\documentclass[12pt]{oxarticle}
\pdfoutput=1
\usepackage{amsmath, amssymb}
\usepackage{graphicx, longtable, afterpage, float}
\usepackage{rangecite, xspace}

\let\SS=\S 
\let\oslash=\o 
\renewcommand{\a}{\alpha}
\renewcommand{\b}{\beta}
\newcommand{\g}{\gamma}
\renewcommand{\d}{\delta}\newcommand{\D}{\Delta}
\newcommand{\e}{\epsilon}
\newcommand{\z}{\zeta}

\renewcommand{\k}{\kappa}
\renewcommand{\l}{\lambda}
\newcommand{\m}{\mu}
\newcommand{\n}{\nu}
\newcommand{\x}{\xi}

\renewcommand{\r}{\rho}
\newcommand{\s}{\sigma}\renewcommand{\S}{\Sigma}
\renewcommand{\t}{\tau}

\newcommand{\ch}{\chi}

\renewcommand{\o}{\omega}

\newcommand{\cA}{\mathcal{A}}
\newcommand{\cB}{\mathcal{B}}
\newcommand{\cC}{\mathcal{C}}

\newcommand{\cE}{\mathcal{E}}

\newcommand{\cM}{\mathcal{M}}

\newcommand{\cP}{\mathcal{P}}

\newcommand{\cU}{\mathcal{U}}

\newcommand{\cX}{\mathcal{X}}
\newcommand{\cY}{\mathcal{Y}}

\newcommand{\IC}{\mathbb{C}}

\newcommand{\IF}{\mathbb{F}}

\newcommand{\IH}{\mathbb{H}}

\newcommand{\IP}{\mathbb{P}}

\newcommand{\IR}{\mathbb{R}}

\newcommand{\IZ}{\mathbb{Z}}

\font\twentyfourrm=cmr12 at 24pt
\font\eightrm=cmr8 at 8pt
\font\csc=cmcsc10

\newcommand{\beq}{\begin{equation}}
\newcommand{\eeq}{\end{equation}}
\newcommand{\bea}{\begin{eqnarray}}
\newcommand{\eea}{\end{eqnarray}}
\newcommand{\bean}{\begin{eqnarray*}}
\newcommand{\eean}{\end{eqnarray*}}
\newcommand{\eref}[1]{(\ref{#1})}

\newcommand{\comment}[1]{}

\newcommand{\pd}[2]{\frac{\partial #1}{\partial #2}}

\newcommand{\defineas}{\buildrel\rm def\over =}

\newcommand{\cy}{Calabi--Yau\xspace}
\newcommand{\cym}{Calabi--Yau manifold\xspace}
\newcommand{\cys}{Calabi--Yau manifolds\xspace}
\newcommand{\hodgenos}{(h^{11},\,h^{21})}
\newcommand{\quotient}[1]{_{\hskip-2pt\lower1pt\hbox{$/$}\lower2pt\hbox{\hskip-1pt$#1$}}}
\newcommand{\symm}[1]{_{\hskip-3pt\lower3pt\hbox{$\left\{#1\right\}$}}}
\newcommand{\fref}[1]{Figure~\ref{#1}}
\newcommand{\tref}[1]{Table~\ref{#1}}

\newcommand{\cicystop}{~\lower8pt\hbox{.}}

\newcommand{\Bigcheck}{\lower3.8pt\hbox{\smash{\hbox{{\twentyfourrm \v{}}}}}}
\newcommand{\Xcheck}{\kern0pt\hbox{\Bigcheck\kern-12.5pt{$X$}}}
\newcommand{\cicy}[2]{\begin{matrix} #1\end{matrix}\!\left[\begin{matrix}#2 \end{matrix}\right]}
\newcommand{\ii}{\text{i}} 
\newcommand{\one}{{\hskip-0.75pt\bf 1}}                                     

\def\place#1#2#3{\vbox to0pt{\kern-\parskip\kern-7pt
                             \kern-#2truein\hbox{\kern#1truein #3}
                             \vss}\nointerlineskip}

\newcommand{\dddarrow}{\hbox{$\begin{matrix}\arrowvert\\[-8pt]%
\arrowvert\\[-8pt] \arrowvert\\[-9pt]\downarrow\end{matrix}$}}

\addtocounter{secnumdepth}{1}
\addtocounter{tocdepth}{1}

\numberwithin{equation}{section}
\proofmodefalse
\begin{document}
\pagestyle{empty}
\begin{center}
\null\vskip0.3in
{\Huge New Calabi-Yau Manifolds\\[1ex]
with Small Hodge Numbers\\[0.5in]}
{\csc Philip Candelas$^1$ and Rhys Davies$^2$\\[0.5in]}
{\it $^1$Mathematical Institute\hphantom{$^1$}\\
Oxford University\\
24-29 St.\ Giles'\\
Oxford OX1 3LB, UK\\[4ex]
$^2$Rudolf Peierls Centre for Theoretical Physics\hphantom{$^2$}\\
Oxford University\\
1 Keble Road\\
Oxford OX1 4NP, UK\\}
\vfill
{\bf Abstract\\[3ex]}
\parbox{6.5in}{\setlength{\baselineskip}{14pt}
It is known that many Calabi-Yau manifolds form a connected web. The question of whether all \cys\ form a single web depends on the degree of singularity that is permitted for the varieties that connect  the distinct families of smooth manifolds. If only conifolds are allowed then, since shrinking two-spheres and three-spheres to points cannot affect the fundamental group, manifolds with different fundamental groups will form disconnected webs. We examine these webs for the tip of the distribution of \cys\ where the Hodge numbers $\hodgenos$ are both small. In the tip of the distribution the quotient manifolds play an important role. We generate via conifold transitions from these quotients a number of new manifolds. These include a manifold with $\ch =-6$, that is an analogue of the $\chi=-6$ manifold found by Yau,  and manifolds with an attractive structure that may prove of interest for string phenomenology. We also examine the relation of some of these manifolds to the remarkable Gross-Popescu manifolds that have Euler number zero.}
\end{center}
\newpage
\tableofcontents
\newpage
\setcounter{page}{1}
\pagestyle{plain}
\section{Introduction and Summary}
\subsection{Preamble}
In a recent paper \cite{Triadophilia} attention was drawn to the fact that there is an interesting region in the distribution of \cys\ where both the Hodge numbers $\hodgenos$ are small. This region contains at least two manifolds that are interesting from the perspective of elementary particle phenomenology. These are the covering space for the three-generation manifold \cite{Yau}, and the split bicubic.
\beq
X^{14,23}=~\cicy{\IP^3\\ \IP^3}{1 & 3 & 0\\ 1 & 0 & 3\\}^{14,23}~,~~~
X^{19,19}=~\cicy{\IP^1\\ \IP^2\\ \IP^2}{1 & 1\\ 3 & 0\\ 0 & 3\\}^{19,19}
\label{twocicys}
\eeq
where the suffices appended to the two configurations denote the Hodge numbers $\hodgenos$. These manifolds belong to a class known as complete intersection \cys\ (CICY's). These are manifolds that can be presented as the complete intersection of polynomials in a product of projective spaces. The covering space of the three generation manifold is the prototypical example and the notation denotes that this manifold can be presented as a submanifold of $\IP^3{\times}\IP^3$ defined by the vanishing locus of three polynomials whose multidegrees are given by the columns of the matrix. One may take, for example, the equations
\beq
\begin{split}
F&= f_{0}\, x_{0}y_{0} + f_{1}\sum_j  x_{j}y_{j}
+ f_{2}\sum_j  x_{j}y_{j+1} +  f_{3}\sum_j  x_{j+1}y_{j}
+ f_{4}\, x_{0}\sum_j  y_{j} +  f_{5}(\sum_j  x_{j})y_{0},\\
G&= x_{0}^{3} -  x_{1}x_{2}x_{3} + g_1\sum_j  x_{j}^{3} + 
g_{2}\, x_{0}\sum_j  x_{j}x_{j+1}~,\\
H&= y_{0}^{3} - y_{1}y_{2}y_{3} + h_1\sum_j  y_{j}^{3} + 
h_{2}\, y_{0}\sum_j  y_{j}y_{j+1}~.
\end{split}
\label{FGH}\eeq
where the $(x_{0},x_{j})$ and $(y_{0},y_{j})$, $j=1,2,3$, are projective coordinates for the two                  
$\IP^{3}$'s and $f_a$, $g_a$ and $h_a$ are coefficients. The separate treatment of the zeroth coordinate anticipates the action of a $\IZ_3$ symmetry group, with generator $S$ that simultaneously permutes the coordinates $x_j$ and $y_k$ cyclically:
$$
S:~(x_0,\,x_j)\times(y_0,\,y_k)~\to ~(x_0,\,x_{j+1})\times(y_0,\,y_{k+1})
$$ 
where the indices $j$ and $k$ are understood to take values in $\IZ_3$. It is easy to see that the action of $S$ is fixed point free so the quotient $X^{14,23}/S$ is smooth and has in fact $\hodgenos=(6,9)$ and hence 
$\chi=-6$, where $\chi$ denotes the Euler number. 

The condition that a configuration corresponds to a \cym, that is has $c_1=0$ is the condition that each row of the matrix sum to one more than the dimension of the corresponding projective space.  A list of the almost 8,000 CICY's was compiled in \cite{CICYsI}; these have Euler numbers in the range $-200\leq\chi\leq 0$, are all simply connected, and have $h^{11}+h^{21}\geq 30$. A review of constructions of \cys\ is given in \cite{Triadophilia} where it is observed that finding manifolds with small Hodge numbers, that is with say 
$h^{11}+h^{21}\leq 24$, is largely synonymous with finding quotients by freely acting groups.

Our aim in this paper is to find such manifolds by finding other CICY's that admit freely acting symmetries and then taking the respective quotient in a manner analogous to that which leads to the three-generation manifold. Manifolds that admit a freely acting symmetry seem to be genuinely rare so our strategy is to try to trace the symmetries through the web of CICY manifolds. To explain this and to keep this account reasonably self contained we digress on the processes of {\em splitting} and {\em contraction} which are fundamental to the strategy.
This is old knowledge and a more detailed account of this than we shall give here, together with many matters pertaining to CICY's, may be found in~\cite{hubsch}.  For a recent interesting reference in which some of the manifolds that are important to us here appear in a different context~see~\cite{Donagi:2008xy}.

\begin{figure}[p]
\begin{center}
\includegraphics[width=6.5in]{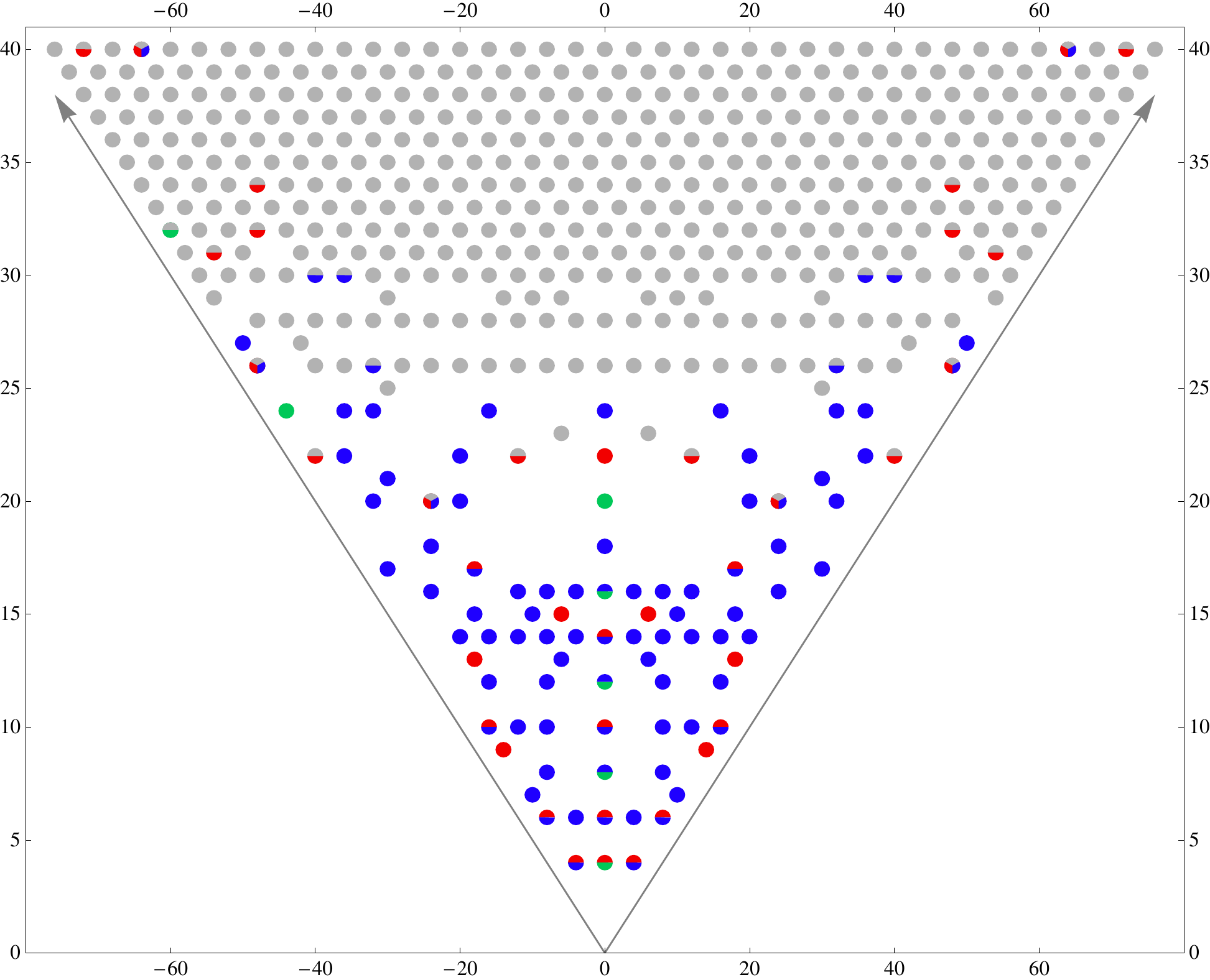}
\vskip10pt
\boxed{
\parbox{5.75in}{\footnotesize\vspace{2pt}
\includegraphics[width=6pt]{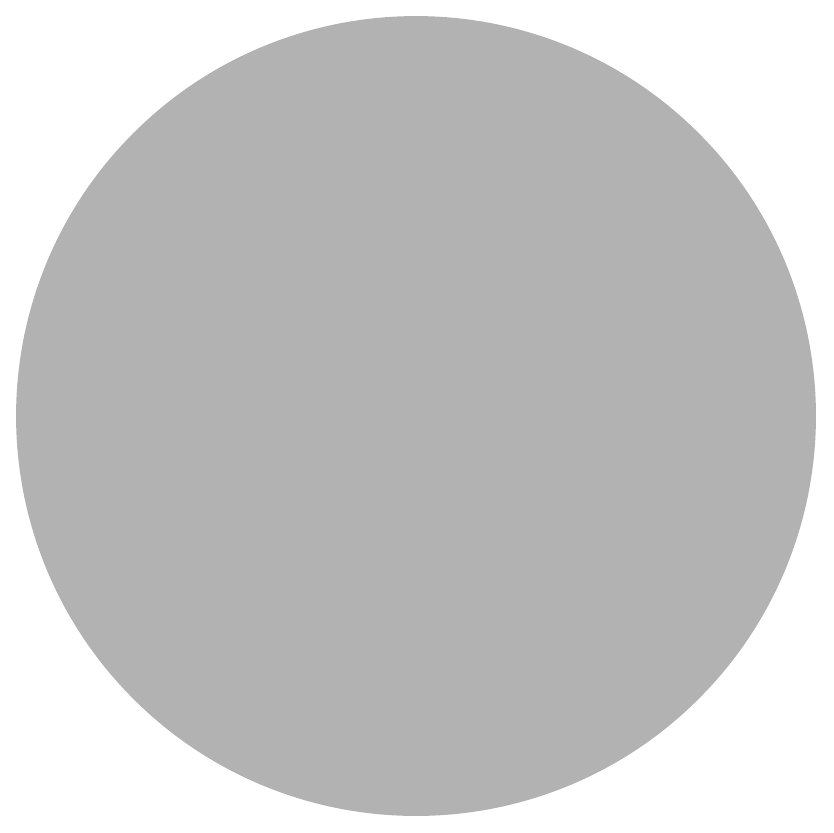}~~The Kreuzer--Skarke list, CICY's, toric CICY's, and  
toric conifolds \cite{CICYsI,KreuzerSkarkeReflexive,KreuzerRieglerSahakyan,%
KlemmKreuzerRieglerScheidegger,BatyrevKreuzerConifolds}, with their mirrors. \\ 
\includegraphics[width=6pt]{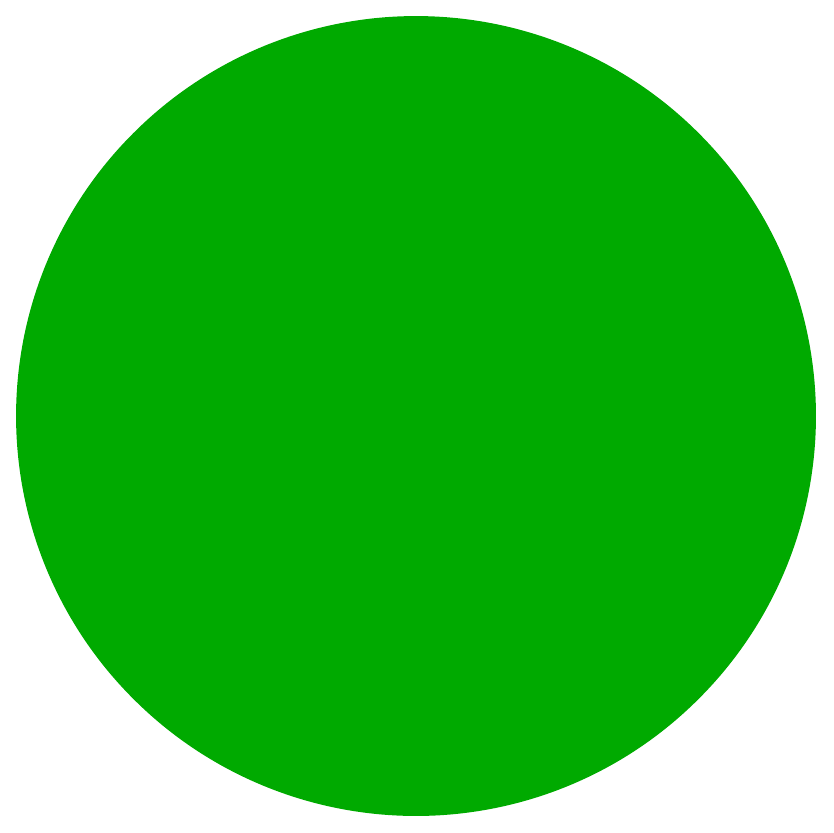}~~The Gross--Popescu,  R{\oslash}dlandand, Tonoli , Borisov-Hua and Hua manifolds \cite{GrossPopescu,Rodland,Tonoli,BorisovHua,Hua}.\\
\includegraphics[width=6pt]{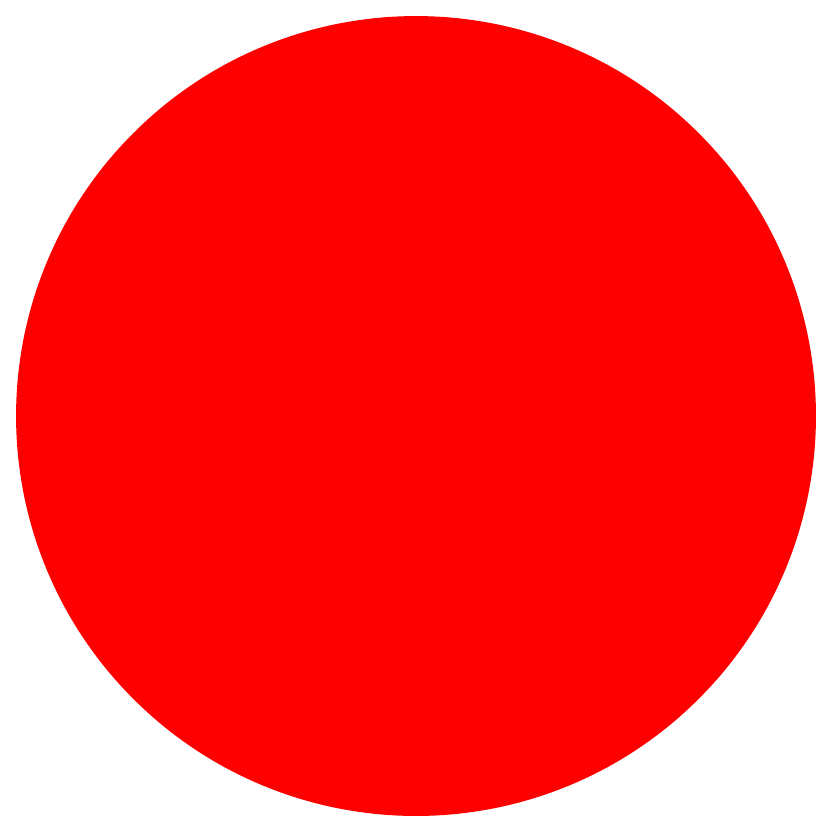}~~Previously known quotients by freely acting groups and their mirrors.\\ 
\includegraphics[width=6pt]{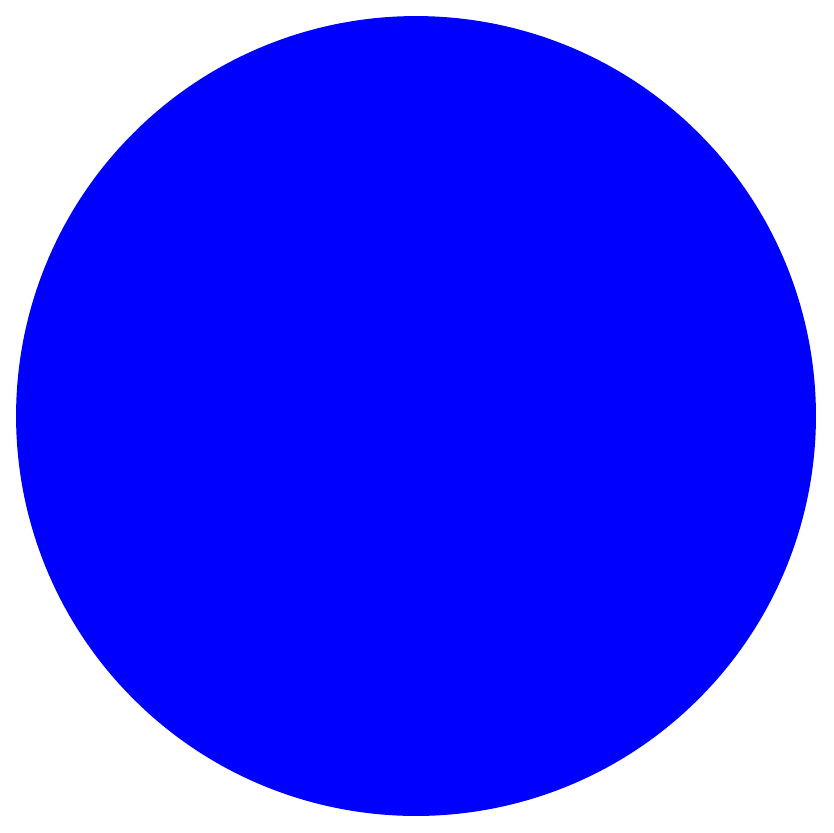}~~New free quotients and resolutions of quotients with fixed points, with their mirrors.\\
\includegraphics[width=6pt]{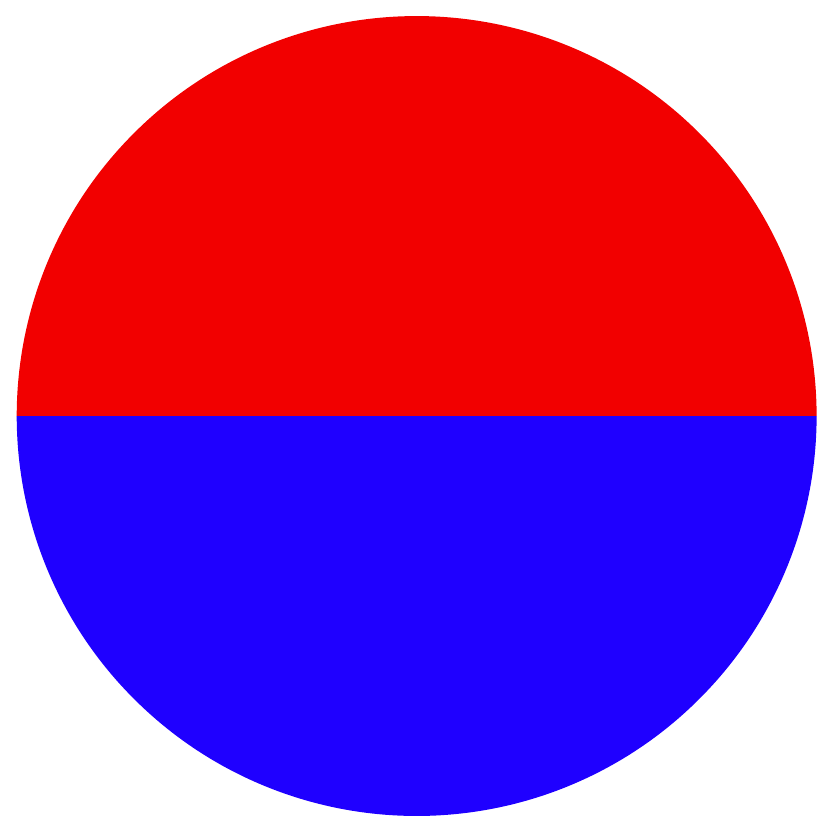}~~Divided dots denote overlays.
\vspace{2pt}}}
\vskip15pt
\parbox{6.0in}{\caption{\label{NewOrbifolds}\small
The tip of the distribution of \cys showing the manifolds that have nontrivial fundamental group, together with their mirrors. The Euler number \hbox{$\ch=2(h^{11}{-}h^{21})$} is plotted horizontally,
$h^{11}{+}h^{21}$ is plotted vertically and the oblique axes bound the region 
$h^{11}\geq 0,\, h^{21}\geq 0$. Manifolds with $h^{11}{+}h^{21}\leq 24$ are identified 
in~\tref{tiptab}.} The constructions that provide the points of the diagram, apart from the new points shown in blue, are reviewed in \cite{Triadophilia}, where references to the original literature are given. The divided dot with Hodge numbers $(1,21)$ corresponds to the single manifold $\IP^4[5]/\IZ_5$. It is recorded with a divided dot owing to the fact that it is a known quotient that appears also in the toric lists.}
\end{center}
\end{figure}
%
\subsection{Splitting and Contraction}
Consider the split bicubic $X^{19,19}$ from~\eref{twocicys}. Taking coordinates $t_a$, for the $\IP^1$, and   $\x_j$ and $\eta_k$, for the two $\IP^2$'s, we may write the two polynomials in the  form
\beq
\begin{split}
t_1 U(\xi) + t_2 W(\xi) ~&=~0 \ , \\
t_1 Z(\eta) + t_2 V(\eta) ~&=~ 0 \ ,
\end{split}
\label{defX1919}\eeq
where $U$, $V$, $Z$ and $W$ are cubics. If we regard $\x$ and $\eta$ as given then we have two equations in the two coordinates $(t_1,t_2)$. These cannot both vanish so it must be the case that
\beq
F_0\defineas U(\xi)V(\eta) - W(\xi)Z(\eta)~=~0~.
\label{Fsharp}\eeq
Now $F_0$ has bidegree $(3,3)$ in $\xi$ and $\eta$ and it is natural to ask how $X^{19,19}$ is related to the bicubic
$$
X^{2,83}~=~\cicy{\IP^2\\ \IP^2}{3\\ 3}~.
$$
The manifolds $X^{19,19}$ and $X^{2,83}$ are clearly different since they have different Hodge numbers. This is due to the fact that $F_0$ necessarily defines a singular bicubic since $F_0$ and all its derivatives vanish at the points where $U=V=W=Z=0$ and, in this example, this will happen in $3^4$ points. Let us denote this singular variety by $X_0$. We may {\em smooth} $F_0$ to $F_s=F_0+s\,K$ which, for suitable polynomial 
$K$, defines a nonsingular manifold for $s$ in a punctured neighbourhood of the origin. Thus we learn that there are smooth bicubics arbitrarily close to $X_0$. 

We have seen that a point $(t,\x,\eta)$ that satisfies \eref{defX1919} must be such that $(\x,\eta)$ satisfy \eref{Fsharp}. Conversely suppose that $(\x,\eta)$ satisfy \eref{Fsharp} and that for these values the cubics $U,V,W,Z$ are not all zero. The equations \eref{defX1919} will then determine a unique ratio $t_1/t_2$ hence a unique point $t\in\IP^1$. If, however, all four of the cubics vanish then the equations \eref{defX1919} are satisfied for all values of  $t\in\IP^1$. For suitable cubics the split manifold defined by \eref{defX1919} is smooth while the conifold defined by $F_0=0$ is singular at a certain number of nodes. The split manifold projects down onto the conifold such that a unique point projects to every nonsingular point of the conifold but an entire $\IP^1$ of the split manifold projects down onto each node. Alternatively we pass from the conifold to the split manifold by blowing up each node to a~$\IP^1$. This is the {\em resolution} of $X_0$. One checks from \eref{defX1919} that the resolution is smooth and since the radii of the resolving $\IP^1$'s can be arbitrarily small there are split bicubics that are also arbitrarily close to $X_0$. Thus the parameter spaces of the bicubic and the split bicubic intersect in points corresponding to the singular variety $X_0$. One can show \cite{comments} that in passing from the smooth bicubic $F_s$ to the singular bicubic $F_0$ an $S^3$ shrinks to zero at each node of the singular variety. Thus we have a process, $F_s\to UV-WZ$, followed by ensuring the vanishing of the determinant by imposing the equations \eref{defX1919}. This shrinks a number of $S^3$'s to zero thereby creating nodes and then resolves the nodes with $\IP^1$'s. This process is realised on the configuration by splitting a column\footnote{This is not a unique process since, in general, a column can be split in different ways. Another way to split the bicubic is
$$\cicy{\IP^2\\ \IP^2}{3\\ 3}~\rightarrow~\cicy{\IP^1\\ \IP^2\\ \IP^2}{1 & 1\\ 2 & 1\\ 1 & 2\\}\cicystop$$} 
as we have done in this example
$$
\cicy{\IP^2\\ \IP^2}{3\\ 3}~\rightarrow~\cicy{\IP^1\\ \IP^2\\ \IP^2}{1 & 1\\ 3 & 0\\ 0 & 3\\}
\cicystop
$$
More generally we may split with a $\IP^n$ in place of the $\IP^1$
$$
\cP[{\bf c}, M]~\rightarrow~
\cicy{\IP^n\\ \cP}{    1       & 1            & \cdots & 1                      & {\bf 0}\\
                             {\bf c}_1&{\bf c}_2& \cdots & {\bf c}_{n+1} & M       }
$$
where $\cP=\IP^{n_1}{\times}\ldots{\times}\IP^{n_k}$ is any product of projective spaces, 
${\bf c}=\sum_{j=1}^{n+1}{\bf c_j}$ is a vector which we decompose as the sum of $n+1$ vectors with nonnegative components, and $M$ is a matrix. On the right we have a configuration with $n+1$ equations that are linear in the $n+1$ coordinates of the $\IP^n$. Since these coordinates cannot all vanish the determinant of the matrix of coefficients, which has multidegree ${\bf c}$, must vanish. Let $A$ denote this matrix of coefficients. Since its determinant vanishes, $A$ cannot have rank $n+1$. If it has rank $n$, which is the generic case, then the coordinates of the $\IP^n$ are determined up to scale, which determines a unique point in the $\IP^n$. A little thought shows that there will be a set of codimension three, in the manifold, that is points, where $A$ will have rank $n-1$ and this will determine a line, that is a $\IP^1$, in the $\IP^n$. In general the rank of $A$ cannot have rank less than $n-1$ since this would occur on a set of codimension greater than three. For the configuration on the left one sees that there is a smoothed equation $F_s=\det A + s\, K$ and that the effect of the limit 
$s\to 0$ is to shrink a finite number of $S^3$'s to nodes. The burden of these comments is that we have the same situation here as previously: we may proceed from a smooth manifold, $X$, corresponding to the configuration on the left, shrink a certain number of $S^3$'s to nodes to arrive at a singular variety 
$X_0$ and then resolve the nodes with $\IP^1$'s to arrive at a smooth manifold corresponding to the configuration on the right that we denote by $\Xcheck$. Now the Euler number of $S^3$ is zero and the Euler number of a $\IP^1$ is 2 so the Euler numbers of $X$ and $\Xcheck$ are related by
$$
\chi\big(\Xcheck\big)~=~\chi(X) + 2\n
$$
where $\n$ is the number of nodes of $X_0$. Thus $\chi\big(\Xcheck\big)\geq \chi(X)$ and there is equality only if the manifold corresponding to $\det A=0$ has, in fact, no nodes. When this is the case 
$\Xcheck=X$. Thus $\Xcheck=X$ if and only if their Euler numbers are equal. This is a useful criterion which we shall use frequently in the following. We refer to the process that we have denoted by $\rightarrow$ as a 
{\em splitting} and we refer to the reversed process as a {\em contraction}.
\subsection{Configurations and diagrams}
Configurations that differ merely by a permutation of their rows or columns determine the same manifold. A more intrinsic representation is given by a diagram that expresses the combinatorics of the degrees of the polynomials in the coordinates of each of the ambient spaces. The individual embedding spaces are represented by open disks
and the polynomials by filled disks. The degree of the polynomial with respect to a given ambient space is encoded by the number of lines connecting the corresponding disks. The diagrams for the quintic, bicubic and split bicubic are as follows:
\begin{gather*}
\IP^4[5]~:~~~\raisebox{-3pt}{\includegraphics[width=37.5pt]{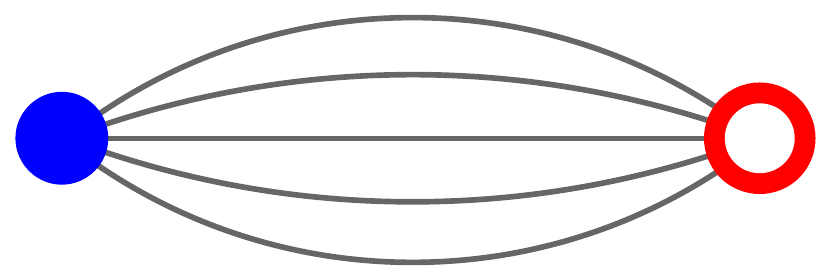}}\hspace{1in}
\cicy{\IP^2\\ \IP^2}{3\\ 3}~:~~~\raisebox{-4pt}{\includegraphics[width=75pt]{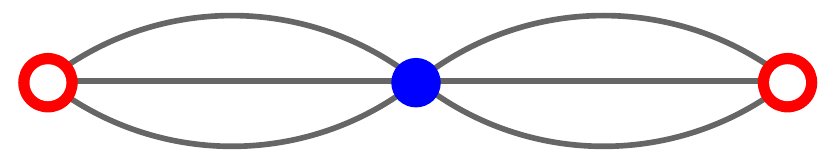}}
\\[20pt]
\cicy{\IP^1\\ \IP^2\\ \IP^2}{1~1\\ 3~0\\ 0~3}~:~~~
\raisebox{-6pt}{\includegraphics[width=150pt]{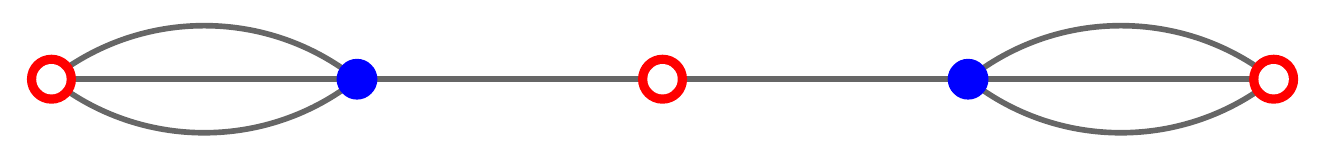}}
\end{gather*}
The notation that we use for the configurations is also redundant since we do not need to write the ambient spaces explicitly owing to the fact that these are determined by the row sums of the matrix. Thus we could just write, for example,
$$
\begin{bmatrix}1~3~0\\ 1~0~3\end{bmatrix}~:~~~
\raisebox{-6pt}{\includegraphics[width=150pt]{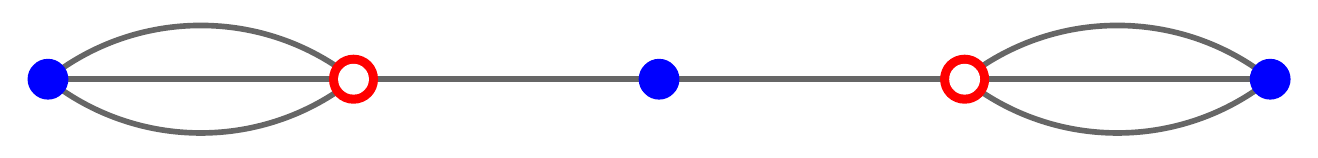}}
$$
for the covering space of the three-generation manifold since the row sums dictate that the two ambient spaces have to be $\IP^3$'s. The transpose of a configuration matrix also defines a CICY. It is an elementary yet surprising fact that this space is also a threefold. This is an elementary combinatorial fact that we leave for the reader to check. 

Transposition interchanges spaces and polynomials so as an operation on the diagram it amounts to simply interchanging the two sorts of disk, as we see, for example, by comparing the diagrams for the covering space of three-generation manifold and the split bicubic above. The transposition of CICY's remains a mysterious process owing to the fact that two different matrices, $M_1$ and $M_2$ say, can represent the same family of manifolds, this is the case, for example, if the matrices differ by an ineffective split. The manifolds corresponding to the transposes $M_1^T$ and $M_2^T$, however, will often be different. There is a converse to this which is that two 
matrices, $M_1$ and $M_2$, may be related by transposition in the sense that the family corresponding to $M_1^T$ is the same family as that corresponding to $M_2$ without it being the case that $M_1^T$ and $M_2$ are equal as matrices. We will come across several examples of this presently; a large but prototypical example is given in \tref{TranspositionId}. Transposition remains, as we say, a mysterious process but it seems to play a role in the webs of manifolds that admit free group~actions.

As mentioned earlier, it sometimes happens that splitting doesn't actually change the manifold. We use two instances of this phenomenon frequently in the following. The first reduces redundancy and is based on the identity
$$
\hskip0.75in\cicy{\IP^1 \\ \IP^1 \\}{1 \\ 1}~\cong~\IP^1 ~~~ \text{or diagrammatically} ~~~
\vcenter{\includegraphics[width=2.5in]{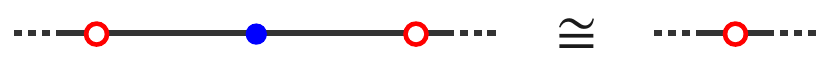}}\hss
$$
This is simply the statement that a bilinear equation in $\IP^1{\times}\IP^1$ is equivalent to 
\hbox{$s_0 t_1=s_1 t_0$}, and this is equivalent to $(s_0,\,s_1)=(t_0,\,t_1)$. This identity prevents us from splitting indefinitely since the splits eventually become ineffective.

A second identity, that arises often in relation to manifolds with a $\IZ_3$ symmetry, is
\beq
\hskip1.0in\cicy{\IP^1 \\ \IP^1 \\ \IP^1 \\ \IP^2 \\}{1~0~0 \\ 0~1~0 \\ 0~0~1 \\ 1~1~1}
~\cong~ \cicy{\IP^1 \\ \IP^1 \\ \IP^1 \\}{1\\1\\1}~~~~\text{or}~~~~~
\vcenter{\includegraphics[width=2.5in]{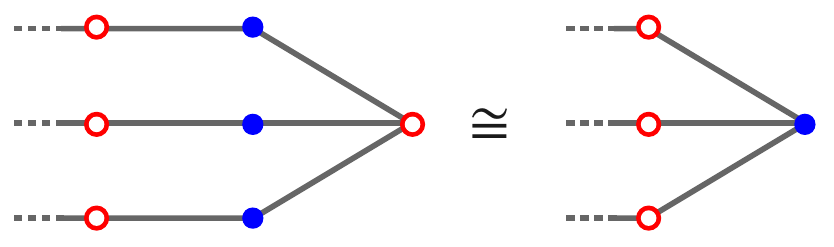}}\hss
\label{dP3id}\eeq
The configuration on the left is a redundant split of the configuration on the right. Both configurations correspond to the del Pezzo surface $dP_6$ given by a $\IP^2$ blown up in three points~\cite{hubsch}. It is natural that this should arise in the context of $\IZ_3$ symmetry since the symmetry can rotate the three exceptional lines into each other. It can happen (as in the CICY's of \eref{twocicys}, for example) that $h^{11}$ can exceed the number of ambient spaces. In suitable cases use of the above identity, read from right to left, can be used to increase the number of ambient spaces and so give explict representation to more of the K\"ahler classes. 
\subsection{Calculation of the Euler characteristic and Hodge numbers}
It is easier to compute the Euler characteristic $\ch=2(h^{11}-h^{21})$ than to compute the Hodge numbers separately since this is computed directly as a function of the matrix. We denote the K\"ahler classes of the ambient spaces by $h_j$, $j=1,\ldots,F$ and the entries of the matrix by $\deg_j(\a)$, where 
$\a=1,\ldots,N$ runs over the polynomials. From these quantities we~form
$$
\x(\a)~=~\sum_{j=1}^F h_j\deg_j(\a)
$$
and a straightforward exercise with Chern classes \cite{CICYsI,hubsch} provides a relation that is easily programmed
\beq
\frac{1}{3}\left(\sum_{j=1}^F (n_j+1)h_j^3 - \sum_{\a=1}^N \x(\a)^3\right)\prod_{\b=1}^N \x(\b)~=~
\chi\,\prod_{j=1}^F h_j^{n_j}~.
\label{eulernum}\eeq
In this relation $n_j$ is the dimension of the $j$'th factor space and we use the fact that
$h_j^{n_j + 1}=0$. As a matter of practical calculation it is important, in order to be able to deal with large matrices, to reduce the product $\prod_\b \x(\b)$ with respect to the ideal generated by the quantities 
$h_j^{n_j+1}$ as the terms are accumulated. This avoids  having to reduce a very large polynomial. 

As for the individual Hodge numbers, these were calculated in \cite{Green:1987cr} for the 7890 CICY matrices constructed in \cite{CICYsI}. This list of CICY's was maintained by Schimmrigk and made available on the 
Calabi-Yau Home Page \cite{CYHomePage}. A version of the list which has the respective Hodge numbers appended, is also available~\cite{DigitalCICYList}. Finding the Hodge numbers corresponding to a given matrix is therefore, in principle, just a matter of looking up the relevant matrix in the list. There are two complications to doing this. The first owes to the fact that an attempt was made to eliminate redundant splits in the compilation of the list so not all CICY's are included.  For the cases considered here, this is not a significant problem since in all cases where the matrix does not occur directly in the list it is related to a space that is listed by redundant splittings and contractions. A second potential problem is that there is no canonical way to write the CICY matrices, so if a given configuration exists in the list it will very likely appear with its rows and columns permuted. 

Consider as an example of this lookup procedure a case that will concern us later. We wish to find the Hodge numbers of the configuration
$$
X~=~~
\cicy{\IP^1\\ \IP^1\\ \IP^1\\ \IP^1\\ \IP^1\\ \IP^1\\ \IP^1\\ \IP^1\\ \IP^1}
{1 & 0 & 0 & 1 & 0 & 0 \\
 0 & 1 & 0 & 1 & 0 & 0 \\
 0 & 0 & 1 & 1 & 0 & 0 \\
 1 & 0 & 0 & 0 & 1 & 0 \\
 0 & 1 & 0 & 0 & 1 & 0 \\
 0 & 0 & 1 & 0 & 1 & 0 \\
 1 & 0 & 0 & 0 & 0 & 1 \\
 0 & 1 & 0 & 0 & 0 & 1 \\
 0 & 0 & 1 & 0 & 0 & 1}\hspace{0.8in}
\raisebox{-0.7in}{\includegraphics[width=1.6in]{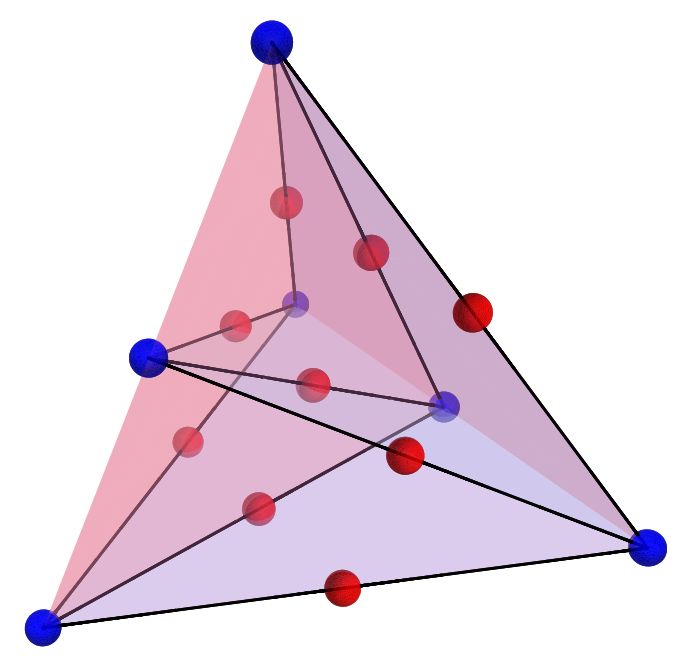}}
$$
By means of \eref{eulernum} we see that the Euler number of the manifold is zero. There are  52 configurations in the list corresponding to $\chi=0$, though only two  with embedding space~$(\IP^1)^9$. These two are configurations 5 and 6 of the list
$$
\cicy{\IP^1\\ \IP^1\\ \IP^1\\ \IP^1\\ \IP^1\\ \IP^1\\ \IP^1\\ \IP^1\\ \IP^1}
 {1 & 1 & 0 & 0 & 0 & 0 \\
 0 & 0 & 1 & 0 & 0 & 1 \\
 0 & 0 & 0 & 1 & 1 & 0 \\
 1 & 0 & 0 & 0 & 0 & 1 \\
 0 & 1 & 1 & 0 & 0 & 0 \\
 0 & 0 & 1 & 0 & 1 & 0 \\
 0 & 0 & 0 & 1 & 0 & 1 \\
 1 & 0 & 0 & 1 & 0 & 0 \\
 0 & 1 & 0 & 0 & 1 & 0}
\hspace{0.3in}\text{and}\hspace{0.3in}
\cicy{\IP^1\\ \IP^1\\ \IP^1\\ \IP^1\\ \IP^1\\ \IP^1\\ \IP^1\\ \IP^1\\ \IP^1}
 {1 & 1 & 0 & 0 & 0 & 0 \\
 0 & 0 & 1 & 0 & 0 & 1 \\
 0 & 0 & 0 & 1 & 1 & 0 \\
 1 & 0 & 0 & 0 & 0 & 1 \\
 0 & 1 & 1 & 0 & 0 & 0 \\
 0 & 0 & 1 & 0 & 1 & 0 \\
 0 & 0 & 0 & 1 & 0 & 1 \\
 0 & 1 & 0 & 1 & 0 & 0 \\
 1 & 0 & 0 & 0 & 1 & 0}
$$
In an attempt to avoid having to draw diagrams for each of these we compute the Euler numbers of their transposes to compare with the Euler number for the transpose of $X$, which is -36. The Euler number for the transposes turns out to be -36 for both of the configurations above, so this shortcut fails on this occasion and we revert to drawing diagrams. A short time spent sketching reveals these as
$$
\raisebox{0.1in}{\includegraphics[width=1.8in]{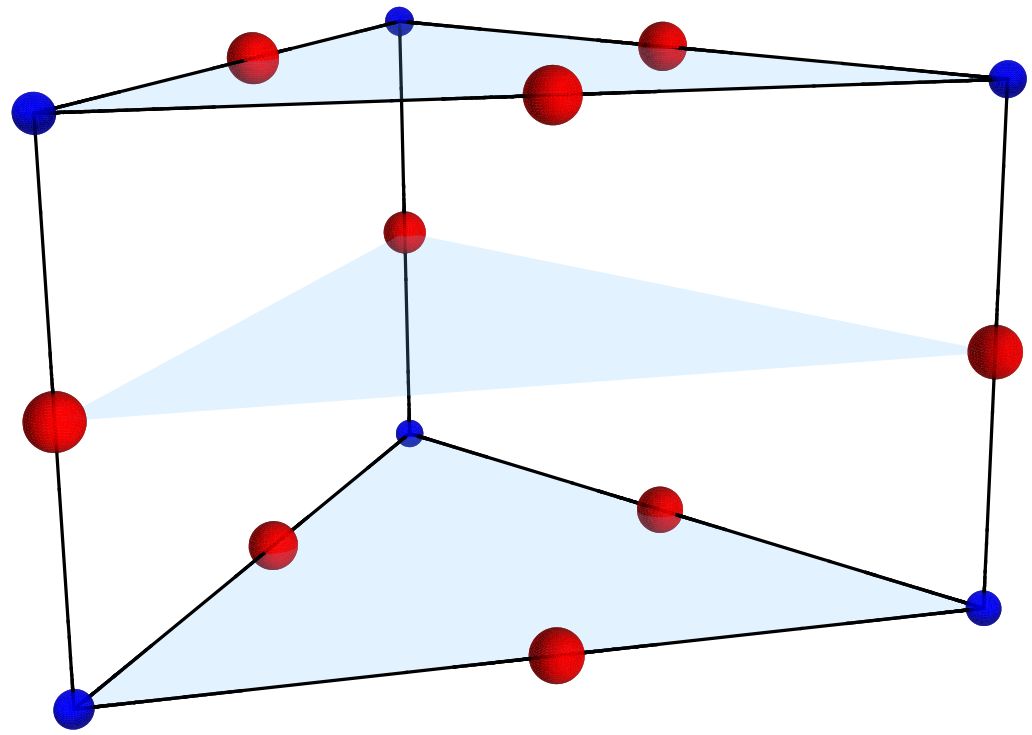}}\hspace{1in}
\raisebox{0in}{\includegraphics[width=1.6in]{fig_15,15_tetrahedron.jpg}}
$$
so $X$ is in fact configuration 6 of the list and has $\hodgenos=(15,15)$.

Determining the Hodge numbers of the quotient manifolds is slightly more challenging.  The easiest quantity to compute is the Euler number, since this simply divides by the order of the freely acting group.  In principle then, we need only calculate one of the individual Hodge numbers, since the other can then be determined from 
$\chi = 2(h^{11} - h^{21})$, however in most cases we calculate both numbers as a check on our results.

The easiest way to determine $h^{11}$ is to find a representation of the CICY in which all non-trivial $(1,1)$-forms on the original manifold arise as the pullbacks of the hyperplane classes of the embedding spaces.  The group action on the second cohomology of the CICY is then determined by the action on the ambient space.  In all cases except one (see \SS\ref{sec:8,44}), each ambient space is either part of an orbit of spaces which are mapped to each other by the symmetry, or the symmetry acts linearly on the homogeneous coordinates within a given $\IP^n$.  In the first case, each orbit contributes 1 to the count for the quotient manifold, and in the second case the space still contributes 1 by itself, since a holomorphic linear action maps hyperplanes bijectively to hyperplanes and preserves orientation.

To calculate $h^{21}$ we count independent parameters in the polynomials defining the symmetric CICY.  There exist cases of CICY's where this does not work, but a theorem from~\cite{Deformations} assures us of the effectiveness of the technique in the majority of cases that we consider here.  Paraphrased, the theorem states that if the diagram for a Calabi-Yau 3-fold $X$ cannot be disconnected by cutting a single leg, and the counting of parameters for the polynomials agrees with $h^{21}$, then the independent parameters in the polynomials act as a basis for $H^{2,1}(X)$.  This means that $h^{21}$ for the quotient manifold will agree with the number of independent parameters in the polynomials after imposing the symmetry.  Examples of how to count the independent parameters can be found throughout this paper, and is guaranteed by the above theorem to work in all cases but two.  For the manifold $X^{19,19}$,  we can indeed count 19 parameters in the defining equations however, in this case, the diagram can be disconnected by cutting a single leg, so the theorem does not apply. We nevertheless assume, for this case, that the coefficients do in fact parametrise the complex structures of $X^{19,19}$.  For the manifold we label as $Y^{5,37}$, the original equations contain far fewer than $37$ independent parameters, and we have to rely on the calculations of the Euler number and $h^{11}$.

In some examples we find group actions which are not free, but can still yield new Calabi-Yau manifolds.  Suppose a group $G$ acts non-freely on a CICY $X$, and denote by $\S$ the set of points on $X$ which are fixed by some element(s) of $G$.  Then we can remove some neighbourhood $U_\S$ of $\S$, leaving a manifold-with-boundary $\widetilde X$ on which $G$ acts freely.  The Euler number is simply
$$
\ch (\widetilde X) = \ch (X) - \ch(U_\S).
$$
Taking the quotient then gives another manifold $\widetilde X/G$ with boundary 
$\partial (\widetilde X/G) = (\partial \overline U_\S)/G$.  In all cases herein we can find a non-compact manifold 
$\cM$ which can be `glued in' along this boundary to give again a smooth Calabi-Yau threefold $X'$.  The resulting Euler number is
$$
\ch (X') = \frac{\ch (X) - \ch(U_\S)}{|G|} + \ch(\cM).
$$
Calculating the individual Hodge numbers depends on the details of the fixed set $\S$, and we will leave the discussion to the relevant parts of the text.  Note that $\S$ and thus $\cM$ will generally have multiple connected components.

\subsection{Checking transversality of the defining equations}
A CICY is constructed as the vanishing locus of $N$ polynomials in an $N+3$ complex dimensional space.  The condition that the resulting variety be of three dimensions and smooth is that the form 
$dp_1\wedge\cdots\wedge dp_N$ be non-vanishing at all points of intersection. When this is so the polynomials are said to have {\it complete intersection\/} or simply to be {\it transverse}. This condition amounts to the following. For each coordinate patch, with coordinates $x_m$, we check that the $(N+3){\times}N$ Jacobian matrix 
$$
H~=~\left(\displaystyle\pd{p_i}{x_{m}}\right)
$$
has rank $N$ on the locus $p_i=0$. This, in turn, requires checking that the equations $p_i=0$ taken together with the vanishing of all the $N{\times}N$ minors of $H$ have no simultaneous solution for general choice of coefficients in the polynomials.  We have a collection of polynomials and the manifold is smooth if the
there is no solution to the entire collection. It is enough to know that there is no solution for a particular choice of parameters since we then know that there is no solution for a general choice. For this it suffices to assign suitable integer values to the coefficients and to perform a Groebner basis calculation. Such a calculation is only practical in finite characteristic since in the process of generating the Groebner basis the coefficients of the basis polynomials, if taken over $\IR$ say, grow very large and, for the computations that we perform here, the computation of the basis will fail to complete. If we choose integer values for the parameters of the defining polynomials of the manifold then the derivatives and determinants, that we take in constructing the ideal, preserve the fact that the coefficients are integers. If there is a simultaneous solution of the equations then there is also a simultaneous solution in characteristic $p$. Such a solution may not exist in $\IF_p$ but it will exist in a finite extension; that is in $\IF_{p^n}$ for sufficiently large $n$. This is the same as saying that the solution will be given by a consistent set of equations with coefficients in $\IF_p$. The Groebner basis calculation finds these if they exist. If, on the other hand, the polynomials do not reduce to a consistent set over $\IF_p$ they cannot have been consistent to start with. The upshot is that if the defining polynomials are transverse over $\IF_p$, for some prime $p$ and choice of integral coefficients, then they are transverse over $\IC$ for generic coefficients.
There can however be an `accidental' solution mod $p$ even if there was no solution to the original equations. There are however only a finite number of these `bad primes'. An example where the variety is singular over  
$\IF_p$ but smooth over $\IC$ is provided by the quintic threefold with equation $\sum_{j=1}^5 x_j^5=0$. This is smooth over $\IC $ but is singular over $\IF_5$ since all the partial derivatives vanish identically. 

We have implemented this procedure directly in Mathematica 6.0 and also in SINGULAR~3.0.4~\cite{Singular3.0.4} which we run from within Mathematica by means of the 
STRINGVACUA package~\cite{Gray:2008zs}. The SINGULAR implementation of the Groebner basis calculation is significantly faster and this is of practical importance since for large matrices the number of minors of the Jacobian matrix grows rapidly and there are also many coordinates and coordinate patches. The number of minors and coordinate patches can be in the hundreds and for these cases the Mathematica implementation seems to be impractical. The number of coordinate patches is not, in itself, as big a problem as it might seem. Consider, as a trivial example, the problem of checking the transversality of $\IP^4[5]$ over the five standard coordinate patches, $U_j=\{x_j\neq 0\}$, of $\IP^4$. Having checked first transversality over $U_0$, we then check transversality over $U_1$ but, since we already know that the polynomial is transverse when $x_0\neq 0$, we may now set $x_0=0$. Similarly when we come to checking transversality over $U_2$ we may set $x_0=x_1=0$, and so on. The complexity of the algorithms grows very rapidly with the number of variables so this simplification, which reduces the number of variables, leads to a very significant increase in speed. 

\begin{table}[p]
\begin{center}
\def\Str{\vrule height0.75in width0pt depth0.65in}
\framebox[6.5in][c]{\hskip0.1in
\begin{tabular}{cccc@{\hspace{-10pt}}c}
\Str\lower0.05in\hbox{\includegraphics[width=0.6in]{fig_1,101_manifold.pdf}} 
&$\IP^4[5]^{1,101}\symm{\IZ_5{\times}\IZ_5}$
&{}\hskip0.7in{}
&\lower0.45in\hbox{\includegraphics[height=1.0in]{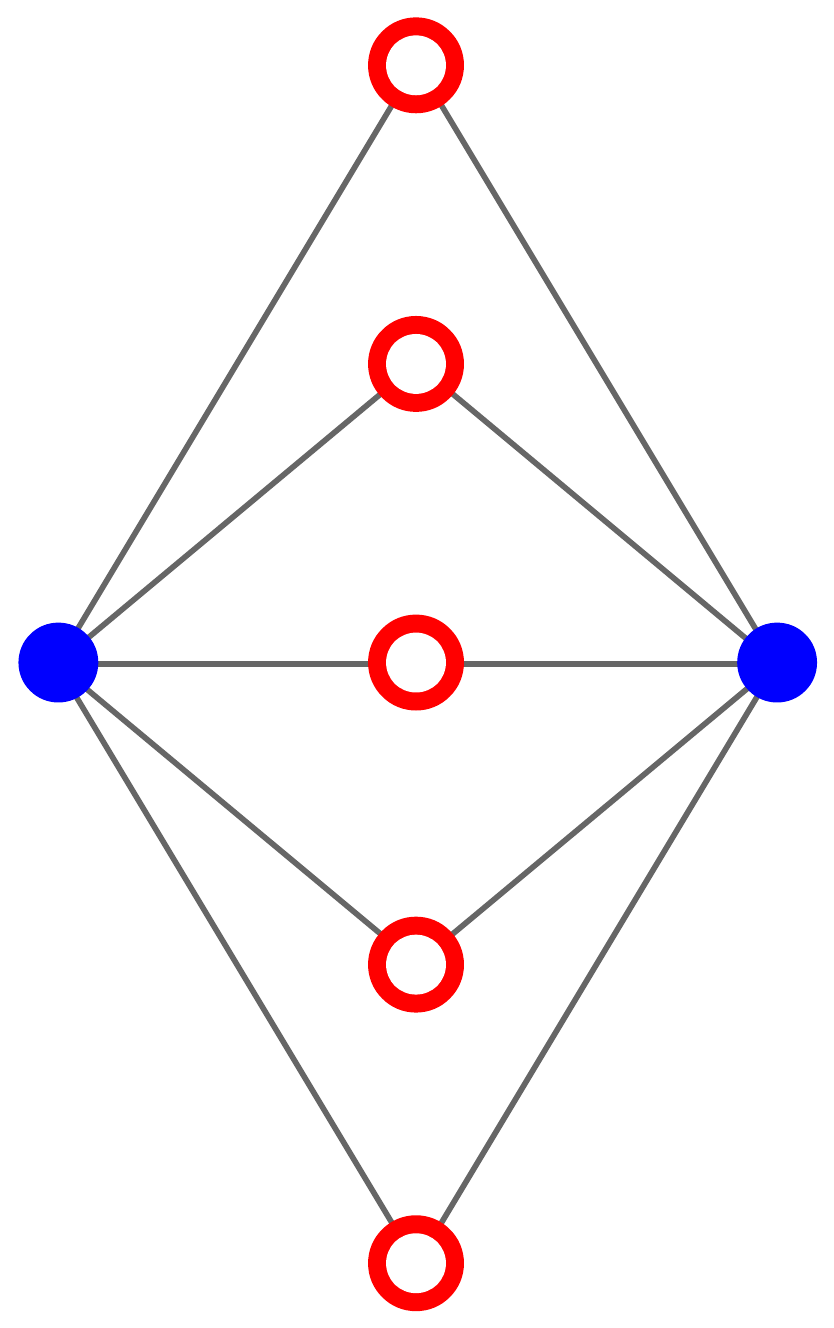}}
&\footnotesize  
\hskip0.55in$\cicy{\IP^1\\\IP^1\\\IP^1\\\IP^1\\\IP^1\\}
{1~1\\1~1\\1~1\\1~1\\1~1\\}^{5,\, 45}\symm{\IZ_5{\times}\IZ_2{\times}\IZ_2}$\\
&\LARGE$\downarrow$&&&\hskip-0.1in\LARGE$\downarrow$\\
\Str\lower0.45in\hbox{\includegraphics[height=1in]{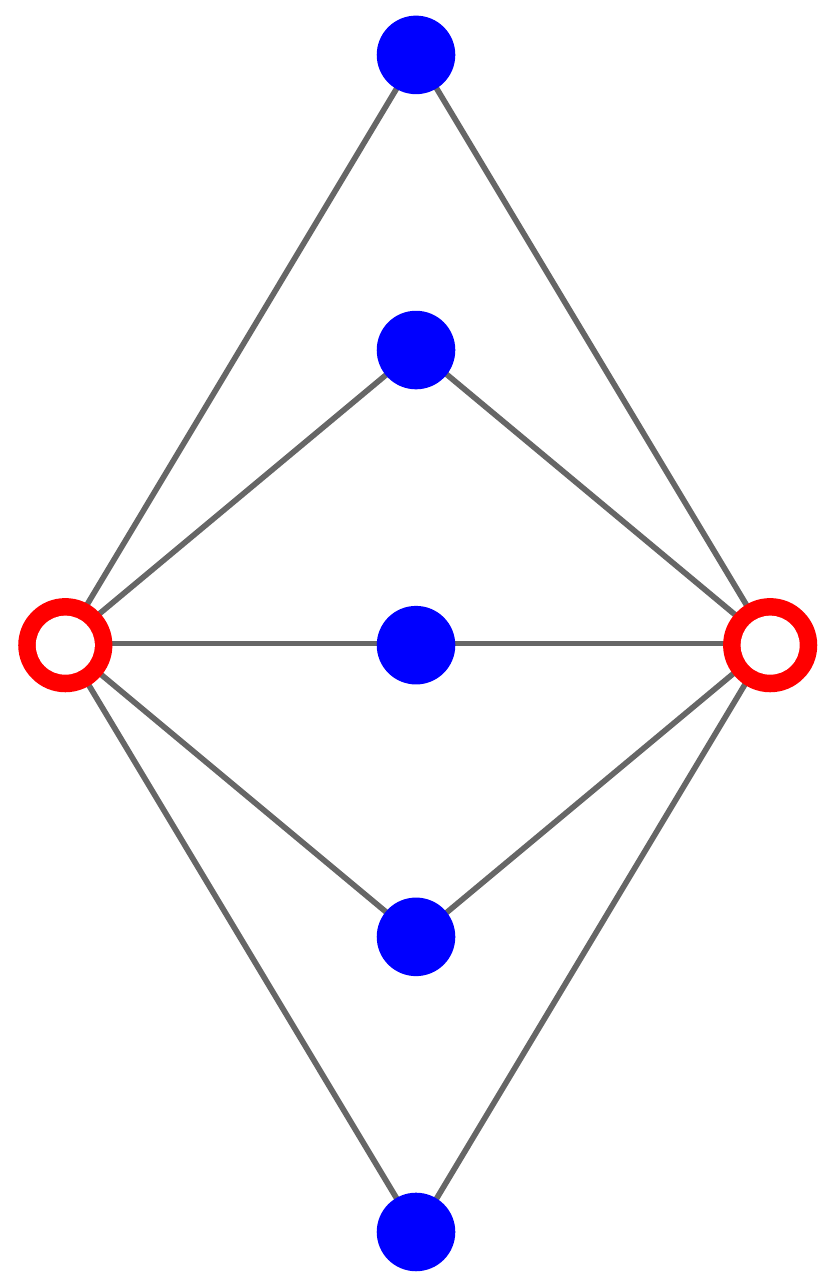}}
&\footnotesize$\cicy{\IP^4\\ \IP^4\\}
{1~1~1~1~1\\ 1~1~1~1~1\\}^{2,52}\symm{\IZ_5{\times}\IZ_2}$
&&\lower0.45in\hbox{\includegraphics[height=1.0in]{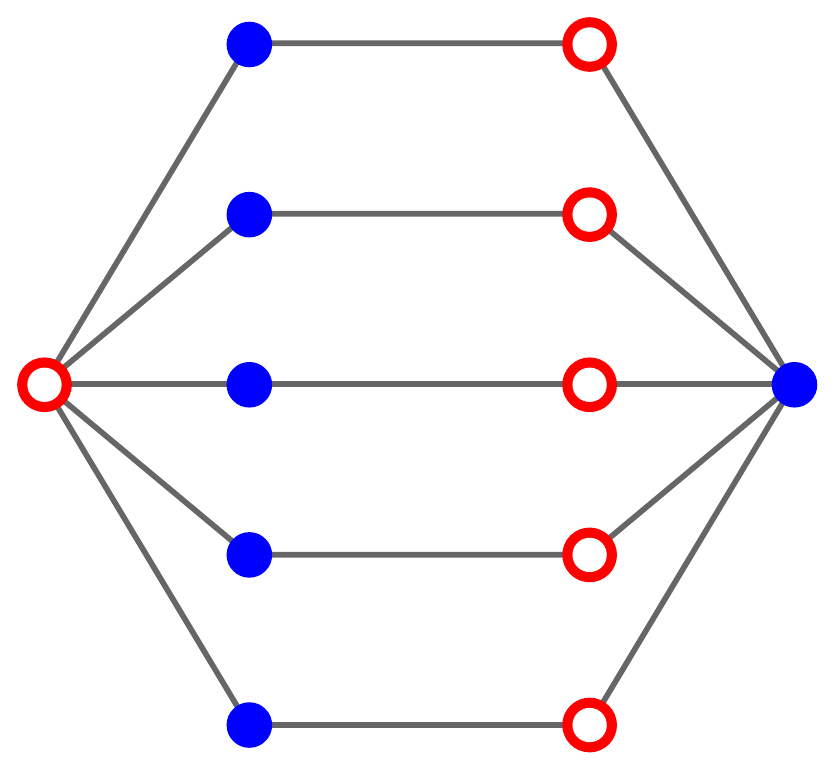}}
&\footnotesize
$\cicy{\IP^1\\ \IP^1\\ \IP^1\\ \IP^1\\ \IP^1\\ \IP^4}
{\one ~ \one ~ 0 ~ 0 ~ 0 ~ 0 \\
 \one ~ 0 ~ \one ~ 0 ~ 0 ~ 0 \\
 \one ~ 0 ~ 0 ~ \one ~ 0 ~ 0 \\
 \one ~ 0 ~ 0 ~ 0 ~ \one ~ 0 \\
 \one ~ 0 ~ 0 ~ 0 ~ 0 ~ \one \\
0~\one~\one~\one~\one~\one}^{6,36}\symm{\IZ_5}$\\
&\multicolumn{2}{c}{\hskip0.5in\LARGE$\searrow$}&&\LARGE\hskip-1.6in$\swarrow$\hskip1in\\
\multicolumn{5}{l}{\Str\hskip0.6in\lower0.45in\hbox{\includegraphics[height=1.0in]{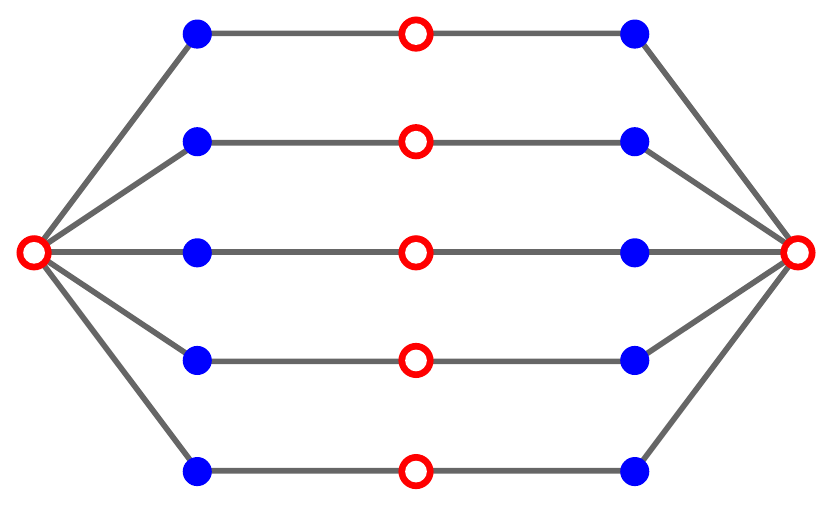}}\hspace{10pt}
\footnotesize$\cicy{\IP^1\\ \IP^1\\ \IP^1\\ \IP^1\\ \IP^1\\ \IP^4\\ \IP^4\\}
{\one ~ 0 ~ 0 ~ 0 ~ 0 ~ \one ~ 0 ~ 0 ~ 0 ~ 0 \\
 0 ~ \one ~ 0 ~ 0 ~ 0 ~ 0 ~ \one ~ 0 ~ 0 ~ 0 \\
 0 ~ 0 ~ \one ~ 0 ~ 0 ~ 0 ~ 0 ~ \one ~ 0 ~ 0 \\
 0 ~ 0 ~ 0 ~ \one ~ 0 ~ 0 ~ 0 ~ 0 ~ \one ~ 0 \\
 0 ~ 0 ~ 0 ~ 0 ~ \one ~ 0 ~ 0 ~ 0 ~ 0 ~ \one \\
 \one ~ \one ~ \one ~ \one ~ \one ~ 0 ~ 0 ~ 0 ~ 0 ~ 0 \\
 0 ~ 0 ~ 0 ~ 0 ~ 0 ~ \one ~ \one ~ \one ~ \one ~ \one \\}^{7,27}\symm{\IZ_5{\times}\IZ_2}$}\\
&\\[-10pt]
\end{tabular}\hskip0.1in}
\vskip5pt
\parbox{5.5in}{\caption{\label{Z5quotients}\small A web of CICY's that admit a freely acting $\IZ_5$ symmetry. The configurations are shown together with their diagrams. The groups that are appended to the configurations are the largest for which we have found a free action of the group.}}


\setlength{\unitlength}{1mm}
\def\str{\vrule height20pt width0pt depth15pt}

\vspace{20pt}
\centerline{
\framebox[3.24in][c]{
\begin{tabular}{l@{\hspace{-5pt}}c@{\hspace{2pt}}c@{\hspace{2pt}}c@{\hspace{2pt}}l}
\str $\left(X^{1,101}/\IZ_5\right)^{1,21}$ &&&& $\left(X^{5,45}/\IZ_5\right)^{1,9}$\\
\hspace{15pt}\LARGE$\downarrow$ &&&& \hspace{15pt}\LARGE$\downarrow$\\
\str $\left(X^{2,52}/\IZ_5\right)^{2,12}$ &&&& $\left(X^{6,36}/\IZ_5\right)^{2,8}$\\
&\hspace{-25pt}\LARGE$\searrow$&&\LARGE$\swarrow$\\
\str&&$\left(X^{7,27}/\IZ_5\right)^{3,7}$\\
&\\[-15pt]
\end{tabular}\hspace{5pt}}
\hspace{3pt}
\framebox[3.24in][c]{
\begin{tabular}{l@{\hspace{-15pt}}c@{}c@{\hspace{-10pt}}c@{\hspace{-3pt}}l}
\str&&&& $\Big(X^{5,45}\quotient{\IZ_5{\times}\IZ_2}\Big)^{1,5}$\\
&&&& \phantom{\LARGE$\downarrow$}\\
\str $\Big(X^{2,52}\quotient{\IZ_5{\times}\IZ_2}\Big)^{1,6}$ &&&&\\
&\hspace{-15pt}\LARGE$\searrow$&&\phantom{\LARGE$\swarrow$}\\
\str&&$\Big(X^{7,27}\quotient{\IZ_5{\times}\IZ_2}\Big)^{2,4}$\smash{
\begin{picture}(0, 40)
\thicklines\put(6, 30){\vector(-1, -2){11}}
\end{picture}}\\
&\\[-15pt]
\end{tabular}\hspace{5pt}}}
\vskip5pt
\parbox{5.5in}{\caption{\label{Z5miniwebs}\small Webs derived from \tref{Z5quotients} with fundamental groups $\IZ_5$ and $\IZ_5{\times}\IZ_2$.}}
\end{center}
\end{table}

%
\subsection{Webs of CICY's with freely acting symmetries}
There is a simple and elegant argument~\cite{Green:1988bp,hubsch} that shows that it is possible to pass from any CICY configuration to any other by a sequence of splittings and contractions. Thus the parameter space of CICY's forms a connected web. The CICY's themselves are all simply connected while the quotient of a CICY by a freely acting group $G$ has fundamental group~$G$. In general splitting will not commute with taking the quotient by a freely acting group since the subvariety of symmetric manifolds, in the parameter space, need not intersect the conifold locus. On the other hand, by suitable choice of splitting, we will sometimes be able to achieve this. When this happens we can move along a web of parameter spaces corresponding to manifolds with fundamental 
group~$G$. Since a conifold transition cannot alter the fundamental group of a manifold there will be webs, not necessarily connected, corresponding to different fundamental groups.

\begin{figure}[t!]
\begin{center}
\includegraphics[width=6.5in]{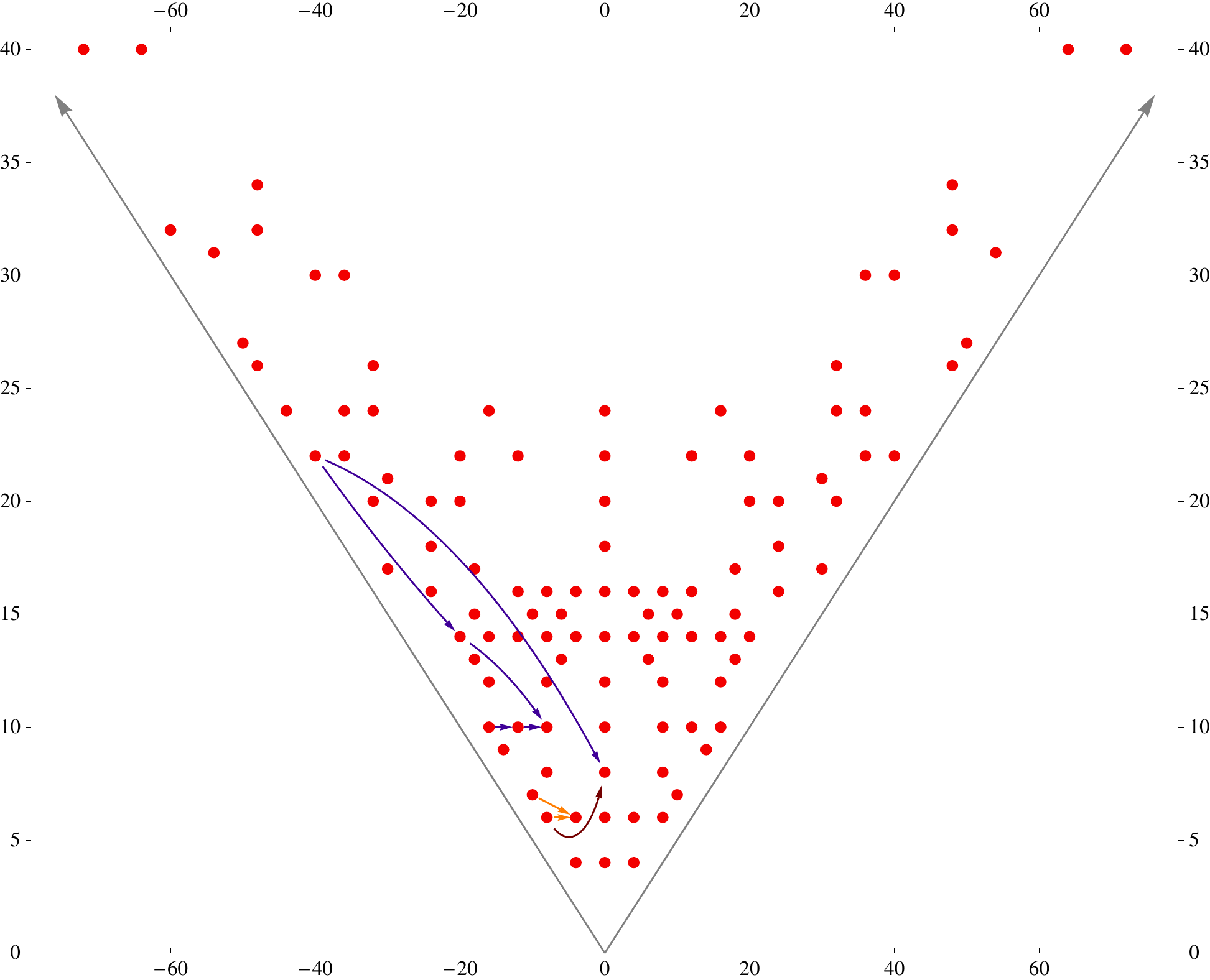}
\vskip10pt
\boxed{
\parbox{4.25in}{\footnotesize\vspace{2pt}
\raisebox{2pt}{\includegraphics[width=20pt]{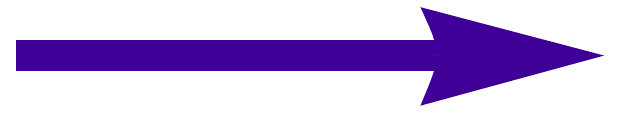}}~~
Splittings between manifolds with fundamental group $\IZ_5$.\\ 
\raisebox{2pt}{\includegraphics[width=20pt]{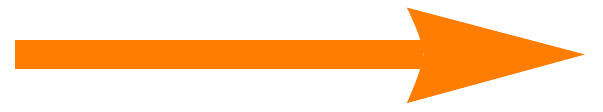}}~~
Splittings between manifolds with fundamental group $\IZ_5{\times}\IZ_2$.\\
\raisebox{2pt}{\includegraphics[width=20pt]{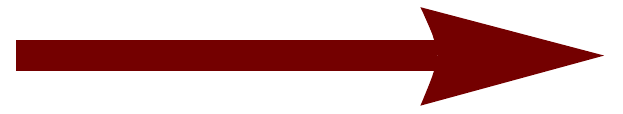}}~~
Splittings between manifolds with fundamental group $\IZ_5{\times}\IZ_5$.
\vspace{2pt}}}
\vskip5pt
\parbox{6.0in}{\caption{\label{Z5WebsInTheTip}\small
The webs of CICY's with fundamental group $\IZ_5$, $\IZ_5{\times}\IZ_2$ and $\IZ_5{\times}\IZ_5$.}}
\end{center}
\end{figure}

We began this investigation with the hypothesis that \cys\ admitting a freely acting group are very rare and although we have found a number of new examples of such manifolds our experience is consistent with a paucity of these. Manifolds admitting free actions by larger groups are seemingly particularly rare. There are manifolds with free actions of groups $G$ of order 64, the largest known~\cite{GrossPopescu,BorisovHua,Hua}. To our knowledge all such manifolds lie at the very tip of the distribution with $\hodgenos=(2,2)$. At order 49 there is a manifold known with a free action of $\IZ_7{\times}\IZ_7$ \cite{GrossPopescu,Rodland} but, to our knowledge, it is the only one such and this also has $\hodgenos=(2,2)$. Only one manifold is known which admits a group of order 36. This has a free action of $\IZ_6{\times}\IZ_6$ and $\hodgenos=(6,6)$. What is more all the cases we have listed thus far belong to the remarkable class of manifolds investigated by 
Gross and Popescu~\cite{GrossPopescu} that are fibered by Abelian surfaces and have Hodge numbers 
$\hodgenos=(n,n)$ for $n=2,4,6,8,10$. There are manifolds admitting free actions of order 32 and these form a short web consisting of manifolds with 
$\hodgenos=(1,3)$ and resolutions of conifolds of these that have $\hodgenos=(2,2)$. At order 25 there is a short web consisting of $\IZ_5{\times}\IZ_5$ quotients of $\IP^4[5]$, the quintic threefold and a resolution of  the Horrocks-Mumford quintic, which is a  highly nodal form of  $\IP^4[5]$. The resolution has 
$\hodgenos=(4,4)$ and is again one of the Gross-Popescu manifolds.

In this paper we find webs for the groups $\IZ_5$, $\IZ_5{\times}\IZ_2$, $\IZ_3$, $\IZ_3{\times}\IZ_3$,
$\IZ_3{\times}\IZ_2$, and $\IH$, with $\IH$ the 
quaternion group. We should explain how $\IH$ arises in this story. Our investigation began with a search for symmetry generators of order 3 and order 5 and so we find groups containing $\IZ_3$ and $\IZ_5$ as subgroups. We have not attempted a search for generators of order 2 since, while there are surely many of these, they are unlikely to give rise to the small Hodge numbers that we seek here. Having sought generators of order 3 
and~5 it seems natural to seek also generators of order 4 particularly in relation to the important configuration 
$\IP^7[2~2~2~2]$ and its close relatives. The diagrams for a selection of these are given below and show a $\IZ_4$ rotational symmetry however, apart from $\IP^7[2~2~2~2]$, this symmetry does not act freely. 
It turns out that there is in fact a freely acting $\IZ_4$ symmetry for each of these manifolds but it does not act as a rotation of the diagram and is contained in $\IH$ as a subgroup.

\begin{gather*}
\IP^7[2~2~2~2]^{1,65}\hskip0.3in\raisebox{-30pt}{\includegraphics[width=0.9in]
{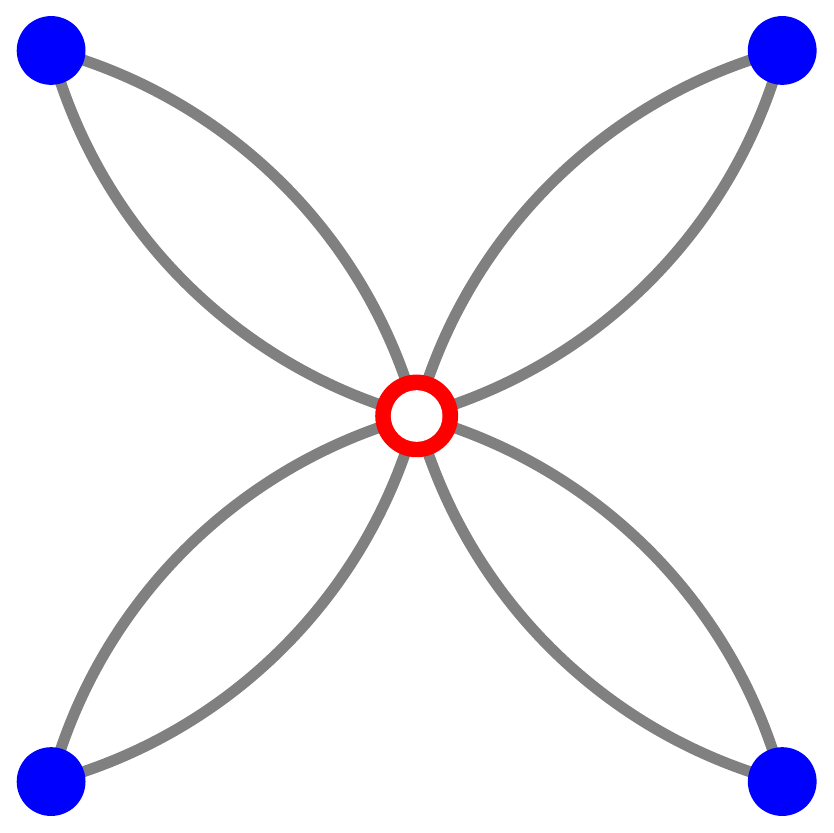}}\hskip1.2in
\cicy{\IP^1\\ \IP^1\\ \IP^1\\ \IP^1}{2\\ 2\\ 2\\ 2}^{4,68}\hskip0.3in
\raisebox{-30pt}{\includegraphics[width=0.9in]{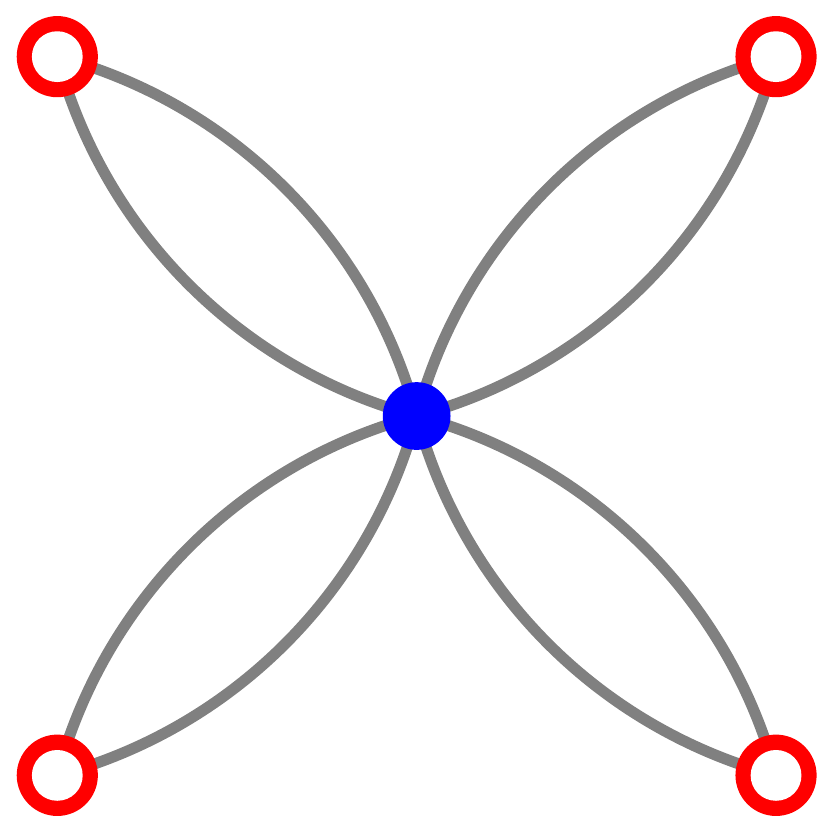}}\\[30pt]
\cicy{\IP^1 \\ \IP^1 \\ \IP^1 \\ \IP^1 \\ \IP^7}
{\one & 0 & 0 & 0 & \one & 0 & 0 & 0 \\
0 & \one & 0 & 0 & 0 & \one & 0 & 0 \\
0 & 0 & \one & 0 & 0 & 0  & \one & 0 \\
0 & 0 & 0 & \one & 0 & 0  & 0 & \one \\
\one & \one & \one & \one & \one & \one & \one & \one}^{5,37}
\hskip0.3in\raisebox{-35pt}{\includegraphics[width=1.1in]
{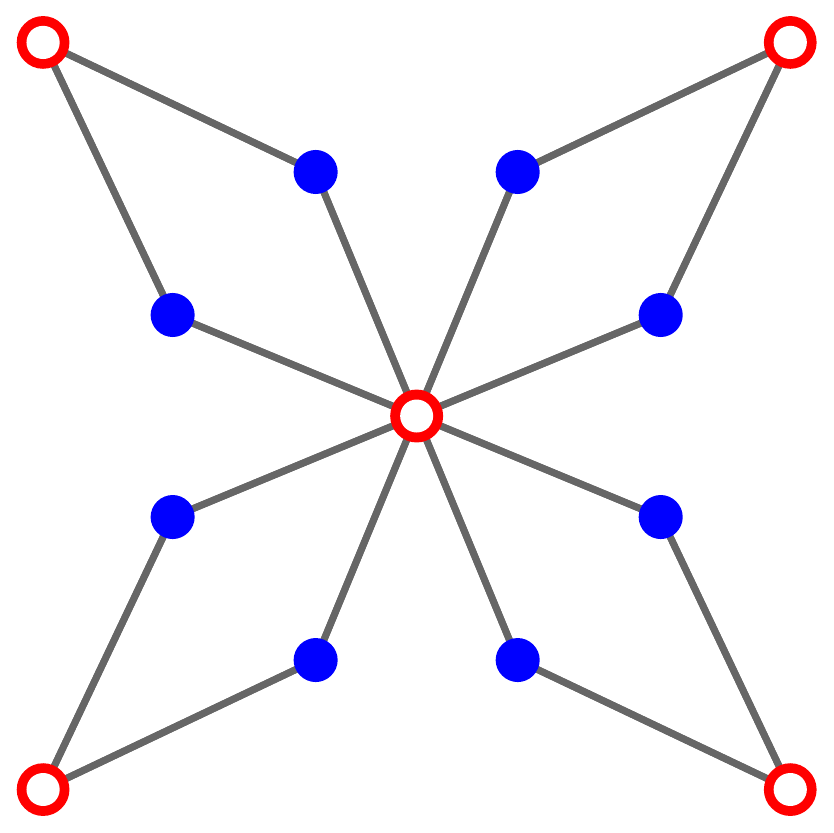}}
\end{gather*}

The process of exploring the webs starts by looking for manifolds with free actions by $\IZ_5$, $\IZ_3$ and $\IH$. The web with group $\IZ_5$ has fewer branches than that for $\IZ_3$ so serves as an introduction. This web is shown in \tref{Z5quotients}. From this table we can form a web of $\IZ_5$ quotients and also a yet smaller web of 
$\IZ_5{\times}\IZ_2$ quotients. These are shown in \tref{Z5miniwebs} and \fref{Z5WebsInTheTip}. In a similar way we obtain from the $\IZ_3$ web smaller webs corresponding to $\IZ_3{\times}\IZ_2$ and 
$\IZ_3{\times}\IZ_3$.

We also find three very small webs, shown in \fref{ResolvedWebs} which consist of resolutions of singular orbifolds.  In one of these the manifolds are simply connected, while the other two contain manifolds with fundamental group $\IZ_3$.  These small webs descend from the large $\IZ_3$ web, since along certain branches an extra $\IZ_3$ or $\IZ_2$ symmetry exists which has fixed points.  We are not sure if the two small $\IZ_3$ webs eventually join with the large $\IZ_3$ web.

A first remark is that the $\IZ_5$, $\IZ_3$ and $\IH$ webs show a striking property with respect to transposition of the matrices. In these tables for each configuration that appears the transposed configuration does so also. In \tref{Z5quotients}, for example, this is clear apart from the configuration with 
$\hodgenos=(7,27)$.
This configuration is however equal to its transpose as we see by performing a further ineffective splitting followed by an inneffective contraction, as illustrated in \tref{TranspositionId}. All three configurations have Euler number -40. It is straightforward to see that by contracting the first five $\IP^1$ rows of the large configuration we return to the configuration with 
$\hodgenos=(7,27)$ whereas if, on the other hand, we contract the rows of the large configuration corresponding to the two $\IP^4$'s then we pass to the transpose of the $(7,27)$ configuration. For the $\IZ_3$-web there are a number of identities analogous to this last one that ensure that the transpose of each configuration is also a configuration of the web. As an illustration of this consider the relation between the $(6,24)$ configuration and the  $(15,15)$. This was written as it is to emphasise that $X^{15,15}$ is a split of $X^{6,24}$. The six $\IP^2$ rows of $X^{15,15}$ can however be contracted without changing the Euler number and when this is done we see that this configuration is also given by the transpose of $X^{6,24}$. The reader wishing to check the remaining identities may do so easily. To find the configuration in the web to which a given configuration is related by transposition: transpose the matrix and compute the new Euler number. This identifies the configuration to which it is related. A~little thought along the lines just presented then shows that the configurations correspond to the same manifold. 
\begin{table}[t!]
\begin{center}
\boxed{\hspace{0.75in}
\begin{tabular}{c@{\hspace{15pt}}c@{\hspace{15pt}}c}
\\[-5pt]
&&\footnotesize$
\cicy{\IP^1\\ \IP^1\\ \IP^1\\ \IP^1\\ \IP^1\\ \IP^4\\ \IP^4\\}
{\one ~ 0 ~ 0 ~ 0 ~ 0 ~ \one ~ 0 ~ 0 ~ 0 ~ 0 \\
 0 ~ \one ~ 0 ~ 0 ~ 0 ~ 0 ~ \one ~ 0 ~ 0 ~ 0 \\
 0 ~ 0 ~ \one ~ 0 ~ 0 ~ 0 ~ 0 ~ \one ~ 0 ~ 0 \\
 0 ~ 0 ~ 0 ~ \one ~ 0 ~ 0 ~ 0 ~ 0 ~ \one ~ 0 \\
 0 ~ 0 ~ 0 ~ 0 ~ \one ~ 0 ~ 0 ~ 0 ~ 0 ~ \one \\
 \one ~ \one ~ \one ~ \one ~ \one ~ 0 ~ 0 ~ 0 ~ 0 ~ 0 \\
 0 ~ 0 ~ 0 ~ 0 ~ 0 ~ \one ~ \one ~ \one ~ \one ~ \one}$\\[0.63in] 
&&\hspace{10pt}\LARGE$\downarrow$\\[8pt]
\footnotesize$
\cicy{\IP^1\\ \IP^1\\ \IP^1\\ \IP^1\\ \IP^1\\ \IP^1\\ \IP^1\\ \IP^1\\ \IP^1\\ \IP^1}
 {\one ~ 0 ~ 0 ~ 0 ~ 0 ~ \one ~ 0 \\
 0 ~ \one ~ 0 ~ 0 ~ 0 ~ \one ~ 0 \\
 0 ~ 0 ~ \one ~ 0 ~ 0 ~ \one ~ 0 \\
 0 ~ 0 ~ 0 ~ \one ~ 0 ~ \one ~ 0 \\
 0 ~ 0 ~ 0 ~ 0 ~ \one ~ \one ~ 0 \\
 \one ~ 0 ~ 0 ~ 0 ~ 0 ~ 0 ~ \one \\
 0 ~ \one ~ 0 ~ 0 ~ 0 ~ 0 ~ \one \\
 0 ~ 0 ~ \one ~ 0 ~ 0 ~ 0 ~ \one \\
 0 ~ 0 ~ 0 ~ \one ~ 0 ~ 0 ~ \one \\
 0 ~ 0 ~ 0 ~ 0 ~ \one ~ 0 ~ \one}$
&\LARGE$\rightarrow$
&\footnotesize
$\cicy{\IP^1\\ \IP^1\\ \IP^1\\ \IP^1\\ \IP^1\\ \IP^1\\ \IP^1\\ \IP^1\\ \IP^1\\ \IP^1\\ \IP^4\\ \IP^4}
{\one~ 0 ~ 0 ~ 0 ~ 0 ~ \one ~ 0 ~ 0 ~ 0 ~ 0 ~ 0 ~ 0 ~ 0 ~ 0 ~ 0 \\
 0 ~ \one~ 0 ~ 0 ~ 0 ~ 0 ~ \one~ 0 ~ 0 ~ 0 ~ 0 ~ 0 ~ 0 ~ 0 ~ 0 \\
 0 ~ 0 ~ \one~ 0 ~ 0 ~ 0 ~ 0 ~ \one~ 0 ~ 0 ~ 0 ~ 0 ~ 0 ~ 0 ~ 0 \\
 0 ~ 0 ~ 0 ~ \one~ 0 ~ 0 ~ 0 ~ 0 ~ \one~ 0 ~ 0 ~ 0 ~ 0 ~ 0 ~ 0 \\
 0 ~ 0 ~ 0 ~ 0 ~ \one~ 0 ~ 0 ~ 0 ~ 0 ~ \one~ 0 ~ 0 ~ 0 ~ 0 ~ 0 \\
 \one~ 0 ~ 0 ~ 0 ~ 0 ~ 0 ~ 0 ~ 0 ~ 0 ~ 0 ~ \one~ 0 ~ 0 ~ 0 ~ 0 \\
 0 ~ \one~ 0 ~ 0 ~ 0 ~ 0 ~ 0 ~ 0 ~ 0 ~ 0 ~ 0 ~ \one~ 0 ~ 0 ~ 0 \\
 0 ~ 0 ~ \one~ 0 ~ 0 ~ 0 ~ 0 ~ 0 ~ 0 ~ 0 ~ 0 ~ 0 ~ \one~ 0 ~ 0 \\
 0 ~ 0 ~ 0 ~ \one~ 0 ~ 0 ~ 0 ~ 0 ~ 0 ~ 0 ~ 0 ~ 0 ~ 0 ~ \one~ 0 \\
 0 ~ 0 ~ 0 ~ 0 ~ \one~ 0 ~ 0 ~ 0 ~ 0 ~ 0 ~ 0 ~ 0 ~ 0 ~ 0 ~ \one\\
 0 ~ 0 ~ 0 ~ 0 ~ 0 ~ \one~ \one~ \one~ \one~ \one~ 0 ~ 0 ~ 0 ~ 0 ~ 0 \\
 0 ~ 0 ~ 0 ~ 0 ~ 0 ~ 0 ~ 0 ~ 0 ~ 0 ~ 0 ~ \one~ \one~ \one~ \one~ \one}$\\[1.15in] 
\end{tabular}
\hspace{0.75in}}
\vskip5pt
\parbox{5.25in}{\caption{\label{TranspositionId}\small The ineffective splits that show that the configuration with Hodge numbers $(7,27)$ from \tref{Z5quotients} corresponds to the same manifold as its transpose.}}
\end{center}
\end{table}
It cannot, however, be simply the case that every web is invariant under transposition since there are many examples of CICY's for which a manifold and its transpose admit different freely acting symmetries. The curious property that characterizes the $\IZ_3$, $\IZ_5$ and $\IH$ webs is that if a manifold of the $\IZ_3$-web, say, admits a freely acting group $G\supset\IZ_3$ then, for all the cases we have studied, the transpose admits a freely acting group $K$ that contains $\IZ_3$ as a subgroup, and a similar statement is true in respect of the 
$\IZ_5$ and~$\IH$~webs.

One last comment, regarding \tref{Z3quotients} concerns the seemingly puzzling split between the manifold with Hodge numbers $(5,50)$ and that with Hodge numbers $(8,44)$ in the fourth column of the table. This becomes clear, however, by splitting the second column of the $(5,59)$-configuration three times in the following~way
$$
\cicy{\IP^1\\ \IP^1\\ \IP^1\\ \IP^2}{\one~\one\\ \one~\one\\ \one~\one\\ 0~\bf 3}^{5,59}_{-108} \rightarrow~~~~~ 
\cicy{\IP^1\\ \IP^1\\ \IP^1\\ \IP^1\\ \IP^1\\ \IP^1\\ \IP^2}
{\one~\one~0~0~0\\ \one~\one~0~0~0\\ \one~\one~0~0~0\\ 0~\one~\one~0~0\\  0~\one~0~\one~0\\  0~\one~0~0~\one\\  0~0~\one~\one~\one}^{8,44}_{-72}~\cong~~~~
\cicy{\IP^1\\ \IP^1\\ \IP^1\\ \IP^1\\ \IP^1\\ \IP^1}
{\one~\one~0\\ \one~\one~0\\ \one~\one~0\\ 0~\one~\one\\  0~\one~\one\\  
0~\one~\one}^{8,44}_{-72}
$$
It is perhaps simplest to see this by contracting the fourth, fifth and sixth $\IP^1$'s of the split manifold, which returns us to the configuration on the left. If, instead, we contract the $\IP^2$ of the split manifold we arrive at the configuration on the right which has the same Euler number and hence is isomorphic. The isomorphism shown here is also an example of the application of the identity \eref{dP3id}.

We have not attempted an automated search for CICY's that admit freely acting groups. We have examined only splits of manifolds that seem likely to admit the desired symmetry. Nevertheless it may be helpful to describe our procedure in the manner of a computer program:
\vbox{
\begin{tabbing}
\hspace*{20pt}\=\bf main loop:~~\=  Split the configurations in all ways likely to admit a $\IZ_5$ or $Z_3$ symmetry.\kill 
\>\bf program:   \>\sl To find CICY's admitting a freely acting $\IZ_3$, $\IZ_5$ or $\IH$ symmetry.\\[2ex]   
\>\bf initialize:   \>\sl Choose a manifold known to admit the required symmetry. \\[2ex]
\>\bf main loop:\>\sl Split the configurations in all ways likely to preserve the symmetry.\\[0.5ex]                     
                     \>\>\sl Transpose all the configurations and add these to the web.\\[0.5ex]
                     \>\>\sl Repeat until the web no longer changes.\\
\end{tabbing}}

\vspace{-20pt}
In this way we have found new quotients as shown in \fref{NewOrbifolds} and listed in \tref{tiptab}. Among these there are some that seem especially interesting. It is apparent from \tref{Z3quotients} that the manifold with Hodge numbers $(15,15)$ is very special and we find $\IZ_2$, $\IZ_3$ and $\IZ_3{\times}\IZ_2$ free quotients with Hodge numbers $(9,9)$, $(7,7)$ and $(3,3)$. This manifold is an elliptic fibration and it would seem to be analogous in a number of ways to $X^{19,19}$ which is at the heart of the heterotic model propounded by the School at the University of Pennsylvania~\cite{Braun:2004xv,Braun:2005nv,Donagi:2000zf,Bouchard:2005ag}. The manifold $X^{9,27}$ is also very special in that it admits a $\IZ_3$ symmetry generator that acts freely so that the quotient $X^{9,27}/S$ has $\chi =-12$ and fundamental group $\IZ_3$.
The manifold admits also a $\IZ_2$ generator $U$ which does not act freely but instead fixes two elliptic curves. On taking the further quotient by $U$ and resolving singularities we find a manifold an analogue of the 
three-generation manifold in that it has Euler number -6 and fundamental group $\IZ_3$ but has now Hodge numbers $(5,8)$ instead of $(6,9)$. We find also a number of interesting manifolds with Hodge numbers near the tip of the distribution as is evident from \fref{NewOrbifolds}.

Although our main aim is to take quotients by freely acting groups, we come upon several cases where there are also symmetry generators with non-trivial fixed point sets, which are often elliptic curves as above. Taking quotients by non-freely acting generators and resolving singularities provides a number of examples of new manifolds. The fundamental groups of the resolved orbifolds which appear in this way can be determined quite simply.  If $X$ is simply-connected and we take its quotient by the group $G$ and resolve fixed points, the fundamental group of the resulting manifold is $G/H$, where $H$ is the normal subgroup generated by elements of $G$ which act with fixed points.  This is intuitive because the spaces we glue in to repair orbifold points are all simply connected, so any loop encircling such a point is homotopically trivial in the resolved manifold.  In the Physics literature this result goes back to \cite{Dixon:1985jw}; for a reference in the mathematical literature see
\cite{BrownHiggins}.
The remaining sections of this paper describe in detail the manifolds that we find. Of necessity, this listing is somewhat repetitive; for each manifold we check that there is a system of polynomials that are transverse and for which the group action is fixed point free. 
\subsection{Webs of transgressions}
\begin{figure}[H]
\begin{center}
\includegraphics[width=6.5in]{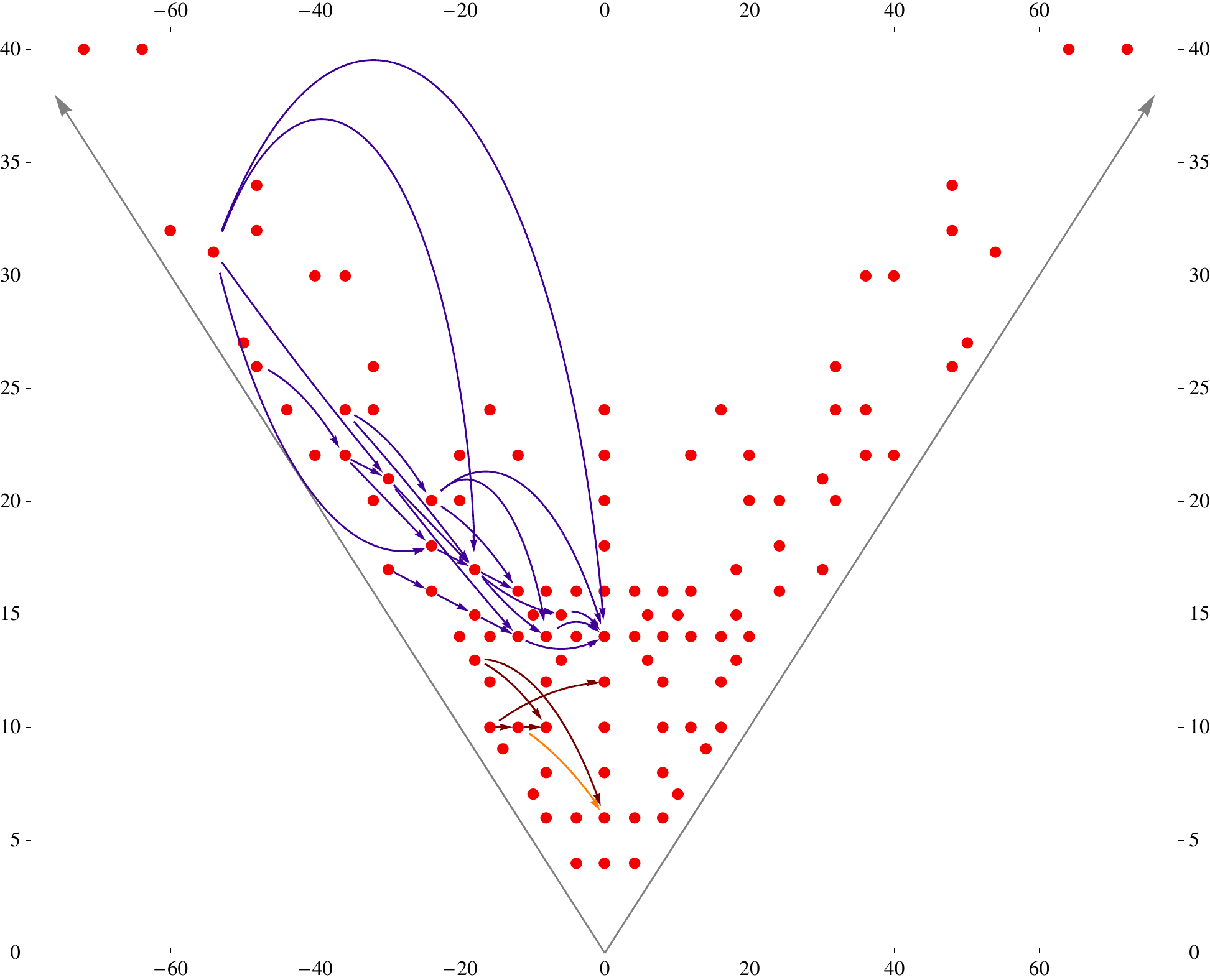}
\vskip10pt
\boxed{
\parbox{4.25in}{\footnotesize\vspace{2pt}
\raisebox{2pt}{\includegraphics[width=20pt]{arrowblue.pdf}}~~
Splittings between manifolds with fundamental group $\IZ_3$.\\ 
\raisebox{2pt}{\includegraphics[width=20pt]{arrowbrown.pdf}}~~
Splittings between manifolds with fundamental group $\IZ_3{\times}\IZ_3$.\\
\raisebox{2pt}{\includegraphics[width=20pt]{arroworange.pdf}}~~
Splittings between manifolds with fundamental group $\IZ_3{\times}\IZ_2$.
\vspace{2pt}}}
\vskip10pt
\parbox{6.0in}{\caption{\label{Z3WebsInTheTip}\small
The webs of CICY's with fundamental group $\IZ_3$, $\IZ_3{\times}\IZ_2$ and $\IZ_3{\times}\IZ_3$.}}
\end{center}
\end{figure}

\begin{figure}[pt!]
\begin{center}
\includegraphics[width=6.5in]{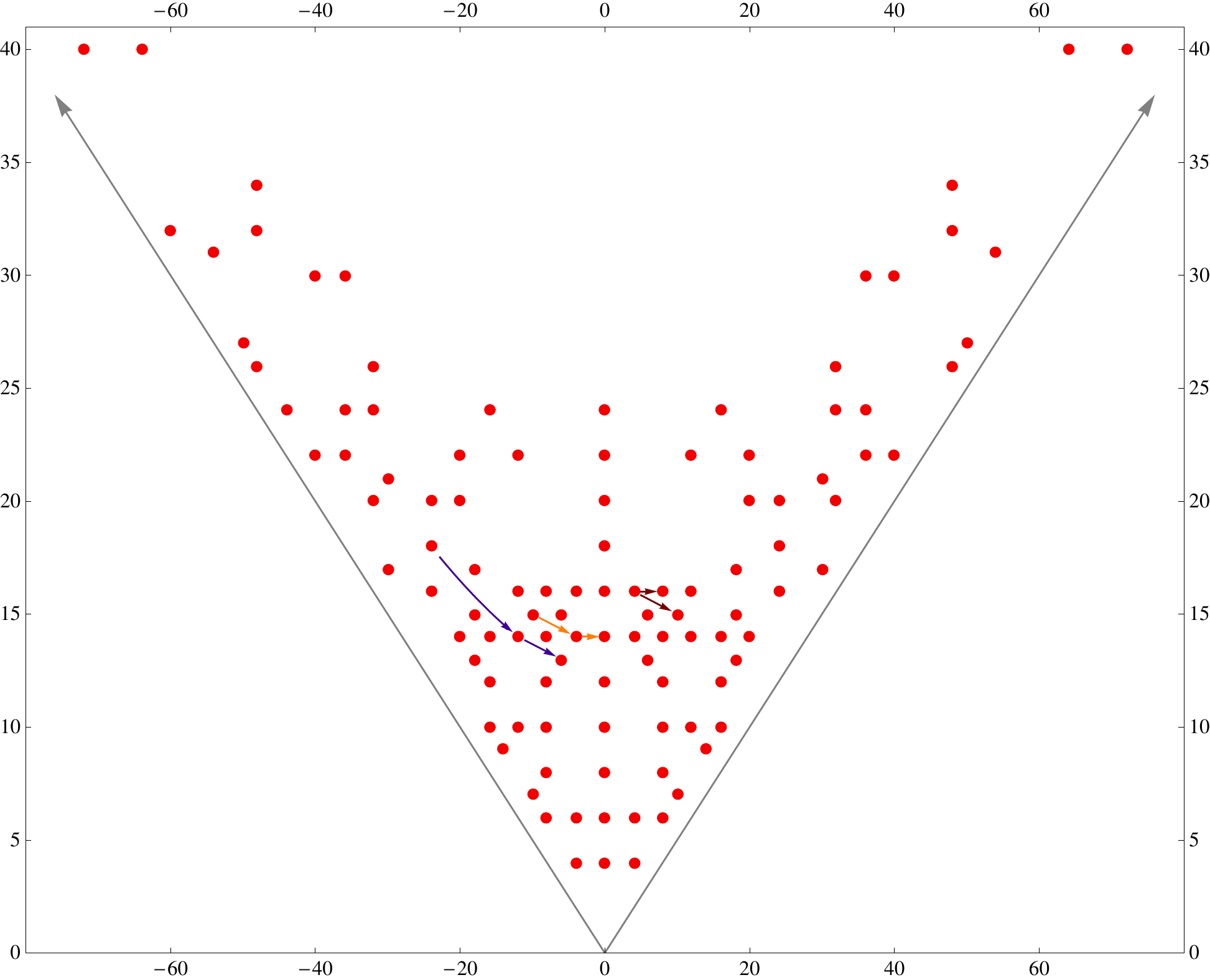}
\vskip10pt
\framebox[6.2in]{
\parbox{6.2in}{\footnotesize\vspace{2pt}
\raisebox{2pt}{\includegraphics[width=20pt]{arrowblue.pdf}}~~Splittings between manifolds with fundamental group $\IZ_3$, from resolutions of $\IZ_3{\times}\IZ_2$ orbifolds.\\[5pt]
\raisebox{2pt}{\includegraphics[width=20pt]{arrowbrown.pdf}}~~Splittings between manifolds with fundamental group $\IZ_3$, from resolutions of $\IZ_3{\times}\IZ_3$ orbifolds.\\[5pt]
\raisebox{2pt}{\includegraphics[width=20pt]{arroworange.pdf}}~~Splittings between simply connected manifolds, from resolutions of $\IZ_3{\times}\IZ_3$ orbifolds.
\vspace{2pt}}}
%
\parbox{6.2in}{\vspace{20pt}\framebox[3.6in][c]{\begin{tabular}{ccccc}
$\left( \widehat{X^{3,48}\quotient{\IZ_3{\times}\IZ_3}} \right)^{5,10}$\hspace*{-30pt}\\
&\Large$\searrow$\hspace*{-20pt}\\
&&$\left( \widehat{X^{6,24}\quotient{\IZ_3{\times}\IZ_3}} \right)^{6,8}$\hspace*{-10pt}
&\Large$\rightarrow$\hspace*{-10pt} 
&$\left( \widehat{X^{15,15}\quotient{\IZ_3{\times}\IZ_3}} \right)^{7,7}$\\
\end{tabular}}
\hfill
\framebox[2.5in][c]{
\begin{tabular}{ccc}
\hskip-5pt$\left( \widehat{X^{5,59}\quotient{\IZ_3{\times}\IZ_3}} \right)^{9,7}$\hspace*{-20pt}
&\Large$\rightarrow$\hspace*{-10pt}
&$\left( \widehat{X^{8,44}\quotient{\IZ_3{\times}\IZ_3}} \right)^{10,6}$\\
&\Large$\searrow$\hspace*{-10pt}\\
&&$\left( \widehat{X^{6,33}\quotient{\IZ_3{\times}\IZ_3}} \right)^{10,5}$
\end{tabular}}}
\vskip2pt
\framebox[6.2in][l]{\vrule width0pt height22pt depth14pt
$\left(\widehat{\IP^5[3,3]^{1,73}\quotient{\IZ_3{\times}\IZ_2}}\right)^{3,15}$\hspace{-5pt}
{\large$\longrightarrow$}
$\left(\widehat{X^{3,39}\quotient{\IZ_3{\times}\IZ_2}}\right)^{4,10}$\hspace{-5pt}
{\large$\longrightarrow$}
$\left(\widehat{X^{9,27}\quotient{\IZ_3{\times}\IZ_2}}\right)^{5,8}$}
\vskip0pt
\parbox{6.0in}{\caption{\label{ResolvedWebs}\small
Webs of resolved orbifolds, that are either simply connected or with fundamental group~$\IZ_3$.
The boxes below the main figure identify the manifolds.}
\vskip-6pt 
}
\end{center}
\end{figure}

One of our principal motivations in studying webs of cicy's with given fundamental groups relates to the process of transgression described in \cite{Triadophilia}. The idea here is to transport vector bundles from one \cym to another across a conifold transition. Subject to important provisos about their stability, these bundles correspond to heterotic string theory vacua. We do not have a clear reason for seeking to minimise some quantity similar to $h^{11}+h^{21}$ though the idea that this is interesting underlies the present investigation. From this perspective, interest attaches to the endpoints of the flow given, for example, by the arrows of \fref{Z3WebsInTheTip}, or more clearly, from the arrows of \tref{Z3andZ3*Z3webs}. The latter table is preferable since it makes clear that there is an endpoint with Hodge numbers $(5,11)$ and that there
are two endpoints with Hodge numbers $(7,7)$, which are the manifolds $X^{19,19}/\IZ_3$ and 
$X^{15,15}/\IZ_3$. Of course, these apparent endpoints might not in fact be true endpoints since there may be other transitions, beyond those that we have found here, that affect this. We do not know, for example, whether the $\IZ_3$-webs of \fref{ResolvedWebs} are, in reality, disjoint from the $\IZ_3$ web of 
\fref{Z3WebsInTheTip}; nor whether the manifold 
$$
\left( \widehat{X^{19,19}\quotient{\IZ_3{\times}\IZ_2{\times}\IZ_2}} \right)^{4,4}~,
$$
which has fundamental group $\IZ_3{\times}\IZ_2$ and appears to be isolated, is in fact so.

\begin{table}[H]
\begin{center}

\boxed{\hskip2pt
\begin{tabular}{ccccc@{\hspace{-5pt}}c@{\hspace{-10pt}}c}
\\[-12pt]
&&\footnotesize$\cicy{\IP^2\\ \IP^2}{3\\ 3}^{2,83}\symm{\IZ_3{\times}\IZ_3}\hspace{-20pt}$
&&\footnotesize$\IP^5[3~3]^{1,73}\symm{\IZ_3{\times}\IZ_3}$\hspace{-20pt}\\[-3pt]
&\LARGE$\swarrow$ 
& \LARGE$\downarrow$ 
&\smash{\begin{picture}(0, -40)\thicklines\put(-10,10){\vector(1, -3){25.0}}\end{picture}}
& \LARGE$\downarrow$\\[-1pt]
\footnotesize$\cicy{\IP^2\\ \IP^3}{3~0\\ 1~3}^{8,35}\symm{\IZ_3}$
&&\footnotesize$\cicy{\IP^2\\ \IP^2\\ \IP^2}
{\one~\one~\one \\\one~\one~\one \\\one~\one~\one}^{3,48}\symm{\IZ_3}$
&&\footnotesize$\cicy{\IP^2\\ \IP^5}{1~1~1~0\\ 1~1~1~3}^{2,56}\symm{\IZ_3{\times}\IZ_3}$\\[-6pt]
\LARGE$\downarrow$ && \LARGE$\downarrow$ && \LARGE$\downarrow$ 
& \LARGE$\searrow$\\[-3pt]
\footnotesize$\cicy{\IP^3\\ \IP^3}{1~3~0\\ 1~0~3}^{14,23}\symm{\IZ_3}$
&&\footnotesize$\cicy{ \IP^2\\ \IP^2\\ \IP^2\\ \IP^2}
{\one~0~0~\one~\one \\
0~\one~0~\one~\one \\
0~0~\one~\one~\one \\
\one~\one~\one~0~0}^{4,40}\symm{\IZ_3}$
&&\footnotesize
$\cicy{\IP^2\\\IP^2\\\IP^5}
{\one~\one~\one~0~0~0\\
0~0~0~\one~\one~\one\\
\one~\one~\one~\one~\one~\one}^{3,39}\symm{\IZ_3{\times}\IZ_3,\IZ_6}$\hspace{-20pt}
&&\hspace{15pt}\footnotesize
$\cicy{\IP^1\\ \IP^1 \\ \IP^1\\ \IP^5}
{\one~\one~0~0~0\\ 
  \one~0~\one~0~0\\ 
  \one~0~0~\one~0\\
  0~\one~\one~\one~3}^{5,50}\symm{\IZ_3}$\\[-5pt]
\LARGE$\downarrow$ && \LARGE$\downarrow$ && \LARGE$\downarrow$ 
& \LARGE$\swarrow$\\[1pt]
\footnotesize$\cicy{\IP^1\\ \IP^2\\ \IP^2}{1~1\\ 3~0\\ 0~3}^{19,19}\symm{\IZ_3{\times}\IZ_3,\IZ_6}$ 
\hspace{-40pt}
&&\footnotesize$\cicy{\IP^2 \\ \IP^2\\ \IP^2\\ \IP^2\\ \IP^2}
{\one~0~0~\one~0~0~\one \\
0~\one~0~0~\one~0~\one \\
0~0~\one~0~0~\one~\one \\
\one~\one~\one~0~0~0~0 \\
 0~0~0~\one~\one~\one~0}^{5,32}\symm{\IZ_3}$\hspace*{-15pt}
&&\footnotesize\smash{
$\cicy{\IP^1\\ \IP^1\\ \IP^1\\ \IP^2\\ \IP^5\\}
{\one~\one~0~0~0~0~0 \\
 \one~0~\one~0~0~0~0 \\
 \one~0~0~\one~0~0~0 \\
 0~0~0~0~\one~\one~\one \\
 0~\one~\one~\one~\one~\one~\one}^{6,33}\symm{\IZ_3}$}
&\LARGE\raisebox{-2pt}{$\leftarrow$}
&\hspace{7pt}\footnotesize
$\cicy{\IP^1\\ \IP^1\\ \IP^1\\ \IP^2}{1~1\\ 1~1\\ 1~1\\ 3~0}^{5,59}\symm{\IZ_3}$\\[-5pt]
&& \LARGE$\downarrow$ && \LARGE$\downarrow$ && \LARGE$\downarrow$\\[-1pt]
&\multicolumn{2}{l}{\footnotesize
$\cicy{\IP^2\\ \IP^2\\ \IP^2\\ \IP^2\\ \IP^2\\ \IP^2\\}
{\one~0~0~\one~0~0~\one~0~0 \\
 0~\one~0~0~\one~0~0~\one~0 \\
 0~0~\one~0~0~\one~0~0~\one \\
 \one~\one~\one~0~0~0~0~0~0 \\
 0~0~0~\one~\one~\one~0~0~0 \\
 0~0~0~0~0~0~\one~\one~\one}^{6,24}\symm{\IZ_3}$}
&&\footnotesize
$\cicy{\IP^1\\ \IP^1\\ \IP^1\\ \IP^1\\ \IP^1\\ \IP^1\\ \IP^5\\}
{0~\one~\one~0~0~0~0~0 \\
  0~\one~0~\one~0~0~0~0 \\
  0~\one~0~0~\one~0~0~0 \\
  \one~0~0~0~0~\one~0~0 \\
  \one~0~0~0~0~0~\one~0 \\
  \one~0~0~0~0~0~0~\one \\
  0~0~\one~\one~\one~\one~\one~\one}^{9,27}\symm{\IZ_3}$
&\LARGE$\leftarrow$
&\footnotesize
$\cicy{\IP^1\\ \IP^1\\ \IP^1\\ \IP^1\\ \IP^1\\ \IP^1\\}
{0~\one~\one \\
  0~\one~\one \\
  0~\one~\one \\
  \one~0~\one \\
  \one~0~\one \\
  \one~0~\one}^{8,44}\symm{\IZ_6}$\\[-12pt]
&&\LARGE$\downarrow$&&
&\smash{\begin{picture}(0, -40)\thicklines\put(3,10){\vector(-2, -3){17}}\end{picture}}
&\smash{\begin{picture}(0, -40)\thicklines\put(-1,10){\vector(0, -1){30}}\end{picture}} \\[-1pt]
\multicolumn{3}{l}{\footnotesize$\cicy{\IP^1\\ \IP^1\\ \IP^1\\ \IP^1\\ \IP^1\\ \IP^1\\ \IP^1\\ \IP^1\\ \IP^1\\ \IP^2\\ \IP^2\\ \IP^2\\ \IP^2\\ \IP^2\\ \IP^2}
{\one~0~0~0~0~0~0~0~0~\one~0~0~0~0~0~0~0~0 \\
 0~\one~0~0~0~0~0~0~0~0~0~0~\one~0~0~0~0~0 \\
 0~0~\one~0~0~0~0~0~0~0~0~0~0~0~0~\one~0~0 \\
 0~0~0~\one~0~0~0~0~0~0~\one~0~0~0~0~0~0~0 \\
 0~0~0~0~\one~0~0~0~0~0~0~0~0~\one~0~0~0~0 \\
 0~0~0~0~0~\one~0~0~0~0~0~0~0~0~0~0~\one~0 \\
 0~0~0~0~0~0~\one~0~0~0~0~\one~0~0~0~0~0~0 \\
 0~0~0~0~0~0~0~\one~0~0~0~0~0~0~\one~0~0~0 \\
 0~0~0~0~0~0~0~0~\one~0~0~0~0~0~0~0~0~\one \\
 \one~\one~\one~0~0~0~0~0~0~0~0~0~0~0~0~0~0~0 \\
 0~0~0~\one~\one~\one~0~0~0~0~0~0~0~0~0~0~0~0 \\
 0~0~0~0~0~0~\one~\one~\one~0~0~0~0~0~0~0~0~0 \\
 0~0~0~0~0~0~0~0~0~\one~\one~\one~0~0~0~0~0~0 \\
 0~0~0~0~0~0~0~0~0~0~0~0~\one~\one~\one~0~0~0 \\
 0~0~0~0~0~0~0~0~0~0~0~0~0~0~0~\one~\one~\one}^{15,15}\symm{\IZ_6}$}
&\hspace*{-10pt}\raisebox{27pt}{\LARGE$\leftarrow$}
&\raisebox{30pt}{\footnotesize$\cicy{\IP^1 \\ \IP^1 \\ \IP^1 \\ \IP^1 \\ \IP^1 \\ \IP^1 \\ \IP^2}
{\one~0~\one~0~0 \\
\one~0~0~\one~0 \\
\one~0~0~0~\one \\
0~\one~\one~0~0 \\
0~\one~0~\one~0 \\
0~\one~0~0~\one \\
0~0~\one~\one~\one}^{9,21}\symm{\IZ_3}$}
&&\raisebox{30pt}{\footnotesize
$\cicy{\IP^1\\ \IP^1\\ \IP^1\\ \IP^1\\ \IP^1\\ \IP^1\\ \IP^1}
{0~0~\one~\one \\
  0~\one~0~\one \\
  0~\one~0~\one \\
  0~\one~0~\one \\
  \one~0~\one~0 \\
  \one~0~\one~0 \\
  \one~0~\one~0}^{19,19}\symm{\IZ_3{\times}\IZ_3,\IZ_6}$}\hspace*{-7pt}\\
\\[-15pt]
\end{tabular}
\hskip1pt}
\vskip-2pt
\parbox{5.5in}{\caption{\label{Z3quotients}\small The web of CICY's that admit a freely acting $\IZ_3$ symmetry. Not all splits are shown and $Y^{6,33}$ of \SS\ref{sec:9,21} is omitted for lack of space. The two configurations with Hodge numbers $(19,19)$ are the same manifold.  }}
\vspace{-15pt} 
\end{center}
\end{table}
\newpage
%

\begin{table}[H]
\begin{center}

\setlength{\tabcolsep}{0pt}
\boxed{
\begin{tabular}
{lc@{\hspace{0pt}}c@{\hspace{-10pt}}c@{\hspace{-20pt}}c@{\hspace{-2pt}}c@{\hspace{-5pt}}c}
\\[-10pt]
&&\hspace{-3pt}\includegraphics[width=50pt]{fig_2,83_manifold.pdf}&&\includegraphics[width=50pt]{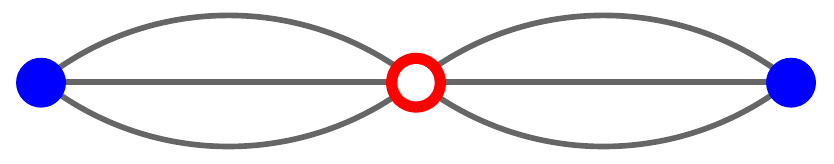}\\[5pt]
& \LARGE$\swarrow$ &\LARGE$\downarrow$ 
&\smash{\begin{picture}(0, -40)\thicklines\put(-15,15){\vector(2, -3){60}}\end{picture}}
& \LARGE$\downarrow$\\[5pt]
\includegraphics[width=75pt]{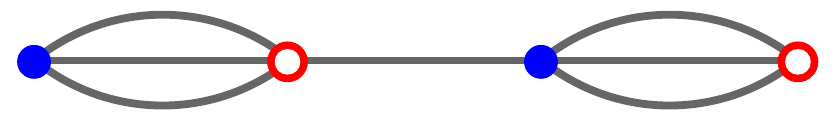}
&&\raisebox{-20pt}{\includegraphics[width=50pt]{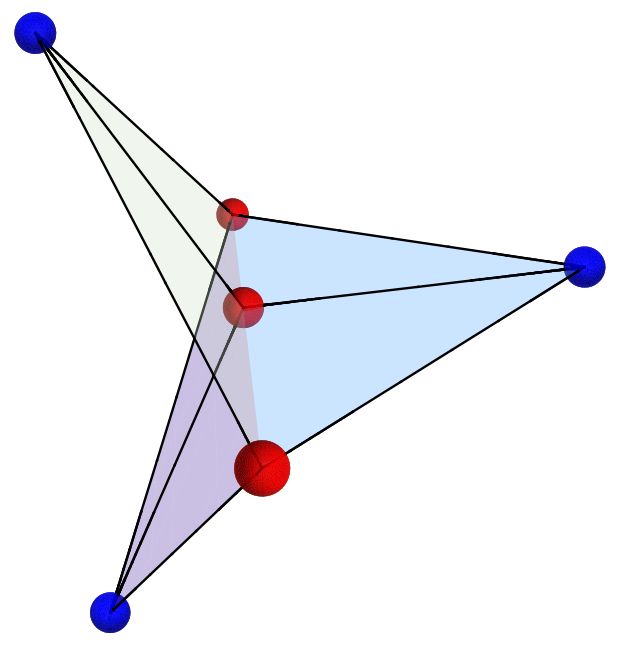}}
&&\hspace{23pt}\raisebox{-10pt}{\includegraphics[width=75pt]{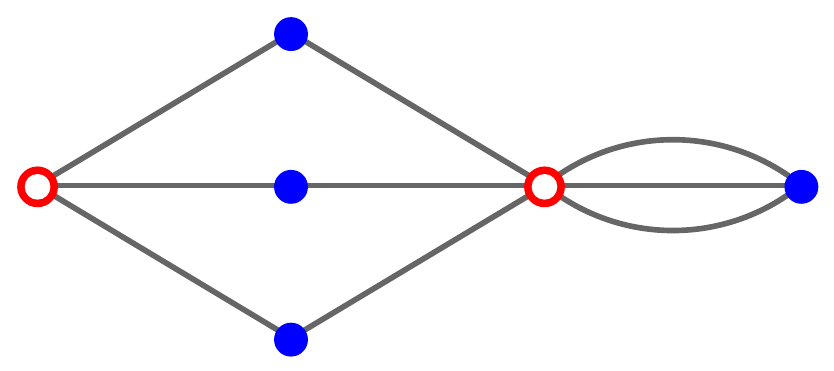}}\\[15pt]
\hspace*{45pt}\LARGE$\downarrow$ && \LARGE$\downarrow$ && \LARGE$\downarrow$ 
& \LARGE$\searrow$\\[1pt]
\raisebox{-5pt}{\includegraphics[width=100pt]{fig_14,23_manifold.pdf}}
&&\raisebox{-40pt}{\includegraphics[width=55pt]{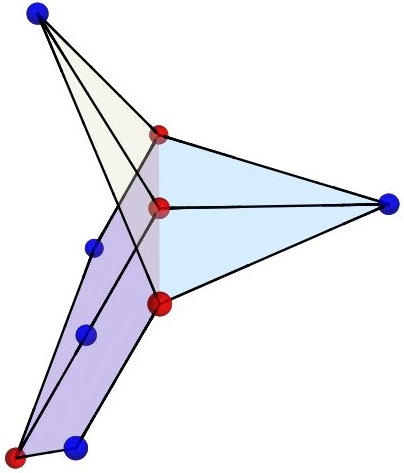}}
&&\raisebox{-13pt}{\includegraphics[width=100pt]{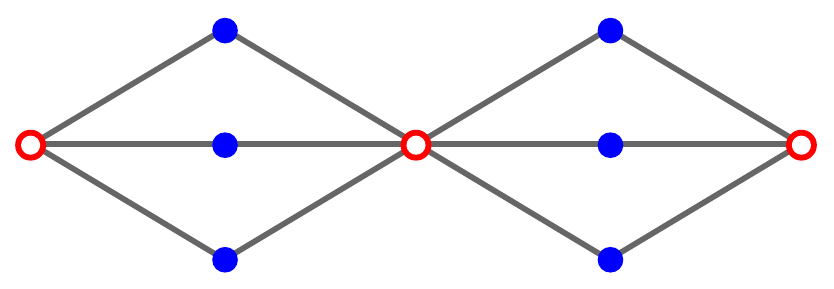}}
&&\raisebox{-13pt}{\includegraphics[width=100pt]{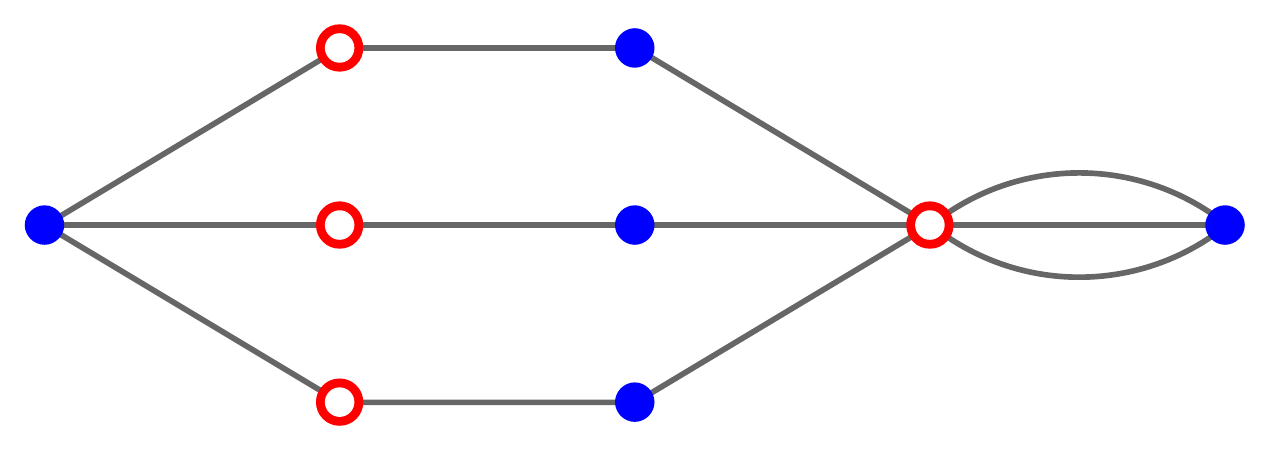}}\\[-15pt]
\hspace*{45pt}\LARGE$\downarrow$ && \LARGE$\downarrow$ && \LARGE$\downarrow$ 
& \LARGE$\swarrow$\\[5pt]
\raisebox{50pt}{\includegraphics[width=100pt]{fig_19,19_manifoldA.pdf}}\hspace{-15pt} 
&&\raisebox{20pt}{\includegraphics[width=65pt]{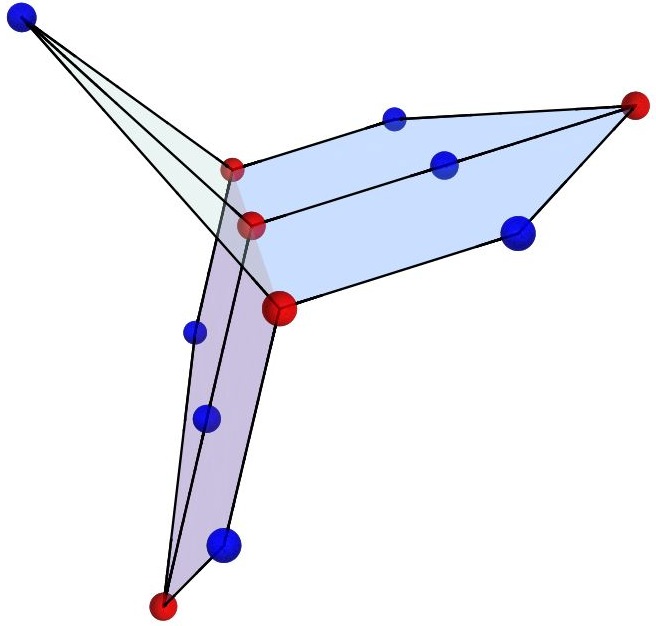}}
&&\hspace{-25pt}\raisebox{40pt}{\includegraphics[width=125pt]{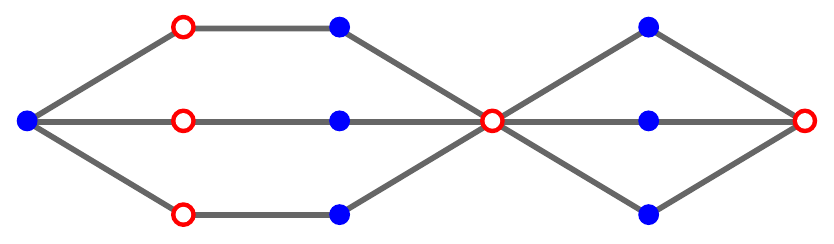}}
&\LARGE\raisebox{54pt}{$\leftarrow$}
&\hspace{-22pt}\raisebox{42pt}{\includegraphics[width=75pt]{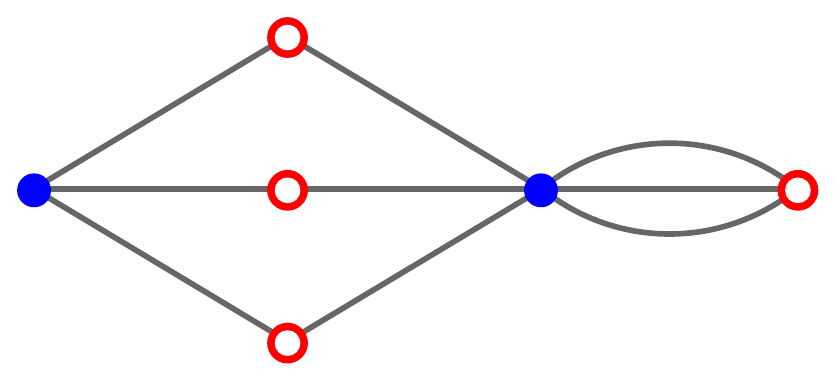}}\\[-40pt]
&& \LARGE$\downarrow$ && \LARGE$\downarrow$ && \hspace{-45pt}\LARGE$\downarrow$\\
&&\hspace*{-15pt}\raisebox{-55pt}{\includegraphics[width=70pt]{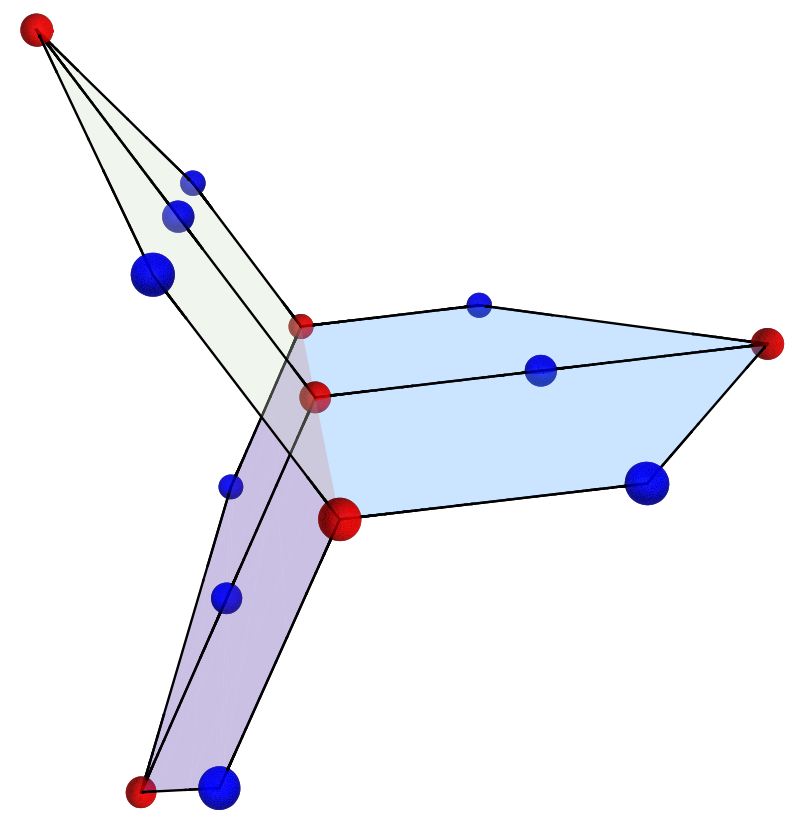}}
&&\raisebox{-14pt}{\includegraphics[width=150pt]{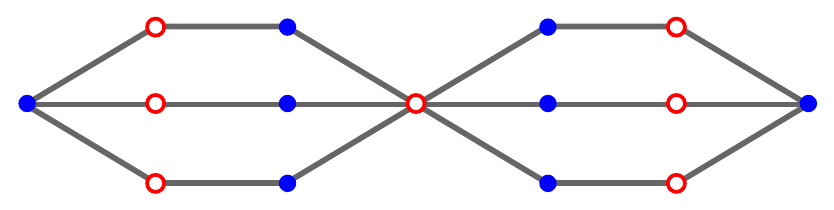}}
&\LARGE$\leftarrow$
&\raisebox{-13pt}{\includegraphics[width=100pt]{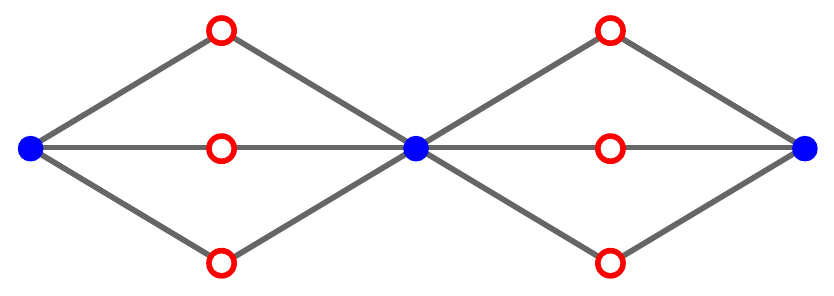}}\\[0pt]
&&\smash{\begin{picture}(0, -40)\thicklines\put(0,15){\vector(0, -1){40}}\end{picture}}
&&& \smash{\begin{picture}(0, -40)\thicklines\put(5,30){\vector(-2, -3){35}}\end{picture}} 
&\smash{\begin{picture}(0, -40)\thicklines\put(-22,30){\vector(0, -1){100}}\end{picture}}\\[20pt]
\multicolumn{4}{l}{\hspace{52pt}\raisebox{-10pt}{\includegraphics[width=140pt]{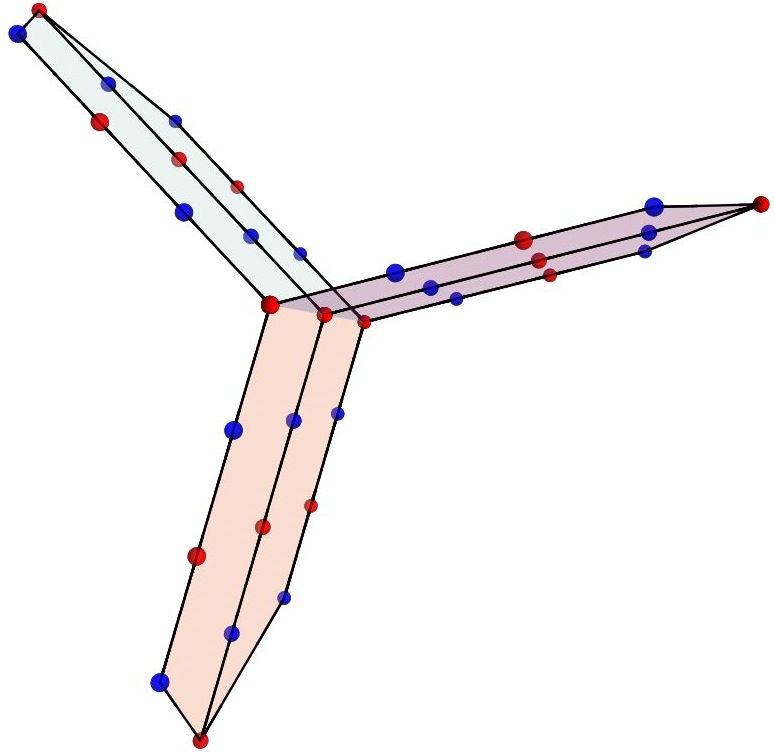}}}
&\hspace{-45pt}\raisebox{83pt}{\LARGE$\longleftarrow$}\hspace*{-5pt}\raisebox{50pt}{\includegraphics[width=70pt]{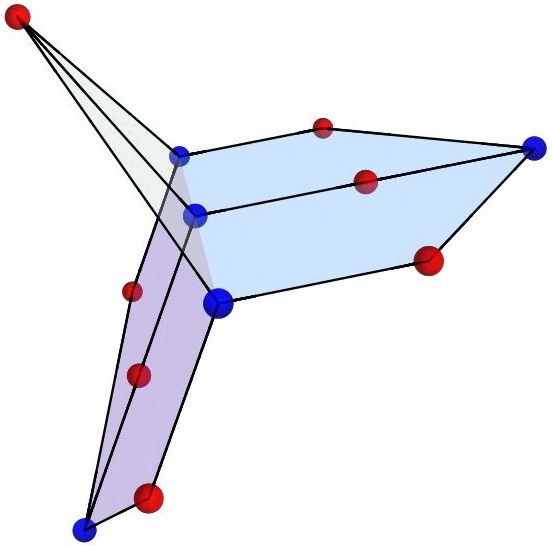}}\hspace{-35pt}
&&\hspace{-43pt}\raisebox{35pt}{\includegraphics[width=150pt]{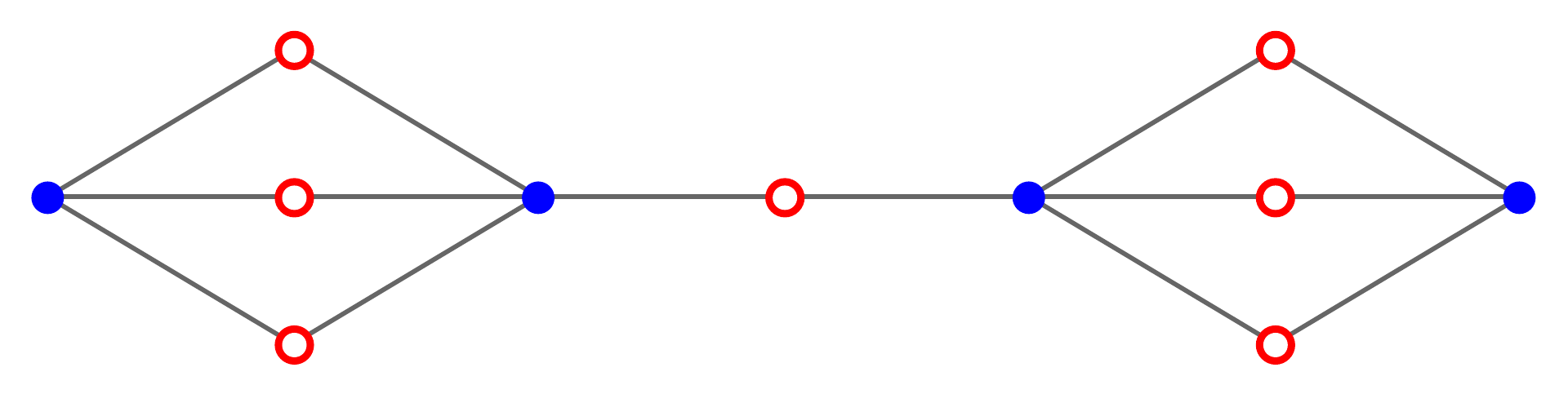}}\\
\\[-10pt]
\end{tabular}
}
\vskip5pt
\parbox{5.8in}{\caption{\label{Z3figs}\small The diagrams for the web of CICY's that admit a freely acting 
$\IZ_3$ symmetry. Again $Y^{6,33}$ is omitted for lack of space.}}
\end{center}
\end{table}
\def\str{\vrule height20pt width0pt depth18pt}

\begin{table}[H]
\begin{center}

\framebox[6.5in][c]{\hskip3pt
\begin{tabular}{ccc@{\hspace{-10pt}}cc@{\hspace{-2pt}}cc@{\hspace{-2pt}}cc}
\\[-12pt]
\str &&&&
\thicklines\smash{\begin{picture}(0,40)\put(-10,-10){\vector(-4,-1){110}}\end{picture}}
$\Big(X^{2,83}\quotient{\IZ_3}\Big)^{2,29}$
\thicklines\smash{\begin{picture}(0,40)\put(-10,-10){\vector(1,-2){42}}\end{picture}}
&&$\Big(X^{1,73}\quotient{\IZ_3}\Big)^{1,25}$\\
&&&& \LARGE$\downarrow$ && \LARGE$\downarrow$\\
\str $\Big(X^{8,35}\quotient{\IZ_3}\Big)^{4,13}$
&&$\Big(X^{3,48}\quotient{\IZ_3}\Big)^{1,16}$
&&$\Big(X^{3,48}\quotient{\IZ_3}\Big)^{3,18}$
&&$\Big(X^{2,56}\quotient{\IZ_3}\Big)^{2,20}$\\
\LARGE$\downarrow$ && \LARGE$\downarrow$ 
&& \thicklines\smash{\begin{picture}(0,40)\put(0,15){\vector(0,-1){140}}\end{picture}}
&& \LARGE$\downarrow$ & \LARGE$\searrow$\\
\str $\Big(X^{14,23}\quotient{\IZ_3}\Big)^{6,9}$
&&  $\Big(X^{4,40}\quotient{\IZ_3}\Big)^{2,14}$
&&&& $\Big(X^{3,39}\quotient{\IZ_3}\Big)^{3,15}$
&&$\Big(X^{5,50}\quotient{\IZ_3}\Big)^{3,18}$\\[4pt]
\LARGE$\downarrow$ && \LARGE$\downarrow$ &&&& \LARGE$\downarrow$ & \LARGE$\swarrow$\\[4pt]
\str $\Big(X^{19,19}\quotient{\IZ_3}\Big)^{7,7}$ 
&&  $\Big(X^{5,32}\quotient{\IZ_3}\Big)^{3,12}$
&&&&$\Big(X^{6,33}\quotient{\IZ_3}\Big)^{4,13}$
&\LARGE\raisebox{-2pt}{$\leftarrow$}
&$\Big(X^{5,59}\quotient{\IZ_3}\Big)^{3,21}$\\
&&& \LARGE$\searrow$ &&& \LARGE$\downarrow$ && \LARGE$\downarrow$ \\[3pt]
\str &&&&$\Big(X^{6,24}\quotient{\IZ_3}\Big)^{4,10}$
&&$\Big(X^{9,27}\quotient{\IZ_3}\Big)^{5,11}$
&\LARGE$\leftarrow$
&$\Big(X^{8,44}\quotient{\IZ_3}\Big)^{4,16}$\\
&&&&\LARGE$\downarrow$&&& \LARGE$\swarrow$ & \LARGE$\downarrow$ \\[3pt]
\str &&&& $\Big(X^{15,15}\quotient{\IZ_3}\Big)^{7,7}$
&\LARGE$\leftarrow$
& $\Big(X^{9,21}\quotient{\IZ_3}\Big)^{5,9}$ && $\Big(X^{19,19}\quotient{\IZ_3}\Big)^{7,7}$\\
\end{tabular}}
\vskip8pt
\framebox[6.5in][l]{\hskip70pt
\begin{tabular}{ccccccc}
\\[-12pt]
\str && $\Big(X^{2,83}\quotient{\IZ_3{\times}\IZ_3}\Big)^{2,11}$ 
&\hspace{10pt}
& $\Big(X^{1,73}\quotient{\IZ_3{\times}\IZ_3}\Big)^{1,9}$\\
&& \thicklines\smash{\begin{picture}(0,40)\put(-5,10){\vector(-1,-2){60}}\end{picture}}
\thicklines\smash{\begin{picture}(0,40)\put(5,10){\vector(2,-3){55}}\end{picture}}
&& \LARGE$\downarrow$\\
\str &&&&$\Big(X^{2,56}\quotient{\IZ_3{\times}\IZ_3}\Big)^{2,8}$\\
&&&& \LARGE$\downarrow$\\
\str &&&& $\Big(X^{3,39}\quotient{\IZ_3{\times}\IZ_3}\Big)^{3,7}$\\
&&&&\phantom{\LARGE$\downarrow$}\\
\str $\Big(X ^{19,19}\quotient{\IZ_3{\times}\IZ_3}\Big)^{3,3}$\\
\end{tabular}
\hskip2pt}
\vskip4pt
\parbox{5.7in}{\caption{\label{Z3andZ3*Z3webs}\small The webs of CICY's obtained, from \tref{Z3quotients} as quotients by freely acting $\IZ_3$ and $\IZ_3{\times}\IZ_3$ symmetries. In the $\IZ_3$ table there are two occurrences of $X^{3,48}/\IZ_3$ owing to the fact that there are distinct quotients by the 
$\IZ_3$-generators $R$ and $S$, see \SS\ref{sec:3,48}.}}
\end{center}
\end{table}
\newcommand{\ddddddarrow}{\hbox{$\begin{matrix}\arrowvert\\[-8pt]\arrowvert\\[-8pt]%
\arrowvert\\[-8pt]\arrowvert\\[-8pt]\arrowvert\\[-8pt]\arrowvert\\[-9pt]\downarrow\end{matrix}$}}

\begin{table}[H] \label{table:H}
\begin{center}

\framebox[6.5in][c]{\hskip3pt
\begin{tabular}{c@{\hspace{-5pt}}ccc@{\hspace{-5pt}}ccc}
\\
\raisebox{-10pt}{\footnotesize$\IP^7[2,\,2,\,2,\,2]^{1,65}\symm{G'}$}\hspace*{-15pt}
&&\footnotesize$\cicy{\IP^1\\ \IP^1\\ \IP^1\\ \IP^1}
{2\\2\\2\\2}^{4,68}\symm{\IH{\times}\IZ_2}$\hspace*{-10pt}
&\LARGE$\rightarrow$\hspace*{20pt}&\footnotesize$\cicy{\IP^1\\ \IP^1\\ \IP^1\\ \IP^1\\ \IP^1}
{1~1\\ 1~1\\ 1~1\\ 1~1\\ 1~1}^{5,45}\symm{\IZ_5{\times}\IZ_2{\times}\IZ_2}$\hspace*{-20pt}
\begin{picture}(-30, 40)
\thicklines\put(-30, -45){\vector(2, -3){83}}
\end{picture}
&&\footnotesize$\cicy{\IP^4\\ \IP^4\\}
{1~1~1~1~1\\ 1~1~1~1~1\\}^{2,52}\symm{\IZ_5{\times}\IZ_2}$
\hspace*{-40pt}\\[30pt]
\hspace*{-5pt}\LARGE$\downarrow$& \LARGE$\swarrow$
&\hspace*{-10pt}\smash{\raisebox{-0.8in}{\LARGE$\ddddddarrow$}}&\hspace*{-45pt} \LARGE$\searrow$
&&&\smash{\raisebox{-0.8in}{\LARGE$\ddddddarrow$}}\hspace*{-15pt}\\[10pt]
\footnotesize$\cicy{\IP^1\\ \IP^1\\ \IP^1\\ \IP^1\\ \IP^7}
{\one~0~0~0~\one~0~0~0 \\
 0~\one~0~0~0~\one~0~0 \\
 0~0~\one~0~0~0~\one~0 \\
 0~0~0~\one~0~0~0~\one \\
 \one~\one~\one~\one~\one~\one~\one~\one}^{5,37}\symm{\IH}$
&&&&\hspace*{-45pt}\footnotesize$\cicy{\IP^1\\ \IP^1\\ \IP^1\\ \IP^1\\ \IP^3}
{2~0~0~0 \\
 0~2~0~0 \\
 0~0~2~0 \\
 0~0~0~2 \\
 \one~\one~\one~\one}^{5,37}\symm{\IH}$
\\[40pt]
&&&&&\\[10pt]
\footnotesize$\cicy{\IP^4\\ \IP^4}{1~2~2~0~0\\ 1~0~0~2~2}^{12,28}\symm{\IH}$
&\LARGE$\rightarrow$
&\footnotesize$\cicy{\IP^1\\ \IP^1\\ \IP^1\\ \IP^1\\ \IP^1}
{1~1\\ 2~0\\ 2~0\\ 0~2\\ 0~2}^{19,19}\symm{\IH}$\hspace*{-10pt}
&&&&\footnotesize$\cicy{\IP^1\\ \IP^1\\ \IP^1\\ \IP^1\\ \IP^1\\ \IP^4\\ \IP^4\\}
{\one ~ 0 ~ 0 ~ 0 ~ 0 ~ \one ~ 0 ~ 0 ~ 0 ~ 0 \\
 0 ~ \one ~ 0 ~ 0 ~ 0 ~ 0 ~ \one ~ 0 ~ 0 ~ 0 \\
 0 ~ 0 ~ \one ~ 0 ~ 0 ~ 0 ~ 0 ~ \one ~ 0 ~ 0 \\
 0 ~ 0 ~ 0 ~ \one ~ 0 ~ 0 ~ 0 ~ 0 ~ \one ~ 0 \\
 0 ~ 0 ~ 0 ~ 0 ~ \one ~ 0 ~ 0 ~ 0 ~ 0 ~ \one \\
 \one ~ \one ~ \one ~ \one ~ \one ~ 0 ~ 0 ~ 0 ~ 0 ~ 0 \\
 0 ~ 0 ~ 0 ~ 0 ~ 0 ~ \one ~ \one ~ \one ~ \one ~ \one \\}^{7,27}\symm{\IZ_{5}{\times}\IZ_2}$
 \hspace*{-10pt}\\
\\
\end{tabular}
\hskip2pt}
\vskip0pt
\parbox{5.5in}{\caption{\label{Hquotients}\small The web of the six CICY's that we consider here that admit a freely acting $\IH$ symmetry, together with some related manifolds that we have met previously.}}
\vskip20pt
\framebox[6.5in][l]{\hskip30pt
\begin{tabular}{ccc@{\hspace*{20pt}}c}
\\
\str $\Big(X^{1,65}\quotient{\IH}\Big)^{1,9}$&& $\Big(X^{4,68}\quotient{\IH}\Big)^{1,9}$&\\
\vspace*{-10pt}
\LARGE$\downarrow$&\LARGE$\swarrow$&\raisebox{-30pt}{\smash{\LARGE$\dddarrow$}}&\hspace{-65pt}\LARGE$\searrow$\vspace*{-10pt}\\
\str $\Big(X^{5,37}\quotient{\IH}\Big)^{2,6}$ &&&$\Big(Y^{5,37}\quotient{\IH}\Big)^{2,6}$\vspace*{15pt}\\
$\Big(X^{12,28}\quotient{\IH}\Big)^{2,4}$&\LARGE$\rightarrow$&$\Big(X^{19,19}\quotient{\IH}\Big)^{3,3}$&\\
&\\

\end{tabular}\hspace{5pt}}
\vskip5pt
\parbox{5.5in}{\caption{\small The web derived from \tref{Hquotients} of manifolds with fundamental group 
$\IH$.}}

\end{center}
\end{table}

\newlength{\myht}
\newlength{\mydp}
\newlength{\mywd}
\newsavebox{\mybox}

\newcommand{\entry}[2]{\settowidth{\mywd}{\footnotesize${}\quotient{#2}$}%
\hspace{\mywd}%
\sbox{\mybox}{\footnotesize$#1\quotient{#2}$}%
\settoheight{\myht}{\usebox{\mybox}}\addtolength{\myht}{6pt}%
\settodepth{\mydp}{\usebox{\mybox}}\addtolength{\mydp}{5pt}%
\vrule height\myht width0pt depth\mydp\usebox{\mybox}}

\newcommand{\simpentry}[1]{%
\sbox{\mybox}{$#1$}%
\settoheight{\myht}{\usebox{\mybox}}\addtolength{\myht}{6pt}%
\settodepth{\mydp}{\usebox{\mybox}}\addtolength{\mydp}{5pt}%
\vrule height\myht width0pt depth\mydp\usebox{\mybox}}

\newpage
\begin{center}

\setlength{\doublerulesep}{3pt}
\def\str{\vrule height15pt width0pt depth9pt}

\begin{longtable}{| c | c | c | c | c |}

\caption{\label{tiptab} The Manifolds of the Tip}\\
\multicolumn{5}{c}{\parbox{6.3in}{\small The manifolds with $y=h^{11}{+}h^{21}\leq 24$ from 
\fref{NewOrbifolds}. In the `Manifold' column $X^{19,19}$ denotes the split bicubic and multiple quotient groups indicates different quotients with the same Hodge numbers. We denote by $\IH$ the quaternion group and the notation $\IP^7[2,2,2,2]^\sharp$ and $\IP^7[2,2,2,2]^{\sharp\sharp}$ denote two different singularizations. The vectors appended to the symmetries of the two weighted CICY's indicate how the generators act. The generator $(\IZ_3:1,2,1,2,0,0,0,0)$, for example, acts by multiplying the first coordinate by $\o$, the second by $\o^2$, etc., with $\o$ a nontrivial cube root of unity. In the entry corresponding to 
$\hodgenos=(6,6)$ we write
$\widehat{\IP^5[3,3]^\sharp}$ to denote a resolution of a conifold of $\IP^5[3,3]$ and ${\bf 1}$ denotes the trivial group.  We also write $\widehat{X^{a,b}/G}$ for the desingularisation of a quotient of 
$X^{a,b}$ by a non-freely acting group $G$.  The column labelled by $\pi_1$ gives the fundamental group.  We only state this explicitly for resolutions of singular quotients; for smooth quotients $\pi_1$ is simply the quotient group.  For each manifold with $\ch<0$, apart possibly from the Tonoli manifold, which is the first entry of the table, there is a mirror which we do not list explicitly. For the Tonoli manifold the mirror is not known to~exist.}}
\vspace{10pt}\\
\hline 
\str $(\chi,\, y)$ & $(h^{11},h^{21})$ & Manifold & $\pi_1$ & Reference \\ 
\hline\hline
\endfirsthead

\multicolumn{5}{c}{\tablename\ \thetable{} \emph{-- Continued from previous page}} \\[1ex]
\hline 
\str $(\chi,\, y)$ & $(h^{11},h^{21})$ & Manifold & $\pi_1$ & Reference \\
\hline\hline 
\endhead

\hline\hline
\multicolumn{5}{|r|}{\str\emph{Continued on the following page}}\\
\hline
\endfoot

\hline\hline
\multicolumn{5}{|c|}{\str}\\
\hline
\endlastfoot
(-44,24) & (1,23) & \simpentry{\text{Degree 17 submanifold of $\IP^6$}} & & \cite{Tonoli}\\
\hline
(-36,24) & (3,21) & \entry{\cicy{\IP^1\\ \IP^1\\ \IP^1\\ \IP^2}{1 ~ 1 \\ 1 ~ 1 \\ 1 ~ 1 \\ 0 ~ 3 \\}}{\IZ_3} & & \SS\ref{sec:5,59} \\
\hline
(-32,24)&(4,20)&\entry{\cicy{\IP^1\\ \IP^1\\ \IP^1\\ \IP^1}{2\\ 2\\ 2\\ 2}}{\IZ_2{\times}\IZ_2} & &
\SS\ref{sec:4,68} \\
\hline
(-16,24)&(8,16)&\entry{\cicy{\IP^4\\ \IP^4\\}
{1~2~2~0~0\\
1~0~0~2~2}}{\IZ_2} & & \SS\ref{sec:12,28} \\
\hline
(0,24) & (12,12) & \simpentry{\widehat{X^{19,19}\quotient{\IZ_2}}}  & $\one$ &
\SS\ref{sec:19,19}\\
\hline
(-44,23) & (10,13) & \simpentry{$\footnotesize$
\IP\!\begin{pmatrix}
0~0~1~1~1~0~0~0~0\\
1~1~0~0~0~1~1~0~0\\
0~0~1~1~0~1~1~1~1\\
3~0~0~1~0~2~2~0~1\end{pmatrix} \begin{bmatrix}3~0\\ 1~3\\ 3~3\\ 3~6\end{bmatrix}}  & & \cite{BatyrevKreuzerConifolds}\\
\hline
(-40,22) & (1,21) & \simpentry{\IP^4[5]/\IZ_5} & & --\\
\hline
 (-36,22)&(2,20)&\entry{\cicy{\IP^2\\ \IP^5\\}{\one~\one~\one~0\\ \one~\one~\one~3\\}}{\IZ_3} & & \SS\ref{sec:2,56} \\
\hline
(-20,22)&(6,16)&\entry{\cicy{\IP^1\\ \IP^1\\ \IP^1\\ \IP^1\\ \IP^1\\ \IP^4\\ \IP^4\\}
{\one~0~0~0~0~\one~0~0~0~0\\ 0~\one~0~0~0~0~\one~0~0~0\\ 0~0~\one~0~0~0~0~\one~0~0\\ 0~0~0~\one~0~0~0~0~\one~0\\ 0~0~0~0~\one~0~0~0~0~\one\\ \one~\one~\one~\one~\one~0~0~0~0~0\\ 0~0~0~0~0~\one~\one~\one~\one~\one}}{\IZ_2} & &
\SS\ref{sec:7,27} \\
\hline
(-12,22)&(8,14)&\simpentry{$\footnotesize$
\IP\!\begin{pmatrix} 4\,2\,2\,2\,1\,1\,0\,0\\ 0\,0\,0\,0\,1\,1\,2\,0\\ 1\,0\,0\,0\,0\,0\,0\,1 \end{pmatrix}\hskip-5pt\begin{bmatrix} 8~4 \cr 0~4 \cr 2~0 \end{bmatrix}\quotient{(\IZ_2:1,1,0,0,1,0,1,0)}$$} &
&\cite{KreuzerRieglerSahakyan,KlemmKreuzerRieglerScheidegger}\\
\hline
(0,22) &(11,11) &\entry{\cicy{\IP^1\\ \IP^2\\ \IP^2}{1~1\\ 3~0\\ 0~3}}{\IZ_2}
& &\cite{BouchardDonagi}\\
\hline
(-30,21)&(3,18)&\entry{\cicy{\IP^1\\ \IP^1\\ \IP^1\\ \IP^5\\}
{\one~\one~0~0~0\\ \one~0~\one~0~0\\ \one~0~0~\one~0\\ 0~\one~\one~\one~3}}{\IZ_3} & & \SS\ref{sec:5,50} \\
\hline
(-30,21)&(3,18)&\entry{\cicy{\IP^2\\ \IP^2\\ \IP^2\\}
{\one~\one~\one\\ \one~\one~\one\\ \one~\one~\one\\}}{\IZ_3} & & \SS\ref{sec:3,48} \\
\hline
(-36,20)&(2,18)&\entry{\cicy{\IP^1\\ \IP^1\\ \IP^1\\ \IP^1}{2\\ 2\\ 2\\ 2}}{\IZ_4} & & 
\SS\ref{sec:4,68} \\
\hline
(-24,20)& (4,16)&\simpentry{$\footnotesize
$\IP\begin{pmatrix}1\,1\,1\,1\,1\,1\,0\\ 1\,1\,0\,0\,0\,0\,1\\ \end{pmatrix}\hskip-3pt
\begin{bmatrix} 3~3\\3~0\\ \end{bmatrix}\quotient{(\IZ_3:1,2,1,2,0,0,0,0)}$$} &
&\cite{KreuzerRieglerSahakyan,KlemmKreuzerRieglerScheidegger}\\
\hline
(-24,20) & (4,16) &\entry{\cicy{\IP^1\\ \IP^1\\ \IP^1\\ \IP^1\\ \IP^1\\ \IP^1\\}
{0~\one~\one\\ 0~\one~\one\\ 0~\one~\one\\ \one~0~\one\\ \one~0~\one\\ \one~0~\one\\}}{\IZ_3} & & \SS\ref{sec:8,44} \\
\hline
(-20,20) & (5,15) &\entry{\cicy{\IP^1\\ \IP^1\\ \IP^1\\ \IP^1\\ \IP^1\\}
{\one~\one\\ \one~\one\\ \one~\one\\ \one~\one\\ \one~\one\\}}{\IZ_2{\times}\IZ_2} & & \SS\ref{sec:5,45} \\
\hline
(0,20) & (10,10) & \simpentry{\widehat{\IP^7[2,2,2,2]^{\sharp\sharp}}} &
&\cite{GrossPopescu,Hua}\\
\hline
(-24,18)&(3,15)&\simpentry{\widehat{\IP^5[3,3]\quotient{\IZ_3{\times}\IZ_2}}}
&$\IZ_3$ & \SS\ref{sec:1,73} \\
\hline
(-24,18)&(3,15)&\entry{\cicy{\IP^2\\ \IP^2\\ \IP^5\\}{\one~\one~\one~0~0~0\\ 0~0~0~\one~\one~\one\\ \one~\one~\one~\one~\one~\one}}{\IZ_3}& 
&\SS\ref{sec:3,39} \\
\hline
(0,18) & (9,9)&\entry{\cicy{\IP^1\\ \IP^1\\ \IP^1\\ \IP^1\\ \IP^1\\ \IP^1\\ \IP^1\\ \IP^1\\ \IP^1\\}
{\one~0~0~\one~0~0 \\
 0~\one~0~\one~0~0 \\
 0~0~\one~\one~0~0 \\
 \one~0~0~0~\one~0 \\
 0~\one~0~0~\one~0 \\
 0~0~\one~0~\one~0 \\
 \one~0~0~0~0~\one \\
 0~\one~0~0~0~\one \\
 0~0~\one~0~0~\one}}{\IZ_2} & & \SS\ref{sec:15,15} \\
\hline
(-30,17)&(1,16)&\entry{\cicy{\IP^2\\ \IP^2\\ \IP^2\\}
{\one~\one~\one\\ \one~\one~\one\\ \one~\one~\one\\}}{\IZ_3} & & \SS\ref{sec:3,48}\\
\hline
(-18,17)&(4,13)&\entry{\cicy{\IP^2\\ \IP^3\\}{3~0\\ 1~3}}{\IZ_3} & & \cite{Schimmrigk} \\
\hline
(-18,17)&(4,13)&\entry{\cicy{\IP^1\\ \IP^1\\ \IP^1\\ \IP^2 \\ \IP^5\\}{\one~\one~0~0~0~0~0\\ \one~0~\one~0~0~0~0\\ \one~0~0~\one~0~0~0\\ 0~0~0~0~\one~\one~\one\\ 0~\one~\one~\one~\one~\one~\one}}{\IZ_3} & & \SS\ref{sec:6,33} \\
\hline
(-18,17)&(4,13)&\entry{\cicy{\IP^1 \\ \IP^1 \\ \IP^1 \\ \IP^2 \\ \IP^2}
{\one ~ \one ~ 0 ~ 0 \\
\one ~ 0 ~ \one ~ 0 \\
\one ~ 0 ~ 0 ~ \one \\
0 ~ \one ~ \one ~ \one \\
0 ~ \one ~ \one ~ \one \\}}{\IZ_3} & & \SS\ref{sec:Y6,33} \\
\hline
(-24,16)&(2,14)&\entry{\cicy{\IP^2\\ \IP^2\\ \IP^2\\ \IP^2\\}
{\one ~ 0 ~ 0 ~ \one ~ \one \\
0 ~ \one ~ 0 ~ \one ~ \one \\
0 ~ 0 ~ \one ~ \one ~ \one \\
\one ~ \one ~ \one ~ 0 ~ 0}}{\IZ_3} & & \SS\ref{sec:4,40} \\
\hline
(-12,16)&(5,11)&\entry{\cicy{\IP^1\\ \IP^1\\ \IP^1\\ \IP^1\\ \IP^1\\ \IP^1\\ \IP^5\\}
{0~\one~\one~0~0~0~0~0\\0~\one~0~\one~0~0~0~0\\ 0~\one~0~0~\one~0~0~0\\ 
\one~0~0~0~0~\one~0~0\\ \one~0~0~0~0~0~\one~0\\ \one~0~0~0~0~0~0~\one\\
0~0~\one~\one~\one~\one~\one~\one\\}}{\IZ_3} & & \SS\ref{sec:9,27} \\
\hline
(-8,16)&(6,10)&\entry{\cicy{\IP^4\\ \IP^4\\}
{1~2~2~0~0\\
1~0~0~2~2}}{\IZ_2{\times}\IZ_2} & & \SS\ref{sec:12,28} \\
\hline
(0,16) & (8,8) & \simpentry{\widehat{X^{19,19}\quotient{\IZ_2{\times}\IZ_2}}}  & $\IZ_2$ 
& \SS\ref{sec:19,19}\\
\hline
(0,16) & (8,8) & \simpentry{\widehat{\IP(1\,1\,1\,1\,4)[8]^\sharp}}  & &\cite{GrossPopescu}\\
\hline
(4,16) & (9,7) & \simpentry{\widehat{X^{5,59}\quotient{\IZ_3{\times}\IZ_3}}}  & $\IZ_3$ & \SS\ref{sec:5,59}\\
\hline
(8,16) & (10,6) & \simpentry{\widehat{X^{8,44}\quotient{\IZ_3{\times}\IZ_3}}}  & $\IZ_3$  & \SS\ref{sec:8,44}\\
\hline
(-18,15)&(3,12)&\entry{\cicy{\IP^2\\ \IP^2\\ \IP^2\\ \IP^2\\ \IP^2 \\}
{\one ~ 0 ~ 0 ~ \one ~ 0 ~ 0 ~ \one \\
0 ~ \one ~ 0 ~ 0 ~ \one ~ 0 ~ \one \\
0 ~ 0 ~ \one ~ 0 ~ 0 ~ \one ~ \one \\
\one ~ \one ~ \one ~ 0 ~ 0 ~ 0 ~ 0 \\
0 ~ 0 ~ 0 ~ \one ~ \one ~ \one ~ 0}}{\IZ_3} & & \SS\ref{sec:5,32} \\
\hline
(-10,15) & (5,10)   &\simpentry{\widehat{X^{3,48}\quotient{\IZ_3{\times}\IZ_3}}}& \one & \SS\ref{sec:3,48} \\
\hline
(-6,15) & (6,9)   &\entry{\cicy{\IP^3\\ \IP^3\\}{1~3~0\\ 1~0~3\\}}{\IZ_3}& & \cite{Yau} \\
\hline
(10,15) & (10,5) &\simpentry{\widehat{X^{6,33}\quotient{\IZ_3{\times}\IZ_3}}} & $\IZ_3$ & \SS\ref{sec:6,33} \\
\hline
(-20,14)&(2,12) &\entry{\cicy{\IP^4\\ \IP^4\\}{\one~\one~\one~\one~\one\\ \one~\one~\one~\one~\one}}{\IZ_5} & &\SS\ref{sec:2,52} \\
\hline
(-16,14)&(3,11) &\entry{\cicy{\IP^1 \\ \IP^1 \\ \IP^1 \\ \IP^1 \\ \IP^7}
{\one ~ 0 ~ 0 ~ 0 ~ \one ~ 0 ~ 0 ~ 0 \\
0 ~ \one ~ 0 ~ 0 ~ 0 ~ \one ~ 0 ~ 0 \\
0 ~ 0 ~ \one ~ 0 ~ 0 ~ 0  ~ \one ~ 0 \\
0 ~ 0 ~ 0 ~ \one ~ 0 ~ 0  ~ 0 ~ \one \\
\one ~ \one ~ \one ~ \one ~ \one ~ \one ~ \one ~ \one}}{\IZ_4} & &\SS\ref{sec:5,37}\\
\hline
(-16,14)&(3,11) &\entry{\cicy{\IP^1 \\ \IP^1 \\ \IP^1 \\ \IP^1 \\ \IP^3}
{2 ~ 0 ~ 0 ~ 0 \\
0 ~ 2 ~ 0 ~ 0 \\
0 ~ 0 ~ 2 ~ 0 \\
0 ~ 0 ~ 0 ~ 2 \\
\one ~ \one ~ \one ~ \one}}{\IZ_4} & &\SS\ref{sec:5,37Y}\\
\hline
(-12,14)&(4,10) & \entry{\cicy{ \IP^2\\ \IP^2\\ \IP^2\\ \IP^2\\ \IP^2\\ \IP^2\\}
{\one~0~0~\one~0~0~\one~0~0 \\
 0~\one~0~0~\one~0~0~\one~0 \\
 0~0~\one~0~0~\one~0~0~\one \\
 \one~\one~\one~0~0~0~0~0~0 \\
 0~0~0~\one~\one~\one~0~0~0 \\
 0~0~0~0~0~0~\one~\one~\one}}{\IZ_3} & & \SS\ref{sec:6,24} \\
\hline
(-12,14)&(4,10) &\simpentry{\widehat{X^{3,39}\quotient{\IZ_3{\times}\IZ_2}}} & $\IZ_3$ & \SS\ref{sec:3,39}\\
\hline
(-8,14)&(5,9) &\entry{\cicy{\IP^1 \\ \IP^1 \\ \IP^1 \\ \IP^1 \\ \IP^1 \\ \IP^1 \\ \IP^2}
{\one ~ 0 ~ \one ~ 0 ~ 0 \\
\one ~ 0 ~ 0 ~ \one ~ 0 \\
\one ~ 0 ~ 0 ~ 0 ~ \one \\
0 ~ \one ~ \one ~ 0 ~ 0 \\
0 ~ \one ~ 0 ~ \one ~ 0 \\
0 ~ \one ~ 0 ~ 0 ~ \one \\
0 ~ 0 ~ \one ~ \one ~ \one \\}}{\IZ_3} &  & \SS\ref{sec:9,21}\\
\hline
(-4,14)&(6,8) &\simpentry{\widehat{X^{6,24}\quotient{\IZ_3{\times}\IZ_3}} }& \one & \SS\ref{sec:6,24} \\
\hline
(0,14) & (7,7) &\entry{\cicy{\IP^1\\ \IP^2\\ \IP^2}{1~1\\ 3~0\\ 0~3}}{\{\IZ_3,\,\IZ_2{\times}\IZ_2\}} & & \cite{BouchardDonagi}\\
\hline
(0,14) & (7,7)&\entry{\cicy{\IP^1\\ \IP^1\\ \IP^1\\ \IP^1\\ \IP^1\\ \IP^1\\ \IP^1\\ \IP^1\\ \IP^1\\}
{\one~0~0~\one~0~0 \\
 0~\one~0~\one~0~0 \\
 0~0~\one~\one~0~0 \\
 \one~0~0~0~\one~0 \\
 0~\one~0~0~\one~0 \\
 0~0~\one~0~\one~0 \\
 \one~0~0~0~0~\one \\
 0~\one~0~0~0~\one \\
 0~0~\one~0~0~\one}}{\IZ_3} & & \SS\ref{sec:15,15} \\
\hline
(0,14) & (7,7) & \simpentry{\widehat{X^{15,15}\quotient{\IZ_3{\times}\IZ_3}}} & \one & \SS\ref{sec:15,15}\\
\hline
(-18,13) & (2,11) &\entry{\cicy{\IP^2\\ \IP^2\\}{3\\ 3\\}}{\IZ_3{\times}\IZ_3}  & & \cite{Triadophilia}\\
\hline
(-6,13) & (5,8) &\simpentry{\widehat{X^{9,27}\quotient{\IZ_3{\times}\IZ_2}}} & $\IZ_3$ & 
\SS\ref{sec:3gen} \\
\hline
(-16,12)&(2,10)&\entry{\cicy{\IP^1\\ \IP^1\\ \IP^1\\ \IP^1}{2\\ 2\\ 2\\ 2}}{\IZ_4{\times}\IZ_2} & & 
\SS\ref{sec:4,68} \\
\hline
(-8,12)&(4,8)&\entry{\cicy{\IP^4\\ \IP^4\\}
{1~2~2~0~0\\
1~0~0~2~2}}{\IZ_4} & & \SS\ref{sec:12,28} \\
\hline
(0,12) & (6,6) & \simpentry{\widehat{X^{19,19}\quotient{\IZ_3{\times}\IZ_2}}}  & $\IZ_3$ & 
\SS\ref{sec:19,19}\\
\hline
(0,12) & (6,6)  &\simpentry{\widehat{\IP^5[3,3]^\sharp}/G~,~~G\subset \IZ_6{\times}\IZ_6} &                
& \cite{GrossPopescu}\\
\hline
(-16,10) & (1,9) &\entry{
\cicy{\IP^1\\ \IP^1\\ \IP^1\\ \IP^1\\ \IP^1\\}{\one~\one\\ \one~\one\\ \one~\one\\ \one~\one\\ \one~\one\\}}{\IZ_5} & & \SS\ref{sec:5,45} \\
\hline
(-16,10)& (1,9) & \simpentry{\IP_5[3,3]\quotient{\IZ_3{\times}\IZ_3}} & & \SS\ref{sec:1,73}\\
\hline
(-16,10)& (1,9) 
&\simpentry{\IP^7[2,2,2,2]/\{\IH,\IZ_8,\IZ_4{\times}\IZ_2,\IZ_2{\times}\IZ_2{\times}\IZ_2\}} & 
&\cite{Strominger:1985it,Hua,Beauville,HjPark}\\
\hline
(-16,10) & (1,9) &\entry{
\cicy{\IP^1\\ \IP^1\\ \IP^1\\ \IP^1}{2\\ 2\\ 2\\ 2}}{\IH} & & \SS\ref{sec:4,68} \\
\hline
(-12,10)&(2,8)&\entry{\cicy{\IP^1\\ \IP^1\\ \IP^1\\ \IP^1\\ \IP^1\\ \IP^4\\}
{\one~\one~0~0~0~0\\ \one~0~\one~0~0~0\\ \one~0~0~\one~0~0\\ \one~0~0~0~\one~0\\
\one~0~0~0~0~\one\\ 0~\one~\one~\one~\one~\one}}{\IZ_5} &
&\SS\ref{sec:6,36} \\
\hline
 (-12,10)&(2,8)&\entry{\cicy{\IP^2\\ \IP^5\\}{\one~\one~\one~0\\ \one~\one~\one~3\\}}{\IZ_3{\times}\IZ_3} & & \SS\ref{sec:2,56} \\
\hline
(-12,10) & (2,8) &\entry{\cicy{\IP^1\\ \IP^1\\ \IP^1\\ \IP^1\\ \IP^1\\ \IP^1\\}{0~\one~\one\\ 0~\one~\one\\ 0~\one~\one\\ \one~0~\one\\ \one~0~\one\\ \one~0~\one\\}}{\IZ_3{\times}\IZ_2} &
& \SS\ref{sec:8,44} \\
\hline
(-8,10)&(3,7)&\entry{\cicy{\IP^1\\ \IP^1\\ \IP^1\\ \IP^1\\ \IP^1\\ \IP^4\\ \IP^4\\}
{\one~0~0~0~0~\one~0~0~0~0\\ 0~\one~0~0~0~0~\one~0~0~0\\ 0~0~\one~0~0~0~0~\one~0~0\\ 0~0~0~\one~0~0~0~0~\one~0\\
0~0~0~0~\one~0~0~0~0~\one\\ \one~\one~\one~\one~\one~0~0~0~0~0\\ 0~0~0~0~0~\one~\one~\one~\one~\one}}{\IZ_5} &
&\SS\ref{sec:7,27}\\
\hline
(-8,10)&(3,7)&\simpentry{\widehat{\IP^5[3,3]\quotient{\IZ_3{\times}\IZ_3{\times}\IZ_2}}}
&$\IZ_3{\times}\IZ_3$ & \SS\ref{sec:1,73} \\
\hline
(-8,10)&(3,7)&\entry{\cicy{\IP^2\\ \IP^2\\ \IP^5\\}{\one~\one~\one~0~0~0\\ 0~0~0~\one~\one~\one\\ \one~\one~\one~\one~\one~\one}}{\IZ_3{\times}\IZ_3} &
& \SS\ref{sec:3,39} \\
\hline
(0,10) & (5,5) &\entry{\cicy{\IP^1\\ \IP^2\\ \IP^2}{1~1\\ 3~0\\ 0~3}}{\IZ_4} & &
\cite{BouchardDonagi}, \SS\ref{sec:19,19H}\\
\hline
(0,10) & (5,5) & \simpentry{\widehat{X^{15,15}\quotient{\IZ_3{\times}\IZ_3{\times}\IZ_2}}} & $\IZ_2$ & \SS\ref{sec:15,15}\\
\hline
(-14,9) & (1,8) &\simpentry{\{\text{Resoln.\ of a Pfaffian CY manifold}\}/\IZ_7} & &\cite{Rodland}\\
\hline
(0,8) & (4,4) & \simpentry{\widehat{X^{19,19}\quotient{\IZ_3{\times}\IZ_2{\times}\IZ_2}}}  & $\IZ_3{\times}\IZ_2$ & \SS\ref{sec:19,19}\\
\hline
(0,8) & (4,4) & \simpentry{\text{Resoln.\ of a Horrocks-Mumford quintic}}  & & \cite[3.2]{GrossPopescu}\\
\hline
(-8,8)&(2,6) &\entry{\cicy{\IP^1 \\ \IP^1 \\ \IP^1 \\ \IP^1 \\ \IP^7}
{\one ~ 0 ~ 0 ~ 0 ~ \one ~ 0 ~ 0 ~ 0 \\
0 ~ \one ~ 0 ~ 0 ~ 0 ~ \one ~ 0 ~ 0 \\
0 ~ 0 ~ \one ~ 0 ~ 0 ~ 0  ~ \one ~ 0 \\
0 ~ 0 ~ 0 ~ \one ~ 0 ~ 0  ~ 0 ~ \one \\
\one ~ \one ~ \one ~ \one ~ \one ~ \one ~ \one ~ \one}}{\IH} & &\SS\ref{sec:5,37}\\
\hline
(-8,8)&(2,6) &\entry{\cicy{\IP^1 \\ \IP^1 \\ \IP^1 \\ \IP^1 \\ \IP^3}
{2 ~ 0 ~ 0 ~ 0 \\
0 ~ 2 ~ 0 ~ 0 \\
0 ~ 0 ~ 2 ~ 0 \\
0 ~ 0 ~ 0 ~ 2 \\
\one ~ \one ~ \one ~ \one}}{\IH} & &\SS\ref{sec:5,37Y}\\
\hline
(-10,7)&(1,6)&\entry{\cicy{\IP^4\\ \IP^4}{\one~\one~\one~\one~\one\\ \one~\one~\one~\one~\one}}{\IZ_5{\times}\IZ_2} & & \SS\ref{sec:2,52} \\
\hline
(-8,6) & (1,5) &\simpentry{\IP^4[5]/\IZ_5{\times}\IZ_5}  & & --\\
\hline
(-8,6) & (1,5)       
& \entry{\cicy{\IP^1\\ \IP^1\\ \IP^1\\ \IP^1\\ \IP^1\\}{\one~\one\\ \one~\one\\ \one~\one\\ \one~\one\\ \one~\one\\}}
{\IZ_5{\times}\IZ_2} & & \SS\ref{sec:5,45} \\
\hline
(-8,6) & (1,5) &\simpentry{\IP^7[2,2,2,2]/G~,~~~|G|=16} & &\cite{Strominger:1985it,BorisovHua,Hua}\\
\hline
(-8,6)&(1,5)&\entry{\cicy{\IP^1\\ \IP^1\\ \IP^1\\ \IP^1}{2\\ 2\\ 2\\ 2}}{\IH{\times}\IZ_2} & & 
\SS\ref{sec:4,68} \\
\hline
(-4,6)&(2,4)
&\entry{\cicy{\IP^1\\ \IP^1\\ \IP^1\\ \IP^1\\ \IP^1\\ \IP^4\\ \IP^4\\}
{\one~0~0~0~0~\one~0~0~0~0\\ 0~\one~0~0~0~0~\one~0~0~0\\ 0~0~\one~0~0~0~0~\one~0~0\\ 0~0~0~\one~0~0~0~0~\one~0\\ 0~0~0~0~\one~0~0~0~0~\one\\ \one~\one~\one~\one~\one~0~0~0~0~0\\ 0~0~0~0~0~\one~\one~\one~\one~\one}}{\IZ_5{\times}\IZ_2} &
&\SS\ref{sec:7,27} \\
\hline
(-4,6)&(2,4)&\entry{\cicy{\IP^4\\ \IP^4\\}
{1~2~2~0~0\\
1~0~0~2~2}}{\IH} & & \SS\ref{sec:12,28} \\
\hline
(0,6) & (3,3)  
&\simpentry{$\footnotesize$\cicy{\IP^1\\ \IP^2\\ \IP^2}{1~1\\ 3~0\\ 0~3}%
\quotient{\{\IZ_3{\times}\IZ_3,\,\IH,\, \IZ_4{\times}\IZ_2,\, \IZ_3{\times}\IZ_2,\, \IZ_5\}}$$} &              
&  \cite{BouchardDonagi}, \SS\ref{sec:19,19H}\\
\hline
(0,6) & (3,3)&\entry{\cicy{\IP^1\\ \IP^1\\ \IP^1\\ \IP^1\\ \IP^1\\ \IP^1\\ \IP^1\\ \IP^1\\ \IP^1\\}
{\one~0~0~\one~0~0 \\
 0~\one~0~\one~0~0 \\
 0~0~\one~\one~0~0 \\
 \one~0~0~0~\one~0 \\
 0~\one~0~0~\one~0 \\
 0~0~\one~0~\one~0 \\
 \one~0~0~0~0~\one \\
 0~\one~0~0~0~\one \\
 0~0~\one~0~0~\one}}{\IZ_3{\times}\IZ_2} & & \SS\ref{sec:15,15}\\
\hline
(-4,4) & (1,3)       
& \entry{\cicy{\IP^1\\ \IP^1\\ \IP^1\\ \IP^1\\ \IP^1\\}
{\one~\one\\ \one~\one\\ \one~\one\\ \one~\one\\ \one~\one\\}}{\IZ_5{\times}\IZ_2{\times}\IZ_2} &
&\SS\ref{sec:5,45} \\
\hline
(-4,4) & (1,3) &\simpentry{\IP^7[2,2,2,2]/G~,~~~|G|=32} & &\cite{Strominger:1985it,BorisovHua,Hua}\\
\hline
(0,4) & (2,2)  &\simpentry{\widehat{\IP^7[2,2,2,2]^\sharp}/G~,~~~|G|~\text{divides}~64} &
&\cite%
{GrossPopescu,BorisovHua,Hua}\\
\hline
(0,4) & (2,2)       &\simpentry{\text{Resoln.\ of Pfaffian CY w.\ 49 nodes}}& & \cite{Rodland}\\
\hline
\end{longtable}
\end{center}
\newpage
\section{Manifolds admitting free actions by $\IZ_5$}
\subsection{The sequence $\IP^4[5]\to X^{2,52}\to X^{7,27}\leftarrow X^{6,36}
\leftarrow X^{5,45}$}
\subsubsection{$X^{2,52}$; a first split of the quintic threefold} \label{sec:2,52}
The quintic family $\IP^4[5]$ contains manifolds that admit a free action by the group $\IZ_5$. Seeking manifolds that are related to $\IP^4[5]$ and which maintain the symmetry it is natural to consider the splitting
\beq
X^{2,52}~=~\cicy{\IP^4\\ \IP^4\\}{1&1&1&1&1\\ 1&1&1&1&1\\}^{2,52}_{-100}
\hskip0.75in
\lower0.5in\hbox{\includegraphics[width=0.75in]{fig_2,52_manifold.pdf}}
\label{(2,52)manifold}
\eeq
We denote the coordinates of the two $\IP^4$'s by $x_j$ and $y_k$ and the five polynomials by $p_i$. The polynomials are bilinear in the coordinates of the two $\IP^4$'s so we may write
\beq
p_i(x,y)~=~\sum_{j,k}A_{ijk}\,x_j y_k
\label{splitquintic}\eeq
with the $A_{ijk}$ constant coefficients. In the present context it is convenient to understand the indices as taking values in $\IZ_5$. Let us choose to make the equations covariant under a generator $S$ by requiring
$$
S:~x_i\to x_{i+1}~;~~y_i\to y_{i+1} ~;~~p_i\to p_{i+1}~.
$$
This requires
$$
A_{ijk}~=~A_{i-1,\,j-1,\,k-1}~~~\text{hence}~~~A_{ijk}~=~A_{0,\,j-i,\,k-i}~.
$$
If we write $a_{jk}$ in place of $A_{0jk}$ and change indices of summation our polynomials take the form
\beq
p_i(x,y)~=~\sum_{j,k}a_{jk}\, x_{j+i}\, y_{k+i}~.
\label{(2,52)polys}
\eeq
The coordinates of the points fixed by $S$ in $\IP^4{\times}\IP^4$ are of the form $x_j=\z^j$ and 
$y_j=\tilde \z^j$ with $\z^5=\tilde \z^5=1$, and it is easy to see that, for general choice of the coefficients $a_{jk}$, none of the 
$p_i$ vanish. We have also checked that the polynomials are transverse, and thus the quotient variety 
$X^{2,52}/S$ is smooth. The Hodge numbers for the quotient follow from the fact that taking the quotient by $S$ does not change $h^{11}$ while the new Euler number is $-100/5=-20$. 

If the coefficient matrix $a_{jk}$ is taken to be symmetric then the polynomials \eref{(2,52)polys} are also invariant under a $\IZ_2$ generator
$$
U:~x_j~\leftrightarrow~y_j~;~~p_i~\to~p_i~.
$$
The fixed points of $U$ are such that $y_k=x_k$ and $p_i(x,x)=0$. The latter equations are 5 quadratic equations acting in a $\IP^4$ which we expect to have no common solution for a generic symmetric matrix $a$. It is easy to check that this is, in fact, the case by means of a Groebner basis calculation. 
For the $U$-quotients $X^{2,52}/U$ and $X^{2,52}/S{\times}U$ we have $h^{11}=1$, since $U$ identifies the two $\IP^4$'s. The value of $h^{2,1}$ follows most simply from the fact that the Euler number divides by the order of the group. Thus we have shown the existence of the quotients of the following table:
\vskip10pt
\begin{table}[H]
\begin{center}
\def\str{\vrule height16pt width0pt depth8pt}
\begin{tabular}{| c | c | c | c |}
\hline 
\str $\hodgenos\left(X^{2,52}/G\right)$ & ~~$(1,6)$~~ & ~~$(2,12)$~~ & ~~$(1,26)$~~ \\
\hline
\str $G$ & ~$\IZ_5{\times}\IZ_2$~ & $\IZ_5$ & $\IZ_2$ \\
\hline
\end{tabular}
\parbox{5.5in}{\caption{\small The Hodge numbers of smooth quotients of $X^{2,52}$.}}
\end{center}
\end{table}
\subsubsection{$X^{7,27}$; a further split of the quintic} \label{sec:7,27}
We can perform a further split of the quintic by splitting $X^{2,52}$ to obtain the configuration
$$
X^{7,27}~=~~
\cicy{\IP^1\\ \IP^1\\ \IP^1\\ \IP^1\\ \IP^1\\ \IP^4\\ \IP^4\\}
{\one & 0 & 0 & 0 & 0 & \one & 0 & 0 & 0 & 0 \\
 0 & \one & 0 & 0 & 0 & 0 & \one & 0 & 0 & 0 \\
 0 & 0 & \one & 0 & 0 & 0 & 0 & \one & 0 & 0 \\
 0 & 0 & 0 & \one & 0 & 0 & 0 & 0 & \one & 0 \\
 0 & 0 & 0 & 0 & \one & 0 & 0 & 0 & 0 & \one \\
 \one & \one & \one & \one & \one & 0 & 0 & 0 & 0 & 0 \\
 0 & 0 & 0 & 0 & 0 & \one & \one & \one & \one & \one \\}^{7,27}_{-40}
\hskip0.75in
\lower0.55in\hbox{\includegraphics[width=2.0in]{fig_7,27_manifold.pdf}}
$$
The structure of the matrix and diagram suggests the possibility of a free $\IZ_5{\times}\IZ_2$ action, and we will show that such a symmetry does in fact exist.  We again take coordinates $(x_i,y_j), i,j \in \IZ_5$ for $\IP^4{\times}\IP^4$, along with coordinates $t_{ia}~,\,i\in \IZ_5,\, a\in \IZ_2$ for the five $\IP^1$'s, so that the defining polynomials are
\beq
\begin{split}
p_i &= \sum_j x_j(A_{ij}\, t_{i0} + B_{ij}\, t_{i1})\\[3pt]
q_i &= \sum_j y_j(C_{ij}\, t_{i0} + D_{ij}\, t_{i1}),
\end{split}
\notag\eeq
We now impose a $\IZ_5{\times}\IZ_2$ symmetry generated by
\beq
\begin{split}
S&:~x_i \to  x_{i+1}~,~~ y_i \to y_{i+1}~,~~ t_{ia} \to t_{i+1,a}~;~ ~
p_i \to p_{i+1}~;~~ q_i \to q_{i+1}\\[3pt]
U&:~x_i \leftrightarrow y_i~,~~t_{i0} \leftrightarrow t_{i1}~;~~p_i \leftrightarrow q_i~.
\end{split}
\notag\eeq
The most general form of the polynomials covariant under these transformations is
\beq
\begin{split}
p_i &= \sum_j x_j(A_{i-j}\, t_{i0} + B_{i-j}\, t_{i1})\\[3pt]
q_i &= \sum_j y_j(B_{i-j}\, t_{i0} + A_{i-j}\, t_{i1}),
\end{split}
\notag\eeq
In order to show that the action is free it suffices to check that both $S$ and $U$ act without fixed points.
A fixed point of $S$  is of the form $x_i = \z^i,~y_i=\tilde\z^i$, where $\z^5 = \tilde\z^5 = 1$, and 
$t_{ia} = t_a$.  At these points, the polynomials are given by
\beq
\begin{split}
p_i&= \z^i \sum_k \z^{-k} (A_k\, t_ 0 + B_k\, t_ 1)\\[3pt]
q_i&= \tilde\z^i \sum_k \tilde\z^{-k} (B_k\, t_ 0 + A_k\, t_ 1)
\end{split}
\notag\eeq
This system reduces to a pair of equations for $(t_ 0, t_ 1) \in \IP^1$, and for general coefficients will have no solution.

A fixed point of $U$ is given by $y_i = x_i$ and $(t_{i0},t_{i1}) = (1,\pm 1)$.  The polynomials then become
\beq
p_i = \sum_j (A_{i-j} \pm B_{i-j} )\, x_j\, t_{i0}, \quad q_i = \pm p_i
\notag\eeq
For general coefficients, regardless of the choice of the signs, the equations $p_i = 0$ place five independent linear constraints on the $x_j$, and therefore have no non-trivial solutions.  We have established that $\IZ_5{\times}\IZ_2$ acts on the manifold without fixed points.  We can use our freedom to change coordinates to isolate the independent coefficients in the polynomials.  The actions of $S$ and $U$ are preserved by any coordinate change of the form
\beq
\begin{split}
x_i \to \sum_k \g_k\, x_{i+k} ~;~~ y_i \to \sum_k \g_k\, y_{i+k}\\[3pt]
(t_{i0},\, t_{i1}) \to (\a\, t_{i0} + \b\, t_{i1}, \b\, t_{i0} + \a\, t_{i1})
\end{split}
\notag\eeq
We can use a transformation of the $x$'s and $y$'s to set $A_i = A_0 \delta_{i0}$.  Then we can transform the $t$'s to enforce $A_0 = B_0$, and absorb $B_0$ into the normalisation of the polynomials.  This leaves us with
\beq
\begin{split}
p_i~&=~ x_i\, t_{i0} + \sum_{j\neq i} B_{i-j}\, x_j\, t_{i1}\\[3pt]
q_i~&= \sum_{j\neq i}B_{i-j}\, y_j \, t_{i0} + y_i\, t_{i1}
\end{split}
\notag\eeq
We have checked that these polynomials are transverse.  The Euler number of the quotient manifold is 
$\chi = -40/10 = -4$, and the group action identifies the five $\IP^1$'s as well as the two $\IP^4$'s, leaving 
$h^{11} = 2$.  Thus $h^{21} = 4$, which agrees with the number of parameters in the polynomials.  In summary we have confirmed the existence of a manifold
$$
\cicy{\IP^1\\ \IP^1\\ \IP^1\\ \IP^1\\ \IP^1\\ \IP^4\\ \IP^4\\}
{\one & 0 & 0 & 0 & 0 & \one & 0 & 0 & 0 & 0 \\
 0 & \one & 0 & 0 & 0 & 0 & \one & 0 & 0 & 0 \\
 0 & 0 & \one & 0 & 0 & 0 & 0 & \one & 0 & 0 \\
 0 & 0 & 0 & \one & 0 & 0 & 0 & 0 & \one & 0 \\
 0 & 0 & 0 & 0 & \one & 0 & 0 & 0 & 0 & \one \\
 \one & \one & \one & \one & \one & 0 & 0 & 0 & 0 & 0 \\
 0 & 0 & 0 & 0 & 0 & \one & \one & \one & \one & \one \\}^{2,4}\quotient{\IZ_5\times\IZ_2}
$$
and hence also of the two- and five-fold covers obtained by taking the separate $\IZ_5$ and $\IZ_2$ quotients. We record the Hodge numbers in the following short table.
\vskip10pt
\begin{table}[H]
\begin{center}
\def\str{\vrule height16pt width0pt depth8pt}
\begin{tabular}{| c | c | c | c |}
\hline 
\str $\hodgenos\left(X^{7,27}/G\right)$ & ~~$(2,4)$~~ & ~~$(3,7)$~~ & ~~$(6,16)$~~ \\
\hline
\str $G$ & ~$\IZ_5{\times}\IZ_2$~ & $\IZ_5$ & $\IZ_2$ \\
\hline
\end{tabular}
\parbox{5.5in}{\caption{\small Hodge numbers of quotients of $X^{7,27}$.}}
\end{center}
\end{table}
\vskip10pt
\subsubsection{$X^{6,36}$; contracting one $\IP^4$} \label{sec:6,36}
As another manifold that is likely to have a freely acting $\IZ_5$ symmetry consider
$$
X^{6,36}~=~~
\cicy{\IP^1\\ \IP^1\\ \IP^1\\ \IP^1\\ \IP^1\\ \IP^4\\}
{\one & \one & 0 & 0 & 0 & 0 \\
 \one & 0 & \one & 0 & 0 & 0 \\
 \one & 0 & 0 & \one & 0 & 0 \\
 \one & 0 & 0 & 0 & \one & 0 \\
 \one & 0 & 0 & 0 & 0 & \one \\
 0 & \one & \one & \one & \one & \one }^{6,36}_{-60}
\hskip0.75in
\lower0.55in\hbox{\includegraphics[width=1.25in]{fig_6,36_manifold.pdf}}
$$
We take coordinates $x_j$, $j\in \IZ_5$ for the $\IP^4$ and $t_{i,a}$, $i\in \IZ_5$, $a\in\IZ_2$, for the five     
$\IP^1$'s. The polynomial corresponding to the first column of the matrix is denoted by $q$ and the remaining columns by $p_i$, $i\in\IZ_5$. We seek to impose a symmetry
$$
S:~x_i\to x_{i+1}~;~~t_{i,a}\to t_{i+1,a}~;~~q\to q~;~~p_i\to p_{i+1}~.
$$
In order to construct a polynomial $q$ we consider the invariants
$$
m_{abcde}~=~\sum_j t_{j,a}\, t_{j+1,b}\, t_{j+2,c}\, t_{j+3,d}\, t_{j+4,e}.
$$
Note that $m_{abcde}$ is invariant under cyclic permutation of its indices. There are therefore 8 linearly independent such terms and the most general invariant polynomial $q$ can be written in the form
$$
q~=~\frac{1}{5}\, m_{00000} + C_0\, m_{00001}+\ldots+ C_5\, m_{01111}+ C_6\, m_{11111}~,
$$
with the first coefficient taken to be $1/5$ since $m_{00000}=5\,t_{00}t_{10}t_{20}t_{30}t_{40}$.

The polynomials $p_i$ take the form
$$
p_i~=~\sum_j x_j (A_{i-j} \,t_{i0} + B_{i-j} \,t_{i1})~.
$$
Now we may change coordinates
$$
(t_{i0},\, t_{i1})~\to~(\a\, t_{i0}+\b\, t_{i1},\, \g\, t_{i0}+\d\, t_{i1})~~~\text{and}~~~
x_i~\to~ \sum_k \s_k\, x_{i+k}
$$ 
and still maintain $S$-covariance. We may use the freedom to redefine the $t_{i a}$ to enforce the conditions
$C_0=C_5=0$ and $C_6=1/5$, say. This brings $q$ to the form
\beq
q~=~\frac{1}{5}\, m_{00000} +  C_1\, m_{00011}+ C_2\, m_{00101}+ C_3\, m_{11010}+ 
C_4\, m_{11100}+\frac{1}{5}\, m_{11111}~.
\label{qeq}\eeq
The effect of an coordinate change of the $x_j$ on the coefficients $A_\ell$ and $B_\ell$ is 
$$
A_\ell\to\tilde{A}_\ell~=~\sum_k \s_{k-\ell}\, A_k~~~\text{and}~~~
B_\ell\to\tilde{B}_\ell~=~\sum_k \s_{k-\ell}\, B_k~.
$$
The determinant $\det( \s_{k-\ell})$ is nonzero for general $\s_k$ so we may use this freedom, say, to set 
$A_\ell=B_0\,\d_\ell$, with $\d_\ell$ an abbreviated form of the Kr\"onecker symbol, and then divide $p_i$ by $B_0$. The upshot is that, without loss of generality, we may rewrite the $p_i$ in the form
\beq
p_i~=~x_i(t_{i0}+ t_{i1}) + \sum_{j \neq i}B_{i-j}\, x_j\, t_{i1}~.
\label{peq}\eeq
The parameter count is that we have 4 free coefficients in Eq~\eref{qeq} and 4 more in Eq~\eref{peq} for a total of 8. The points in the embedding space that are fixed by $S$ are of the form 
\hbox{$t_{ia}=t_a, x_i=\z^i$} where $\z^5 = 1$, and it is easy to see that, for sufficiently general coefficients, these points do not satisfy the equations $q=0$ and $p_i=0$.  We have also checked the transversality of the polynomials by means of a Gr\"obner basis calculation. Thus we have shown the existence of the quotient manifold
$$
\cicy{\IP^1\\ \IP^1\\ \IP^1\\ \IP^1\\ \IP^1\\ \IP^4}
{\one & \one & 0 & 0 & 0 & 0 \\
 \one & 0 & \one & 0 & 0 & 0 \\
 \one & 0 & 0 & \one & 0 & 0 \\
 \one & 0 & 0 & 0 & \one & 0 \\
 \one & 0 & 0 & 0 & 0 & \one \\
 0 & \one & \one & \one & \one & \one }^{2,8}\quotient{\IZ_5}{~\lower35pt\hbox{.}}
$$
The Hodge number $h^{11}$ is now 2, since $S$ identifies the 5 $\IP^1$'s. The Euler number divides by the order of the group and so is now $\ch= -60/5= -12$. It follows that $h^{21}=8$, which agrees with our parameter count.
\subsubsection{$X^{5,45}$; contracting the second $\IP^4$} \label{sec:5,45}
Consider now the manifold
$$
X^{5,45}~=~~
\cicy{\IP^1\\\IP^1\\\IP^1\\\IP^1\\\IP^1\\}{1&1\\1&1\\1&1\\1&1\\1&1\\}_{-80}^{5,\, 45}
\hskip0.75in
\lower0.5in\hbox{\includegraphics[width=0.7in]{fig_5,45_manifold.pdf}}
$$
The matrix corresponding to this manifold is the transpose of that of the split quintic. It arises naturally by contracting the $\IP^4$ row in the matrix of the manifold of the previous subsection and it seems likely to admit a freely acting $\IZ_5$ symmetry. We will show in fact that there is a freely acting group isomorphic to 
to $\IZ_5{\times}\IZ_2{\times}\IZ_2$.

As previously we denote by $t_{ia}~,\,i\in \IZ_5,\, a\in \IZ_2$, the coordinates of the five          
$\IP^1$'s and again write
$$
m_{abcde}~=~\sum_j t_{j,a}\, t_{j+1,b}\, t_{j+2,c}\, t_{j+3,d}\, t_{j+4,e}~.
$$
For the two polynomials that define $X^{5,45}$ we take
\beq
\begin{split}
p_1~&=~\frac{1}{5}m_{00000} + A_1\, m_{00011}+A_2\, m_{00101}+A_3\, m_{01111}\\[5pt]
p_2~&=~\frac{1}{5}m_{11111} + A_1\, m_{11100}+A_2\, m_{11010}+A_3\, m_{10000}
\end{split}\label{polys}
\eeq
We have confirmed that these polynomials are transverse.  The locus $p_1=p_2=0$ is invariant under the symmetries
\beq
\begin{split}
S:~t_{ia}&\to t_{i+1,a}\\[3pt]
U:~t_{ia}&\to (-1)^a t_{ia}\\[3pt]
V:~t_{i0}& \leftrightarrow t_{i1}~.
\end{split}\label{symms}
\eeq
We shall now show that the group $G\cong \IZ_5{\times}\IZ_2{\times}\IZ_2$ generated by $S$, $U$ and     $V$ acts without fixed points. Note first that if $S^k$ has a fixed point
then so has $S^{k\ell}$ for all $\ell$ and, since $\IZ_5$ is a field, so has $S$. If $S^k U$ has a fixed point then so has its square $S^{2k}$, and so also $S$. The same argument applies also to $S^k V$ and $S^k UV$. Thus it suffices to check that $S$, $U$, $V$ and $UV$ act without fixed points.

A fixed point of $S$ in the embedding space, $(\IP^1)^5$, has the form $t_{ia}=t_a$, independent of $i$. The constraints 
$p_1$ and $p_2$ become two quintics acting in $\IP^1$, which have no common solution for generic values of the parameters. One sees, in fact, that there is no solution unless either $A_3=1$, or $A_3=-1$ and 
$A_1+A_2=0$.

A fixed point of $U$ in the embedding space is one of the 32 points obtained as the choices
$\big\{(1,0),\,(0,1)\big\}$ for each $\IP^1$. Each fixed point is such that precisely one of the monomials $m_{abcde}$ is nonzero there. Each monomial occurs once in either $p_1$ or $p_2$ so, provided none of the parameters $A_1,A_2,A_3$ vanishes, one of the two constraints fails to vanish.

Fixed points of $V$ in the embedding space are the 32 choices corresponding to choosing a point $(1,\pm 1)$ for each $\IP^1$. On each such point each monomial takes the value $\pm 1$. In particular the leading monomials in $p_1$ and $p_2$ take these values and, for generic values of the parameters, cannot be cancelled by the remaining terms. The fixed points of $UV$ are the 32 choices of points, each of the form $(1,\,\pm \ii)$, and an argument parallel to the previous one shows that the constraints do not vanish for generic choices of the parameters.

For the simply connected manifold the Euler number is -80 and $h^{11}=5$ since $H^2$ is spanned by the hyperplane sections of the 5 $\IP^1$'s. It follows that $h^{21}=45$. Under $S$ the 5 hyperplane sections of the $\IP^1$'s are identified so 
$h^{11}\left(X/G\right)=1$ and the Euler number of the quotient is $-80/20=-4$. It follows that $h^{21}\left(X/G\right)=3$, which coincides with our count of the free parameters in the polynomials \eref{polys}. Since the manifold $X/\IZ_5{\times}\IZ_2{\times}\IZ_2$ exists so too do the covering spaces
$X/K$ with $K$ a subgroup of $\IZ_5{\times}\IZ_2{\times}\IZ_2$. The Hodge numbers of these manifolds are given in the following table:
\vskip10pt
\begin{table}[H]
\begin{center}
\def\str{\vrule height16pt width0pt depth8pt}
\begin{tabular}{| c | c | c | c | c | c |}
\hline 
\str $\hodgenos\left(X^{5,45}/G\right)$ &$(1,3)$ &~$(1,5)$~ & ~$(1,9)$~ &$(5,15)$ &$(5,25)$\\
\hline
\str $G$ &$\IZ_5{\times}\IZ_2{\times}\IZ_2$ &$\IZ_5{\times}\IZ_2$ &$\IZ_5$ 
   &$\IZ_2{\times}\IZ_2$ &$\IZ_2$\\
\hline
\end{tabular}
\parbox{5.5in}{\caption{\small Hodge numbers of quotients of $X^{5,45}$.}}
\end{center}
\end{table}
\vskip10pt
\newpage
\section{Manifolds admitting free actions by $\IZ_3$}
\subsection{The branch $X^{2,83}\to X^{8,35}\to X^{14,23}\to X^{19,19}$}
The relation between these manifolds is discussed in detail in \SS2 of \cite{Triadophilia}. The manifold $X^{19,19}$ occurs also in the lower right of \tref{Z3quotients} and its $\IZ_6$-quotient will be discussed when we come to that branch.
\subsection{The branch $X^{2,83}\to X^{3,48}\to X^{4,40}\to X^{5,32}\to X^{6,24}\to X^{15,15}$}
\subsubsection{$X^{3,48}$; a $\IP^2$ split of the bicubic} \label{sec:3,48}
A very symmetrical-looking split of the bicubic is given by splitting with a single $\IP^2$:
$$
X^{3,48}~=~~
\cicy{\IP^2\\ \IP^2\\ \IP^2}
{1 & 1 & 1 \\
  1 & 1 & 1 \\
  1 & 1 & 1}_{-90}^{3,48}
$$
This has two equivalent diagrams. The first suggests a $\IZ_3$ symmetry while the second makes it apparent, in the diagram, that the matrix is identical to its transpose
$$
\includegraphics[width=1.75in]{fig_3,48_manifold.jpg}\hskip1in
\lower0pt\hbox{\includegraphics[width=1.50in]{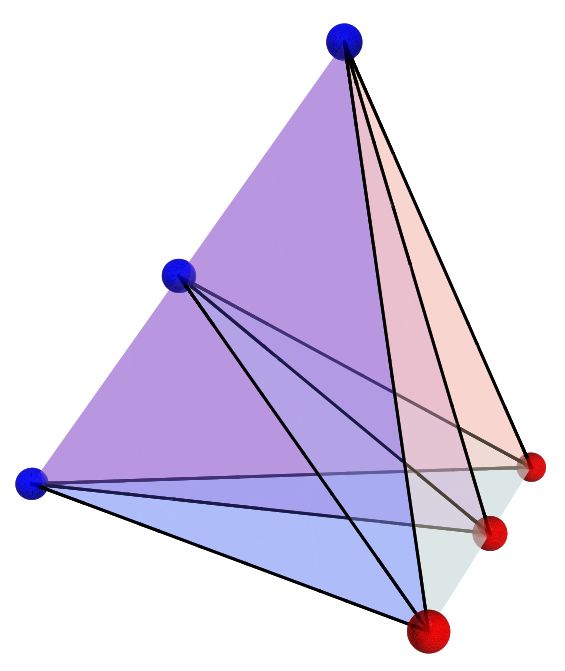}}\hskip0.5in
$$
Let us denote the coordinates of the three $\IP^2$'s by $x_{ij}$ where $i$ labels the space and $j$ its coordinate, we understand the labels to take values in $\IZ_3$. The polynomials take the form
\beq
p_i~=~\sum_{jkl}A_{ijkl}\, x_{0,j}\, x_{1,k}\, x_{2,\ell}~.
\label{(3,48)polys}
\eeq  
The counting of parameters, before we impose any symmetries, is that there are $3^4$ coefficients. There are 
$3{\times}8$ degrees of freedom corresponding to redefinitions of the coordinates, up to scale, and a further 9 degrees of freedom corresponding to redefinitions of the polynomials
$$
x_{ij}~\to~\sum_k\a_{ijk}\, x_{ik}~;\hspace{0.5in} p_i~\to~\sum_k \b_{ik}\, p_k~.
$$
The count is that there are $81-24-9=48$ free parameters in the polynomials which agrees with the value that we have for $h^{21}$. 

We impose first an internal symmetry
$$
S:~x_{i,j}~\to~x_{i,j+1}~;~~~p_i~\to~p_{i+1}
$$
which requires 
$$
A_{ijk\ell}~=~A_{0,j-i,k-i,\ell-i}
$$
so, setting $A_{0ijk}=C_{ijk}$, we have
\beq
p_i~=~\sum_{jk\ell}C_{jk\ell}\, x_{0,j+i}\, x_{1,k+i}\, x_{2,\ell+i}~.
\label{(3,48)Sinvar}\eeq
Now there are 27 coefficients and we are allowed redefinitions 
$$
x_{i,j}~\to~\sum_k \a_{ik}\, x_{i,j+k}~~~\text{and}~~~p_i~\to~\sum_k \b_k\, p_{i+k}~.
$$
with 6 degrees of freedom, up to scale, originating in the redefinition of the coordinates and 3 in the redefinition of the polynomials. The count is now that there are now $27-6-3=18$ free parameters in the equations. Fixed points of $S$, in the embedding space, are of the form $x_{ij}^S=\x_i^j$, with the $\x_i$ cube roots of unity. Evaluating on the fixed points we have
$$
p_i\big(x^S\big)~=~(\x_0 \x_1 \x_2 )^i\, p_0\big(x^S\big)
$$
and $p_0(x^S)\neq 0$ for generic choices of coefficients. The polynomials are also transverse for generic coefficients though this is easiest to show for the more symmetric polynomials we shall consider shortly.
The quotient $X^{3,48}/S$ will have $\chi = -90/3 = -30$, and $h^{11} = 3$ since $S$ only acts internally on each $\IP^2$.  This implies that $\hodgenos=(3,18)$ so we have agreement between $h^{21}$ and the degrees of freedom in the equations. 

We return to \eref{(3,48)polys} and impose the symmetry
$$
R:~x_{i,j}~\to~x_{i+1,j}~;~~p_i~\to~p_i~.
$$
This requires $A_{ijk\ell}$ to be invariant under cyclic permutation of the last three indices. A~tensor 
$B_{jk\ell}$ that is invariant under cyclic permutation of its labels has 11 degrees of freedom so our coefficients $A_{ijk\ell}$ have 33. Now there are a total of 8 degrees of freedom in the allowed redefinition of coordinates and 9 in the redefinition of the polynomials
$$
x_{i,j}~\to~\sum_k \a_{jk}\, x_{i,k}~;\hspace{0.5in}p_i~\to~\sum_k \b_{ik}\, p_k~.
$$
so the count is that there are $33-8-9=16$ degrees of freedom in the polynomials. The generator $R$ identifies the three $\IP^2$'s so $h^{11}=1$, for the quotient. Since the Euler number again divides we find 
$\hodgenos=(1,16)$ and the value we find for $h^{21}$ confirms our parameter count. Fixed points of $R$ are of the form
$$
x_{ij}^R~=~w_j
$$
The $w_j$ parametrise a $\IP^2$ so the three equations $p_i(w)=0$ will, generically, have no solution. this can be checked explicitly, together with the fact that the polynomials are transverse, for the simple polynomials given below.

If we impose invariance under both $R$ and $S$ then the polynomials are as in \eref{(3,48)Sinvar} but with the coefficient tensor invariant under cyclic permutation of its indices. A counting of parameters analogous to that above reveals that there are $11-2-3=6$ free parameters in the equations. A choice that exhibits these is 
\beq
\begin{split}
p_i &= x_{0,i}\,x_{1,i}\,x_{2,i} + \\
&\hspace{0pt} \sum_{s=\pm 1}\!\left\{ \! E_s\, x_{0,i+s}\,x_{1,i+s}\,x_{2,i+s}
+ F_s\!\sum_{k=0}^2 x_{k,i}\, x_{k+1,i+s}\, x_{k+2,i+s}
+ G_s\!\sum_{k=0}^2 x_{k,0}\, x_{k+s,1}\, x_{k+2s,2}\right\}
\end{split}
\label{(3,48)RSinvar}\eeq
and it is straightforward to check that these equations are transverse, for generic choice of the coefficients.

We have seen that $R$ and $S$ act without fixed points. The diagonal generators, however, do have fixed points.
We have $RS\,x_{i,j}=x_{i+1,j+1}$ and this action has fixed points that satisfy
$$
x^{RS}_{i,j}~=~x^{RS}_{i-1,j-1}~~~\text{hence}~~~x^{RS}_{i,j}~=~x^{RS}_{0,j-i}~.
$$
If we regress to writing the polynomials as in \eref{(3,48)Sinvar} with cyclically invariant $C_{jk\ell}$ then we see that, when evaluated on the fixed points,
\beq
p_{i+1}~=~\sum_{jk\ell}C_{jk\ell}\, x_{1,j+i}^{RS}\, x_{2,j+i-1}^{RS}\, x_{0,j+i-2}^{RS}
                =~p_i~.
\notag\eeq
Thus the three equations $p_i$ are all equivalent to $p_0$, say, and this is a cubic in $x_{0,k}^{RS}$ so the fixed point set is an elliptic curve that we denote by $\cE^{RS}$. In a similar way we see that $R^2S$ has fixed points 
$$
x^{R^2S}_{i,j}~=~x^{R^2S}_{0,j+i}
$$
and that these points make up a second elliptic curve $\cE^{R^2S}$. A point of intersection of the two elliptic curves is a simultaneous fixed point of $RS$ and $R^2S$ and hence also of $R$ and $S$. We have seen above that $R$ and $S$ act without fixed points so the two elliptic curves cannot intersect. The fixed point set is resolved by replacing a neighbourhood of each elliptic curve by a bundle of open sets taken from the interior of an $A_2$ surface resolution.  An $A_2$ surface has two $(1,1)$ forms $\r_{1,2}$, so each resolution increases $h^{11}$ by two.  Furthermore, an elliptic curve has a holomorphic $(1,0)$-form $\eta$, so we also get two new $(2,1)$-forms $\r_{1,2}\wedge\eta$.  For the resolved manifold then we have $h^{11}=1+2{\times}2=5$ and $h^{21}=6+2{\times}2=10$.  Since $RS$ and $R^2S$ both have fixed points, and these two elements generate the whole quotient group, the manifold obtained in this way is simply connected. The manifolds we have found in this section are summarised in the following table
\vskip10pt
\begin{table}[H]
\begin{center}
\def\str{\vrule height17pt width0pt depth10pt}
\begin{tabular}{| l | c | c | c |}
\hline 
\str Hodge numbers  & $(1,\,16)$ & $(3,\,18)$ & $(5,10)$ \\
\hline
Manifold            &~~$X^{3,48}/R$~~&~~$X^{3,48}/S$~~&~~$\simpentry{\widehat{X^{3,48}/R{\times}S}}$~~\\       
\hline
\str Fundamental group           & & & \one \\
\hline
\end{tabular}
\parbox{4.5in}{\caption{\small Hodge numbers of smooth quotients, and the resolution of the singular quotient, of $X^{3,48}$.}}
\end{center}
\end{table}
\vskip10pt
\subsubsection{$X^{4,40}$; a second split of the bicubic} \label{sec:4,40}
We can introduce a fourth $\IP^2$ and split one column of the above configuration to obtain
$$
X^{4,40}~=~~
\cicy{ \IP^2\\ \IP^2\\ \IP^2\\ \IP^2\\}
{\one & 0 & 0 & \one & \one \\
0 & \one & 0 & \one & \one \\
0 & 0 & \one & \one & \one \\
\one & \one & \one & 0 & 0 \\}_{-72}^{4,40}
$$
with diagrams
$$
\includegraphics[width=1.5in]{fig_4,40_manifold.jpg}
\hskip1in
{\includegraphics[width=1.6in]{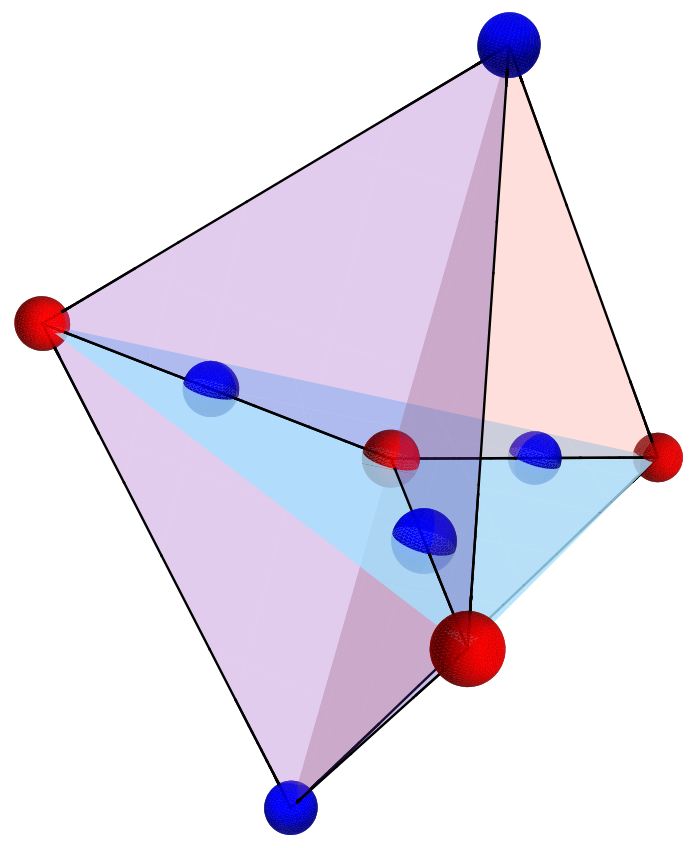}}
\vspace{10pt}
$$
\vskip-20pt
Take coordinates $x_{ij}$ on the first three spaces as in the previous subsection, and coordinates $u_i$ on the last space.
Denote by $q_i$ the first three polynomials, and by $p_1,p_2$ the last two.  We can then extend the definition of
the $\IZ_3$ generator $R$ to
$$
R:~x_{i,j}~\to~x_{i+1,j}~,~~ u_i~ \to~u_{i+1}~;~~q_i~\to~q_{i+1}.
$$
The most general polynomials covariant under this action are
\beq
\begin{split}
q_i &= \sum_{j,k} A_{jk}\, x_{ij}\, u_{i+k} \\
p_1 &= \sum_{i,j,k} B_{ijk}\, x_{0,i}\, x_{1,j}\, x_{2,k} \\
p_2 &= \sum_{i,j,k} C_{ijk}\, x_{0,i}\, x_{1,j}\, x_{2,k}
\end{split}
\eeq
where $B_{ijk}$ and $C_{ijk}$ are cyclic in their indices.  We have checked that these polynomials are transverse. Fixed points of $R$ in the ambient space occur when $u_i = \o$, where $\o^3 = 1$, and 
$x_{ij} = x^\ast_j$.  At these points, the equations $q_0 = p_1 = p_2 = 0$ impose three independent homogeneous constraints in the $\IP^2$ parametrised by $x^\ast_j$, so there are no solutions, in general.  Therefore the quotient is smooth.

Since $R$ identifies three of the four ambient spaces, $h^{11}$ will be reduced to 2 for the quotient.  The Euler number will be $-72/3 = -24$, and this implies that $h^{21} = 14$.  We can confirm this with a parameter count.  We start with the general non-symmetric case, in which there are 81 terms in the polynomials.  There are $4{\times}8 = 32$ parameters in coordinate changes, and $3$ parameters in rescaling the $q_i$.  Next we use an observation first made in~\cite{Deformations} in the context of the manifold $X^{2,56}$ 
(see \SS\ref{sec:2,56}).  For a solution to $q_i = 0$ to exist for all $i$, we must have 
$0 = \det(\partial q_i/\partial u_j) \equiv D$.  But $D$ is a homogeneous trilinear polynomial in the
first three spaces, so the most general redefinition of the polynomials $p_1, p_2$ is
$$
p_1 \to \k_{11}\, p_1 + \k_{12}\, p_2 + K_1 D~,~~p_2 \to \k_{21}\, p_1 + \k_{22}\, p_2 + K_2 D
$$
which contains 6 more parameters.  Therefore the number of meaningful parameters in the defining polynomials is
$81 - (32 + 3 + 6) = 40$, which agrees with $h^{21}$.  Now impose covariance under $R$.  Demanding that $C_{ijk}$ is cyclic leaves $11$ independent components, so the polynomials now contain $9 + (2\times 11) = 31$ coefficients.  The coordinate changes compatible with the action of $R$ are $x_{ij} \to \sum_{k} \a_{jk}\, x_{ik}$ and $u_{i} \to \sum_{j} \b_{ij}\, u_{j}$, which contain $8+2 = 10$ parameters, up to irrelevant scaling.  Finally, we can rescale the $q_i$ by a common factor, and redefine
$p_1$ and $p_2$ exactly the same way as in the non-symmetric case, so altogether this gives 7 more parameters, leaving $31 - (10 + 7) = 14$ independent coefficients.  This agrees with our previous determination of $h^{21}$, so we have found a quotient manifold with fundamental group $\IZ_3$ and $\hodgenos = (2,14)$.
\subsubsection{$X^{5,32}$; a third split of the bicubic} \label{sec:5,32}
Splitting a second column with a $\IP^2$ leads to the configuration
$$
X^{5,32}~=~~
\cicy{\IP^2 \\ \IP^2\\ \IP^2\\ \IP^2\\ \IP^2\\}
{\one & 0 & 0 & \one & 0 & 0 & \one \\
0 & \one & 0 & 0 & \one & 0 & \one \\
0 & 0 & \one & 0 & 0 & \one & \one \\
\one & \one & \one & 0 & 0 & 0 & 0 \\
 0 & 0 & 0 & \one & \one & \one & 0 \\}_{-54}^{5,32}
$$
with diagrams
$$
\includegraphics[width=1.75in]{fig_5,32_manifold.jpg}
\hskip 1.0in
{\includegraphics[width=1.05in]{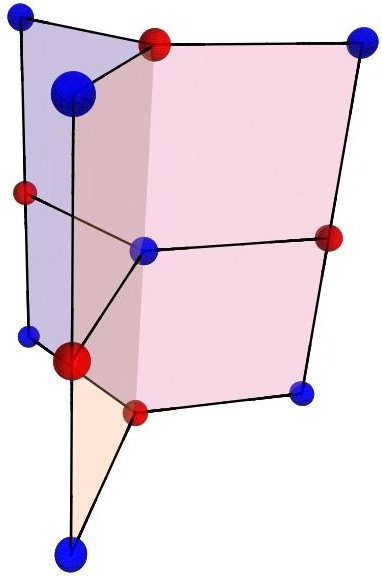}}\hskip0.7in
$$
Again take coordinates $x_{ij}$ for the first three spaces, $u_i$ for the fourth, and introduce coordinates $v_i$ for the last space.  Denote by $p_i$ the first three polynomials, by $q_i$ the next three, and by $r$ the last.  Then we can again extend the definition of the $\IZ_3$ generator $R$ to
$$
R:~x_{i,j}~\to~x_{i+1,j}~,~~ u_i~ \to~u_{i+1}~,~~ v_i~\to~v_{i+1}~;~~p_i~\to~p_{i+1}~;~~
q_i~\to~q_{i+1}.
$$
The most general polynomials covariant under this action are
\beq
\begin{split}
p_i &= \sum_{j,k} A_{jk}\, x_{ij}\, u_{i+k} \\
q_i &= \sum_{j,k} B_{jk}\, x_{ij}\, v_{i+k} \\
r &= \sum_{i,j,k} C_{ijk}\, x_{0,i}\, x_{1,j}\, x_{2,k}
\end{split}
\eeq
where $C_{ijk}$ is again cyclic in its indices.  We have checked that these polynomials are transverse and the action of $R$ is fixed point free, so we obtain a smooth quotient manifold.  Three of the five ambient spaces get identified, leaving $h^{11} = 3$.  The Euler number of the quotient is simply $-54/3 = -18$, and together these numbers imply $h^{21} = 12$.

The analysis of fixed points and the parameter counting is very similar to the previous case, so we do not present it in detail, but we note that there are now two determinants which can be used to redefine 
$r$: $\det(\partial p_i/\partial u_j)$ and $\det(\partial q_i/\partial v_j)$.  Indeed the counting confirms that $h^{21} = 12$, so we have found a manifold with fundamental group $\IZ_3$ and $\hodgenos = (3,12)$.
\subsubsection{$X^{6,24}$; a fourth split of the bicubic} \label{sec:6,24}
We can introduce one final $\IP^2$ and arrive at the very symmetrical configuration represented by the following matrix
$$
X^{6,24}~=~~
\cicy{\IP^2\\ \IP^2\\ \IP^2\\ \IP^2\\ \IP^2\\ \IP^2\\}
{\one & 0 & 0 & \one & 0 & 0 & \one & 0 & 0 \\
 0 & \one & 0 & 0 & \one & 0 & 0 & \one & 0 \\
 0 & 0 & \one & 0 & 0 & \one & 0 & 0 & \one \\
 \one & \one & \one & 0 & 0 & 0 & 0 & 0 & 0 \\
 0 & 0 & 0 & \one & \one & \one & 0 & 0 & 0 \\
 0 & 0 & 0 & 0 & 0 & 0 & \one & \one & \one \\}_{-36}^{6,24}
$$
with diagrams
$$
\includegraphics[width=1.75in]{fig_6,24_manifold.jpg}\hskip1in
\raise10pt\hbox{\includegraphics[width=1.75in]{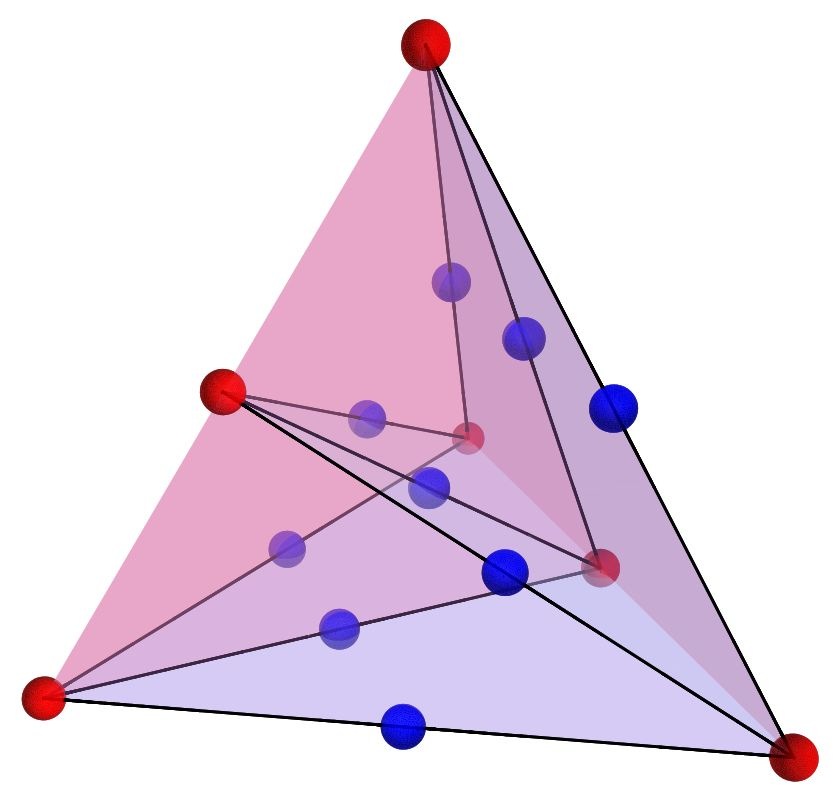}}\hskip0.5in
$$
We will find a family of such manifolds with a freely acting $\IZ_3$ symmetry.  Take coordinates $x_{i j}$ on the first three spaces, where the first index labels the spaces and the second the coordinates.  Take similarly coordinates $y_{ij}$ on the last three spaces.  In the second diagram the spaces corresponding to the $x_{ij}$ lie on one edge of the tetrahedron and spaces corresponding to the $y_{ij}$ on the opposite edge. We may impose the $\IZ_3$ symmetry generated by
$$
R:~x_{ij}~\to~ x_{i,j+1}~;~~y_{ij}~\to~y_{i+1,j}~.
$$
This symmetry is suggested by the first diagram where we think of the $x$-spaces as lying along the axis and the $y$-spaces as being permuted cyclically. There is a symmetry under the exchange of the $x$ and $y$ spaces so we may impose a second symmetry
$$
R':~x_{ij}~\to~ x_{i+1,j}~;~~y_{ij}~\to~y_{i,j+1}~.
$$
Polynomials that are symmetric under these generators are of the form
$$
p_{ij}~=~\sum_{k,\ell} C_{k\ell}\, x_{i,j+k}\, y_{j,i+\ell}
$$
For generic coefficients these polynomials are transverse and they are fixed point free under the action of $R$, to avoid undue repetition we will forego an explicit demonstration of this here. Since $R$ identifies the three $y$-spaces we have $h^{11}=4$ for the quotient. In this way we find a quotient $X^{6,24}/R$ with Hodge numbers $\hodgenos=(4,10)$. 

The action of $R'$ is also fixed point free however, in a manner similar to the previous case, the diagonal generator $RR'$ and $R^2R'$ have fixed points. These take the form
$$
x^{RR'}_{i,j}~=~x^{RR'}_{0,j-i}~,~~y^{RR'}_{i,j}~=~y^{RR'}_{0,j-i}~~~\text{and}~~~
x^{R^2R'}_{i,j}~=~x^{R^2R'}_{0,j+i}~,~~y^{R^2R'}_{i,j}~=~y^{R^2R'}_{0,j+i}~.
$$
The defining polynomials give the constraints
$$
\sum_{k,\ell} C_{k\ell}\, x^{RR'}_{k+j}\, y^{RR'}_{\ell-j}~=~0~~~\text{and}~~~
\sum_{k,\ell} C_{k\ell}\, x^{R^2R'}_{k+j}\, y^{R^2R'}_{\ell+j}~=~0~;~~k=0,1,2,
$$
corresponding, in each case, to the configuration 
$$
\cicy{\IP^2\\ \IP^2\\}{1&1&1\\ 1&1&1\\}
$$
indicating that the fixed point sets are one-dimensional CICY's and hence elliptic curves. Intersection points of the two elliptic curves would be simultaneously fixed points of $RR'$ and of $R^2R'$ and hence of $R$ and $R'$. Since $R$ and $R'$ act freely we see that the elliptic curves do not intersect.  Repeating a construction analogous to that of a previous section we see that the resolution, $\widehat{X^{6,24}/R{\times}R'}$ has $\hodgenos=(6,8)$ and trivial fundamental group. We summarize the hodge numbers of the manifolds we have found with a short table.
\vskip10pt
\begin{table}[H]
\begin{center}
\def\str{\vrule height17pt width0pt depth10pt}
\begin{tabular}{| l | c | c |}
\hline 
\str Hodge numbers  & $(4,\,10)$ & $(6,\,8)$  \\
\hline
\str Manifold            &~~$X^{6,24}/R$~~&~~$\simpentry{\widehat{X^{6,24}/R{\times}R'}}$~~\\       
\hline
\str Fundamental group           & & \one \\
\hline
\end{tabular}
\parbox{3.5in}{\caption{\small Hodge numbers of smooth quotients, and resolutions of singular quotients, of $X^{6,24}$.}}
\end{center}
\end{table}
\vskip10pt
\subsubsection{$X^{15,15}$; a fifth split of the bicubic} \label{sec:15,15}
We can also consider the final split shown in \tref{Z3quotients}. As can be shown in a similar way to the example in \tref{TranspositionId} this is also the transpose of the CICY of the previous subsection
$$
X^{15,15}~=~~
\cicy{\IP^1\\ \IP^1\\ \IP^1\\ \IP^1\\ \IP^1\\ \IP^1\\ \IP^1\\ \IP^1\\ \IP^1\\}
{\one & 0 & 0 & \one & 0 & 0 \\
 0 & \one & 0 & \one & 0 & 0 \\
 0 & 0 & \one & \one & 0 & 0 \\
 \one & 0 & 0 & 0 & \one & 0 \\
 0 & \one & 0 & 0 & \one & 0 \\
 0 & 0 & \one & 0 & \one & 0 \\
 \one & 0 & 0 & 0 & 0 & \one \\
 0 & \one & 0 & 0 & 0 & \one \\
 0 & 0 & \one & 0 & 0 & \one}_0^{15,15}
$$
This manifold has diagrams
$$
\includegraphics[width=1.75in]{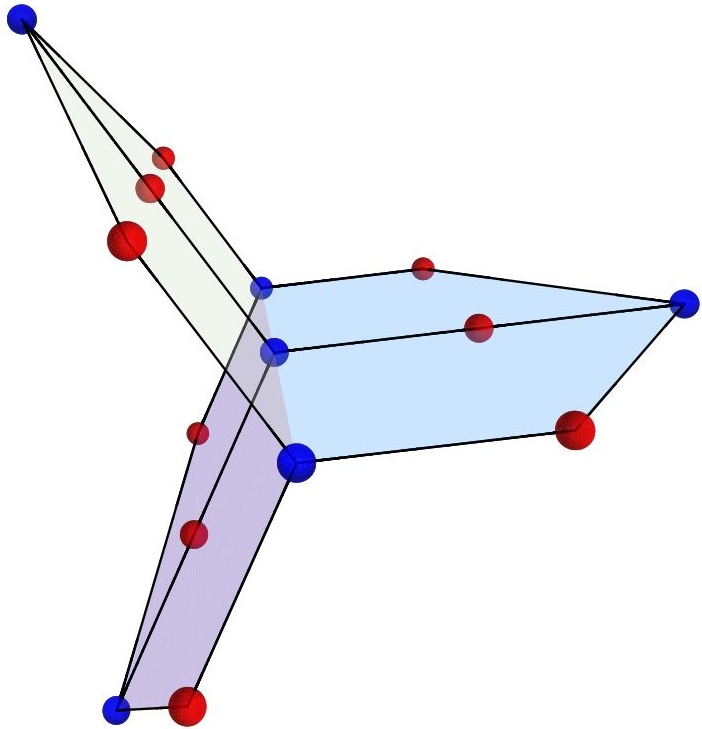}\hskip1in
\raise10pt\hbox{\includegraphics[width=1.75in]{fig_15,15_tetrahedron.jpg}}\hskip0.5in
$$
The second diagram in particular suggests that we take three polynomials $p_i$, corresponding to one edge of the tetrahedron, and three other polynomials $q_j$ corresponding to the opposite edge. The indices $i,j$ take values in $\IZ_3$. We take the $\IP^1$ lying on the line that connects $p_i$ to $q_j$ to have coordinates $x_{ija}$, $a\in\IZ_2$.  The polynomials are
\beq
\begin{split}
p_i ~&=~ \sum_{abc} A_{iabc}\, x_{0,i,a}\, x_{1,i,b}\, x_{2,i,c} \\[3pt]
q_i ~&=~\sum_{abc} B_{iabc}\, x_{i,0,a}\, x_{i,1,b}\, x_{i,2,c}
\end{split}
\notag\eeq
We will first demand that the manifold be invariant under the action of $\IZ_3$ generated by
$$
S:~ x_{i,j,a} ~\to~ x_{i,j+1,a}~;~~~p_i~\to~p_{i+1}~,~~~q_j~\to~q_j
$$ 
Covariance under $S$ requires the $B_{iabc}$ to be invariant under cyclic permutations of the indices $abc$ and the $A_{iabc}$ to be independent of $i$. Thus there are $3{\times}3$ independent coefficients in the $q_i$, since an overall scale is irrelevant. There are also 9 parameters, up to scale in the coordinate transformations consistent with~$S$ 
$$
(x_{ij0},\, x_{ij1}) \to (\a_i\, x_{ij0} + \b_i\, x_{ij1}, \, \g_i\, x_{ij0} + \d_i\, x_{ij1})~.
$$
We may use this freedom to bring the $q_i$ to the form
$$
q_i~=~\n_{i\,001}+\frac{1}{3}\,\n_{i\,111}
$$
where
$$
\n_{iabc} = \sum_{j} x_{ija}\, x_{i,j+1,b}\, x_{i,j+2,c}~.
$$
We have eliminated the free coefficients from the $q_i$ but we are left with the 8 coefficients, hence 7 parameters in the $p_i$. These polynomials are transverse though, again, it is easier to check this for the more symmetric polynomials that will be given shortly. It is also straightforward to check that the $S$ action on the manifold is fixed point free.  The fact that there are seven free parameters in the polynomials suggests that $h^{21}=7$ and so, since the quotient manifold must still have Euler number zero, that the quotient has Hodge numbers $(7,7)$.  
A~way to check this is to consider an `extended' representation as in \tref{tab(15,15)Extended}, with fifteen ambient spaces providing the fifteen $(1,1)$ forms, and note that exactly seven of them will still be independent after the $\IZ_3$ identification.
\begin{table}[H]
\begin{center}
\boxed{\hskip2pt
\begin{tabular}{cc}
\footnotesize
$\cicy{\IP^1\\ \IP^1\\ \IP^1\\ \IP^1\\ \IP^1\\ \IP^1\\ \IP^1\\ \IP^1\\ \IP^1\\ \IP^2\\ \IP^2\\ \IP^2\\ \IP^2\\ \IP^2\\ \IP^2}
{\one ~ 0 ~ 0 ~ 0 ~ 0 ~ 0 ~ 0 ~ 0 ~ 0 ~ \one ~ 0 ~ 0 ~ 0 ~ 0 ~ 0 ~ 0 ~ 0 ~ 0 \\
 0 ~ \one ~ 0 ~ 0 ~ 0 ~ 0 ~ 0 ~ 0 ~ 0 ~ 0 ~ 0 ~ 0 ~ \one ~ 0 ~ 0 ~ 0 ~ 0 ~ 0 \\
 0 ~ 0 ~ \one ~ 0 ~ 0 ~ 0 ~ 0 ~ 0 ~ 0 ~ 0 ~ 0 ~ 0 ~ 0 ~ 0 ~ 0 ~ \one ~ 0 ~ 0 \\
 0 ~ 0 ~ 0 ~ \one ~ 0 ~ 0 ~ 0 ~ 0 ~ 0 ~ 0 ~ \one ~ 0 ~ 0 ~ 0 ~ 0 ~ 0 ~ 0 ~ 0 \\
 0 ~ 0 ~ 0 ~ 0 ~ \one ~ 0 ~ 0 ~ 0 ~ 0 ~ 0 ~ 0 ~ 0 ~ 0 ~ \one ~ 0 ~ 0 ~ 0 ~ 0 \\
 0 ~ 0 ~ 0 ~ 0 ~ 0 ~ \one ~ 0 ~ 0 ~ 0 ~ 0 ~ 0 ~ 0 ~ 0 ~ 0 ~ 0 ~ 0 ~ \one ~ 0 \\
 0 ~ 0 ~ 0 ~ 0 ~ 0 ~ 0 ~ \one ~ 0 ~ 0 ~ 0 ~ 0 ~ \one ~ 0 ~ 0 ~ 0 ~ 0 ~ 0 ~ 0 \\
 0 ~ 0 ~ 0 ~ 0 ~ 0 ~ 0 ~ 0 ~ \one ~ 0 ~ 0 ~ 0 ~ 0 ~ 0 ~ 0 ~ \one ~ 0 ~ 0 ~ 0 \\
 0 ~ 0 ~ 0 ~ 0 ~ 0 ~ 0 ~ 0 ~ 0 ~ \one ~ 0 ~ 0 ~ 0 ~ 0 ~ 0 ~ 0 ~ 0 ~ 0 ~ \one \\
 \one ~ \one ~ \one ~ 0 ~ 0 ~ 0 ~ 0 ~ 0 ~ 0 ~ 0 ~ 0 ~ 0 ~ 0 ~ 0 ~ 0 ~ 0 ~ 0 ~ 0 \\
 0 ~ 0 ~ 0 ~ \one ~ \one ~ \one ~ 0 ~ 0 ~ 0 ~ 0 ~ 0 ~ 0 ~ 0 ~ 0 ~ 0 ~ 0 ~ 0 ~ 0 \\
 0 ~ 0 ~ 0 ~ 0 ~ 0 ~ 0 ~ \one ~ \one ~ \one ~ 0 ~ 0 ~ 0 ~ 0 ~ 0 ~ 0 ~ 0 ~ 0 ~ 0 \\
 0 ~ 0 ~ 0 ~ 0 ~ 0 ~ 0 ~ 0 ~ 0 ~ 0 ~ \one ~ \one ~ \one ~ 0 ~ 0 ~ 0 ~ 0 ~ 0 ~ 0 \\
 0 ~ 0 ~ 0 ~ 0 ~ 0 ~ 0 ~ 0 ~ 0 ~ 0 ~ 0 ~ 0 ~ 0 ~ \one ~ \one ~ \one ~ 0 ~ 0 ~ 0 \\
 0 ~ 0 ~ 0 ~ 0 ~ 0 ~ 0 ~ 0 ~ 0 ~ 0 ~ 0 ~ 0 ~ 0 ~ 0 ~ 0 ~ 0 ~ \one ~ \one ~ \one}_0^{15,15}$
&\lower115pt\hbox{\includegraphics[width=3.4in]{fig_15,15_Extended.jpg}}
\end{tabular}}
\vskip5pt
\parbox{5.5in}{\caption{\label{tab(15,15)Extended}\small The matrix and diagram for the extended representation of $X^{15,15}$ for which all 15 K\"ahler classes are represented by ambient spaces.}}
\end{center}
\end{table}
We now impose invariance under a second $\IZ_3$, generated by
$$
S':~ x_{i,j,a}\to x_{i+1,j,a}~;~~~p_i~\to~p_{i}~,~~~q_j~\to~q_{j+1}
$$
The most general polynomials consistent with this additional symmetry are obtained by specialising those given previously. Now the coefficients $A_{iabc}$ must be independent of $i$ and cyclically symmetric in the indices $abc$. Equivalently the polynomials may be written in the form
\beq
p_i ~=~ A\, \m_{i000} + B\, \m_{i001} + C\, \m_{i011} + D\, \m_{i111}~,~~~
q_i ~= ~\n_{i000} + \frac{1}{3}\, \n_{i111} 
\label{eq:15,15 Z3Z3}\eeq
where we have defined also $\m_{iabc} = \sum_j\, x_{j,i,a}\, x_{j+1,i,b}\, x_{j+2,i,c}$.  There are no further coordinate changes, consistent with $S$ and $S'$, that preserve the form of the $q_i$. So we are left with the three degrees of freedom that are visible in the coefficients of the $p_i$.
We have checked that these equations are transverse and that the actions of both $S$ and $S'$ on the resulting manifold are fixed point free. As in previous cases the elements $SS'$ and $S^2S'$ have fixed points.  These are subject to the relations
$$
x^{SS'}_{i,j,a}~=~x^{SS'}_{0,j-i,a}~~~\text{and}~~~
x^{S^2S'}_{i,j,a}~=~x^{S^2S'}_{0,j+i,a}~.
$$
In each case, putting these conditions into the polynomials gives the configuration
$$
\cicy{\IP^1\\ \IP^1\\ \IP^1 \\}
{1~1 \\
 1~1 \\
 1~1 \\}
 $$
which again is an elliptic curve.  By exactly the same arguments as previously, we see that the two elliptic curves do not intersect.  We can therefore construct a new manifold  by beginning with the smooth quotient of $X^{15,15}$ by $S$, and orbifolding again by the action of~$S'$.  This fixes the two elliptic curves, which can be resolved, as before, with the aid of the $A_2$ surface.  The resulting manifold will have Euler number zero and trivial fundamental group.  Inspecting the diagram for the extended representation of $X^{15,15}$, we see that there will be three independent $(1,1)$-forms after identification under $S$ and $S'$. We will also have three $(2,1)$-forms coming from the original space, corresponding to the three undetermined co-efficients
in~\eref{eq:15,15 Z3Z3}.  The resolution of each elliptic curve contributes two $(1,1)$-forms and two $(2,1)$-forms, so the Hodge numbers of the resolution are $(7,7)$.

An alternative to the second $\IZ_3$ is to impose a $\IZ_2$ symmetry generated by
$$
U:~ x_{ija} ~\to~ (-1)^a x_{ija}~;~~~p_i ~\to~ p_i~,~~~ q_i ~\to~ -q_i~.
$$
Specialising the $S$-covariant polynomials to this case we have
\beq
\begin{split}
p_i ~&=~ A\, x_{0i0}\, x_{1i0}\, x_{2i0} + B\, x_{0i1}\, x_{1i1}\, x_{2i0} + C\, x_{0i1}\, x_{1i0}\, x_{2i1} + D\, x_{0i0}\, x_{1i1}\, x_{2i1} \\[3pt]
q_i ~&=~ \n_{i100} + \frac{1}{3}\, \n_{i111}
\end{split}
\notag\eeq
It is again the case that there are no further coordinate changes, consistent with the symmetries, that preserve the form of the $q_i$ and that we are left with 3 free parameters that are visible in the coefficients of the $p_i$. We have checked that these equations are also transverse and the generators act without fixed points.

Since a $\IZ_3{\times}\IZ_2$ quotient exists, so too must the three-fold cover which is a $\IZ_2$ quotient of the original simply-connected manifold.  A straightforward counting of parameters suffices to determine its Hodge numbers.  Before imposing any symmetries, there are $2{\times}2{\times}2=8$ terms in each of the six equations, half even under the action of $U$, and half odd.  Thus after imposing that the $q_i$ be odd and the $p_i$ even, we have $6{\times}(8/2) = 24$ terms.  The remaining allowed coordinate transformations are $(x_{ij0}, x_{ij1}) \to (x_{ij0}, \b_{ij} x_{ij1})$, with nine parameters $\b_{ij}$, and we can rescale each of the six polynomials independently.  We conclude that the number of parameters in the 
$\IZ_2$ covariant polynomials is $24{-}9{-}6 = 9$.  They will be transverse since the subfamily which is also 
$\IZ_3$ covariant are transverse, so we get a smooth quotient with Hodge numbers $(9,9)$.

Finally we can consider polynomials covariant under all three generators $S,\, S',\, U$
\beq
p_i = A\, \m_{i000} + B\, \m_{i110}~,~~ q_i = C\, \n_{i100} + D\, \n_{i111}
\notag\eeq
The only coordinate change consistent with all three symmetries is $(x_{ij0},x_{ij1})\to(x_{ij0},\b x_{ij1})$, which allows us to remove a single
coefficient.  After also scaling the $p_i$'s and $q_i$'s we are left~with
\beq
p_i = \frac 13 \m_{i000} + B \m_{i110}~,~~ q_i = \n_{i100} + \frac 13 \n_{i111}~.
\notag\eeq
These are transverse, and again the action of $\IZ_3{\times}\IZ_2$ generated by $S$ and $U$ is fixed point free.  As before, the elements $SS'$ and $S^2S'$ each fix an elliptic curve, and since these do not intersect they can be resolved independently.  Note that there is no need to consider products of these elements with $U$ since, for example, $(USS')^3 = U$, and we know $U$ does not have fixed points.  The
Hodge numbers for the resolved orbifold can be determined by noting that the polynomials have only a single free parameter, corresponding to a single $(2,1)$-form, while the resolution of each fixed curve contributes another 2.  Therefore the manifold obtained in this way has fundamental group $\IZ_2$ and Hodge
numbers $(5,5)$.

We summarise the various manifolds we have found in the following table:
\vskip10pt
\begin{table}[H]
\begin{center}
\def\str{\vrule height16pt width0pt depth8pt}
\begin{tabular}{| l | c | c | c | c | c |}
\hline 
\str Hodge numbers & ~~$(3,3)$~~ & ~~$(7,7)$~~ &  ~~$(9,9)$~~ & ~~$(5,5)$~~ & ~~$(7,7)$~~ \\
\hline
Manifold & $X^{15,15}\quotient{\IZ_3{\times}\IZ_2}$ & $X^{15,15}\quotient{\IZ_3}$
 & $X^{15,15}\quotient{\IZ_2}$ & \simpentry{\widehat{X^{15,15}\quotient{\IZ_3{\times}\IZ_3{\times}\IZ_2}}}  & \simpentry{\widehat{X^{15,15}\quotient{\IZ_3{\times}\IZ_3}}} \\
\hline
\str Fundamental group           & & & & $\IZ_2$ & \one \\
\hline
\end{tabular}
\parbox{5.0in}{\caption{\small The Hodge numbers of the smooth quotients, and resolutions of singular quotients, which have been found of $X^{15,15}$.}}
\end{center}
\end{table}
\vskip10pt
\subsection{Two transposes}
\subsubsection{$X^{9,21}$; the transpose of $X^{5,32}$} \label{sec:9,21}
Taking the transpose of the matrix representing $X^{5,32}$ leads us to the configuration
$$
X^{9,21}~=~~
\cicy{\IP^1 \\ \IP^1 \\ \IP^1 \\ \IP^1 \\ \IP^1 \\ \IP^1 \\ \IP^2}
{\one & 0 & \one & 0 & 0 \\
\one & 0 & 0 & \one & 0 \\
\one & 0 & 0 & 0 & \one \\
0 & \one & \one & 0 & 0 \\
0 & \one & 0 & \one & 0 \\
0 & \one & 0 & 0 & \one \\
0 & 0 & \one & \one & \one \\}^{9,21}_{-24}
$$
with corresponding diagrams
$$
\includegraphics[width=1.75in]{fig_9,21_manifold.jpg}\hskip1.0in
\includegraphics[width=1.3in]{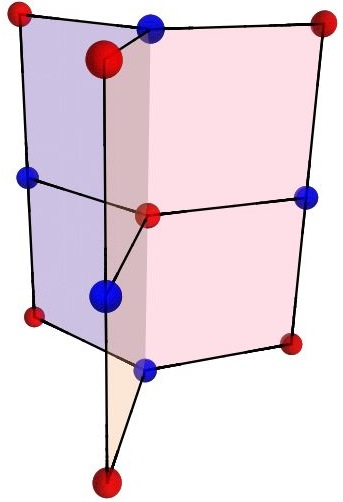}
$$
Introduce coordinates $s_{ia}$ on the first three $\IP^1$'s, $t_{ia}$ on the next three, and $u_i$ on the $\IP^2$.  Denote by $r, q$ the
polynomials trilinear in the $s$'s and $t$'s respecively, and by $p_i$ the three trilinear polynomials.  Define a 
$\IZ_3$ action with generator
$$
S:~s_{ia} \to s_{i+1,a}~,~~t_{ia} \to t_{i+1,a}~,~~u_i \to u_{i+1}~;~~p_i \to p_{i+1}
$$
If we define the $S$-invariants
$$
m_{abc} = \sum_i s_{ia}\, s_{i+1,b}\, s_{i+2,c}~,~~ n_{abc} = \sum_i t_{ia}\, t_{i+1,b}\, t_{i+2,c}
$$
then by a suitable choice of coordinates we can bring the symmetric polynomials to the form
\begin{eqnarray} \label{eq:9,21polys}
r =~ \frac 13 m_{000} + m_{110}~,~~ q =~ \frac 13 n_{000} + n_{110} ~~~~~~~~~~~\\[5pt] \nonumber
p_i =~ s_{i0}\, t_{i0}\, u_i + \sum_{j} (D_j s_{i0}\, t_{i1} + E_j s_{i1}\, t_{i0} + F_j s_{i1}\, t_{i1})\, u_{i+j}
\end{eqnarray}
These polynomials are transverse, and $S$ acts on the corresponding manifold without fixed points.  The quotient is thus smooth, and its Euler number is given by $-24/3 = -8$.  To determine the Hodge numbers we can pass to a representation which exhibits all the $(1,1)$-forms as hyperplane sections:
$$
X^{9,21}~=~~
\cicy{\IP^1 \\ \IP^1 \\ \IP^1 \\ \IP^1 \\ \IP^1 \\ \IP^1 \\ \IP^2 \\ \IP^2 \\ \IP^2}
{\one & 0 & 0 & 0 & 0 & 0 & \one & 0 & 0 \\
0 & \one & 0 & 0 & 0 & 0 & 0 & \one & 0 \\
0 & 0 & \one & 0 & 0 & 0 & 0 & 0 & \one \\
0 & 0 & 0 & \one & 0 & 0 & \one & 0 & 0 \\
0 & 0 & 0 & 0 & \one & 0 & 0 & \one & 0 \\
0 & 0 & 0 & 0 & 0 & \one & 0 & 0 & \one \\
\one & \one & \one & 0 & 0 & 0 & 0 & 0 & 0 \\
0 & 0 & 0 & \one & \one & \one & 0 & 0 & 0 \\
0 & 0 & 0 & 0 & 0 & 0 & \one & \one & \one \\}^{9,21}_{-24}
$$
The action of $S$ is the same on the $\IP^1$'s, and now acts internally on each $\IP^2$ separately.  This leaves $5$ independent hyperplane sections after taking the quotient, giving $h^{11} = 5$.  Along with $\chi = -8$, this determines $h^{21} = 9$, which corresponds to the 9 coefficients visible in~\eqref{eq:9,21polys}.  In summary we have found a manifold with fundamental group
$\IZ_3$ and Hodge numbers $\hodgenos = (5,9)$.
\subsubsection{$Y^{6,33}$; the transpose of $X^{4,40}$} \label{sec:Y6,33}
The transpose of $X^{4,40}$ has Hodge numbers $\hodgenos = (6,33)$, which are the same as those of the manifold $X^{6,33}$ in section \ref{sec:6,33}.  We do not know if these two manifolds are the same, so we include both, and distinguish the new one by labelling it $Y^{6,33}$:
$$
Y^{6,33}~=~~
\cicy{\IP^1 \\ \IP^1 \\ \IP^1 \\ \IP^2 \\ \IP^2}
{\one & \one & 0 & 0 \\
\one & 0 & \one & 0 \\
\one & 0 & 0 & \one \\
0 & \one & \one & \one \\
0 & \one & \one & \one \\}^{6,33}_{-54}
$$

The corresponding diagrams are
$$
\includegraphics[width=1.75in]{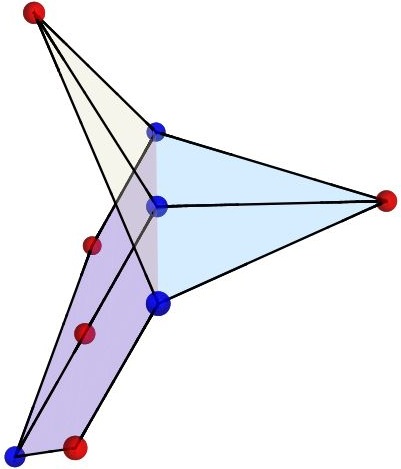}\hskip1in
\includegraphics[width=1.75in]{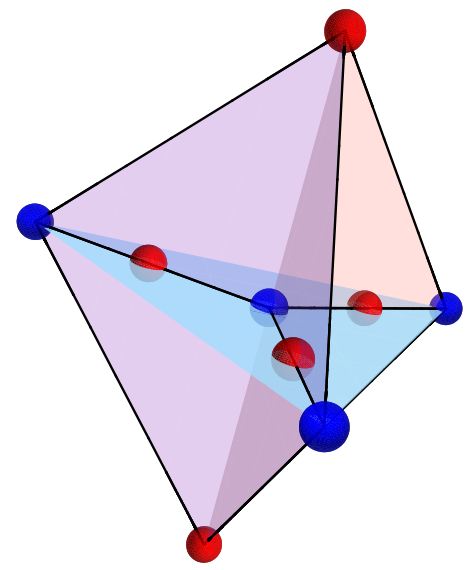}
$$
This manifold is clearly a split of $X^{3,48}$, but it is also a split of $X^{5,59}$, which can be seen by contracting one of the $\IP^2$'s.  Since $X^{6,33}$ is also a (similar) split of $X^{5,59}$, this can be considered circumstantial evidence that these two manifolds are in fact the same.

Introduce coordinates $s_{ia}$ on the three $\IP^1$'s, as well as $u_i$ and $v_i$ on the two $\IP^2$'s.  Denote by~$r$ the polynomial trilinear in the $s$'s, and by $p_i$ the three trilinear polynomials.  Define also a 
$\IZ_3$ action with generator
$$
S:~s_{ia} \to s_{i+1,a}~,~~u_i \to u_{i+1}~,~~v_i \to v_{i+1}~;~~p_i \to p_{i+1}
$$
A specific choice of coordinates allows us to put polynomials consistent with the symmetry in the form
\beq \label{eq:Y6,33}
\begin{split}
r =&~ \frac 13 m_{000} + m_{110} \\
p_i =&~ s_{i0}\, u_i\, v_i +  \sum_{j,k=1}^2 B_{jk}\, s_{i0}\, u_{i+j}\, v_{i+k} + \sum_{j,k} C_{jk}\, s_{i1}\, u_{i+j}\, v_{i+k}
\end{split}
\eeq
where $m_{abc}$ is defined as before.  These polynomials are transverse, and $S$ acts freely on the corresponding manifold.  Therefore we obtain a smooth quotient with Euler number $-54/3 = -18$.  To determine the individual Hodge numbers we again examine an equivalent representation in which all $(1,1)$-forms arise from the pullbacks of hyperplane sections:
$$
Y^{6,33}~=~~
\cicy{\IP^1 \\ \IP^1 \\ \IP^1 \\ \IP^2 \\ \IP^2 \\ \IP^2}
{\one & 0 & 0 & \one & 0 & 0 \\
0 & \one & 0 & 0 & \one & 0 \\
0 & 0 & \one & 0 & 0 & \one \\
\one & \one & \one & 0 & 0 & 0 \\
0 & 0 & 0 & \one & \one & \one \\
0 & 0 & 0 & \one & \one & \one \\}^{6,33}_{-54}
$$
The action of $S$ on the new $\IP^2$ is the same as its action on the others.  We see that since the three $\IP^1$'s get identified, the quotient will have $h^{1,1} = 4$.  Since $\chi = -18$, we deduce that 
$h^{21} = 13$, and indeed this agrees with the number of parameters in the equations~\eqref{eq:Y6,33}.  In summary, the quotient has fundamental group $\IZ_3$ and Hodge numbers $\hodgenos = (4,13)$.  These data agree with those for the $\IZ_3$ quotient of $X^{6,33}$, increasing our suspicion that this may be the same manifold.
\subsection{The branches $\IP^5[3,\,3]\to X^{2,56}\to X^{3,39}\to X^{6,33}\to X^{9,27}$
and $X^{2,56}\to X^{5,50}$}
\subsubsection{$\IP^5[3,3]$} \label{sec:1,73}
The fact that $\IP^5[3,3]$ admits a freely acting $\IZ_3{\times}\IZ_3$ symmetry has been known for some time. We recall this in order to be able to take a further quotient by a $\IZ_2$ generator, $V$, that does not act freely. This generator reappears in splits of $\IP^5[3,3]$ including in a configuration that we study in \SS\ref{sec:9,27} and \SS\ref{sec:3gen} whose quotient by $V$ leads, after resolution, to a manifold with $\chi=-6$.

We take coordinates $(x_0,x_1,x_2,y_0,y_1,y_2)$ for $\IP^5$ and impose two cubics, $p_1$ and $p_2$, that we take to be invariant under the symmetries generated by
\beq
\begin{split}
S:&~x_i~\to~x_{i+1}~,~~y_i~\to~y_{i+1}~,\\[5pt]
T:&~x_i \to \zeta^i x_i~,~~y_i \to \zeta^i y_i
\end{split}
\notag\eeq
with $\z$ a non-trivial cube root of unity. To write suitable invariant equations define the following invariant cubics
\begin{align*}
\cX_0~&=~\sum_i x_i^3~,\hskip1in & \cY_0~&=~\sum_i y_i^3~,\\
\cX_1~&=~\sum_i x_i y_i^2~,& \cY_1~&=~\sum_i x_i^2 y_i~,\\
\cX_2~&=~\sum_i x_i y_{i+1} y_{i+2}~,& \cY_2~&=~\sum_i x_i x_{i+1} y_{i+2}~,\\
\cX_3~&=~x_0 x_1 x_2~,& \cY_0~&=~y_0 y_1 y_2~.
\end{align*}
In terms of these we form invariant polynomials
$$
p_1~=~\sum_{\a=0}^3\big( A_\a\,\cX_\a + B_\a\,\cY_\a \big)~,~~~
p_2~=~\sum_{\a=0}^3\big( C_\a\,\cX_\a + D_\a\,\cY_\a \big)~,
$$
and one can check that these equations are transverse for generic values of the coefficients, though this is easier to check for the more symmetric polynomials that we shall write below.
To show that the group generated by $S$ and $T$ acts without fixed points it is sufficient to show that $ST^k$ acts without fixed points for $k=0,1,2$, and this is straightforward. There are 16 coefficients in $p_1$ and $p_2$ and, up to scale, a three-parameter freedom to redefine the coordinates in a way that preserves the symmetry
$$
x_i~\to~\a\, x_i + \b\, y_i~,~~~y_i~\to~\g\, x_i + \d\, y_i~.
$$
There is also a four-parameter freedom to redefine $p_1$ and $p_2$
$$
p_1~\to~ a\, p_1 + b\, p_2~,~~~p_2~\to~ c\, p_1 + d\, p_2~.
$$
Thus there are $16-3-4=9$ true degrees of freedom in the equations. For the quotient it is still the case that $h^{11}=1$ and $\chi=-144/9=-16$ so $h^{21}=9$, confirming our counting of parameters.

Consider now the the specialisation of the defining equations with $C_\a=B_\a$ and $D_\a=A_\a$
\beq
p_1~=~\sum_{\a=0}^3\big( A_\a\,\cX_\a + B_\a\,\cY_\a \big)~,~~~
p_2~=~\sum_{\a=0}^3\big( B_\a\,\cX_\a + A_\a\,\cY_\a \big)~.
\label{eq:1,73Z3Z3V}\eeq
These equations are covariant under the $\IZ_2$-generator 
$$
V:~x_i~\leftrightarrow~y_i~;~~p_1~\leftrightarrow~p_2~.
$$
We have checked that the equations \eref{eq:1,73Z3Z3V} are transverse.
Now the counting is that there are 8 coefficients, a one-parameter family of coordinate redefinitions
$x_i\to \a x_i + \b y_i$, $y_i\to \b x_i + \a y_i$ and a two-parameter family of redefinitions of the defining equations $p_1\to a p_1 + b p_2$, $p_2\to b p_1 + a p_2$. Thus there are now $8-1-2=5$ degrees of freedom in the defining equations. Checking for fixed points we see that if $S^k T^\ell V$ has fixed points then so has 
$V$. The fixed points of $V$, in the embedding space, are of the form $(x_i, x_j)$ and $(x_i, -x_j)$ and these correspond to two disjoint elliptic curves
$$
\cE_\pm~:~~~\sum_\a \big( A_\a \pm B_\a \big) \cX_\a(x,x)~=~0~.
$$
The fixed curves are resolved by replacing a tubular neighborhood of each by bundle of ball neighborhoods of the origin in $EH_2$ which is the $A_1$ surface resolution. Each elliptic curve therefore introduces an extra 
$(1,1)$-form and an extra $(2,1)$-form. Thus $h^{1,1}=1+2=3$ and $h^{2,1}=5+2=7$. Since the elliptic curves have Euler number zero the resolutions do not affect the fact that the Euler number divides so the resolved manifold has $\chi=-16/2=-8$ which is a check on our computation of the Hodge numbers.

We summarize with a table the manifolds that we have obtained from quotients of the manifold 
$X^{1,73}=\IP^5[3,3]$.
\vskip10pt
\begin{table}[H]
\begin{center}
\def\str{\vrule height16pt width0pt depth8pt}
\begin{tabular}{| l | c | c | c | c | c | c |}
\hline 
\str Hodge numbers & ~$(1,9)$~ & ~$(1,25)$~  & ~$(1,73)$~ & ~$(3,7)$~ & ~$(3,15)$~ & ~$(3,39)$~ \\
\hline
\str Manifold & $X^{1,73}\quotient{\IZ_3{\times}\IZ_3}$ & $X^{1,73}\quotient{\IZ_3}$
& $X^{1,73} $ & \simpentry{\widehat{X^{1,73}\quotient{\IZ_3{\times}\IZ_3{\times}\IZ_2}}}
& \simpentry{\widehat{X^{1,73}\quotient{\IZ_3{\times}\IZ_2}}} 
& \simpentry{\widehat{X^{1,73}\quotient{\IZ_2}}}\\
\hline
\str Fundamental group           & & & & $\IZ_3{\times}\IZ_3$ & $\IZ_3$ & $\bf 1$\\
\hline
\end{tabular}
\parbox{5.5in}{\caption{\small Hodge numbers for quotients of $X^{1,73}$.}}
\end{center}
\end{table}
\vskip10pt
\subsubsection{$X^{2,56}$; splitting the first polynomial} \label{sec:2,56}
By splitting one of the cubics which define $\IP^5[3,\,3]$ we obtain the Calabi-Yau matrix
$$
X^{2,56}~=~~
\cicy{\IP^2\\ \IP^5}
{1 & 1 & 1 & 0 \\
 1 & 1 & 1 & 3 \\}^{2,56}_{-108}
\hskip0.5in
\lower0.3in\hbox{\includegraphics[width=1.5in]{fig_2,56_manifold.pdf}}
$$
We take coordinates $u_i$ for the $\IP^2$, and write $(x_0,x_1,x_2,y_0,y_1,y_2)$ for those of the $\IP^5$, allowing us to define a $\IZ_3{\times}\IZ_3$ action with generators
\beq
\begin{split}
S:&~u_i \to u_{i+1}~,~~x_i \to x_{i+1}~,~~y_i \to y_{i+1} \\[5pt]
T:&~u_i \to \zeta^i u_i~,~~x_i \to \zeta^i x_i~,~~y_i \to \zeta^i y_i
\end{split}
\notag\eeq
where $\zeta$ is a non-trivial cube root of unity.  If we denote by $p_i$ the three bilinear equations, and by
$q$ the cubic, we can demand that $q$ be invariant under the group action, and that the $p_i$ transform as
\beq
S:~p_i \to p_{i+1}~,~~T:~p_i \to \zeta^{-i}p_i
\notag\eeq
The most general polynomials which satisfy these conditions are
\beq
\begin{split}
p_i &= \sum_k\, (A_k\, x_{i+k} + B_k\, y_{i+k}) u_{i-k} \\
q &= \sum_i\sum_{k=0}^1 \, (C_k\, x_i\, x_{i-k}\, x_{i+k} + D_k\, x_i\, x_{i-k}\, y_{i+k} + E_k\, x_i\, y_{i-k}\, y_{i+k} + F_k\, y_i\, y_{i-k}\, y_{i+k} )
\end{split}
\notag\eeq
We have checked that the elements $S,T,ST,ST^2$ all act on the resulting manifold without fixed points, and this is enough to guarantee the same for the entire group.  Furthermore, the above polynomials are transverse, so the $\IZ_3{\times}\IZ_3$ quotient is smooth.  The Euler number of the quotient is $-12$.  The group action is trivial on the second cohomology, so $h^{11}$ will not change upon taking the quotient.  This is enough to determine that the Hodge numbers of the resulting manifold are $(2,8)$.  The threefold cover of this manifold, obtained by taking the quotient by just a single $\IZ_3$ factor, will have Euler number $-36$ and Hodge numbers $(2,20)$.

We can confirm the values of $h^{21}$ by counting the parameters in the defining equations.  We begin with the original manifold.  There are $3{\times}3{\times}6 + 6{\times}7{\times}8/3! = 110$ terms in the most
general polynomials.  From this we subtract $45$ for coordinate redefinitions, $9$ for redefining 
$p_i \to \sum_j \k_{ij}\, p_j$ and $1$ for rescaling $q$, giving $57$.  Finally, for a solution to exist we must have $\det (\partial p_i/\partial u_j) = 0$, and the freedom to redefine $q$ by multiples of this determinant eliminates one more coefficient, leaving us with $56$ true degrees of freedom.  For the polynomials symmetric under only $S$, the $p_i$ contain $3{\times}6 = 18$ independent coefficients, while a quick computer calculation shows that $q$ contains $20$.  The allowed coordinate changes are 
$x_i \to \sum_j (\a_j\, x_{i+j} + \b_j\, y_{i+j})$, $y_i \to \sum_j (\g_j\, x_{i+j} + \d_j\, y_{i+j})$,
$u_i \to \sum_j \e_j\, u_{i+j}$,
which have $13$ parameters, neglecting scale.  There is a 3-parameter freedom to redefine the $p_i$ according to $p_i \to \sum_j \k_j\, p_{i+j}$, and finally we can rescale $q$ as well as applying the determinant constraint, eliminating two more coefficients. This leaves us with $38-(13+3+2) = 20$ meaningful coefficients, agreeing with the value of $h^{21}$ determined above.

The $\IZ_3{\times}\IZ_3$ symmetric polynomials start with $2{\times}3 + 4{\times}2 = 14$ terms.  The allowed coordinate changes are $x_i \to \a x_i + \b y_i,~y_i \to \g x_i + \d y_i$, which have three parameters, neglecting scale.  We can now only rescale the $p_i$ by a common factor, as well as rescaling $q$ and taking account of the freedom to redefine $q$ by multiples of $\det (\partial p_i/\partial u_j)$, which altogether eliminates three more coefficients, leaving $14 - 6 = 8$, which agrees with the previous determination of $h^{21}$.

In summary, we have found two smooth quotients:
\vskip10pt
\begin{table}[H]
\begin{center}
\def\str{\vrule height16pt width0pt depth8pt}
\begin{tabular}{| c | c | c |}
\hline 
\str $\hodgenos\left(X^{(2,56)}/G\right)$ & ~~$(2,8)$~~ & ~~$(2,20)$~~ \\
\hline
\str $G$ & ~$\IZ_3{\times}\IZ_3$~ & $\IZ_3$ \\
\hline
\end{tabular}
\parbox{5.5in}{\caption{\small Hodge numbers of smooth quotients of $X^{2,56}$.}}
\end{center}
\end{table}
\vskip10pt
\subsubsection{$X^{3,39}$; splitting the second polynomial} \label{sec:3,39}
The next manifold we consider is a split of both  $\IP^5[3,3]$ and its transpose.
$$
X^{3,39}~=~~
\cicy{\IP^2\\ \IP^2\\ \IP^5}
{\one & \one & \one & 0 & 0 & 0 \\
  0 & 0 & 0 & \one & \one & \one \\
  \one & \one & \one & \one & \one & \one \\}^{3,39}_{-72}
\hskip0.75in
\lower0.3in\hbox{\includegraphics[width=2.0in]{fig_3,39_manifold.pdf}}
$$
Clearly contracting the two $\IP^2$-rows brings us back to $\IP^5[3,3]$ while contracting the $\IP^5$-row takes us to the transpose. Note also a first parallel with our split of the quintic: as in that case the Euler number of the split manifold is half the Euler number of the original manifold. 

Let us take coordinates $u_j$ and $v_k$ for the two $\IP^2$'s and write 
$(x_0,\,x_1,\,x_2,\,y_0,\,y_1,\,y_2)$ for the coordinates of the $\IP^5$. We denote by $p_i$ the first three polynomials and by $q_i$ the remaining three. With these conventions the polynomials have the form
\beq
\begin{split}
p_i~&=~\sum_{j,k}(A_{ijk}\,x_k + B_{ijk}\,y_k)\, u_j \\
q_i~&=~\sum_{j,k}(C_{ijk}\,x_k + D_{ijk}\,y_k)\, v_j \\
\end{split}\notag\eeq
where the indices $i,j,k$ are understood to take values in $\IZ_3$. We take the equations to be covariant under the action of a generator
$$
S:~(x_i,\, y_j)~\to~(x_{i+1},\, y_{j+1})~;~~(u_i,\, v_j)~\to~(u_{i+1},\, v_{j+1})~;~~
(p_i,\, q_j)~\to~(p_{i+1},\, q_{j+1})~.
$$
Covariance of the polynomials again requires the relations $A_{i,j,k}~=~A_{i-k,\,j-k,\,0}$
with analogous relations for $B$, $C$ and $D$. By setting $k=2(i+j+m)$ and writing 
$a_{\ell m}=A_{-\ell+m,\, \ell+m,\,0}$ we bring the polynomials to the form
\beq
\begin{split}
p_i~&=~\sum_{j,m}\big(a_{i-j,\, m}\,x_{2(i+j+m)} + b_{i-j,\, m}\,y_{2(i+j+m)}\big)\, u_j \\
q_i~&=~\sum_{j,m}\big(c_{i-j,\, m}\,x_{2(i+j+m)} + d_{i-j,\, m}\,y_{2(i+j+m)}\big)\, v_j
\end{split}\label{Zthreesplit}
\eeq 
The parameter count is that there $4{\times}3{\times}3=36$ coefficients and from this we need to subtract the number of degrees of freedom in making changes of coordinates that preserve the symmetry under $T$, and the number of degrees of freedom in redefining the polynomials in a way that preserves the symmetry.
Neglecting an overall change of scale, there is a two parameter freedom in changing 
$u_j\to\sum_k g_k u_{j+k}$ and similarly for $v$. Again neglecting an overall scale, there is an 11 parameter freedom to redefine the coordinates $x$ and $y$
$$
x_i~\to~\sum_k (\a_k\, x_{i+k} + \b_k\, y_{i+k})~~~,~~~
y_i~\to~\sum_k (\g_k\, x_{i+k} + \d_k\, y_{i+k})~,
$$
(in this context the $\d_k$ denote parameters, rather than the Kr\"onecker symbol). Finally there is a 3 parameter freedom to redefine $p_i\to \sum_k \k_k\, p_{i+k}$ and similarly for the $q_i$, where we here do permit a scaling of the polynomials. The count is that we have a total of $36 - 21=15$ free parameters.

It is straightforward to show that the equations \eref{Zthreesplit} are fixed point free and transverse. Hence we have checked the existence of the quotient
$$
\cicy{\IP^2\\ \IP^2\\ \IP^5}
{\one & \one & \one & 0 & 0 & 0 \\
  0 & 0 & 0 & \one & \one & \one \\
  \one & \one & \one & \one & \one & \one \\}^{3,15}\quotient{\IZ_3}{~\lower15pt\hbox{.}}
$$
For the quotient $h^{11}=3$ and $\chi=-24$, so $h^{21}=15$, confirming our parameter~count.

If we now seek to impose the second symmetry
$$
T:~(x_i,\,y_j)~\to~(\o^i x_i,\,\o^j y_j)~;~~(u_i,\,v_j)~\to~(\o^i u_i,\,\o^j v_j)~;~~
(p_i,\,q_j)~\to~(\o^{2i} p_i,\,\o^{2j} q_j)~.
$$
where again $\o$ is a non-trivial cube root of unity, then we must take $a_{\ell, m}=0$ for $m\neq 0$ and similarly for $b$, $c$ and $d$. Dropping the second index on the coefficients, we have
\beq
\begin{split}
p_i~&=~\sum_{j}\big( a_{i-j}\,x_{2(i+j)} + b_{i-j}\,y_{2(i+j)} \big)\, u_j \\
q_i~&=~\sum_{j}\big( c_{i-j}\,x_{2(i+j)} + d_{i-j}\,y_{2(i+j)} \big)\, v_j
\end{split}\label{ZthreeZthreesplit}
\eeq 
In contradistinction to the quintic case these equations are fixed point free and transverse for general values of the coefficients. Now there are 12 coefficients, a three-parameter family of coordinate redefinitions up to scaling
$$
x_i~\to~\a\, x_{i} + \b\, y_{i}~~~,~~~
y_i~\to~\g\, x_{i} + \d\, y_{i}~,
$$
and two rescalings, $p_i\to\tilde\a\, p_i$ and $q_i\to\tilde\b\, q_i$ of the polynomials. Thus the polynomials \eref{ZthreeZthreesplit} contain $12-3-2=7$ parameters. The Euler number is now $-24/3=-8$ and $h^{11}=3$, as previously, so $h^{21}=7$, in agreement with our counting of parameters. A convenient way to represent the 7 parameters is to demand that $c_j=b_j$ and that $a_0=1$ and $b_0=0$, leaving the two free components of $a_j$, the two free components of $b_j$ and the three $d_j$.

We have thus checked the existence of the $\IZ_3{\times}\IZ_3$ quotient
$$
\cicy{\IP^2\\ \IP^2\\ \IP^5}
{\one & \one & \one & 0 & 0 & 0 \\
  0 & 0 & 0 & \one & \one & \one \\
  \one & \one & \one & \one & \one & \one \\}^{3,7}\quotient{\IZ_3{\times}\IZ_3}{~\lower15pt\hbox{.}}
$$

Another possibility is to impose, instead of the second $\IZ_3$ symmetry, a $\IZ_2$ symmetry, generated by
$$
U:~u \leftrightarrow v~,~~x_i \leftrightarrow y_i
$$
Clearly this must act on the polynomials as $p_i \leftrightarrow q_i$, which forces them to take the form
\beq
\begin{split}
p_i~&=~\sum_{j,m}\big(a_{i-j,\, m}\,x_{2(i+j+m)} + b_{i-j,\, m}\,y_{2(i+j+m)}\big)\, u_j \\
q_i~&=~\sum_{j,m}\big(b_{i-j,\, m}\,x_{2(i+j+m)} + a_{i-j,\, m}\,y_{2(i+j+m)}\big)\, v_j
\end{split}
\eeq 
Again $S$ acts without fixed points, but $U$ fixes two curves in the manifold.  To see this, note that fixed points of $U$ in the ambient space are those for which $y_i = \l\, x_i$, with $\l=\pm 1$, and $v_i = u_i$.  At these points, we have $q_i = \l\, p_i$, and
$$
p_i = ~\sum_{j,m}\big(a_{i-j,\, m} + \l\, b_{i-j,\, m}\big)\,x_{2(i+j+m)}\, u_j
$$
We see then that the two fixed curves $\cE_\pm$ are in fact CICY's corresponding to 
$$
\cicy{\IP^2\\ \IP^2}
{1&1&1 \\
  1&1&1}
$$
and thus elliptic curves.  The $U$-quotient is resolved by removing a neighbourhood of each of the fixed curves, taking the quotient and replacing the missing sets by bundles of 4-balls taken from the center of $E\! H_2$, the two dimensional Eguchi-Hanson space, which is asymptotic at infinity to $S^3/\IZ_2$. Since the Euler numbers of the $\cE_\pm$ are zero the Euler number of the resolved quotient is simply $-72/6=-12$. The generator $U$ identifies the two $\IP^2$'s so the embedding spaces now contribute only 2 to $h^{11}$. An $E\! H_2$ has a single $(1,1)$-form, $\r$, and an elliptic curve has a holomorphic $(1,0)$-form, $\eta$. In this way we obtain two additional $(1,1)$-forms $\r_\pm$ and two additional $(2,1)$-forms $\r_\pm\wedge\eta_\pm$  that arise from the resolution of the fixed curves.  The resolved orbifold therefore has fundamental group $\IZ_3$, $\chi = -12$, and Hodge numbers $(4,10)$.
 
We can also obtain $h^{21} = 10$ more directly.  There are 18 parameters in the polynomials, but we are free to make the following coordinate changes consistent with the action of the~group:
\beq
\begin{split}
 x_i \to \sum_j (\a_j\, x_{i+j} + \b_j\, y_{i+j})~,&~~y_i \to \sum_j (\b_j\, x_{i+j} + \a_j\, y_{i+j}) \\
 u_i \to \sum_j \g_j\, u_{i+j}~,&~~ v_i \to \sum_j \g_j\, v_{i+j}
\end{split}
\notag\eeq
We neglect overall scaling, so this is a 7 parameter freedom.  We can also redefine the polynomials, in a way consistent with the symmetries:
\beq
p_i \to \sum_j \k_j\, p_{i+j}~,~~ q_i \to \sum_j \k_j\, q_{i+j}~.
\notag\eeq
Here the overall scale is significant, so we have another 3 parameters, giving us 10 in total.  Subtracting this from the 18 parameters in the original equations leaves 8, which together with the two $(2,1)$-forms coming from the resolutions gives an independent determination of $h^{21}=10$.

We summarise the manifolds found in this section in the following table:
\vskip10pt
\begin{table}[H]
\begin{center}
\def\str{\vrule height16pt width0pt depth8pt}
\begin{tabular}{| l | c | c | c|}
\hline 
\str Hodge numbers & ~~$(3,7)$~~ & ~~$(3,15)$~~ & ~~$(4,10)$~~ \\
\hline
Manifold & ~$X^{3,39}\quotient{\IZ_3{\times}\IZ_3}$~ & ~$X^{3,39}\quotient{\IZ_3}$~
&~ $\simpentry{\widehat{X^{3,39}\quotient{\IZ_3{\times}\IZ_2}}}$ \\
\hline
\str Fundamental group           & & & $\IZ_3$ \\
\hline
\end{tabular}
\parbox{5.5in}{\caption{\small Hodge numbers of smooth quotients of $X^{3,39}$.}}
\end{center}
\end{table}
\subsubsection{$X^{6,33}$; a second split of the first cubic} \label{sec:6,33}
We may proceed to a manifold which is an analogue of the split quintic of \SS\ref{sec:6,36}. We do this in two steps; the first consists in splitting the first $\IP^2$. Consider the manifold
\beq
X^{(6,33)}~=~
\cicy{\IP^1\\ \IP^1\\ \IP^1\\ \IP^2\\ \IP^5\\}
{\one & \one & 0 & 0 & 0 & 0 & 0 \\
 \one & 0 & \one & 0 & 0 & 0 & 0 \\
 \one & 0 & 0 & \one & 0 & 0 & 0 \\
 0 & 0 & 0 & 0 & \one & \one & \one \\
 0 & \one & \one & \one & \one & \one & \one \\}^{6,33}_{-54}
\hskip0.4in
\lower0.35in\hbox{\includegraphics[width=2.5in]{fig_6,33_manifold.pdf}}\hskip0.2in
\label{(6,33)manifold}\eeq
Take coordinates $s_{ia}$ for the three $\IP^1$'s, $v_i$ for the $\IP^2$ and $(x_i,y_j)$ for the 
$\IP^5$.  If we impose the $\IZ_3$ symmetry generated by
\beq
S:~ s_{ia} \to s_{i+1,a} ~;~ v_i \to v_{i+1} ~;~ (x_i,y_j) \to (x_{i+1}, y_{j+1})~,
\notag\eeq
then, by choice of coordinates, the defining polynomials may be brought to the form
\beq
\begin{split}
P    ~&=~\frac{1}{3}(m_{000}+m_{111})~=~s_{00}s_{10}s_{20} + s_{01}s_{11}s_{21} \\[5pt]
p_i ~&=~x_i\, s_{i0} + y_i\, s_{i1}\\[5pt]
q_i ~&=~\sum_{j,k}(A_{i-j,i-k}\, x_j + B_{i-j,i-k}\, y_j)\, v_k~,
\end{split}\label{shorteqs}
\eeq
where $m_{a,b,c} = \sum_{i}s_{ia}s_{i+1,b}s_{i+2,c}$.  These equations are transverse,  
$S$ acts on the resulting manifold without fixed points and the quotient has Euler number 
$-54/3 = -18$. There are 18 coefficients in the polynomials and we still have freedom to make the transformations \hbox{$v_i\to\sum_{j} \a_{i-j} v_j$}, in which there are two parameters (up to scale), and to redefine $q_i\to\sum_{j} \k_{i-j} q_j$, in which there are three parameters;  this suggests $h^{21}=18-5=13$. 
The $S$-action identifies the three $\IP^1$'s suggesting that $h^{11}=6-2=4$. The values of the Hodge numbers that are suggested are in fact correct, as we will show below by considering a representation for $X^{(6,33)}$ that has 6 embedding spaces so that the action of $S$ on $H^2$ can be seen explicitly. In this way we will establish the existence of the quotient
\beq
\cicy{\IP^1\\ \IP^1\\ \IP^1\\ \IP^2\\ \IP^5\\}
{\one & \one & 0 & 0 & 0 & 0 & 0 \\
 \one & 0 & \one & 0 & 0 & 0 & 0 \\
 \one & 0 & 0 & \one & 0 & 0 & 0 \\
 0 & 0 & 0 & 0 & \one & \one & \one \\
 0 & \one & \one & \one & \one & \one & \one \\}^{4,13}\quotient{\IZ_3}
\notag\eeq
 
In order to exhibit all of $H^2$ as the K\"ahler forms of ambient spaces we consider the representation 
$$
X^{(6,33)}~=~~~
\cicy{\IP^1\\ \IP^1\\ \IP^1\\ \IP^2\\ \IP^2\\ \IP^5\\}
{\one & 0 & 0 & \one & 0 & 0 & 0 & 0 & 0 \\
 0 & \one & 0 & 0 & \one & 0 & 0 & 0 & 0 \\
 0 & 0 & \one & 0 & 0 & \one & 0 & 0 & 0 \\
 \one & \one & \one & 0 & 0 & 0 & 0 & 0 & 0 \\
 0 & 0 & 0 & 0 & 0 & 0 & \one & \one & \one \\
 0 & 0 & 0 & \one & \one & \one & \one & \one & \one \\}^{6,33}_{-54}
$$
We see that this manifold is indeed isomorphic to that given by \eref{(6,33)manifold} by noting that contracting the first $\IP^2$ returns us to \eref{(6,33)manifold} while the fact that the Euler numbers are the same ensures that the contraction does not introduce any nodes. This representation corresponds to the diagram
\vskip20pt
\vbox{
$$
\includegraphics[width=3.0in]{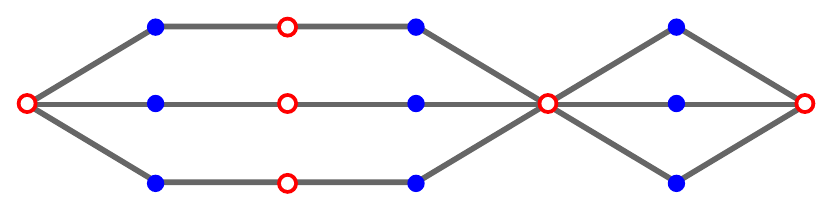}
$$
\vskip0pt
\place{1.65}{0.57}{$u$}
\place{4.75}{0.57}{$v$}
\place{2.25}{1.08}{$P_i$}
\place{2.75}{1.05}{$s_i$}
\place{3.20}{1.05}{$p_i$}
\place{3.55}{0.85}{$(x,y)$}
\place{4.15}{1.05}{$q_i$}
}
We assign coordinates $u_i$ to the new $\IP^2$. The polynomials $p_i$ and $q_i$ are as before but 
$P$ has been replaced by three equations which, by suitable choice of $u$, can be brought to the form~shown
\beq
\begin{split}
P_i ~&=~u_i\, s_{i0} + \Big(\sum_j D_{i-j}\, u_j\Big)\,s_{i1} \\
p_i ~&=~x_i\, s_{i0} + y_i\, s_{i1}\\[7pt]
q_i ~&=~\sum_{j,k}(A_{i-j,i-k}\, x_j + B_{i-j,i-k}\, y_j)\, v_k,
\end{split}\label{extendedeqs}
\eeq
The form of these equations is preserved by the transformations
\beq
\begin{split}
s_{i0}~&\to~\a\, s_{i0} + \b\, s_{i1}\\[5pt]
x_i      ~&\to~\d\, x_i - \g\, y_i\\[7pt]
&\hskip35pt u~\to~\big(\a\, {\bf 1}+\g\, D\big)^{-1}\, u
\end{split}
\hskip-30pt
\begin{split}
s_{i1}~&\to~\hphantom{-}\g\, s_{i0} + \d\, s_{i1}\\[5pt]
y_i      ~&\to~-\b\, x_i + \a\, y_i\\[7pt]
           ~&
\end{split}\notag
\eeq
where $D$ denotes the matrix $(D_{i-j})$. The upshot is that the matrix $D$ in the first of 
Eqs~\eref{extendedeqs} can be transformed
$$
D~\to~\widetilde{D}~=~\big(\b\,{\bf 1} + \d\, D\big)\big(\a\,{\bf 1} + \g\, D\big)^{-1}
$$
and we may use this freedom to set, say, $D_0=D_1=0$ and $D_2=1$. With this choice the equations $P_i$ become
$$
P_i~=~u_i\,s_{i0} + u_{i+1}\,s_{i1}~.
$$
These are three equations in the three coordinates $u_i$ so the determinant of the matrix of coefficients must vanish yielding the condition
$$
\det\left(\begin{matrix}
s_{00} & s_{01} & 0        \\
        0 & s_{10} & s_{11}\\
s_{21} &     0     & s_{20}\\
\end{matrix}\right) ~=~ s_{00}s_{10}s_{20} + s_{01}s_{11}s_{21}~=~0
$$
and, in this way, we recover the equation $P=0$ of \eref{shorteqs}. If we take 
$S:~u_i\to u_{i+1}$ then the equations \eref{extendedeqs} are covariant, transverse and fixed point free. Notice that in passing from $P$ to the three equations $P_i$ no new parameters have been introduced so the number of free parameters in the equations remains 13. Since $S$ identifies the three $\IP^1$'s we now see explicitly that, for the quotient, $h^{11}=4$ and since $\chi=-18$ we have $h^{21}=13$ confirming the parameter count.

The polynomials $P$ and $p_i$ exhibit a second $\IZ_3$ symmetry and by specialising the polynomials $q_i$ so that the polynomials \eref{shorteqs} now take the form
\beq
\begin{split}
P    ~&=~s_{00}s_{10}s_{20} + s_{01}s_{11}s_{21} \\[5pt]
p_i ~&=~x_i\, s_{i0} + y_i\, s_{i1}\\[5pt]
q_i ~&=~\sum_j (a_{i-j}\, x_j\, v_{2i+2j} + b_{i-j}\, y_j\, v_{2i+2j+1})
\end{split}\label{Tshorteqs}
\eeq
we see that we have a symmetry
$$
T:~(s_{i,0},\,s_{i,1}) \to (s_{i,0},\,\o s_{i,1})~~;~(x_i,\, y_j) \to (\o^i\, x_i,\, \o^{j+2}\, y_j)~~;~
v_i \to \o^i\, v_i~~;~q_i \to \o^{2i}\,q_i~,
$$
with $\o^3=1$. The polynomials \eref{Tshorteqs} are transverse, indeed checking this is the simplest way to check the transversality of \eref{shorteqs}. The action of $T$ however has fixed points. The fixed points for 
$(x,y)$ are $\IP^1$'s equivalent under $S$ to 
$$
(x_*,\,y_*)~=~(\l,0,0,0,\m,0)
$$
and the fixed point of $v$ is given by $v_{*,i}=\d_{i,m}$ for some $m$.  At these points, the only nontrivial equation among the $q_i$ is
$$
q_{2m}~=~a_{2m}\,\l + b_{2m-1}\,\m~=~0
$$
which fixes a point in the $\IP^1$. The fixed points of the $s_i$ are $s_{i,*}=(0,1)$ or $(1,0)$ for each $i$.
The choice of fixed points is constrained by the $p_i$, for which the nontrivial equations are
$$
p_0~=~\l\, s_{00}~=~0~~~\text{and}~~~p_1~=~\m\, s_{11}~=~0~.
$$
For generic coefficients, these are consistent with the previous conditions only if $s_{00}=s_{11}=0$. We are left with two choices for the fixed points of the $s_i$
$$
(s_{ia})~=~\Big\{\,\big( (0,1),\,(1,0),\,(1,0)\big),\,\big( (0,1),\,(1,0),\,(0,1)\big)\,\Big\}
$$
both of which satisfy the equation $P=0$. Combining these choices with the three choices for the fixed points of  $v$ gives a total of 6 fixed points on the manifold $X^{(6,33)}/S$. The fixed points of the $T$-quotient are resolved by replacing a ball neighborhood of each of the fixed points by a ball neighborhood of the origin of the three-dimensional Eguchi-Hanson
 manifold $E\!H_3$, which has boundary $S^5/\IZ_3$. Each $E\!H_3$ contributes 1 to the count of $(1,1)$-forms so, for the resolution, $h^{11}=4+6=10$. There are 5 parameters visible in \eref{Tshorteqs} and the resolutions of the singularities do not introduce new complex structure parameters, so we expect $h^{21}=5$ for the resolution. The Euler number of $E\!H_3$ is 3 so the Euler number of the resolution is given by
$$
\chi~=~(-18-6)/3 + 6\times 3~=~10
$$
which confirms the parameter count. To summarise: we have constructed a manifold
$$
\widehat{X}^{10,5}~=~\widehat{X^{6,33}/S{\times}T}~~~\text{with}~~~\hodgenos=(10,5)~.
$$
Since $T$ has fixed points this manifold has fundamental group $\IZ_3$.  The manifolds constructed in this section are presented in
the following table
\vskip10pt
\begin{table}[H]
\begin{center}
\def\str{\vrule height16pt width0pt depth8pt}
\begin{tabular}{| l | c | c |}
\hline 
\str Hodge numbers & ~~$(4,13)$~~ & ~~$(10,5)$~~ \\
\hline
Manifold & $X^{6,33}\quotient{S}$~&~ $\simpentry{\widehat{X^{6,33}\quotient{S{\times}T}}}$ \\
\hline
\str Fundamental group           & & $\IZ_3$ \\
\hline
\end{tabular}
\parbox{5.5in}{\caption{\small Hodge numbers of manifolds constructed from quotients of $X^{6,33}$.}}
\end{center}
\end{table}
\vskip10pt
\subsubsection{$X^{9,27}$; a second split of both cubics} \label{sec:9,27}
We now split the other $\IP^2$ in $X$ to obtain the manifold
\beq
X^{9,27}~=~~
\cicy{\IP^1\\ \IP^1\\ \IP^1\\ \IP^1\\ \IP^1\\ \IP^1\\ \IP^5\\}
{0 & \one & \one & 0 & 0 & 0 & 0 & 0 \\
  0 & \one & 0 & \one & 0 & 0 & 0 & 0 \\
  0 & \one & 0 & 0 & \one & 0 & 0 & 0 \\
  \one & 0 & 0 & 0 & 0 & \one & 0 & 0 \\
  \one & 0 & 0 & 0 & 0 & 0 & \one & 0 \\
  \one & 0 & 0 & 0 & 0 & 0 & 0 & \one \\
  0 & 0 & \one & \one & \one & \one & \one & \one \\}^{9,27}_{-36}
\hskip0.3in
\lower0.35in\hbox{\includegraphics[width=3.0in]{fig_9,27_manifold.pdf}}
\label{(9,27)TwoSidedExtension}\eeq
We take coordinates $s_{ia}$ for the first three $\IP^1$'s, $t_{ia}$ for the remaining three, and $(x_i, y_j)$ for 
$\IP^5$ with $i, j \in \IZ_3$ and $a \in \IZ_2$.  We first impose the $\IZ_3$ symmetry generated by
\beq
S:~ s_{ia} \to s_{i+1,a} ~;~ t_{ia} \to t_{i+1,a} ~;~ x_i \to x_{i+1} ~;~ y_i \to y_{i+1}.
\notag\eeq
By appropriate choice of coordinates we can bring the polynomials to the form
\beq
\begin{split}
P ~&=~\frac{1}{3}( m_{000} + m_{111} )\\[5pt]
Q~&=~\frac{1}{3}( n_{000} + n_{111} )\\[5pt]
p_i~&=~x_i\, s_{i0} + \sum_j \big( A_{i-j}\, x_j + B_{i-j}\, y_j\big) s_{i1}\\[3pt]
q_i~&=~y_i\, t_{i0} + \sum_j \big( C_{i-j}\, x_j + D_{i-j}\, y_j\big) t_{i1}
\end{split}
\label{(9,27)smalleqs}\eeq
where $m_{abc}=\sum_i s_{ia}s_{i+1,b}s_{i+2,c}$ and $n_{abc}=\sum_i t_{ia}t_{i+1,b}t_{i+2,c}$.
There is a scaling $(x_i,y_j)\to (x_i,\m\, y_j)$ that preserves the form of the polynomials. The effect of the scaling is to change the coefficients $B_k\to \m\, B_k$ and $C_k\to \m^{-1}C_k$, with $A_k$ and $D_k$ remaining unchanged. This freedom can be used to set $B_0=C_0$, for example, so there are 11 free parameters in the equations. For generic coefficients the polynomials are transverse, though this is easier to check for the $\IZ_6$-invariant subfamily that we will come to shortly. 

The solution set has no fixed points under $S$.  The fixed point analysis goes as follows.  Fixed points of $S$ are of the form 
$$
s_{ia} = s_a~,~~ t_{ia} = t_ a~,~~ (x_{i},\,y_{j}) = (\l \o^i,\,\m \o^j)
$$
where $\o^3 = 1$, and $(\l,\m)$ parametrise a $\IP^1$.  The equations $P=0$ and $Q=0$ become cubics in 
$s_a$ and $t_ a$ respectively and restrict $s_a$ and $t_ a$ to discrete values.  The $p_i$ and $q_i$ then become two independent equations for the variables $(\l,\m)$, which in general will have no solution for $(\l,\m)$ a point of a $\IP^1$.

As in the previous example we may pass to an extended representation for which the hyperplane sections of the embedding spaces generate the second cohomology and this enables us to calculate the Hodge numbers of the quotient. To this end consider the representation
\beq
X^{9,27}~=~~~
\cicy{\IP^1\\ \IP^1\\ \IP^1\\ \IP^1\\ \IP^1\\ \IP^1\\ \IP^2\\ \IP^2\\ \IP^5}
{\one & 0 & 0 & 0 & 0 & 0 & \one & 0 & 0 & 0 & 0 & 0 \\
 0 & \one & 0 & 0 & 0 & 0 & 0 & \one & 0 & 0 & 0 & 0 \\
 0 & 0 & \one & 0 & 0 & 0 & 0 & 0 & \one & 0 & 0 & 0 \\
 0 & 0 & 0 & \one & 0 & 0 & 0 & 0 & 0 & \one & 0 & 0 \\
 0 & 0 & 0 & 0 & \one & 0 & 0 & 0 & 0 & 0 & \one & 0 \\
 0 & 0 & 0 & 0 & 0 & \one & 0 & 0 & 0 & 0 & 0 & \one \\
 0 & 0 & 0 & 0 & 0 & 0 & \one & \one & \one & 0 & 0 & 0 \\
 0 & 0 & 0 & 0 & 0 & 0 & 0 & 0 & 0 & \one & \one & \one \\
 \one & \one & \one & \one & \one & \one & 0 & 0 & 0 & 0 & 0 & 0}^{9,27}_{-36}
\label{(9,27)Extended}\eeq

corresponding to the diagram
\vskip15pt
$$
\hbox{\includegraphics[width=4.0in]{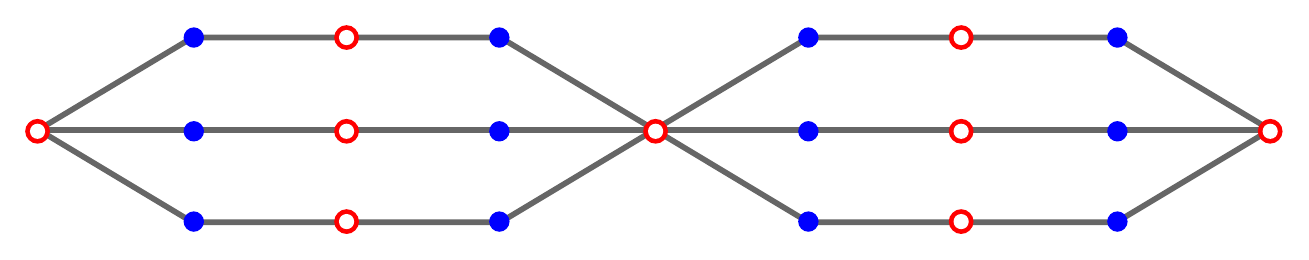}}
$$
\vskip0pt
\place{1.15}{0.57}{$u$}
\place{1.8}{1.05}{$P_i$}
\place{2.25}{1.05}{$s_{ja}$}
\place{2.75}{1.05}{$p_i$}
\place{3.1}{0.9}{$(x,y)$}
\place{3.7}{1.05}{$q_i$}
\place{4.15}{1.05}{$t_{ja}$}
\place{4.6}{1.05}{$Q_i$}
\place{5.25}{0.57}{$v$}

We extend the action of $S$ to the new $\IP^2$'s
$$
S:~u_i~\to~u_{i+1}~;~~ v_i~\to~ v_{i+1}~.
$$
By applying twice the process of the previous subsection we may, without loss of generality, take equations for the extended manifold to be of the form
\beq\begin{split}
P_i~&=~u_i\, s_{i0} +  u_{i+1}\, s_{i1}\\[9pt]
Q_i~&=~v_i\, t_{i0} +  v_{i+1}\, t_{i1}\\[7pt]
p_i~&=~x_i\, s_{i0} + \sum_j (A_{i-j}\, x_j + B_{i-j}\, y_j)\, s_{i1}\\
q_i~&=~y_i\, t_{i0} + \sum_j (C_{i-j}\, x_j + D_{i-j}\, y_j)\, t_{i1}
\end{split}\label{bigmatrixeqs}
\eeq
where, as in \eref{(9,27)smalleqs}, we take $B_0=C_0$. These equations are also transverse and fixed point free, though, again, it is easiest to check the transversality for the $\IZ_6$ invariant subfamily that follows. 
 
The six $\IP^1$'s form two orbits under the action of $S$ so it is now clear that, for the quotient, $h^{11}=5$. For the Euler number we have $\chi=-12$ hence $h^{21}=11$ which agrees with our parameter count. Thus we have found a quotient

\beq
\cicy{\IP^1\\ \IP^1\\ \IP^1\\ \IP^1\\ \IP^1\\ \IP^1\\ \IP^5\\}
{0 & \one & \one & 0 & 0 & 0 & 0 & 0 \\
  0 & \one & 0 & \one & 0 & 0 & 0 & 0 \\
  0 & \one & 0 & 0 & \one & 0 & 0 & 0 \\
  \one & 0 & 0 & 0 & 0 & \one & 0 & 0 \\
  \one & 0 & 0 & 0 & 0 & 0 & \one & 0 \\
  \one & 0 & 0 & 0 & 0 & 0 & 0 & \one \\
  0 & 0 & \one & \one & \one & \one & \one & \one \\}^{5,11}\quotient{\IZ_3}\lower40pt\hbox{.}
\notag\eeq
It is of interest to take a further quotient by a $\IZ_2$-generator, $U$, that fixes two elliptic curves. The resolution of this quotient has $\ch=-6$. We defer discussion of this further quotient to~\SS\ref{sec:3gen}.
\subsubsection{$X^{5,50}$; another split of $X^{2,56}$} \label{sec:5,50}
The manifold $X^{2,56}$, which earlier we split to $X^{3,39}$, can be split another way while still retaining the
$\IZ_3$ symmetry. If we leave the cubic equation intact and instead split the $\IP^2$ we~obtain
$$
X^{5,50}~=~~
\cicy{\IP^1\\ \IP^1\\ \IP^1\\ \IP^5\\}
{1 & 1 & 0 & 0 & 0 \\
1 & 0 & 1 & 0 & 0 \\
1 & 0 & 0 & 1 & 0 \\
0 & 1 & 1 & 1 & 3 \\}^{5,50}_{-90}
\hskip0.75in
\lower0.33in\hbox{\includegraphics[width=2.0in]{fig_5,50_manifold.pdf}}
$$
We take coordinates $(x_i,y_j)$ for the $\IP^5$ and $t_{i,a}$ for the $\IP^1$'s, where $a\in\IZ_2$, and define
a $\IZ_3$ action generated by
$$
S:~x_i\to x_{i+1}~,~y_j\to y_{j+1}~,~t_{i,a}\to t_{i+1,a}
$$
We will denote the trilinear equation by $r$, the three bilinear equations by $p_i$, and the cubic by $q$.  We demand that $q$ and $r$ be invariant under the action of $S$, and that the $p_i$ transform as 
$p_i \to p_{i+1}$.  If, as previously, we define the $S$-invariant quantities $m_{abc} = \sum_i t_{i,a}\, t_{i+1,b}\, t_{i+2,c}$
then the polynomials take the form
\beq \label{eq:5,50}
\begin{split}
p_i &= \sum_j\Big( (A_j\, x_{i+j} + B_j\, y_{i+j})\,t_{i,0} 
+ (C_j\, x_{i+j} + D_j\, y_{i+j})\,t_{i,1}\Big) \\[5pt]
r &= \sum_{ijk} \Big(G_{jk}\, x_i\, x_{i+j}\, x_{i+k} + H_{jk}\, x_i\, x_{i+j}\, y_{i+k} 
+ J_{jk}\, x_i\, y_{i+j}\, y_{i+k} + K_{jk}\, y_i\, y_{i+j}\, y_{i+k} \Big) \\[5pt]
q &= E_0\, m_{000} + E_1\, m_{100} + E_2\, m_{110} + E_3\, m_{111}
\end{split}
\eeq
We have checked that these equations are transverse, and that the action of $S$ is fixed point free.  To find the Hodge numbers of the quotient, we consider the extended representation in which all the $(1,1)$-forms are all pullbacks from the ambient spaces:
$$
X^{5,50}~=~~
\cicy{\IP^1\\ \IP^1\\ \IP^1\\ \IP^2\\ \IP^5\\}
{1 & 0 & 0 & 1 & 0 & 0 & 0 \\
 0 & 1 & 0 & 0 & 1 & 0 & 0 \\
 0 & 0 & 1 & 0 & 0 & 1 & 0 \\
 1 & 1 & 1 & 0 & 0 & 0 & 0 \\
 0 & 0 & 0 & 1 & 1 & 1 & 3 \\}^{5,50}_{-90}
\hskip0.5in
\lower0.3in\hbox{\includegraphics[width=2.5in]{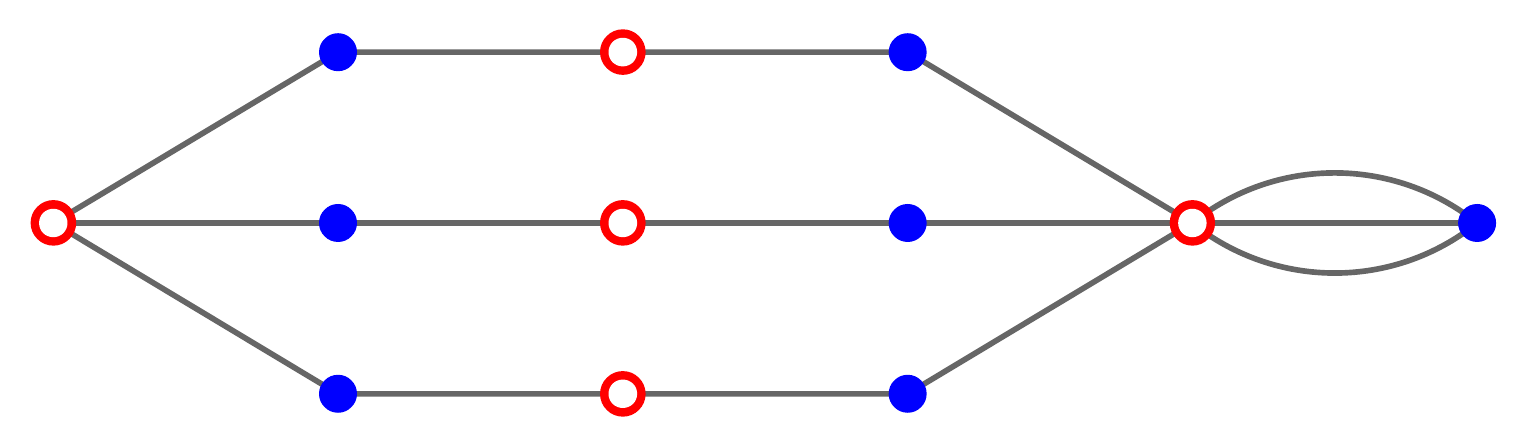}}
$$
Here we take coordinates $u_i$ on the $\IP^2$, and extend the $\IZ_3$ action to include $u_i \to u_{i+1}$.  Thus the action identifies the three $\IP^1$'s, but only acts internally on the other spaces, leaving the quotient with $h^{11}=3$.  Since the Euler number will be $\chi = -90/3 = -30$, it must be that $h^{21}=18$.  We can as usual check the value of $h^{21}$ by counting coefficients, which we do first for the manifold before imposing the symmetry.

Consider the first configuration.  The most general polynomial $q$ has $2^3 = 8$ coefficients, each $p_i$ has $2\times6 = 12$, and $r$ is characterised by a symmetric rank three $SO(6)$ tensor, which has $6\times7\times8/1\times2\times3 = 56$ components, giving altogether $100$ coefficients.  A general change of coordinates has $3\times3 + 35 = 44$ parameters, neglecting an overall scaling in each space, and we can also rescale each of the five polynomials.  Finally, there is one non-trivial consistency condition between the polynomials.  The general form of the $p_i$ is
$$
p_i = \cA_i(x,y)\, t_{i0} + \cB_i(x,y)\, t_{i1}
$$
where $\cA$ and $\cB$ are linear, so $p_i = 0$ then gives
$$
t_{i1} = -\frac{\cA_i(x,y)}{\cB_i(x,y)}\, t_{i0}.
$$
If we substitute this into $q=0$ (and multiply by $\cB_0\,\cB_1\,\cB_2$) we obtain a cubic constraint on $x,y$, which can be used to eliminate a single coefficient from the polynomial $r$.  Our count of independent coefficients is therefore $100 - 50 = 50$, which is indeed the value of $h^{21}$.

Repeating the above analysis for the $\IZ_3$-symmetric case has only one subtlety; $G_{jk}$ and $K_{jk}$ have only four independent components each, while $H_{jk}$ and $J_{jk}$ have six.  This gives a total of $36$ coefficients in \eqref{eq:5,50}.  Coordinate changes consistent with the symmetry are
\beq
\begin{split}
x_i \to \sum_j (\a_j\, x_{i+j} + &\b_j\, y_{i+j})~,~~y_i \to \sum_j (\g_j\, x_{i+j} + \d_j\, y_{i+j}) \\
(t_{i0},t_{i1}) & \to (\m\,t_{i0} + \n\,t_{i1}, \r\,t_{i0} + \s\,t_{i1})
\end{split}
\eeq
Neglecting overall scaling there are $14$ parameters here.  We can also rescale all the $p_i$ by the same factor, and rescale $r$ and $q$, giving three more.  The consistency condition discussed above eliminates one more coefficient, leaving $36 - 18 = 18$, which agrees with our previously-determined value of $h^{21}$.  In summary, the quotient manifold has fundamental group $\IZ_3$ and Hodge numbers $\hodgenos = (3,18)$.

\subsection{The branch $X^{5,59} \to X^{8,44} \to X^{19,19}$}
\subsubsection{$X^{5,59}$; a contraction of $X^{6,33}$} \label{sec:5,59}
If we contract the $\IP^5$ in the definition of the manifold we have labelled $X^{6,33}$, we are led
to a matrix which is the transpose of that for $X^{2,56}$:
$$
X^{5,59}~=~~
\cicy{\IP^1\\ \IP^1\\ \IP^1\\ \IP^2}
{1 & 1 \\
  1 & 1 \\
  1 & 1 \\
  0 & 3 \\}^{5,59}_{-108}
\hskip0.75in
\lower0.3in\hbox{\includegraphics[width=1.5in]{fig_5,59_manifold.pdf}}
$$
We take coordinates $u_i$ for the $\IP^2$ and $t_{ia}$ for the three $\IP^1$'s, and define a $\IZ_3$ generated by
$$
S:~u_i \to u_{i+1}~,~~t_{ia}\to t_{i+1,a}
$$
Writing the most general $S$-invariant form of the sixth degree polynomial is complicated, but we do not need to work in the most general case.  We again define $m_{abc} = \sum_i t_{i,a}\, t_{i+1,b}\, t_{i+2,c}$ and take
$S$-invariant cubics in the $u_i$:
$$
\cU_0 = u_0^3 + u_1^3 + u_2^3~,~~
\cU_1 = u_0 u_1^2 + u_1 u_2^2 + u_2 u_0^2~,~~
\cU_2 = u_0^2 u_1 + u_1^2 u_2 + u_2^2 u_0~,~~
\cU_3 = u_0 u_1 u_2~.
$$
Then we can take our equations to be:
\beq \label{(5,59)polys}
\begin{split}
r &= A\, m_{000} + B\, m_{100} + C\, m_{110} + D\, m_{111} \\
q &= \sum_{I=0}^3\Big( E_I\, \cU_I\, m_{000} +  F_I\, \cU_I\, m_{100}
 +  G_I\,\cU_I\, m_{110} +  H_I\, \cU_I \,m_{111} \Big)~.\\
\end{split}
\eeq
We have checked that these equations are transverse.
The fixed points of $S$ are of the form $u_i = \o^i,~t_{i,a} = t_a$ where $\o^3 = 1$, and it is easy to see that at all such points, $r$ and $q$ constitute two independent equations in the $\IP^1$ parametrised by $t_a$, for which there are no solutions.  Therefore the $\IZ_3$ acts freely on the variety, and the quotient is smooth, with
Euler number $-108/3 = -36$.  The easiest way to find the Hodge numbers is again to examine a representation
with one ambient space for each non-trivial $(1,1)$ form:
$$
X^{5,59}~=~~
\cicy{\IP^1\\ \IP^1\\ \IP^1\\ \IP^2\\ \IP^2}
{1 & 0 & 0 & 1 \\
 0 & 1 & 0 & 1 \\
 0 & 0 & 1 & 1 \\
 1 & 1 & 1 & 0 \\
 0 & 0 &  0 & 3 \\}^{5,59}_{-108}
\hskip0.75in
\lower0.33in\hbox{\includegraphics[width=2.0in]{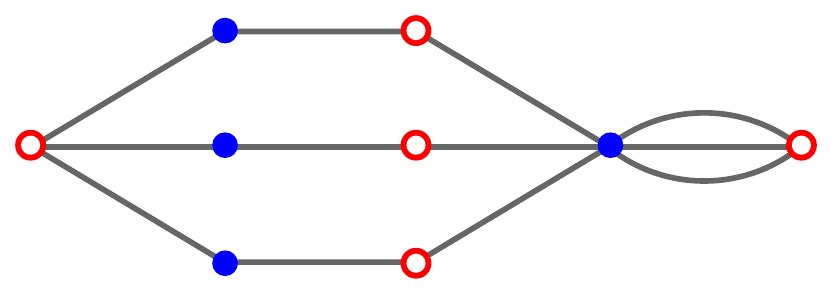}}
$$
The $Z_3$ action is extended to include $v_i \to v_{i+1}$, where the $v_i$ are coordinates on the extra $\IP^2$.  We see then that the quotient will have one independent $(1,1)$-form coming from the triplet of $\IP^1$'s, and one from each of the $\IP^2$'s, giving $h^{11} = 3$.  Since $\chi = -36$, we conclude that the Hodge numbers are $(3,21)$.

We can now impose a second $\IZ_3$ symmetry
$$
T:~t_{ia}\to \o^a\, t_{ia}~,~~u_i\to \o^i\, u_i~,
$$
with $\o^3=1$.
Under $T$ the following special cases of the polynomials in~\eqref{(5,59)polys} are invariant
\beq
\begin{split}
r &= \frac13 m_{000} + D\, m_{111} \\[3pt]
q &=  (A_0\, \cU_0 + A_1\, \cU_3)\, m_{000} + A_2\, \cU_1\, m_{100} +  A_3\, \cU_2\, m_{110}
 + (A_4\, \cU_0 + A_5\, \cU_5)\, m_{111}
\end{split}
\notag\eeq
We have checked that these are transverse; however the action of $T$ has fixed points, which we now analyse.
Points fixed by $T$ in the ambient space are given by $u_i = \d_{i,k}$ for some fixed $k$, and 
$(t_{i0} ,t_{i1}) \in \{(0,1),(1,0)\}$ for each $i$.  All such points are equivalent under $S$ to one with 
$k = 0$, so we consider this case.  The only non-zero terms in $q$ will be those proportional to $u_0^3$, and these must be multiplied by either $m_{000}$ or $m_{111}$ to be $T$-invariant.  Our polynomials above are therefore general enough to capture the full fixed-point structure, and since $m_{000}$ and $m_{111}$ both vanish unless we choose $(t_{i0},t_{i1}) = (0,1)$ or $(t_{i0},t_{i1}) = (1,0)$ for all values of $i$, there will be six fixed points, corresponding to the six other fixed points in the ambient space.  The generators $ST$ and $S^2T$ do not have fixed points (it can easily be seen that at their fixed points in the ambient space $t_{ia} = \d_{a,0}$ or $t_{ia} = \d_{a,1}$, but $r$ clearly does not vanish at these points), so this is the full set of points on $X^{5,59}/S$ fixed by $T$.  To obtain a smooth manifold we remove a neighbourhood of each of these points before taking the quotient by $T$, then glue in six copies of the Eguchi-Hanson space $E\!H_3$.  Since $\chi(E\!H_3) = 3$, the Euler number of the resulting manifold is $(-36-6)/3 + 6\times3 = 4$, and it has fundamental group $\IZ_3$. Since $T$ acts internally on each projective space, we still expect to have 3 $(1,1)$-forms from the ambient space, and, with another from each $E\!H_3$, to obtain $h^{11}=9$.  This implies $h^{21} = 7$, which we can check by parameter counting.

First we will see how $h^{21}=59$ comes about for the original manifold, before any symmetries are imposed.  The trilinear polynomial $r$ has $2^3=8$ terms, and $q$ has $2^3\times10 = 80$ terms, where $10$ is the number of independent monomials $u_i u_j u_k$. There are $3\times3 + 8 = 17$ parameters in coordinate changes (neglecting scale). There are 10 degrees of freedom corresponding to redefining $q$ by a multiple of $r$.

Finally, we can rescale $r$ and $q$, so the number of independent parameters is $88-17-10-2 = 59$, which agrees with the value of $h^{21}$.  We can therefore expect to obtain the correct value by repeating the counting for the symmetric manifold.  A short piece of computer algebra shows that there are $12$ independent terms in the $\IZ_3{\times}\IZ_3$-symmetric polynomials.  The only remaining coordinate change consistent with the symmetry is $(t_{i0},t_{i1})\to(\a t_{i0},\d t_{i1})$, which gives one parameter.  We eliminate two coefficients by substituting $r=0$ into $q$, and two by rescaling the polynomials.  Therefore we obtain 
$h^{21} = 12-1-2-2=7$, which agrees with the previously determined value.

We summarise the manifolds found in this section in a short table
\vskip10pt
\begin{table}[H]
\begin{center}
\def\str{\vrule height16pt width0pt depth8pt}
\begin{tabular}{| l | c | c |}
\hline 
\str Hodge numbers & ~~$(3,21)$~~& ~~$(9,7)$~~ \\
\hline
Manifold & ~$X^{5,59}\quotient{\IZ_3}$~&~ \simpentry{\widehat{X^{5,59}\quotient{\IZ_3{\times}\IZ_3}}} \\
\hline
\str Fundamental group           & & $\IZ_3$ \\
\hline
\end{tabular}
\parbox{4.0in}{\caption{\small Hodge numbers of manifolds constructed from quotients of $X^{5,59}$.}}
\end{center}
\end{table}
\vskip10pt
\subsubsection{$X^{8,44}$; splitting the $\IP^2$} \label{sec:8,44}
Splitting the $\IP^2$ in $X^{5,59}$ yields a manifold given by the transpose of the matrix corresponding to $X^{3,39}$
\beq
X^{8,44}~=~
\cicy{\IP^1\\ \IP^1\\ \IP^1\\ \IP^1\\ \IP^1\\ \IP^1\\}
{0 & \one & \one \\
  0 & \one & \one \\
  0 & \one & \one \\
  \one & 0 & \one \\
  \one & 0 & \one \\
  \one & 0 & \one \\}^{8,44}_{-72}
\hskip0.75in
\lower0.33in\hbox{\includegraphics[width=2.0in]{fig_8,44_manifold.pdf}}
\label{(8,44)cicy}\eeq
Fortunately the transposes of \eref{(6,33)manifold} and \eref{(9,27)TwoSidedExtension} as well as their `overextended' matrices all reduce to this case since they are all splits of $X^{8,44}$ and all have $\chi=-72$.

Again take coordinates $s_{ia}$ for the first three $\IP^1$'s and $t_{ia}$ for the remaining three, with 
$i \in \IZ_3$ and $a \in \IZ_2$.  We define an action of $\IZ_3$ generated by
$$
S:~ s_{ia} \to s_{i+1,a} ~;~ t_{ia} \to t_{i+1,a}
$$
In order to write polynomials invariant under $S$ it is again useful to consider the invariant quantities
$$
m_{abc}=\sum_j s_{ja}\, s_{j+1,b}\, s_{j+2,c}~~,~~~
n_{abc}=\sum_j t_{ja}\, t_{j+1,b}\, t_{j+2,c}
$$ 
and
$$
\ell_{abcdef}~=~\sum_j s_{ja}\, s_{j+1,b}\, s_{j+2,c}\, t_{jd}\, t_{j+1,e}\, t_{j+2,f}~.
$$
The $S$-invariant polynomials can then be written as
\beq \label{(8,44)Z3polys}
p= \sum_{abc} A_{abc}\, m_{abc} ~,~~ q= \sum_{abc} B_{abc}\, n_{abc} ~,~~ 
r= \sum_{abcdef} C_{abcdef}\, \ell_{abcdef}
\eeq
Note that the individual terms obey the symmetries $m_{abc} = m_{cab}$, $n_{abc} = n_{cab}$, 
$\ell_{abcdef} = \ell_{cabfde}$, and therefore we may require the same symmetry in the coefficients.  The polynomials are transverse and the action of $S$ is fixed point free, but this is easier to check for the more symmetric polynomials which follow.

We now impose also a $\IZ_2$ symmetry generated by
$$
U:~ s_{ia} \to (-1)^a s_{ia} ~~;~ t_{ia} \to (-1)^a t_{ia}
$$
The polynomials~\eqref{(8,44)Z3polys} are also invariant under $U$ if the indices of each term sum to $0$ in 
$\IZ_2$.  By choosing coordinates suitably we may take the first two polynomials to be
\beq
p~=~\frac{1}{3}\, m_{000} + m_{110}~~~\text{and}~~~q~=~\frac{1}{3}\, n_{000} + n_{110}
\label{(8,44)PandQ}
\eeq
We shall want to discuss the general form of the third polynomial in some detail but a simple choice suffices to show that the equations can be chosen so as to be transverse and fixed point free. Such a choice, for $r$, is 
\beq
r~=~\frac{1}{9}\,m_{000}\, n_{000} + A\, m_{100}\, n_{100} + B\, m_{111}\, n_{100} + 
C\, m_{100}\, n_{111} + D\, m_{111}\, n_{111}~.
\label{simplepoly}
\eeq 

To show that the polynomials are fixed point free it suffices to check the fixed points of $S$ and of $U$. A fixed point of $S$ is such that $s_{ja}=s_a$ and $t_{ja}=t_a$, independent of $j$. The polynomials $p$ and $q$ impose the conditions
$$
s_0(s_0^2 + 3 s_1^2)~=~0~~~\text{and}~~~t_0(t_0^2 + 3 t_1^2)~=~0
$$
and the nine solutions to these equations do not satisfy \eref{simplepoly} for generic values of the coefficients.
Fixed points of $U$, in the embedding space, consist of the 64 points with $s_{ja}$ and $t_{ja}$ given by independent choices of $\{(0,1),\,(1,0)\}$. For each of the 8 choices of fixed points for $s_{ja}$ there is precisely one of the polynomials $m_{abc}$ that is nonzero and similarly there is precisely one of the polynomials $n_{abc}$ that is nonzero for the fixed points of $t_{ja}$. Our three polynomials then cannot vanish if all the coefficients shown are nonzero.

We have shown that the $\IZ_3{\times}\IZ_2$ quotient exists. In this representation of the space we see only 6 of the 8 independent cohomology classes of $H^2$ among the embedding spaces. 
As in previous examples we may here also pass to a representation for which all of $H^2$ is generated by the hyperplane classes of the ambient spaces. Such a representation is given by the transpose of \eref{(9,27)TwoSidedExtension}
$$
X^{8,44}~=~~
\cicy{\IP^1\\ \IP^1\\ \IP^1\\ \IP^1\\ \IP^1\\ \IP^1\\ \IP^2\\ \IP^2\\}
{\one & \one & 0 & 0 & 0 & 0 & 0 \\
 \one & 0 & \one & 0 & 0 & 0 & 0 \\
 \one & 0 & 0 & \one & 0 & 0 & 0 \\
 \one & 0 & 0 & 0 & \one & 0 & 0 \\
 \one & 0 & 0 & 0 & 0 & \one & 0 \\
 \one & 0 & 0 & 0 & 0 & 0 & \one \\
 0 & \one & \one & \one & 0 & 0 & 0 \\
 0 & 0 & 0 & 0 & \one & \one & \one \\}^{8,44}
\hskip0.25in\lower25pt\hbox{\includegraphics[width=3.0in]{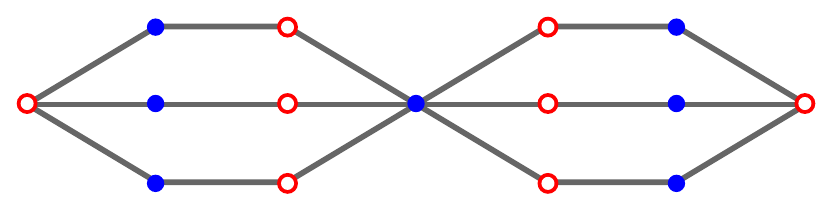}}
\label{(8,44)Extended}
$$
The polynomials $p$ and $q$ are each replaced by three equations involving the coordinates of the additional 
$\IP^2$'s
\beq
\begin{split}
p_i~&=~s_{i0}(u_{i+1} + u_{i+2}) + s_{i1}(u_{i+1} - u_{i+2}) \\[3pt]
q_i~&=~t_{i0}(v_{i+1} + v_{i+2}) + t_{i1}(v_{i+1} - v_{i+2})~.
\end{split}
\label{(8,44)pandq}\eeq
It is convenient to think of these as matrix equations $p_i=\sum_j M_{ij}(s)u_j$ and 
$q_i=\sum_j M_{ij}(t)v_j$ with 
\beq
M_{ij}(s)~=~ \begin{pmatrix}
0 & s_{00} + s_{01} & s_{00} - s_{01} \\
s_{10} - s_{11} & 0 & s_{10} + s_{11} \\
s_{20} + s_{21} & s_{20} - s_{21} & 0 \\
\end{pmatrix}
\label{Mmatrix}\eeq
Eliminating the new coordinates from \eref{(8,44)pandq} yields the conditions $\det M(s)= \det M(t)=0$ and returns us to \eref{(8,44)PandQ}. The fact that the Euler numbers for the two configurations are the same shows that eliminating the new coordinates does not introduce any nodes so that the manifolds are indeed isomorphic. We have taken the equations $p_i$ and $q_i$ above to be covariant under $S$ and it is straightforward to take the $S$-quotient since this identifies the $\IP^1$'s in two groups of three while leaving the $\IP^2$'s invariant. It follows that $X^{8,44}/S$ has Hodge numbers $\hodgenos=(4,16)$. 

It is, at first sight, puzzling how to describe the action of $U$ in the extended representation.  
For while the polynomials $p$, $q$ and $r$ are invariant under $U$ it is easy to see that there is no 
{\em linear\/} transformation of the coordinates $u_i$ and $v_i$ that renders the polynomials $p_i$ and $q_i$ covariant.

Consider the equations $p_i=\sum_j M_{ij}(s)u_j =0$. Clearly the matrix $M_{ij}(s)$ cannot have rank 3. It is immediate, from the explicit form \eref{Mmatrix}, that $M_{ij}(s)$ can never have rank 1. Thus the matrix always has rank 2. Hence given three points $(s_{i0},s_{i1})$ the equations $p_i=0$ determine a unique point
$u_i\in \IP^2$. Conversely given a point $u_i\in \IP^2$, that is not one of the three special points 
$u_i = (1,0,0),\, (0,1,0),\, (0,0,1)$, and the general form of $M_{ij}$ from \eref{Mmatrix} we determine three
points $(s_{i0},s_{i1})\in(\IP^1)^3$. For the special points, however, an entire $\IP^1$ is left undetermined by this process. For $u_i=(1,0,0)$, for example $(s_{00},s_{01})$ is undetermined. Thus the surface determined by the equations $p_i=0$ is the del Pezzo surface $dP_6$ given by a $\IP_2$ blown up in the three special points. 

We know that, given the points $(s_{i0},s_{i1})$, there is a unique solution for $u_i$, up to scale so, since we know the action of $U$ on the $s_{ia}$ we may deduce the action on the $u_i$.\footnote{We are grateful to Duco van Straten for explaining to us the geometry of this situation, and the associated Cremona transformation of $\IP^2$.}.  To this end we write down an explicit solution and act on it with $U$:
\beq 
\begin{split}
\begin{pmatrix}
u_0 \\
u_1 \\
u_2
\end{pmatrix} ~&=~ 
\begin{pmatrix}
-(s_{10} + s_{11})(s_{20} - s_{21})\\
\phantom{-}(s_{10} + s_{11})(s_{20} + s_{21})\\
-(s_{10} - s_{11})(s_{20} - s_{21})
\end{pmatrix} \\[10pt]
 ~&\stackrel{U}{\to}\hskip3pt\begin{pmatrix}
-(s_{10} - s_{11})(s_{20} + s_{21})\\
\phantom{-}(s_{10} - s_{11})(s_{20} - s_{21})\\
-(s_{10} + s_{11})(s_{20} + s_{21})
\end{pmatrix} \\[10pt]
~&=~ \frac{1}{(s_{10} + s_{11})(s_{20} - s_{21})}
\begin{pmatrix}
u_1 u_2 \\
u_0 u_2 \\
u_0 u_1
\end{pmatrix} 
\end{split}
\label{eq:delPezzo}\eeq
Since an overall scale is irrelevant, we see that the action of $U$ is related to a Cremona transformation of 
$\IP^2$, given by $u_i \to u_{i+1} u_{i+2}$, or equivalently $u_i \to 1/u_i$.  This is defined everywhere except at the three special points and squares to the identity away from the special lines $u_0 = 0,~u_1 = 0,~u_2 = 0$. Furthermore it maps the special lines to the special points.  There is a nice way to understand this in terms of the del Pezzo surface.  Since this is just the blow up of $\IP^2$ at the three special points it contains six interesting lines -- the three special lines $u_i = 0$, just described, and the three exceptional lines arising as the blow ups -- these six lines form a hexagon.  The actions of $S$ and $U$ can be thought of as rotations of this hexagon of order 3 and 2 respectively.  The reason the action of $U$ cannot be realised as an isomorphism of $\IP^2$ is that the projection blows down three sides of the hexagon to points.
\begin{figure}[H]
\begin{center}
\framebox[6.5in][c]{\parbox{6.0in}{\vspace{10pt}
$$
\raisebox{15pt}{\includegraphics[width=1in]{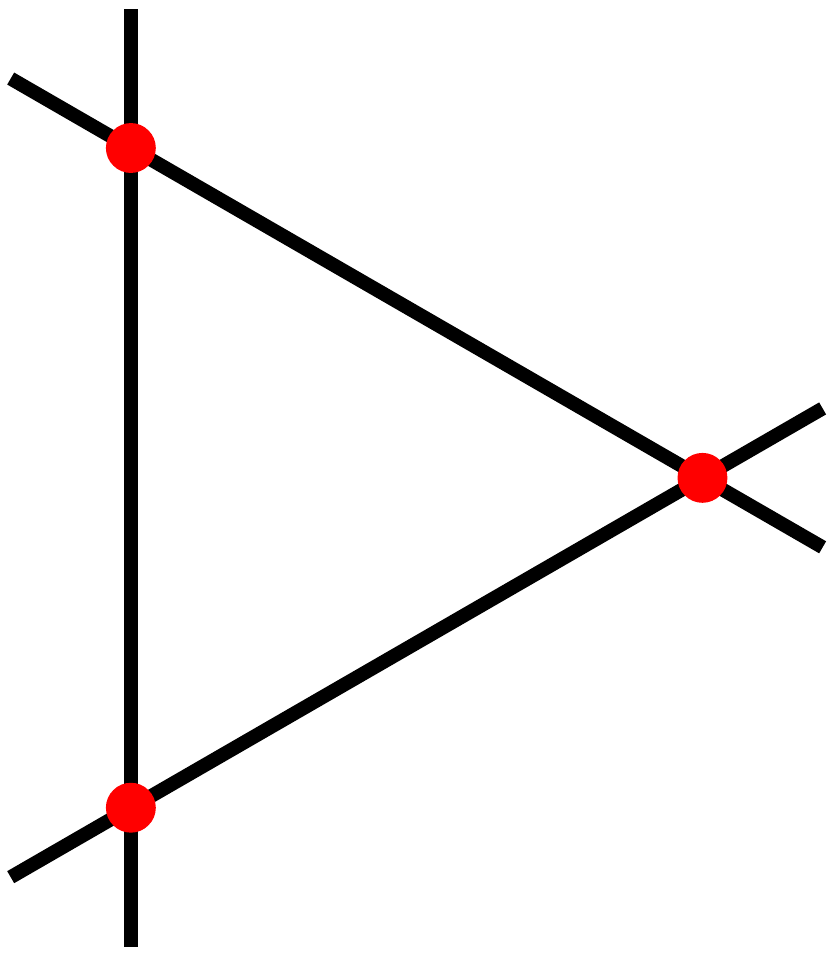}}\hspace{0.8in}
\includegraphics[width=1.5in]{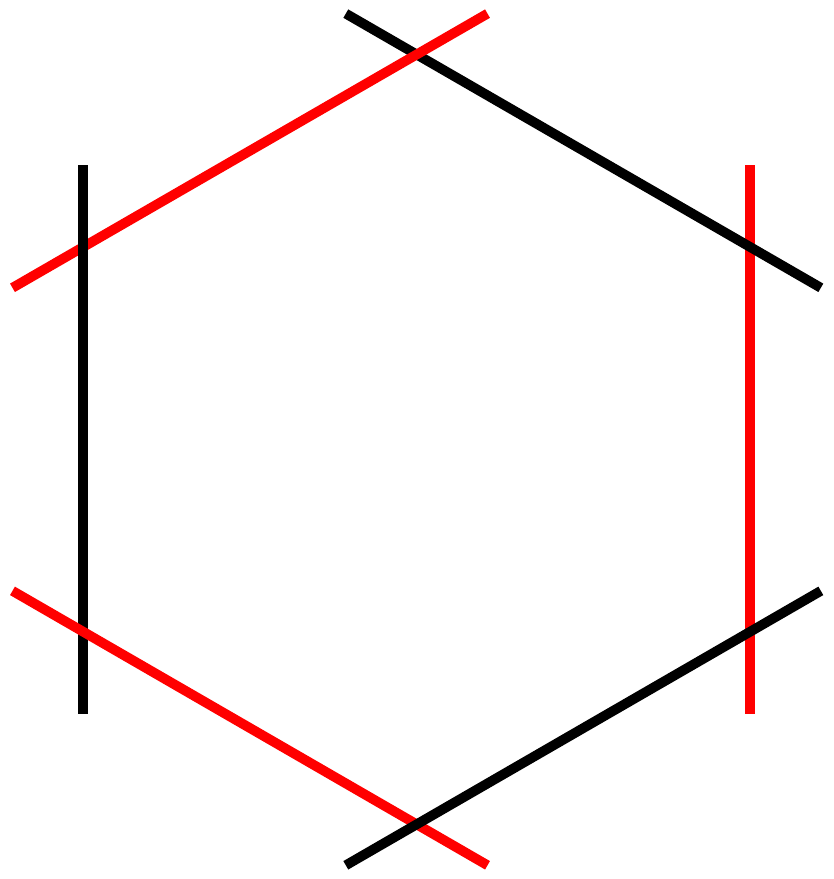}\hspace{0.8in}
\raisebox{15pt}{\includegraphics[width=1in]{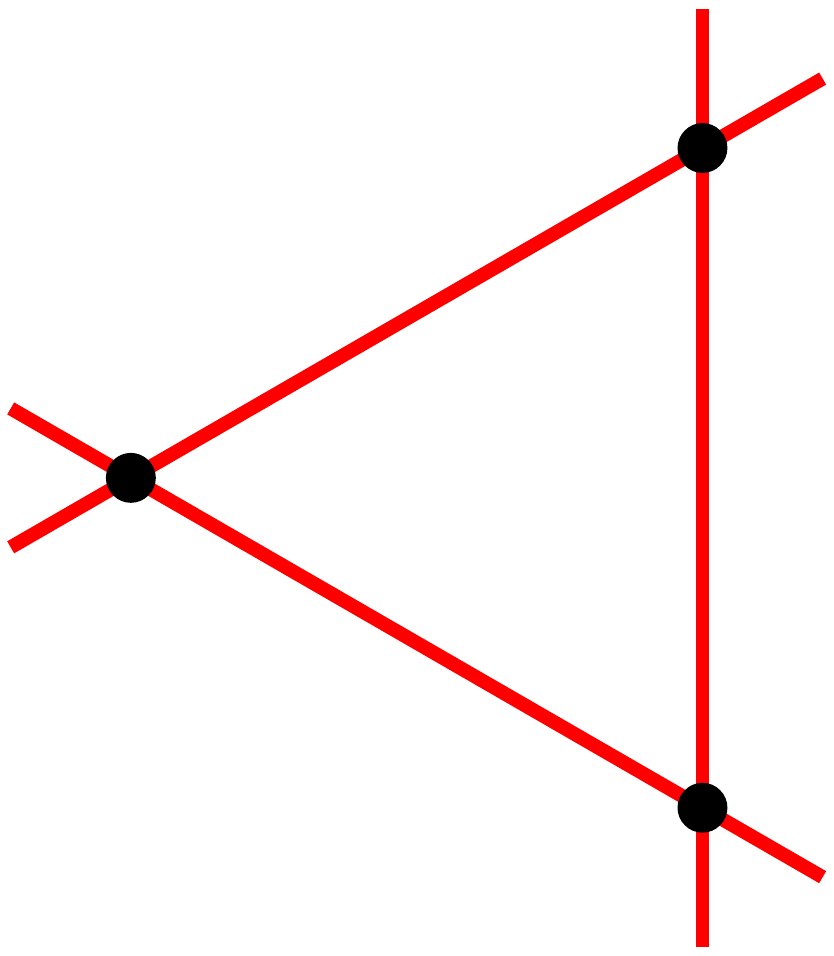}}
$$
\vspace{-10pt}}}
\vskip5pt
\parbox{5.5in}{\caption{\label{dp6figs}\small The surface $dP_6$ is 
obtained by blowing up three points in
$\IP^2$. The three exceptional lines together with the three lines 
that passed through pairs of the points that were blown up form a 
hexagon. The effect of $U$ in $dP_6$ is to make a rotation of order 2 
of the hexagon. The effect of $U$ in $\IP^2$ is to blow up the special 
points and blow down the special lines and so `maps' the triangle on 
the left to the triangle on the right, and cannot be realised as an 
isomorphism of $\IP^2$.}}
\end{center}
\end{figure}
The reader may worry that the above map is ill-defined on our manifold just as on $\IP^2$, but this is not the case.  Consider the point $(1,0,0) \in \IP^2$, at which the Cremona transformation is not defined.  Inspection of~\eref{Mmatrix} shows that this point occurs when $s_{11} = s_{10}$ and $s_{21} = -s_{20}$.  The value of $s_{01}/s_{00}$ is however undetermined, as we have just seen.  The second line of~\eref{eq:delPezzo}  shows that this $\IP^1$ is mapped bijectively to the line $u_1 = 0$.  The same analysis applies to the other two special points. The action of $U$ is thus well-defined on the del Pezzo surface, and hence also on our manifold, which is embedded in the Cartesian product of two copies of this surface.

To determine $h^{11}$ for the quotient manifolds we need to calculate the action of $U$ on the pullback of the hyperplane class of $\IP^2$. Alternatively we can determine $h^{21}$ for the quotients involving $U$ by a careful counting of parameters in the polynomials.  Let us return to the polynomials~\eqref{(8,44)Z3polys} that define $X^{8,44}$. We may choose coordinates $s_{ja}$ and $t_{ja}$ such that the polynomials $P$ and $Q$ take the form shown in \eref{(8,44)PandQ}. Fixing the form of these polynomials does not completely exhaust the freedom to make redefinitions. For the $s_{ja}$ we may consider the 
$S\!L(2,\IC)$ transformations
$$
\begin{pmatrix}s_{j0}\\ s_{j1}\end{pmatrix}~\to~
\begin{pmatrix} a_j & b_j \\ c_j & d_j \end{pmatrix} \begin{pmatrix} s_{j0}\\ s_{j1} \end{pmatrix}~.
$$
There are a total of 9 parameters in these transformations. There are 8 monomials $s_{1a}\,s_{2b}\,s_{3c}$, and an overall scale is irrelevant, so fixing the form of $P$ imposes 7 constraints. We are left with a two parameter freedom to make redefinitions of the $s_{ja}$. If we restrict to $S\!L(2,\IC)$ transformations that are infinitesimally
close to the identity then we can check this quite explicitly. The transformations that preserve $P$ take the form
\beq
\begin{pmatrix}s_{j0}\\ s_{j1}\end{pmatrix}~\to~
\begin{pmatrix} 1 & \e\b_j \\ \e\b_j & 1 \end{pmatrix} \begin{pmatrix} s_{j0}\\ s_{j1} \end{pmatrix}~.
\label{sparams}
\eeq
with the $\b_j$ subject to the constraint $\sum_j \b_j=0$. Consider now the monomials 
$$
s_{1a}\,s_{2b}\,s_{3c}\,t_{1d}\,t_{2e}\,t_{3f}
$$
that can be included in $r$. There are 64 of these, of which 32 are even under $U$ and 32 odd. Terms of the form
$$
m_{011}\,t_{1d}\,t_{2e}\,t_{3f}~~~\text{and}~~~
s_{1a}\,s_{2b}\,s_{3c}\,n_{011}~,
$$
of which there are 7 that are even and 8 that are odd, may be eliminated from $r$ through the equations $P=0$ and $Q=0$ and this does not change the parity of the terms. There is also the freedom to redefine $r$ by an overall scale which we use to, say, set the coefficient of the monomial $m_{000}n_{000}$ to unity. We still dispose of a two parameter freedom to redefine the $s_{ja}$ and another two parameter freedom to redefine the $t_{ja}$. Consider the effect of making the redefinition \eref{sparams} in the leading monomial $m_{000}n_{000}$
$$
m_{000}\,n_{000}~\to~m_{000}\,n_{000} + \e\big(\b_0\, s_{01}\,s_{10}\,s_{20} + 
\b_1\, s_{00}\,s_{11}\,s_{20} + \b_2\, s_{00}\,s_{10}\,s_{21}\big)n_{000}~.
$$
We may use this freedom, together with the corresponding freedom for the $t_{ja}$, to eliminate the 4 odd monomials
$$
s_{01}\,s_{10}\,s_{20}\,n_{000}~;~~s_{00}\,s_{11}\,s_{20}\,n_{000}~;~~ 
m_{000}\, t_{00}\,t_{10}\,t_{21}~;~~m_{000}\,t_{00}\,t_{11}\,t_{20}~.$$
The counting, so far, is that we have $32-7-1=24$ free parameters in $r$ associated with even monomials and 
$32-8-4=20$ associated with odd monomials. The total count is 44 which agrees with the value of $h^{21}$.
Since we have a complete description of the parameter space we know that the even parameters are the parameters of the quotient $X^{8,44}/U$ which therefore has Hodge numbers $\hodgenos=(6,24)$. A more formal argument proceeds via the Lefschetz fixed point theorem with the same result. 

If we turn now to the $S$-quotient. We start with the following complete list of the labels for the polynomials 
$\ell_{abcdef}$.
$$
\def\topstr{\vrule height14pt width0pt depth3pt}
\def\botstr{\vrule height12pt width0pt depth6pt}
\def\str{\vrule height12pt width0pt depth3pt}
\setlength{\doublerulesep}{3pt}
\begin{tabular}{| @{\hspace{5pt}}l @{\hspace{4pt}}c@{\hspace{4pt}}
@{\hspace{3pt}}c@{\hspace{4pt}} r@{\hspace{5pt}} |}
\hline 
\multicolumn{4}{|c|}{Even polynomials}\\
\hline\hline
\topstr (000\,000) & (001\,001) & (011\,000) & (111\,001) \\
\str      (000\,011) & (001\,010) & (011\,011) & (111\,111) \\
\str                        & (001\,100) & (011\,101) & \\
\botstr                   & (001\,111) & (011\,110) & \\
\hline
\end{tabular}
\quad
\begin{tabular}{| @{\hspace{5pt}}l @{\hspace{4pt}}c@{\hspace{4pt}}
@{\hspace{3pt}}c@{\hspace{4pt}} r@{\hspace{5pt}} |}
\hline 
\multicolumn{4}{|c|}{Odd polynomials}\\
\hline\hline
\topstr (000\,001) & (001\,000) & (011\,001) & (111\,000) \\
\str      (000\,111) & (001\,011) & (011\,010) & (111\,011) \\
\str                        & (001\,101) & (011\,100) & \\
\botstr                   & (001\,110) & (011\,111) & \\
\hline
\end{tabular}
$$
\vskip10pt
We again take $r$ to have leading term $\ell_{000000}=m_{000}n_{000}$ with coefficient unity, which removes one parameter, and we should remove also 3 even parameters owing to the freedom to eliminate terms of the form
$$
m_{011}\,n_{000}=3\ell_{011000}~,~~m_{000}\,n_{011}=3\ell_{000011}~~~\text{and}~~~
m_{011}\,n_{011} = \ell_{011011} + \ell_{011101} + \ell_{011110}
$$
by means of $P$ and $Q$. Elimination of terms of the form
$$
m_{011}\,n_{001}~,~~m_{011}\,n_{111}~,~~m_{001}\,n_{011}~,~~m_{111}\,n_{011}
$$
removes 4 odd parameters. Thus we are left with $12-3-1=8$ even parameters and $12-4=8$ odd parameters. The total number is 16 which is indeed the value of $h^{21}$ for $X^{8,44}/S$. We learn also that $h^{21}=8$ for the quotient $X^{8,44}/U{\times}S$ and hence $h^{11}=2$. Again we could deduce this more formally by applying the Lefschetz fixed point theorem.

Instead of the $\IZ_2$ we can try to impose a second $\IZ_3$ generated by
$$
T:~s_{ia}\to \o^a s_{ia}~;~t_{ia}\to \o^a t_{ia}~,
$$
with $\o^3=1$.
The most general polynomials invariant under both $S$ and $T$ are given by
\beq
\begin{split}
p ~&=~ A_0\, m_{000} + A_1\, m_{111} \\[5pt]
q ~&=~ B_0\, n_{000} + B_1\, n_{111} \\[5pt]
r ~&=~ C_0\, \ell_{000000} + C_1\, \ell_{100110} + C_2\, \ell_{100101} + C_3\, \ell_{100011} 
 + C_4\, \ell_{110100}\\
&\hskip10pt + C_5\, \ell_{110010} + C_6\, \ell_{110001} + C_7\, \ell_{111000}  + C_8\, \ell_{000111} 
 + C_9\, \ell_{111111}
\end{split}
\notag\eeq
The only coordinate changes which are allowed are $(s_{i0}, s_{i1})\to (\a_s s_{i0}, \b_s s_{i1})$ and 
$(t_{i0}, t_{i1})\to (\a_t t_{i0}, \b_t t_{i1})$. We can use this along with the freedom to rescale $p$ and $q$ to set $A_a = B_a = \frac 13$, for $a=0,1$.  We can also use the freedom to redefine $r$ by multiples of $p$ and $q$ to eliminate $C_7, C_8$ and $C_9$ in favour of $C_0$, and then rescale $r$ so that $C_0 = \frac 19$.  This leaves us with
\beq
\begin{split}
p =~ & \frac 13 (m_{000} + m_{111}) \\[3pt]
q =~ & \frac 13 (m_{000} + m_{111}) \\[3pt]
r =~ & \frac 19 \ell_{000000} + C_1\, \ell_{100110} + C_2\, \ell_{100101} + C_3\, \ell_{100011}
 + C_4\, \ell_{110100} + C_5\, \ell_{110010} + C_6\, \ell_{110001}
\end{split}
\notag\eeq
These are transverse, but the action of $T$ has fixed points.  The elements $ST$ and $S^2T$ however act freely, which can be demonstrated the same way as for the manifold in \SS\ref{sec:5,59}.  The fixed points in the ambient space are given by $(s_{i0}, s_{i1}), (t_{i0}, t_{i1})\in \{(0,1),(1,0)\}$ for all $i$, corresponding to $8\times8 = 64$ points.  However, if the same choice is made for all values of $i$ for either $s$ or $t$, the
polynomials are easily seen not to vanish, and the remaining choices for (say) $s$ are all equivalent under $S$ to either $\{(1,0),(0,1),(0,1)\}$ or $\{(1,0),(1,0),(0,1)\}$.  Therefore there are actually only $2\times6 = 12$ points to consider.  At each of these points we have $p=q=0$, but a moment's thought shows that in exactly half the cases, $r\neq0$.  Therefore the action of $T$ on $X^{8,44}/S$ has six isolated fixed points.  We can resolve these as usual by deleting a ball neighbourhood of each before taking the quotient, and then gluing in six copies of $E\!H_3$.  The Euler number of the resulting manifold is $(-24-6)/3 + 6\times3 = 8$ and it will have fundamental group $\IZ_3$.  We expect $h^{21}=6$, as that is the number of free parameters in the above polynomials, and since the $T$ acts only internally on each ambient space we will get 4 $(1,1)$-forms from these plus 6 from the blow ups of the fixed points, giving $h^{11}=10$.  Indeed this gives the correct Euler number.

We summarize with a table the manifolds that we have obtained from quotients of $X^{8,44}$.
\vskip10pt
\begin{table}[H]
\begin{center}
\def\str{\vrule height16pt width0pt depth8pt}
\begin{tabular}{| l | c | c | c | c |}
\hline 
\str Hodge numbers & ~~$(2,8)$~~ & ~~$(4,16)$~~ & ~~$(6,24)$~~ & ~~$(10,6)$~~ \\
\hline
\str Manifold & ~$X^{8,44}\quotient{\IZ_3{\times}\IZ_2}$~ & $X^{8,44}\quotient{\IZ_3}$
& $X^{8,44}\quotient{\IZ_2}$~ & ~\simpentry{\widehat{X^{8,44}\quotient{\IZ_3{\times}\IZ_3}}} \\
\hline
\str Fundamental group           & & & & $\IZ_3$ \\
\hline
\end{tabular}
\parbox{5.5in}{\caption{\small Hodge numbers for quotients of $X^{8,44}$.}}
\end{center}
\end{table}
\vskip10pt
\subsection{Some quotients of $X^{19,19}$} \label{sec:19,19}
The split bicubic $X^{19,19}$ is well known and its quotients have been studied, not least in connection with 
a heterotic model of elementary particle interactions, and a listing of its quotients, which has been incorporated into \tref{tiptab}, is given in \cite{BouchardDonagi}. Our aim here is to recover the $\IZ_2$, $\IZ_3$ and          $\IZ_2{\times}\IZ_3$ quotients of $X^{19,19}$ by our methods and to point out that there are `extended' representations of these spaces that exhibit the full number of $(1,1)$-forms as ambient spaces. The diagram for this space exhibits a compelling left-right symmetry (see \eref{fig(19,19)B} below). Taking the quotient by this symmetry and resolving singularities we find manifolds with Hodge numbers $(12,12)$, $(8,8)$, $(6,6)$ and $(4,4)$.

We start with $X^{19,19}$ represented in the form
\beq
X^{19,19}~=~~
\cicy{\IP^1\\ \IP^1\\ \IP^1\\ \IP^1\\ \IP^1\\ \IP^1\\ \IP^1}
{\one ~ \one ~ 0 ~ 0 \\
 \one ~ 0 ~ \one ~ 0 \\
 \one ~ 0 ~ \one ~ 0 \\
 \one ~ 0 ~ \one ~ 0 \\
 0 ~ \one ~ 0 ~ \one \\
 0 ~ \one ~ 0 ~ \one \\
 0 ~ \one ~ 0 ~\one}^{19,19}_0
\label{(19,19)B}\eeq
and we shall see presently that this space is indeed the one that is often represented as the split bicubic. We name the polynomials and coordinates for this manifold as in the diagram
\vskip25pt
\vbox{
\beq
\includegraphics[width=3.0in]{fig_19,19_manifoldB.pdf}
\label{fig(19,19)B}\eeq
\vskip0pt
\place{1.6}{0.57}{$q_0$}
\place{4.77}{0.57}{$q_1$}
\place{2.25}{1.08}{$s_i$}
\place{2.75}{0.75}{$p_0$}
\place{3.22}{0.75}{$x$}
\place{3.68}{0.75}{$p_1$}
\place{4.15}{1.08}{$t_i$}
}
\vskip-15pt
We first impose a symmetry generated by
$$
S:~s_{i,a}~\to~s_{i+1,a}~,~~~t_{i,a}~\to~t_{i+1,a}~;~~~p_a~\to~p_a~,~~~q_b~\to~q_b~.
$$
By a process that is, by now, very familiar coordinates may be chosen such that, without loss of generality, polynomials $q_0$ and $q_1$ take the form
$$
q_0~=~m_{001} + \frac{1}{3}\, m_{111}~;~~~q_1~=~n_{001} + \frac{1}{3}\, n_{111}~,
$$
where 
$$
m_{abc}~=~\sum_i s_{i,a}\, s_{i+1,b}\, s_{i+2,c}~~~\text{and}~~~
n_{abc}~=~\sum_i t_{i,a}\, t_{i+1,b}\, t_{i+2,c}~.
$$
The polynomials $p_0$ and $p_1$ can be written in the general form
\beq
\begin{split}
p_0~&=~\sum_{abc}\left(x_0\, A_{abc}\, m_{abc} + x_1\, B_{abc}\, m_{abc}\right)\\[3pt]
p_1~&=~\sum_{abc}\left(x_0\, C_{abc}\, n_{abc} + x_1\, D_{abc}\, n_{abc}\right)
\end{split}
\notag\eeq
where the coefficients are cyclically symmetric in their indices. By means of the equations $q_a$, we may remove terms containing $m_{111}$ or $n_{111}$ from the $p_a$ thus the coefficients $A_{abc},\ldots, D_{abc}$ each correspond to three degrees of freedom. There are two degrees of freedom in choosing a scale for $p_0$ and $p_1$ and three degrees of freedom in making an $S\!L(2,\IC)$ transformation on the coordinates $x_a$. Thus we are left with $4{\times}3-2-3=7$ parameters in the polynomials.

We now impose a second generator
\beq
U:~s_{i,a}\to (-1)^a\, s_{i,a}~,~~t_{i,b}\to (-1)^b\, t_{i,b}~,~~x_a\to x_a
~;~~p_a\to p_a~,~~q_b\to -q_b~.
\label{(19,19)Udef}\eeq
The polynomial $p_0$ can now only contain the $s$-coordinates through the terms $m_{000}$ and $m_{011}$
and $p_1$ can only contain the $t$'s through the terms $n_{000}$ and $n_{011}$. So now there are 
$4{\times}2-2-3=3$ parameters in the equations which we may write in the form
\begin{align*}
p_0~&= \left(\frac{1}{3}\, m_{000}+ a\, m_{011}\right) x_0 + c\, m_{011}\, x_1
&  q_0~&=~m_{001} + \frac{1}{3}\, m_{111} \\[3pt]
p_1~&=~~c\, n_{011}\, x_0 + \left(\frac{1}{3}\, n_{000}+ b\, n_{011}\right)\, x_1
&  q_1~&=~\, n_{001} + \frac{1}{3}\, n_{111}~.
\end{align*}
It is straightforward to check that these equations are fixed point free and transverse. 

Since the $\IZ_3{\times}\IZ_2$ free quotient exists so too does the $\IZ_2$ quotient corresponding to $U$. It is instructive to count parameters for this case but before doing so we return to the case with no symmetries to account for the 19 parameters. For the general case we may take equations
\beq
\begin{aligned}
p_0~&=\sum_{abc}\left(x_0\, A_{abc}\,s_{0a}\,s_{1b}\,s_{2c} + 
x_1\, B_{abc}\,s_{0a}\,s_{1b}\,s_{2c}\right)~,\hskip15pt
& q_0~&=~\sum_{abc}E_{abc}\,s_{0a}\,s_{1b}\,s_{2c}~,\\[3pt]
p_1~&=\sum_{abc}\left(x_0\, C_{abc}\,t_{0a}\,t_{1b}\,t_{2c} + 
x_1\, D_{abc}\,t_{0a}\,t_{1b}\,t_{2c}\right)~,
& q_1~&=~\sum_{abc}F_{abc}\,t_{0a}\,t_{1b}\,t_{2c}~.
\end{aligned}
\label{(19,19)geneqs}\eeq
There are a total of $6{\times}8=48$ coefficients in these equations. We have $7{\times}3$ degrees of freedom to make $S\!L(2,\IC)$ transformations on the coordinates $s_{ia}$, $t_{ia}$ and $x_a$. In addition we may redefine $p_0$ and $p_1$ by adding multiples of $q_0$ and $q_1$
$$
p_0~\to~p_0 + (\a\, x_0 + \b\, x_1)\, q_0~,~~~~p_1~\to~p_1 + (\g\, x_0 + \d\, x_1)\, q_1~;
$$
there are 4 degrees of freedom here. Finally we may rescale the 4 polynomials so the count is
$48-21-4-4=19$ parameters.

If we impose $U$-covariance as in, \eref{(19,19)Udef}, then there are 24 coefficients. Allowed redefinitions of coordinates correspond to an $S\!L(2,\IC)$ transformation for $x_a$ but merely
scalings 
$$
(s_{i0},\, s_{i1})~\to~ (s_{i0},\, \l_i\, s_{i1})~,~~~~(t_{i0},\, t_{i1})~\to~ (t_{i0},\, \m_i\, t_{i1})
$$
for the other coordinates, so these correspond to a total of 9 parameters. Since the $q_a$ are odd under $U$ we cannot use them to redefine the $p_b$, which are even. There remain the 4 degrees of freedom corresponding to changing the scale of the polynomials. So now the count is $24-9-4=11$.

We summarise the discussion, thus far, with a table
\vskip10pt
\begin{table}[H]
\begin{center}
\def\str{\vrule height16pt width0pt depth8pt}
\begin{tabular}{| c | c | c | c |}
\hline 
\str $\hodgenos\left(X^{19,19}/G\right)$ & ~~$(3,3)$~~ & ~~$(7,7)$~~ & ~~$(11,11)$~~ \\
\hline
\str $G$ & ~$\IZ_3{\times}\IZ_2$~ & $\IZ_3$ & $\IZ_2$ \\
\hline
\end{tabular}
\parbox{5.5in}{\caption{\small Hodge numbers for the smooth quotients of $X^{19,19}$.}}
\end{center}
\end{table}

Given the symmetry of the diagram \eref{fig(19,19)B} under a reflection in a vertical axis it is natural to seek to impose also an external $\IZ_2$ symmetry
$$
V:~s_{ia}~\leftrightarrow~t_{ia}~,~~~x_0~\leftrightarrow~x_1~;~~~p_0~\leftrightarrow~p_1~,
~~~q_0~\leftrightarrow~q_1~.
$$
Covariance under $V$ forces the relations
$$
A_{abc}~=~D_{abc}~,~~~B_{abc}~=~C_{abc}~,~~~E_{abc}~=~F_{abc}~.
$$
so that we now have just two constraints $p_0$ and $q_0$. The $\IP^1$ with coordinates $x$ has two fixed points 
$$
x_a^V~=~\left\{(1,1),\,(1,-1)\right\}
$$
and substituting these values into the constraints we find that the fixed point set consists of two non-intersecting elliptic curves, 
$\cE_\pm$, with the equations and configuration
\vspace{3pt}
\beq
\begin{split}
\sum_{abc}\left( A_{abc} \pm B_{abc} \right)\,s_{0a}\,s_{1b}\,s_{2c}~&=~0 \\
\sum_{abc}E_{abc}\,s_{0a}\,s_{1b}\,s_{2c}~&=~0
\end{split}
\hspace{1in}\raisebox{3pt}{$\cicy{\IP^1\\ \IP^1\\ \IP^1}{1~1\\ 1~1\\ 1~1}$}
\notag\eeq
Returning to the equations that describe the $V$-quotient, which we can take to be those of \eref{(19,19)geneqs}, we count parameters. Between $A_{abc}$, $B_{abc}$ and $E_{abc}$ there are 24 coefficients. There is a 3 parameter freedom to redefine each of the coordinates $s_{i}$ and a 1 parameter freedom to redefine $x$, in a manner consistent with $V$, so a total of 10 parameters corresponding to coordinate redefinitions. In addition we have a 2 parameter freedom to redefine 
\hbox{$p_0\to (\a\, x_0 + \b\, x_1)\, q_0$}, and finally another 2 parameter freedom to scale $p_0$ and $q_0$. Thus the equations have $24-10-2-2=10$ free parameters. Two further parameters arise when we repair the two curves of $\IZ_2$ fixed points with $A_1$ spaces. In this way we arrive at a manifold with 
$\hodgenos=(12,12)$, which we denote by $X^{12,12}$,
$$
X^{12,12}~=~\widehat{X^{19,19}/V}~.
$$
We may repeat this analysis by taking and resolving $V$-quotients of the free quotients $X^{19,19}$ that we have discussed above. The parameter counting, for these cases, is summarised by~\tref{tab(12,12)quotients}.

\begin{table}[H]
\begin{center}
\def\str{\vrule height16pt width0pt depth8pt}
\def\Str{\vrule height18pt width0pt depth10pt}

\begin{tabular}{| l | c | c | c | c |}
\hline 
\str \hfil$G$ & $V$ & $V{\times}U$ & $V{\times}S$ & $V{\times}U{\times}S$ \\ \hline\hline
\str Coefficients                                 & 24 & 12 & 12 & 6 \\ \hline
\str Coordinate changes                     & 10 & 4   & 4   & 2 \\ \hline
\str Modification of $p_0$ by $q_0$  & 2   & 0   & 2   & 0 \\ \hline
\str Scaling of $p_0$ and $q_0$       & 2   & 2   & 2   & 2 \\ \hline\hline
\Str $\hodgenos(\widehat{X^{19,19}/G})$ & ~$(12,12)$~ & ~~$(8,8)$~~ & ~~$(6,6)$~~ & $(4,4)$ \\ \hline
\end{tabular}
\vskip5pt
\parbox{5.5in}{\caption{\label{tab(12,12)quotients}\small The parameter count for the resolutions of the singular quotients of $X^{19,19}$. The Hodge numbers $h^{21}$ are obtained by subtracting rows two, three and four from the first row and then adding two to account for the resolution of the two elliptic curves of $A_1$ singularities.}}
\end{center}
\end{table}
It is also of interest to enquire to what extent one can extend the configuration of $X^{19,19}$ so as to represent more of the 19 dimensions of the space of $(1,1)$-forms by generators corresponding to ambient spaces. There does not seem to be a CICY configuration that represents all 19 in this way however we may represent 11 by taking $\IP^3$ splits of the columns of the representation \eref{(19,19)B} that contain four $1$'s and taking $\IP^2$ splits of the columns with three $1$'s. In this way we arrive at the configuration
\beq
X^{19,19}~=~~\mbox{\footnotesize$
\cicy{\IP^1\\ \IP^1\\ \IP^1\\ \IP^1\\ \IP^1\\ \IP^1\\ \IP^1\\ \IP^2\\ \IP^2\\ \IP^3\\ \IP^3}
{\one ~ 0 ~ 0 ~ 0 ~ \one ~ 0 ~ 0 ~ 0 ~ 0 ~ 0 ~ 0 ~ 0 ~ 0 ~ 0 \\
 0 ~ \one ~ 0 ~ 0 ~ 0 ~ 0 ~ 0 ~ 0 ~ \one ~ 0 ~ 0 ~ 0 ~ 0 ~ 0 \\
 0 ~ 0 ~ \one ~ 0 ~ 0 ~ 0 ~ 0 ~ 0 ~ 0 ~ \one ~ 0 ~ 0 ~ 0 ~ 0 \\
 0 ~ 0 ~ 0 ~ \one ~ 0 ~ 0 ~ 0 ~ 0 ~ 0 ~ 0 ~ \one ~ 0 ~ 0 ~ 0 \\
 0 ~ 0 ~ 0 ~ 0 ~ 0 ~ \one ~ 0 ~ 0 ~ 0 ~ 0 ~ 0 ~ \one ~ 0 ~ 0 \\
 0 ~ 0 ~ 0 ~ 0 ~ 0 ~ 0 ~ \one ~ 0 ~ 0 ~ 0 ~ 0 ~ 0 ~ \one ~ 0 \\
 0 ~ 0 ~ 0 ~ 0 ~ 0 ~ 0 ~ 0 ~ \one ~ 0 ~ 0 ~ 0 ~ 0 ~ 0 ~ \one \\
 0 ~ 0 ~ 0 ~ 0 ~ 0 ~ 0 ~ 0 ~ 0 ~ \one ~ \one ~ \one ~ 0 ~ 0 ~ 0 \\
 0 ~ 0 ~ 0 ~ 0 ~ 0 ~ 0 ~ 0 ~ 0 ~ 0 ~ 0 ~ 0 ~ \one ~ \one ~ \one \\
 \one ~ \one ~ \one ~ \one ~ 0 ~ 0 ~ 0 ~ 0 ~ 0 ~ 0 ~ 0 ~ 0 ~ 0 ~ 0 \\
 0 ~ 0 ~ 0 ~ 0 ~ \one ~ \one ~ \one ~ \one ~ 0 ~ 0 ~ 0 ~ 0 ~ 0 ~ 0}$}
\notag\eeq
Since we began with a configuration with Euler number zero the Euler number of every split must also be zero so the split configuration represents the same manifold.
A first observation is that if we contract all the $\IP^1$-rows, apart from the first, and then contract the two 
$\IP^3$-rows then we recover the representation
$$
X^{19,19}~=~~\cicy{\IP^1\\ \IP^2\\ \IP^2}{1~1\\ 3~0\\ 0~3}
$$
which identifies the space as being indeed $X^{19,19}$. 

The diagram corresponding to the extended configuration is
$$
\includegraphics[width=5in]{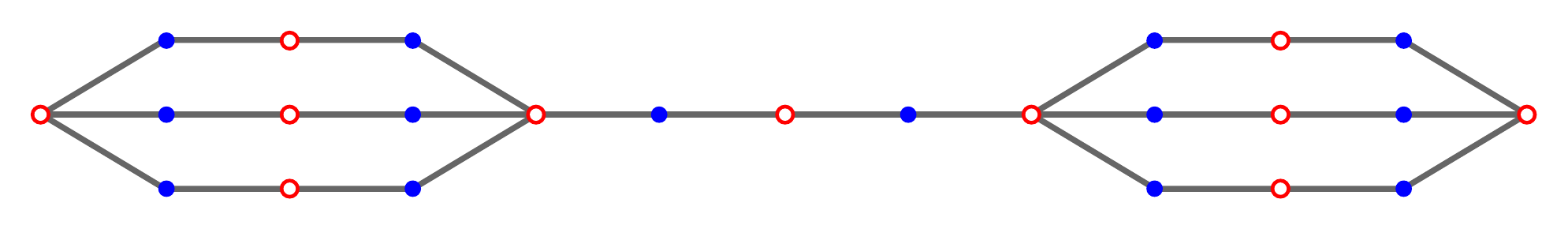}
$$
This has 11 ambient spaces so can represent the full number of $(1,1)$-forms for $X^{19,19}/U$. The 
$S$-quotient identifies the three $\IP^1$'s in each lobe of the diagram. This reduces the number of independent ambient spaces to 7, which is sufficient to represent the $(1,1)$-forms for $X^{19,19}/S$. For the 3 
$(1,1)$-forms of the $S{\times}U$ quotient the small configuration above suffices.
\subsection{A  resolved quotient with $\chi=-6$}\label{sec:3gen}
Let us consider further the family $X^{9,27}/S$ of \SS\ref{sec:9,27}. In particular the subfamily of 
\eref{bigmatrixeqs} that is invariant under the $\IZ_2$ generator
$$
U:~x_i~\leftrightarrow~ y_i~;~~s_{ia}~\leftrightarrow~t_{ia}~;~~u_i~\leftrightarrow~v_i~.
$$
The equations are now
\beq\begin{split}
P_i~&=~u_i\, s_{i0} +  u_{i+1}\, s_{i1}\\[2ex]
Q_i~&=~v_i\, t_{i0} +  v_{i+1}\, t_{i1}\\[2ex]
p_i~&=~x_i\, s_{i0} + \sum_j (A_{i-j}\, x_j + B_{i-j}\, y_j)\, s_{i1}\\
q_i~&=~y_i\, t_{i0} + \sum_j (B_{i-j}\, x_j + A_{i-j}\, y_j)\, t_{i1}~,
\end{split}\label{bigmatrixUeqs}\eeq
which are transverse for generic choices of coefficients. The constraint between $B_0$ and $C_0$ is now irrelevant so the equations now contain 6 parameters. The generator $U$ does not act freely: the fixed point set consists of two disjoint curves, $\cE_\pm$, for which $y_i=\l\, x_i$, with $\l=\pm 1$, $t_{ia}=s_{ia}$ and $v_i=u_i$, and with the coordinates subject to the independent constraints $p_i=P_i=0$. The curves $\cE_\pm$ correspond to the one dimensional CICY's
$$
\cicy{\IP^1\\ \IP^1\\ \IP^1\\ \IP^2\\ \IP^2\\}
{1 & 0 & 0 & 1 & 0 & 0\\
  0 & 1 & 0 & 0 & 1 & 0\\
  0 & 0 & 1 & 0 & 0 & 1\\
  1 & 1 & 1 & 0 & 0 & 0\\
  0 & 0 & 0 & 1 & 1 & 1\\}{~\lower30pt\hbox{.}}
$$
In this way we see that $\cE_\pm$ are elliptic curves, and can be resolved as in \SS\ref{sec:3,39}. Since the Euler numbers of the $\cE_\pm$ are zero the Euler number of the resolved quotient is simply $-36/6=-6$. The generators $S$ and $U$, between them, identify the six $\IP^1$'s and the two $\IP^2$'s so the embedding spaces contribute 3 to $h^{11}$. Two additional $(1,1)$-forms and two additional $(2,1)$-forms arise from the resolution of the fixed curves. In this way we find a manifold with fundamental group $\IZ_3$, $\chi=-6$ and $\hodgenos=(5,8)$.  We summarise the new manifolds found in this section in a short table
\vskip10pt
\begin{table}[H]
\begin{center}
\def\str{\vrule height16pt width0pt depth8pt}
\begin{tabular}{| l | c | c |}
\hline 
\str Hodge numbers & ~~$(5,11)$~~ & ~~$(5,8)$~~ \\
\hline
Manifold & $X^{9,27}\quotient{\IZ_3}$ ~&~ \simpentry{\widehat{X^{9,27}\quotient{\IZ_3{\times}\IZ_2}}}  \\
\hline
\str Fundamental group           & & $\IZ_3$ \\
\hline
\end{tabular}
\parbox{5.0in}{\caption{\small Hodge numbers of manifolds constructed from quotients of $X^{9,27}$, including a new ``three
generation" manifold, with Euler number $-6$.}}
\end{center}
\end{table}
\newpage
\section{Six manifolds with quaternionic symmetry}
\subsection{The manifold $\IP^7[2~2~2~2]$}
The manifold $X^{1,65} = \IP^7[2~2~2~2]$ admits free actions by groups, both abelian and nonabelian, of order 32 \cite{Strominger:1985it,BorisovHua,Hua}, see also \cite{Triadophilia} for a brief review. We repeat some of the salient points here since we will want to split the manifold. The diagram for the manifold is:
$$
X^{1,65} = \IP^7[2~2~2~2]\hskip1in\raisebox{-30pt}{\includegraphics[width=0.9in]{fig_1,65_manifold.pdf}}
$$
If the four quadrics are taken in the form 
\beq
p_m~= ~z_m^2 + z_{m+4}^2 + a\, z_{m+2} z_{m+6} + b\, (z_{m+1} z_{m-1} + z_{m+3} z_{m-3})
+ c\,(z_{m+1}z_{m+3}+z_{m+5}z_{m-1}) ~,
\label{SWquadrics}\eeq
where the indices take values in $\IZ_8$, and it is seen that $p_{m+4}=p_m$, then the manifold admits a freely acting $\IZ_8{\times}\IZ_4$ symmetry generated by $A:~z_m\to z_{m+1}$ and $V:~z_m\to (-\ii)^m z_m$.
These same equations are also invariant under a nonabelian group of order 32, that is denoted by $G'$ in \cite{Strominger:1985it}, that contains the quaternionic group as a subgroup. To see the action of this group it is convenient to rename the coordinates $x_\a$ where the index $\a$ now takes values in the quaternion group
$$
\IH=\big\{\,1,~i,~j,~k,\,-1,\,-i,\,-j,\,-k\,\big\}~.
$$
The quadrics above, which we now index with the elements of $\IH$, take the form
$$
p_\a~=~x_\a^2 + x_{-\a}^2 + a\, x_{\a j}\, x_{-\a j} + b\, ( x_{\a i}\, x_{-\a k} +  x_{-\a i}\, x_{\a k} )
+ c\, ( x_{\a i}\, x_{\a k} +  x_{-\a i}\, x_{-\a k} )
$$
and now $p_{-\a}=p_\a$. The symmetries $U_\g,\,\g\in\IH$ act by
$$
U_\g:~x_\a~\to~x_{\g\a}~;~~p_\a~\to~p_{\g\a}~.
$$
It is straightforward to check that
$$
V^4~=~1~,~~V\,U_{-1}~=~U_{-1}\,V~,~~V\,U_i~=~U_i\,V^3~,~~V\,U_j~=~U_j\,V~,~~
V\,U_k~=~U_k\,V^3~.
$$
It follows that the elements of the group $G'$ generated by $V$ and the $U_\a$ can be enumerated as 
$U_\a\,V^r$ with $\a\in\IH$ and $r=0,\ldots,3$. In this way it is seen that the group has order 32.
\subsection{$X^{5,37}$, a symmetrical split of $\IP^7[2~2~2~2]$}\label{sec:5,37}
We can split this in a symmetrical-looking way to obtain a manifold that admits the quaternionic group as a freely acting symmetry. The split and its diagram are given below
\vskip10pt
\vbox{
$$
X^{5,37} = \cicy{\IP^1 \\ \IP^1 \\ \IP^1 \\ \IP^1 \\ \IP^7}
{\one & 0 & 0 & 0 & \one & 0 & 0 & 0 \\
0 & \one & 0 & 0 & 0 & \one & 0 & 0 \\
0 & 0 & \one & 0 & 0 & 0  & \one & 0 \\
0 & 0 & 0 & \one & 0 & 0  & 0 & \one \\
\one & \one & \one & \one & \one & \one & \one & \one}_{-64}^{5,37}
\hskip0.6in\raisebox{-55pt}{\includegraphics[width=1.6in]{fig_5,37_manifold.pdf}}
$$
\vskip0pt
\place{5.75}{1.60}{$\s$}
\place{5.05}{1.4}{$p_\s$}
\place{5.55}{1.0}{$q_\s$}
}
\vskip-10pt
It is useful to denote by $\IH_{+}$ the set of `positive' unit quaternions
$$
\IH_{+}=\big\{\,1,~i,~j,~k\,\big\}~.
$$
and use these elements to label the four $\IP^1$'s. The coordinates of the $\IP^1$ labelled by $\s\in \IH_{+}$ are taken to be $(s_\s,s_{-\s})$ and the polynomials `connected' to this $\IP^1$
are denoted by $p_\s$ and~$q_\s$. We also take coordinates $x_\a,~\a\in\IH$ for the $\IP^7$ as above.   

Consider the polynomials
\beq
\begin{split}
p_\a~&=~s_\a\, x_\a + s_{-\a}\, x_{-\a}\\[5pt]
q_\a~&=~\sum_{\b\in\IH}a_\b\,(s_\a x_{\a\b} - s_{-\a} x_{-\a\b})
\end{split}\label{(2,6)polys}
\eeq
where the index $\a$ runs over $\IH$. This is a harmless extension of the indexing on the polynomials since 
$p_{-\a}=p_\a$  and $q_{-\a}=-q_\a$.
For $\g\in\IH$ these equations are covariant under the action
$$
U_\g:~x_\a~\to~x_{\g\a}~,~~s_\a~\to~s_{\g\a}~;~~p_\a~\to~p_{\g\a}~,~~
q_\a~\to~q_{\g\a}~.
$$
To check that the action is fixed point free it is sufficient to check that $U_{-1}$ acts without fixed points since the elements $\pm i,\pm j,\pm k$ all square to $-1$. A fixed point of $U_{-1}$ has the form
$$
x_{-\a}^\ast~=~\l\,x_\a^\ast~,~~(s_\s^\ast,\,s_{-\s}^\ast)~=~(1,\,s_{-\s}^\ast)~~
\text{with}~~\l^2=(s_{-\s}^\ast)^2=1
$$
with the $x_\a^\ast$ not all zero. Imposing the constraints $p_\s=q_\s=0$ requires
$$
(1+\l s_{-\s}^\ast)\, x_\s^\ast~=~0~~~\text{and}~~~
(1-\l s_{-\s}^\ast)\sum_{\b\in\IH} a_\b\, x_{\s\b}^\ast~=~0~.
$$
and these require, for generic choice of the coefficients $a_\b$, that the $x_\s^\ast$ all vanish. 
 
Generic invariant polynomials with the parity $p_{-\a}=p_\a$  and $q_{-\a}=-q_\a$ can be brought to the form \eref{(2,6)polys} by suitable redefinition of the coordinates $x_\a$. This does not completely fix the coordinates since there remains a one-parameter freedom to make the redefinition
$$
x_\a~\to~\l\, x_\a + \m\, x_{-\a}~,~~s_\a~\to~\l\, s_\a - \m\, s_{-\a}~;~~\l^2 - \m^2=1
$$
which preserves the form of the $p_\a$. Such a redefinition changes the coefficients in the $q_\a$
$$
a_\b~\to~(\l^2 + \m^2)\, a_\b + 2\l\m\, a_{-\b}
$$
and we may use this freedom to require $a_{-1}=0$, for example. In this way we see that there are 6 free parameters in equations \eref{(2,6)polys}. We have checked that these polynomials are generically transverse.

The Euler number for the quotient is $-64/8=-8$ and $h^{11}=2$, since the four $\IP^1$'s are identified under the action of $\IH$, so the Hodge numbers for the quotient are $\hodgenos=(2,6)$ and $h^{21}$ agrees with our counting of parameters. The group $\IH$ has subgroups $\IZ_4$ and $\IZ_2$ generated by (say) $i$ and $-1$ respectively, and we give the Hodge numbers of the corresponding quotients in the table below.
\vskip10pt
\begin{table}[H]
\begin{center}
\def\str{\vrule height16pt width0pt depth8pt}
\begin{tabular}{| c | c | c | c |}
\hline 
\str $\hodgenos\left(X^{5,37}/G\right)$ & ~~$(2,6)$~~ & ~~$(3,11)$~~ & ~~$(5,21)$~~ \\
\hline
\str $G$ & ~$\IH$~ & $\IZ_4$ & $\IZ_2$ \\
\hline
\end{tabular}
\parbox{5.5in}{\caption{\small The Hodge numbers of smooth quotients of $X^{5,37}$.}}
\end{center}
\end{table}
\subsection{$X^{4,68}$; contracting the $\IP^7$}\label{sec:4,68}
Contracting the $\IP^7$ of the configuration above brings us to the tetraquadric, which is the transpose of $\IP^7[2,2,2,2]$.
$$
X^{4,68}~=~\cicy{\IP^1\\ \IP^1\\ \IP^1\\ \IP^1}{2\\ 2\\ 2\\ 2}^{4,68}\hskip1.25in
\raisebox{-30pt}{\includegraphics[width=0.9in]{fig_4,68_manifold.pdf}}
$$
For this manifold it is possible to write a defining polynomial that is transverse and also invariant and and fixed point free under the group $\IH{\times}\IZ_2$. We again choose coordinates 
$(s_\s,\,s_{-\s}),~\s\in\IH_{+}$ for the four $\IP^1$'s and define symmetry generators $U_\g,~\g\in\IH$, as before, together with a new generator, $W$,
$$
U_\g:~s_\a~\to~s_{\g\a}~~\text{for}~~\a\in\IH~,~~
W:~(s_\s,\,s_{-\s})~\to~(s_\s, -s_{-\s})~~\text{for}~~\s\in\IH_{+}~.
$$
There are $3^4=81$ tetraquadric monomials in the $s_\a$. One of these is the fundamental monomial,
$\prod_{\a\in\IH} s_\a$, that is invariant under the full group. Of the other 80 monomials 40 are even under  $W$ and 40 odd. The 40 even monomials fall into 5 orbits of length 8 under the action of $\IH$. Thus there is a 5 parameter family of invariant polynomials. The symmetry $\IH{\times}\IZ_2$ does not permit any redefinition of the coordinates so the number of parameters in the polynomials is also the number of parameters of the manifold.  For the quotient $h^{11}=1$, since the four $\IP^1$'s are identified, and the Euler number is $-128/16=-8$. Hence $\hodgenos=(1,5)$, confirming the parameter count.

It is straightforward to check that the generic member of this family is transverse and fixed point free. In order to check that the group action is fixed point free it suffices to check that 
$U_{-1}$, $W$ and $U_{-1}W$ act without fixed points. Each of these symmetries has a set of 16 fixed points in the embedding space and it is simple to check that the generic polynomial does not vanish on any of these~points.
\vskip10pt
\begin{table}[H]
\begin{center}
\def\str{\vrule height16pt width0pt depth8pt}
\begin{tabular}{| c | c | c | c | c | c | c | c |}
\hline 
\str $\hodgenos\left(X^{4,68}/G\right)$ & ~~$(1,5)$~~ & ~~$(1,9)$~~ & ~~$(2,10)$~~ 
& ~~(2,18)~~ & ~~$(4,20)$~~ & ~~$(4,36)$~~\\
\hline
\str $G$ & ~$\IH{\times}\IZ_2$~ & $\IH$ & $\IZ_4{\times}\IZ_2$ & $\IZ_4$ & $\IZ_2{\times}\IZ_2$
& $\IZ_2$\\
\hline
\end{tabular}
\parbox{5.5in}{\caption{\small The Hodge numbers of smooth quotients of $X^{4,68}$.}}
\end{center}
\end{table}
\subsection{An $\IH$ quotient of the manifold with $\hodgenos=(19,19)$}\label{sec:19,19H}
The split of the tetraquadric, $X^{4,68}$, has $\hodgenos=(19,19)$ and, as we saw in the previous section is another presentation of the split bicubic. We will see that, presented as the split tetraquadric, the manifold admits a free action by $\IH$. We start by labelling four of the $\IP^1$'s by the positive unit quaternions as indicated in the diagram. The coordinates of these spaces are taken to be $(s_\s,\, s_{-\s}),~\s\in\IH_{+}$. The $\IP^1$ introduced by the splitting is taken to have coordinates $(t_1,\,t_i)$. Generators $U_\b,~\b\in\IH$ act on these coordinates and on polynomials $p_\a$, to be given below, as follows
$$
U_\b:~s_\a~\to~s_{\b\a}~,~~t_\a~\to~t_{\b\a}~;~~p_\a~\to~p_{\b\a}~,
$$
where we will understand the $t_\a$ and the $p_\a$ as subject to the identifications
$$
t_{\a}~=~t_{j\a}~~~\text{and}~~~p_{\a}~=~p_{j\a}~.
$$
it follows from these identifications that the only independent values for the coordinates $t_\a$ are $t_1$ and $t_i$ and similarly for the polynomials $p_\a$.
\vskip10pt
\vbox{
$$
X^{19,19}~=~\cicy{\IP^1\\ \IP^1\\ \IP^1\\ \IP^1\\ \IP^1}
{1&1\\2&0\\ 2&0\\ 0&2\\ 0&2}^{19,19}\hskip1.0in
\raisebox{-25pt}{\includegraphics[width=1.75in]{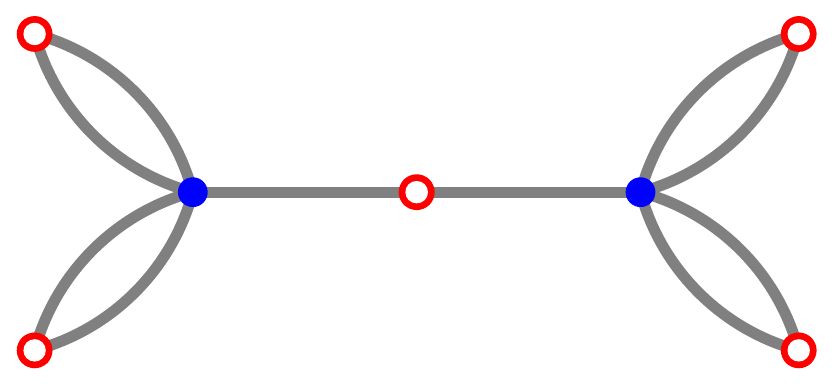}}
$$
\vskip0pt
\place{3.65}{0.92}{$1$}
\place{3.65}{0.25}{$j$}
\place{5.55}{0.92}{$i$}
\place{5.55}{0.25}{$k$}
\place{4.2}{0.45}{$p_1$}
\place{4.93}{0.45}{$p_i$}
\place{4.4}{0.73}{$(t_1,\, t_i)$}
}
Consider the following set of three polynomials, which we denote by $m_a,~a=1,2,3$,
$$
(s_1^2+s_{-1}^2)(s_j^2+s_{-j}^2)~,~~
(s_1^2+s_{-1}^2) s_j s_{-j} + s_1 s_{-1} (s_j^2+s_{-j}^2)~,~~~
s_1 s_{-1} s_j s_{-j}
$$
These are biquadratic in the variables $(s_1,s_{-1})$ and $(s_j,s_{-j})$, and invariant under $U_j$. We form linear combinations
$$
f_1~=~\sum_{a=1}^3 B_a\,m_a~~~\text{and}~~~g_i~=~\sum_{a=1}^3 C_a\,m_a
$$
and define
$$
f_\a~=~U_\a\, f_1~~~\text{and}~~~g_\a~=~U_{-\a i}\,g_i~.
$$
Note that, by construction, $f_\a=f_{\a j}=f_{-\a}=f_{j\a}$ and similarly for $g_\a$. Thus $f_\a$ and $g_\a$ each take only two independent values as $\a$ ranges over $\IH$ and these can be taken to be the values corresponding to $\a=1$ and $\a=i$. The defining polynomials can be written in terms of the $f_\a$ and $g_\a$.
$$
p_\a~=t_\a\, f_\a + t_{\a i}\, g_{\a i}
$$
where the $t_\a$ are also understood to be subject to the identifications $t_\a=t_{\a j}=t_{-\a}=t_{j\a}$. 

To check that the action of $\IH$ is fixed point free it is, again, only necessary to check that $U_{-1}$ acts freely. In the embedding space, a fixed point of $U_{-1}$ has $(s_\s^\ast,\,s_{-\s}^\ast)=(1,\pm 1)$ for each $\s\in\IH_{+}$. For each of the 16 fixed points, the independent polynomials $p_1$ and $p_i$ give two equations for $(t_1,t_i)$ and these have no solution apart from $t_1=t_i=0$ for a general choice of coefficients. It is straightforward to check that the polynomials $p_1$ and $p_i$ are transverse.

The parameter count is that there are 6 free coefficients in the definition of $f_1$ and $g_i$. There is a two-parameter freedom to redefine coordinates $s_\a\to \l s_\a +\m s_{-\a}$ and $t_\a\to \n t_\a + \r t_{\a i}$ and there is a one-parameter freedom to rescale the polynomials $p_\a\to \t p_\a$. This suggests that the manifold has $6-3=3$ parameters. We do not have a presentation of the manifold that accounts for all of $H_2$ in terms of the embedding spaces and we shall simply assume that our count of parameters is correct in this case.
The Hodge numbers for $X^{19,19}/\IH$ are, subject to this assumption, $\hodgenos=(3,3)$.
\subsection{$X^{12,28}$; the transpose of the split tetraquadric} \label{sec:12,28}
Taking the transpose of the configuration considered in the last section brings us to the manifold given by
\vskip10pt
\vbox{
\beq
X^{12,28}~=~\cicy{\IP^4\\ \IP^4\\}
{1~2~2~0~0 \\ 1~0~0~2~2}^{12,28}\hskip0.6in
\raisebox{-25pt}{\includegraphics[width=1.75in]{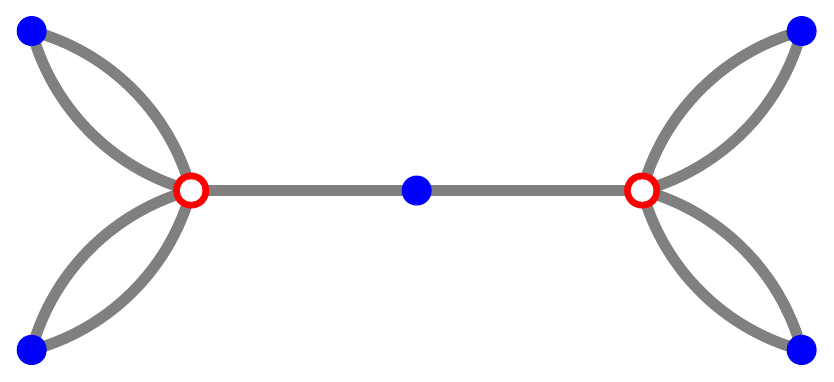}}
\label{(12,28)manifold}\eeq
\vskip0pt
\place{3.55}{0.72}{$p_1$}
\place{3.55}{0.05}{$p_j$}
\place{5.45}{0.72}{$p_i$}
\place{5.45}{0.05}{$p_k$}
\place{4.15}{0.25}{$1$}
\place{4.92}{0.25}{$i$}
\place{4.55}{0.53}{$r$}
}
This is an interesting manifold in as much as it is an analogue of $X^{14,23}$, the covering space of the 
three-generation manifold. Like $X^{14,23}$ it is both the transpose of a representation of $X^{19,19}$ and has $X^{19,19}$ as a split. This fact is perhaps not immediately obvious but can be seen by noting that
$$
X^{12,28}~\longrightarrow~
\cicy{\IP^1\\ \IP^4\\ \IP^4}{1~1~0~0~0~0\\ 1~0~2~2~0~0\\ 0~1~0~0~2~2}^{19,19}
\hskip-5pt\cong\hskip15pt
\cicy{\IP^1\\ \IP^1\\ \IP^1\\ \IP^1\\ \IP^1}{1~1\\ 2~0\\ 2~0\\ 0~2\\ 0~2}^{19,19}
$$
The equivalence of the two configurations above is demonstrated by the fact that they both have Euler number zero and both split to the following configuration, which also has Euler number zero
$$
\cicy{\IP^1\\ \IP^1\\ \IP^1\\ \IP^1\\ \IP^1\\ \IP^4\\ \IP^4}
{\one~\one~0~0~0~0~0~0~0~0 \\
 0~0~\one~\one~0~0~0~0~0~0 \\
 0~0~0~0~\one~\one~0~0~0~0 \\
 0~0~0~0~0~0~\one~\one~0~0 \\
 0~0~0~0~0~0~0~0~\one~\one \\
 \one~0~\one~\one~\one~\one~0~0~0~0 \\
 0~\one~0~0~0~0~\one~\one~\one~\one}^{19,19}\raisebox{-60pt}{.}
$$
Returning to \eref{(12,28)manifold},
we take coordinates $(w_1, x_1, x_{-1}, x_j, x_{-j})$ and $(w_i, x_i, x_{-i}, x_k, x_{-k})$ for the two projective spaces.  The bilinear polynomial is denoted by $r$, the two quadratic polynomials on the first space by $p_1, p_j$, and those on the second by $p_i, p_k$.  We can then define an action of $\g \in \IH$ on the system~by
$$
U_\g :~ w_\a \to w_{\g \a}~,~~ x_\a \to x_{\g \a}~;~~r \to r~,~~p_\a \to p_{\g \a}
$$
where $w_\a$ and $p_\a$ are identified according to $w_\a = w_{j\a}$ and $p_\a = p_{-\a}$.
The polynomials which are covariant under this action are given by
\beq
\begin{split}
r ~&=~ A_0\, w_1 w_i 
+ A_1\,\Big( w_1 (x_i+ x_{-i} + x_k + x_{-k}) + w_i (x_1 + x_{-1} + x_j + x_{-j}) \Big) \\
 &\hskip40pt + A_2\, \Big( (x_1+ x_{-1})( x_i +x_{-i}) +(x_j+ x_{-j})( x_k+x_{-k}) \Big) \\
 &\hskip40pt+ A_3\, \Big( (x_1+ x_{-1})( x_k +x_{-k}) +(x_j+ x_{-j})( x_i+x_{-i}) \Big) \\[5pt]
p_1 ~&=~ B_0\,w_1^2 + B_1\,w_1 (x_1 + x_{-1}) + B_2\, w_1 (x_j+ x_{-j}) 
+ B_3\,(x_1^2 + x_{-1}^2) + B_4\,(x_j^2 + x_{-j}^2) \\
 &\hskip40pt + B_5\,x_1\, x_{-1} + B_6\, x_j\, x_{-j} + B_7\,(x_1\, x_j + x_{-1}\, x_{-j}) 
 + B_8\,(x_1\, x_{-j} + x_{-1}\, x_j)
\end{split}
\notag\eeq
with $p_i,\, p_j,\, p_k$ being obtained by acting on $p_1$ with $U_i,\, U_j, U_k$ respectively.  To check that the action of $\IH$ on the resulting manifold is fixed point free it again suffices to check that $U_{-1}$ acts without fixed points. Fixed points in the ambient space are given by
\beq
\begin{split}
(w_1,\,  x_1,\,  x_{-1},\,  x_j,\, x_{-j}) ~&=~ \left(\frac{(1+\eta)}{2}\,a\,, \, b,\, \eta\, b,\, c\, ,\, \eta\, c\,\right)~;~~\eta=\pm1 \\[5pt]
(w_i,\, x_i,\, x_{-i},\, x_k,\, x_{-k})      ~&=~ \left(\frac{(1+\eta')}{2}d, \, e,\, \eta' e,\, f,\, \eta' f\right)~;~~\eta'=\pm1~.
\end{split}
\notag\eeq
Therefore the dimension of the set of fixed points is (at most) four, and since the five polynomials are independent, they will not simultaneously vanish anywhere on this set for a general choice of coefficients.  We have checked that the polynomials are transverse, so we conclude that the quotient variety is smooth, with Euler number $-32/8 = -4$. 

We can redefine the coordinates and polynomials, while maintaining the symmetry,  as follows
\begin{alignat*}{2}
w_1 &\to \l_0\, w_1 + \l_1\, (x_1 + x_{-1} + x_j + x_{-j})~,~~&
x_1  &\to\m_0\, w_1 + \m_1\, x_1 + \m_{-1}\, x_{-1} + \m_j\, x_j + \m_{-j}\, x_{-j} \\[3pt]
r &\to K\, r~,              & p_1 &\to L\, p_1 + M\, p_j~.
\end{alignat*}
with all other transformations determined by the action of $\IH$ on the ones given.  Neglecting an overall scaling of the coordinates on each space, this is a 9 parameter freedom.  The original polynomials had 13 independent coefficients, so this suggests $h^{21} = 13-9 = 4$ for the quotient, which thus has Hodge numbers $\hodgenos = (2, 4)$.  We have  no representation which explicitly displays the 12 $(1,1)$-forms, so we do not have our usual check but we will nevertheless assume that the calculation of the Hodge numbers is correct in this case.

We can also consider quotients by the subgroups $\IZ_2$ and $\IZ_4$ of $\IH$.  In this case the action of $\IZ_4$ on the ambient space depends on the generator we take ($i$ and $k$ interchange the two $\IP^4$'s, whereas $j$ does not), but the Hodge numbers come out the same.  This suggests that the actions are in fact the same when restricted to the manifold.
\subsubsection{A quotient by the Klein group, $\IZ_2{\times}\IZ_2$}
There is an alternative way to try to define a quaternionic action on $X^{12, 28}$, which is to take coordinates $(x_0, x_1, x_i, x_j, x_k)$ on e.g. the first $\IP^4$ and consider $x_0 \to x_0,~x_\a \to x_{\g\a}$ for $\g \in \IH$, where $x_{-\a}$ is considered to be the same as $x_\a$.  Obviously $-1 \in \IH$ fixes the whole manifold in this case, so this is really an action of the quotient group $\IH/\{1,-1\}$, which is the Klein group, isomorphic to $\IZ_2{\times}\IZ_2$.  It will be convenient to abandon the quaternion notation and simply consider an action by $\IZ_2{\times}\IZ_2$.

Take coordinates $(w, x_m),~ m = 0, \ldots, 3$ for the first $\IP^4$, and $(z, y_m),~ m = 0, \ldots, 3$ for the second.  Again denote by $r$ the bilinear polynomial, but this time label the quadratics in $x$ by $p_1, p_2$, and those in $y$ by $q_1, q_2$.  Define the generators of two commuting $\IZ_2$ actions by
\beq
\begin{split}
U:~ x_0 \leftrightarrow x_1 ~,~~ x_2 \leftrightarrow x_3 ~,~~ y_0 \leftrightarrow y_1 ~,~~ y_2 \leftrightarrow y_3 \\
V:~ x_0 \leftrightarrow x_2 ~,~~ x_1 \leftrightarrow x_3 ~,~~ y_0 \leftrightarrow y_2 ~,~~ y_1 \leftrightarrow y_3 \\
\end{split}
\notag\eeq
We are taking all the polynomials to be invariant under both $U$ and $V$, which determines their form to be
\beq
\begin{split}
r &~=~ A_0\, w\, z + \sum_{m=0}^3 (A_1\, w\, y_m + A_2\, z\, x_m) + A_3 \sum_{m=0}^3 x_m\, y_m \\
 &\hskip40pt + A_4 \sum_{m=0}^3 x_m\, y_{m+2}  + A_5 \sum_{m=0}^3 x_m\, y_{1-m}  + A_6 \sum_{m=0}^3 x_m\, y_{3-m} \\[5pt]
p_a &~=~ B_{a,0}\, w^2 + B_{a,1} \sum_{m=0}^3 w\, x_m + B_{a,2} \sum_{m=0}^3 x_m^2 + B_{a,3} (x_0\, x_1 + x_2\, x_3) \\
 &\hskip40pt + B_{a,4} (x_0\, x_2 + x_1\, x_3) + B_{a,5} (x_0\, x_3 + x_1\, x_2) \\[8pt]
q_a &~=~ C_{a,0}\, z^2 + C_{a,1} \sum_{m=0}^3 z\, y_m + C_{a,2} \sum_{m=0}^3 y_m^2 + C_{a,3} (y_0\, y_1 + y_2\, y_3) \\
 &\hskip40pt + C_{a,4} (y_0\, y_2 + y_1\, y_3) + C_{a,5} (y_0\, y_3 + y_1\, y_2)
\end{split}
\notag\eeq
where index addition is performed in $\IZ_4$.  It is easy to check that $U$, $V$ and $UV$ all act on the variety without fixed points, and we have confirmed that the polynomials are transverse.  The quotient will have Euler number $-32/4 = -8$.

We may perform the following coordinate transformations which are consistent with the symmetry
$$
w \to \m_0\, w + \m_1 (x_0 + x_1 + x_2 + x_3)~,~~x_0 \to \l_w\, w + \l_0\,x_0 + \l_1\, x_1 + \l_2\, x_2 + \l_3\, x_3
$$
with the transformations of the other $x_m$ being determined from the above by the action of $U$ and $V$, and similar transformations for the $y$ coordinates.  We may also redefine our polynomials
$$
r \to F\, r~,~~
p_1 \to K\, p_1 + L\, p_2 ~,~~ p_2 \to M\, p_1 + N\, p_2~,~~
q_1 \to K'\, q_1 + L'\, q_2 ~,~~ q_2 \to M'\, q_1 + N'\, q_2~.
$$
Neglecting overall scaling of the coordinates of each $\IP^4$, we have a 21 parameter freedom to redefine coordinates and polynomials.  Subtracting this from the 31 independent coefficients in the polynomials determines that $h^{21} = 10$ for the quotient.  We again have no independent check on the value of $h^{11}$, but we will assume that the counting is correct, so that the quotient manifold has $\hodgenos = (6,10)$.

We list the Hodge numbers for the manifolds constructed in this section in the table below
\vskip10pt
\begin{table}[H]
\begin{center}
\def\str{\vrule height16pt width0pt depth8pt}
\begin{tabular}{| c | c | c | c | c | c | c | c |}
\hline 
\str $\hodgenos\left(X^{12,28}/G\right)$ & ~~$(2,4)$~~ & ~~$(4,8)$~~ & ~~$(6,10)$~~ 
& ~~(8,16)~~\\
\hline
\str $G$ & ~$\IH$~ & $\IZ_4$ & $\IZ_2{\times}\IZ_2$ & $\IZ_2$ \\
\hline
\end{tabular}
\parbox{5.5in}{\caption{\small The Hodge numbers of smooth quotients of $X^{12,28}$.}}
\end{center}
\end{table}

\subsection{$Y^{5,37}$; another split of the tetraquadric} \label{sec:5,37Y}
We have seen earlier that the tetraquadric $X^{4,68}$ can, with the introduction of a $\IP^7$, to $X^{5,37}$.  If we instead split it by introducing a $\IP^3$ we obtain another manifold with the same Hodge numbers, which we will call $Y^{5,37}$, given by the following configuration.
$$
Y^{5,37}~=~\cicy{\IP^1\\ \IP^1\\ \IP^1\\ \IP^1\\ \IP^3}{2 & 0 & 0 & 0\\0 & 2 & 0 & 0\\0 & 0 & 2 & 0\\0 & 0 & 0 & 2\\ 1 & 1 & 1 & 1}^{5,37}\hskip1.25in
\raisebox{-50pt}{\includegraphics[width=1.5in]{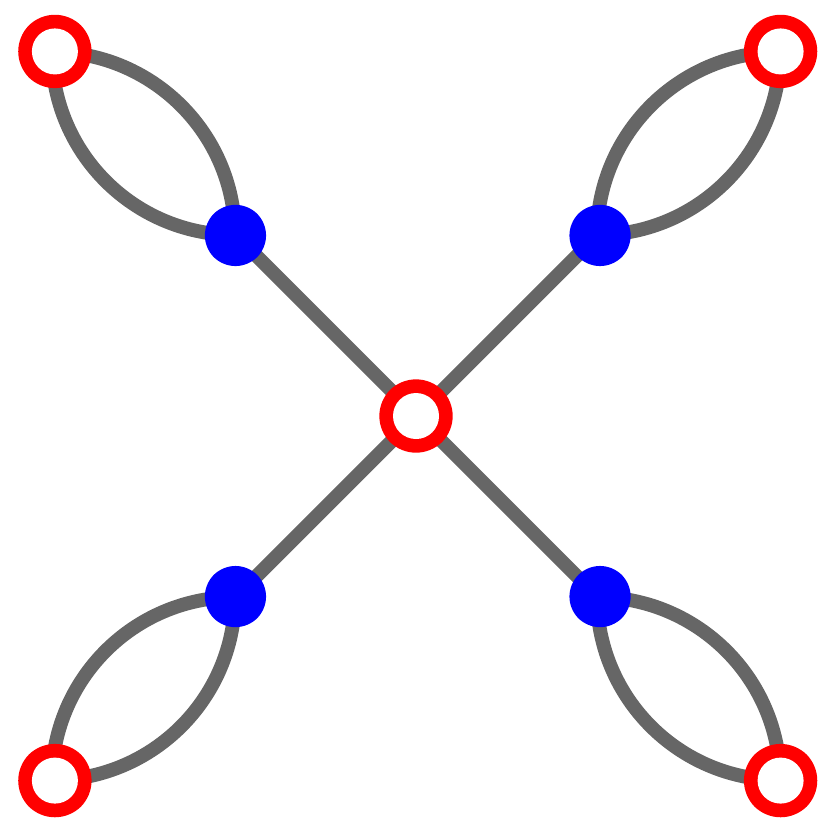}}
$$
This also admits a free action by $\IH$.  Proceeding as in previous sections, we can introduce coordinates $(s_\s, s_{-\s})$ on the four $\IP^1$'s, where $\s \in \IH_+$.  We also take coordinates $x_\a$ on the $\IP^3$ and label the four polynomials by $p_\a$, where $\a \in \IH$ and we identify $x_\a \equiv x_{-\a}, p_\a \equiv p_{-\a}$.  The action of $\g \in \IH$ is defined by
$$
U_\g: s_\a \to s_{\g\a}~,~~x_\a \to x_{\g\a}~;~~p_\a \to p_{\g\a}
$$
which determines the polynomials to be
$$
p_\a = \sum_{\s\in\IH_+}\Big(A_\s (s_\a^2+s_{-\a}^2) + B_\s\, s_\a\, s_{-\a}\Big)\, x_{\a\s}
$$
We have shown that these polynomials are transverse and it is easy to see, as in previous sections, that $U_{-1}$ acts without fixed points on the resulting manifold, guaranteeing the same for the whole group.  The quotient manifold has Euler number $-64/8 = -8$, and since the four $\IP^1$'s get identified, $h^{11} = 2$.  We conclude from this that $h^{21} = 6$.  We cannot verify this by counting coefficients, since the most general polynomials for the original manifold contain far fewer than $37$ independent coefficients, which is easy to see.  Since this manifold and its quotients have the same Hodge numbers as $X^{5,37}$ and its quotients, it seems likely that they are the same manifold, although this cannot be seen explicitly.

We should also consider the transpose of $Y^{5,37}$, which is given by
$$
\cicy{\IP^2\\ \IP^2\\ \IP^2\\ \IP^2}
{2 & 0 & 0 & 0 & 1 \\
0 & 2 & 0 & 0 & 1 \\
0 & 0 & 2 & 0 & 1 \\
0 & 0 & 0 & 2 & 1}
$$
Due to the identity $\IP^2[2~ 1] \cong \IP^1[2]$, this is just another representation of the tetraquadric, and so does not need to be treated separately.  The identity can be established by embedding $\IP^1$ in $\IP^2$ via the map $(z_0, z_1, z_2) = (t_0 t_1, t_0^2, t_1^2)$, the image of which is determined by $z_0^2 - z_1 z_2 = 0$.  This shows that $\IP^2[2] \cong \IP^1$ and that a polynomial of degree  $n$ in the $z$'s becomes one of degree $2n$ in the $t$'s.

The equivalence can equivalently be seen by noting that, analogous to earlier examples, both the tetraquadric and the transpose of $Y^{5,37}$ can be split to the following configuration, which has the same Euler number as both of them.
$$
\cicy{\IP^1\\ \IP^1\\ \IP^1\\ \IP^1\\ \IP^2\\ \IP^2\\ \IP^2\\ \IP^2}
{1 & 1 & 0 & 0 & 0 & 0 & 0 & 0 & 0 \\
0 & 0 & 1 & 1 & 0 & 0 & 0 & 0 & 0 \\
0 & 0 & 0 & 0 & 1 & 1 & 0 & 0 & 0 \\
0 & 0 & 0 & 0 & 0 & 0 & 1 & 1 & 0 \\
1 & 1 & 0 & 0 & 0 & 0 & 0 & 0 & 1 \\
0 & 0 & 1 & 1 & 0 & 0 & 0 & 0 & 1 \\
0 & 0 & 0 & 0 & 1 & 1 & 0 & 0 & 1 \\
0 & 0 & 0 & 0 & 0 & 0 & 1 & 1 & 1}
$$
Contracting the four $\IP^1$'s leads to the transpose of $Y^{5,37}$, whereas contracting the four $\IP^2$'s returns us to the tetraquadric.
\newpage
\section{Comments on the Gross-Popescu manifolds}
Calabi-Yau manifolds admitting abelian surface fibrations have been 
studied by Gross and Popescu \cite{GrossPopescu}. These manifolds have 
many remarkable properties among which are the fact that they have   
Euler number zero and admit free actions by groups of the form $
\IZ_n{\times}\IZ_n$. These manifolds are not themselves CICY's however 
they are closely related to some of the manifolds that we consider 
here. In some cases the Gross-Popescu manifolds may be realised as the 
resolutions of singular CICY's that have many nodes and the very 
symmetric split manifolds that we have studied here may be thought of 
as partial resolutions of the  nodal varieties that give rise to the 
Gross-Popescu manifolds. There are Gross-Popescu manifolds that are 
related to three manifolds that are basic to us here. These three 
cases relate to the manifolds
$$
\IP^4[5]^{1,101}_{-200}~,~~\IP[3,\,3]^{1,73}_{-144}~,~~\IP^7[2,\,2,\,
2,\,2]^{4,68}_{-128}
$$
In each case there is a very symmetric singularization of the variety 
that has a number of nodes that is half the absolute value of the 
respective Euler number, so 100, 72 and 64 in these three cases. 
Remarkably these nodes can be simultaneously resolved to yield a \cym 
which then has Euler number zero. This fact is intimately related to 
the property that the Gross-Popescu manifolds are fibered by abelian 
surfaces. It is the fact that all the nodes lie on a single fiber that 
allows them to be simultaneously resolved while retaining the 
vanishing of the first Chern class.

For the quintic we have the split
$$
\IP_4[5]^{1,101}~\longrightarrow~\cicy{\IP^4\\ \IP^4}{1~1~1~1~1\\ 
1~1~1~1~1}^{2,52}
$$
One way of looking on this is that there is a determinantal quintic 
with 50 nodes (to account for the difference in the Euler numbers) and 
the split manifold is the resolution of this. The split manifold 
admits a freely acting
$\IZ_5$ symmetry but not a free $\IZ_5{\times}\IZ_5$ symmetry. At 
least this is so for values of the parameters such the variety is 
smooth. We can however find a family of split manifolds that have 50 
nodes that does admit a free action by $\IZ_5{\times}\IZ_5$. These 
singular varieties are closely related to the
Horrocks-Mumford quintic and can be regarded as partial resolutions of 
these. The 50 nodes of the split variety can be simultaneously 
resolved and this yields the Gross-Popescu manifold with Hodge numbers 
$(4,4)$.

For $\IP^7[2,\,2,\,2,\,2]$ we have the equations \eref{SWquadrics} 
which, for generic values of the parameters, admit free actions by 
groups of order 32 and are transverse. By specialising the parameters 
by setting $c=0$ one finds free actions also by groups of order 64, 
one of which is $\IZ_8{\times}\IZ_8$. The polynomials are, however, no 
longer transverse and the variety has now 64 nodes. These may be 
simultaneously resolved and in this way one arrives at the Gross-
Popescu manifold with Hodge numbers $(2,2)$.
\subsection{$X^{2,52}$ and the Horrocks-Mumford quintic}
{\label{HMquintic}}
We return to the split quintic of \SS\ref{sec:2,52} prior to imposing 
the symmetry generated by $U$ and impose a second $\IZ_5$ symmetry
$$
T:~x_i\to\z^i\, x_i~;~~y_i\to\z^{2i}\, y_i~;~~p_i\to \z^{3i}\, p_i~.
$$
where $\z$ is a non-trivial fifth root of unity.  We have, from Eq.~
\eref{splitquintic},
$$
T\,p_i~=~\z^{3i}\sum_{jk} a_{jk}\,x_{j+i}\, y_{k+i}\,\z^{j+2k}
$$
so covariance requires $a_{jk}=0$ for $j\neq 3k$. We set $c_k=a_{3k,k}
$ and make the replacement
$k\to k-i$ in the sum so that the polynomials become
\beq
p_i~=~\sum_k c_{k-i}\,x_{3(k+i)}\,y_k~.
\label{z5z5bilinears}\eeq
In order to check that the group generated by $S$ and $T$ is fixed 
point free it is sufficient to check that $S$,
$T$ and $ST^\ell$, $\ell=1,\ldots,4$ act without fixed points. The 
fixed point set of each of these generators, in the embedding space, 
consists of a number of isolated points and for generic values of the 
coefficients $c_k$ these do not satisfy the equations $p_i=0$. In 
particular, the fixed points of $T$ are avoided if none of the $c_k$ 
vanish.

We have seen that the polynomials are fixed point free; the 
coefficients, however, are now so constrained that the polynomials are 
no longer transverse. We learn from a Groebner basis calculation that 
the variety defined by \eref{z5z5bilinears} has 50 nodes. An 
alternative way to look at this is to return to \eref{splitquintic} 
which we write in the form
$$
p_i~=~A_{ij}(x)\,y_j~~~;~~~A_{ij}(x)~=~A_{ijk}\, x_k~.
$$
Since the $y_j$ cannot all be zero the equations $p_i=0$ require $
\det(A_{ij}(x))=0$. The variety corresponding to the vanishing of the 
determinant is a conifold. For generic choices of the coefficients 
$A_{ijk}$ the conifold has 50 nodes and has \eref{(2,52)manifold} as 
its resolution. The 50 nodes account for the difference in the Euler 
number between the quintic and the split quintic. For our very 
symmetric polynomials \eref{z5z5bilinears} the coefficients are no 
longer generic. The determinantal equation is now
$$
q(x,c)~=~\det\big(c_{k-i}\, x_{3(k+i)}\big)~=~0.
$$
In particular the variety corresponding to $q$ has 100 nodes. The 
resolution of this variety is the
Horrocks-Mumford quintic which has many remarkable properties 
\cite{HorrocksMumford,Borcea:HMquintics,Fryers}. We may think of this 
manifold, equivalently, as the resolution of the variety corresponding 
to $q=0$ or as the resolution of the variety corresponding to 
\eref{z5z5bilinears}.

It is simple to see that $q$ is invariant under the replacement
$$
x\to Ix~~~,~~~Ix_j~=~x_{-j}~.
$$
Of the 100 nodes, 50 occur at points $x=Gc$, where $G$ denotes the 
group generated by $S$, $T$ and $I$. There exists also another vector 
$c'\notin Gc$, unique up to the action of $G$, such that 
$q(x,c)=q(x,c')$ and the other 50 nodes occur at $x=Gc'$. The Horrocks-
Mumford quintic is an abelian surface fibration and the nodes all lie 
on the same fiber. The nodes may be simultaneously resolved by blowing 
up this abelian surface and this gives rise to a manifold which we 
shall here denote by $X^{4,4}$ which is simply connected and has
$\hodgenos=(4,4)$. The four complex structure parameters entailed by 
$h^{21}=4$ are visible in $q$ as the components of $c_k$ taken up to 
scale.
\subsection{$X^{3,39}$ and the Gross-Popescu manifold with $
\hodgenos=(6,6)$}{\label{GP(6,6)}}
We continue the discussion of the manifold $X^{3,39}$ of \SS\ref{sec:3,39}.
In order to discuss extensions of $\IZ_3{\times}\IZ_3$, and facilitate 
a comparison\footnote{Notation: our
$z_\a$ is the coordinate $x_\a$ of \cite{GrossPopescu}. The 
coefficients which we can here denote by
$c_\a=(a_0,\,b_2,\,a_1,\,b_0,\,a_2,\,b_1)$ are the $y_\a$ of 
\cite{GrossPopescu}. Finally the symmetry generators $\s$ and $\t$ 
that we will introduce shortly are the inverses of the corresponding 
generators of \cite{GrossPopescu}.} with \cite{GrossPopescu}, let us 
relate the $x_i$ and the $y_i$ to the coordinates
$z_\a$, $\a\in\IZ_6$ of the $\IP^5$ by
$$
z_\a~=~(x_0,\, y_2,\, x_1,\, y_0,\, x_2,\, y_1)
$$
and let $\s$ and $\t$ denote generators of the group $\IZ_6{\times}
\IZ_6$ that have the action
$$
\s:~z_\a~\to~z_{\a+1}~;~~~ \t:~z_\a~\to~(-\z^2)^\a\, z_\a~.
$$
In terms of the $x$ and $y$ coordinates we have
\beq
\begin{split}
\s:&~(x_i,\, y_i)~\to~(y_{i+2},\,x_{i+2})~~~ ;~~~(u_i,\, v_i)~\to~(v_{i
+2},\, u_{i+2}) \\[1ex]
\t:&~(x_i,\, y_i)~\to~(\z^i\,x_i,-\z^i\, y_i)~;~~~(u_i,\, v_i)~
\to~(\z^i\, u_i,\,\z^i\, v_i)
\end{split} \notag
\eeq
where we extend the action also to the coordinates $u$ and $v$. We see 
that $\s^2=S$ and $\t^4=T$. The group $\IZ_3{\times}\IZ_3$ is extended 
to $\IZ_6{\times}\IZ_6$ by adding the two elements of order two
$U=\s^3$ and $V=\t^3$
\beq
\begin{split}
U&~:~~(x_i,\,y_j)~\to~(y_i,\, x_j)~~~;~~(u_i,\,v_j)~\to~(v_i,\, u_j)\\[1ex]
V&~:~~(x_i,\,y_j)~\to~(x_i,-y_j)~;~~(u_i,\,v_j)~\to~(u_i,\, v_j)~.
\end{split} \notag
\eeq
For the generators $\s$ and $\t$ we have the relations $\s=S^2 U$ and $
\t=TV$.

Let us now return to the discussion of the counting of parameters 
following
eq.~\eref{ZthreeZthreesplit}. The general form of the equations that 
are invariant under $\IZ_3{\times}\IZ_3$ can be taken to be
\beq
\begin{split}
p_i~&=~\sum_{j}\big( a_{i-j}\,x_{2(i+j)} + b_{i-j}\,y_{2(i+j)} \big)\, 
u_j \\
q_i~&=~\sum_{j}\big( b_{i-j}\,x_{2(i+j)} + d_{i-j}\,y_{2(i+j)} \big)\, 
v_j
\end{split}\label{ZthreeZthreesplitnew}
\eeq
with $a_0=1$ and $b_0=0$. The 7 parameters being visible as the two 
free components of each of $a_j$ and $b_j$ and the three components of 
$d_j$. We may further impose invariance under $U$. This requires 
$d_j=a_j$
and reduces the number of parameters in the equations to four. The 
action of $U$ is not, however, free. The fixed points satisfy $y_j=\pm 
x_j$ and $v_j=u_j$ and these lie on two elliptic curves that we denote 
$\cE^U_\pm$. These fixed curves do not trouble us so long as we do not 
take the quotient by $U$. A more immediate point is that the equations 
are no longer transverse and, generically have 36 nodes. The 
determinants that result from the elimination of $u_j$ and $v_j$
\beq
\D_1~=~\det\!\big( a_{i-j}\,x_{2(i+j)} + b_{i-j}\,y_{2(i+j)} \big)~~~
\text{and}~~~
\D_2~=~\det\!\big( b_{i-j}\,x_{2(i+j)} + a_{i-j}\,y_{2(i+j)} \big)
\label{ZthreeZsixsplit}\eeq
define a variety (see \cite[Thm 5.2]{GrossPopescu}) whose resolution 
is $G\!P^{6,6}$, the Gross-Popescu manifold with Hodge numbers 
$(6,6)$. The resolution of the variety corresponding to 
\eref{ZthreeZthreesplitnew}, with $d_j=a_j$, is a four-parameter 
subfamily of the six-parameter family of $G\!P^{6,6}$ manifolds. This 
suggests that the generic $G\!P^{6,6}$ manifold has symmetry $
\IZ_3{\times}\IZ_3$ and a four-parameter subfamily has symmetry
$\IZ_3{\times}\IZ_6$. The fact that \eref{ZthreeZthreesplitnew} is a 
seven-parameter family suggests that there is a single constraint, $
\cC(a,b,c,d)$, on the coefficients, such that the six-parameter family 
$\cC=0$ is a family of varieties each of which has 36 nodes and whose 
resolutions are $G\!P^{6,6}$ manifolds. We have tested this for the 
line in the parameter space given by $d_j=(1+h,\,a_1,\,a_2)$. We find 
that there are 36 nodes when $h$ satisfies a certain cubic of the form 
$h(h^2 + \m h + \n)$, indicating that $\cC$ is a cubic in the 
coefficients. It is shown in \cite{GrossPopescu} that there are $G\!
P^{6,6}$ manifolds with symmetry $\IZ_6{\times}\IZ_6$. It is not clear 
how such manifolds can arise in our construction owing to the fact 
that the $\IZ_6{\times}\IZ_6$-invariant specialization of 
\eref{ZthreeZthreesplitnew} has $b_k=0$ and is very singular.
\subsection{A comment on the split of $\IP^7[2~2~2~2]$}
As recalled above the manifold $\IP^7[2~2~2~2]$ is closely related to 
the Gross-Popescu manifold with Hodge numbers $(2,2)$. It is also the 
case that the split $X^{5,37}$ of \SS\ref{sec:5,37} has  $\chi=-64$ so 
the determinant has half the number of nodes of the singular member of 
the $\IP^7[2~2~2~2]$ family whose resolution is $G\!P^{2,2}$. On the 
other hand a resolution of a nodal form of $X^{5,37}$ would have to 
have $h^{11}\geq 6$ and so cannot be directly related to $G\!P^{2,2}$.
\newpage
\section*{Acknowledgements}
It is a pleasure to acknowledge fruitful conversations with Mark Gross, James Gray, Yang-Hui He, Maximilian Kreuzer, Andre Lukas, Duco van Straten and Bal\'{a}zs Szendr\H{o}i. In addition, James Gray and Andre Lukas provided 
much assistance with the installation and application of the STRINGVACUA package, which proved very useful in this work. One of the authors, PC, wishes to thank the Tata Institute for Fundamental Research for hospitality during part of this project.
\newpage


\begin{thebibliography}{99}
\frenchspacing

\bibitem{Triadophilia}
P.~Candelas, X.~de la Ossa, Y.~H.~He and B.~Szendroi,
``Triadophilia: A Special Corner in the Landscape,'' Adv.~Theor.~Math.~Phys.~{\bf 12} 429 (2008), 
arXiv:0706.3134 [hep-th].
  
\bibitem{Yau}
S.-T. Yau,
``Compact three-dimensional K\"ahler manifolds with zero Ricci curvature'',
in: Proc. of Symposium on Anomalies, Geometry, Topology, 395, World Scientific, Singapore, 1985. \\
G. Tian and S.-T. Yau, 
``Three-dimensional algebraic manifolds with $c_1=0$ and $\chi=-6$'', in: Mathematical aspects of string theory (San Diego, Calif., 1986),  543, Adv. Ser. Math. Phys., 1, World Scientific, Singapore, 1987.

\bibitem{CICYsI}
P.~Candelas, A.~M.~Dale, C.~A.~Lutken and R.~Schimmrigk,
``Complete Intersection Calabi-Yau Manifolds,''
Nucl.\ Phys.\  B {\bf 298} (1988) 493.

\bibitem{hubsch}
T.~Hubsch ``Calabi-Yau Manifolds --- A Bestiary for Physicists'',
World Scientific, Singapore, 1994.

\bibitem{Donagi:2008xy}
R.~Donagi and K.~Wendland, 
``On orbifolds and free fermion constructions,''\\
arXiv:0809.0330 [hep-th].

\bibitem{KreuzerSkarkeReflexive}
M.~Kreuzer and H.~Skarke,
``Complete classification of reflexive polyhedra in four dimensions'',
Adv.\ Theor.\ Math.\ Phys.\  {\bf 4} (2002) 1209 
[arXiv:hep-th/0002240].

\bibitem{KreuzerRieglerSahakyan}
M.~Kreuzer, E.~Riegler and D.~A.~Sahakyan,
``Toric complete intersections and weighted projective space'',
J.\ Geom.\ Phys.\  {\bf 46} (2003) 159
[math.ag/0103214].

\bibitem{KlemmKreuzerRieglerScheidegger}
  A.~Klemm, M.~Kreuzer, E.~Riegler and E.~Scheidegger,
``Topological string amplitudes, complete intersection Calabi-Yau spaces  and threshold corrections'', 
JHEP {\bf 0505} (2005) 023
[arXiv:hep-th/0410018].

\bibitem{BatyrevKreuzerConifolds}
V.~Batyrev and M.~Kreuzer,
``Constructing new Calabi-Yau 3-folds and their mirrors via conifold transitions,''
arXiv:0802.3376 [math.AG].

\bibitem{GrossPopescu}
M.~Gross and S.~Popescu, ``Calabi-Yau Threefolds and Moduli of Abelian Surfaces I'',
 Compositio Math. 127  (2001) 169 [math.ag/0001089].
 
\bibitem{Rodland}
E.~A.~R\oslash dland, 
``The Pfaffian CalabiÐYau, its Mirror, and their Link to the Grassmannian G(2,7)'',
Compositio. Math. {\bf 122} (2000) 135, ArXiv:math.AG/9801092

\bibitem{Tonoli}
F. Tonoli, ``Construction of \cy\ 3-folds in $\IP^6$'',
J. Algebraic Geom. {\bf 13} (2004) 209.

\bibitem{BorisovHua}
L. Borisov and Z. Hua, ``On Calabi-Yau Threefolds with Large Nonabelian Fundamental Group'',
arXiv:math.AG/0609728

\bibitem{Hua}
Z. Hua, ``Classification of Free Actions on Complete Intersections of Four Quadrics'',
arXiv:math.AG/0707.4339


\bibitem{comments}
P.~Candelas and X.~C.~de la Ossa,
``Comments on Conifolds,''
Nucl.\ Phys.\  B {\bf 342} (1990) 246.

\bibitem{Green:1987cr}
 P.~S.~Green, T.~Hubsch and C.~A.~Lutken,
 ``All Hodge Numbers Of All Complete Intersection Calabi-Yau Manifolds,''
 Class.\ Quant.\ Grav.\  {\bf 6} (1989) 105.
 
\bibitem{CYHomePage}
{\tt http://thp.uni-bonn.de/Supplements/cy.html~}.
 
\bibitem{DigitalCICYList}
There is a link to a Mathematica list, that contains the Hodge numbers, on\\ 
{\tt http://www-thphys.physics.ox.ac.uk/user/AndreLukas~}.
 
 \bibitem{Deformations}
P.~Green and T.~H\"ubsch,
``Polynomial Deformations And Cohomology Of Calabi-Yau Manifolds",
Commun.\ Math.\ Phys.\  {\bf 113} (1987) 505.

\bibitem{Singular3.0.4} G.-M. Greuel, G. Pfister, and H. Sch\"onemann.
{\sc Singular} 3.0.4. A Computer Algebra System for Polynomial
Computations. Centre for Computer Algebra, University of
Kaiserslautern (2007). {\tt http://www.singular.uni-kl.de}.
 
\bibitem{Gray:2008zs}
J.~Gray, Y.~H.~He, A.~Ilderton and A.~Lukas,
``STRINGVACUA: A Mathematica Package for Studying Vacuum Configurations in String Phenomenology,''
arXiv:0801.1508.
 
\bibitem{Green:1988bp}  P.~S.~Green and T.~Hubsch, 
``Connecting Moduli Spaces of Calabi-Yau Threefolds'', 
Comm. Math. Phys. {\bf 119} (1988) 431.

\bibitem{Braun:2004xv}
V.~Braun, B.~A.~Ovrut, T.~Pantev and R.~Reinbacher,
 ``Elliptic Calabi-Yau threefolds with Z(3) x Z(3) Wilson lines'',
JHEP {\bf 0412} (2004) 062
[arXiv:hep-th/0410055].

\bibitem{Braun:2005nv}
  V.~Braun, Y.~H.~He, B.~A.~Ovrut and T.~Pantev,
  ``The exact MSSM spectrum from string theory'',
  JHEP {\bf 0605} (2006) 043 
  [arXiv:hep-th/0512177].\\
V.~Braun, Y.~H.~He and B.~A.~Ovrut,
  ``Stability of the minimal heterotic standard model bundle'',
  JHEP {\bf 0606} (2006) 032 
  [arXiv:hep-th/0602073].
    
\bibitem{Donagi:2000zf}
  R.~Donagi, B.~A.~Ovrut, T.~Pantev and D.~Waldram,
  ``Standard-model bundles on non-simply connected Calabi-Yau threefolds'',
  JHEP {\bf 0108} (2001) 053 
  [arXiv:hep-th/0008008].\\
--, ``Standard-model bundles'', math/0008010.\\
R.~Donagi, Y.~H.~He, B.~A.~Ovrut and R.~Reinbacher,
  ``The particle spectrum of heterotic compactifications'',
  JHEP {\bf 0412} (2004) 054 
  [arXiv:hep-th/0405014].\\
--, ``The spectra of heterotic standard model vacua'',
  JHEP {\bf 0506} (2005) 070 
  [arXiv:hep-th/0411156].\\
V.~Braun, Y.~H.~He, B.~A.~Ovrut and T.~Pantev,
  ``A heterotic standard model'',
  Phys.\ Lett.\  B {\bf 618} (2005) 252 
  [arXiv:hep-th/0501070].

\bibitem{Bouchard:2005ag}
V.~Bouchard and R.~Donagi,
``An SU(5) heterotic standard model'',
Phys.\ Lett.\  B {\bf 633} (2006) 783 
[arXiv:hep-th/0512149].

\bibitem{HorrocksMumford}
G. Horrocks and D. Mumford, 
``A rank 2 vector bundle on $\IP^4$ with 15,000 symmetries'', 
Topology 12 (1973), 63Ð-81.

\bibitem{Borcea:HMquintics}
 C. Borcea, 
 ``On desingularized HorrocksÐMumford quintics'', 
 J. Reine Angew. Math. 421 (1991), 23Ð41. 

\bibitem{Fryers}
M. J. Fryers,
``The Movable Fan of the Horrocks-Mumford Quintic'',\\
ArXiv:math.AG/0102055

\bibitem{BouchardDonagi}
V. Bouchard and R. Donagi,
``On a Class of Non-Simply Connected \cy\ Threefolds'',
arXiv:0704.3096.

\bibitem{Schimmrigk}
R. Schimmrigk,
``A new construction of a three-generation Calabi-Yau manifold'',
Phys.\ Lett.\ {bf B193} (1987) 175.

\bibitem{Strominger:1985it}
A.~Strominger and E.~Witten,
``New Manifolds For Superstring Compactification,''
Commun.\ Math.\ Phys.\  {\bf 101} (1985) 341.

\bibitem{Beauville}
A. Beauville,
``A Calabi-Yau Threefold with Non-Abelian Fundamental Group'',
in: New trends in algebraic geometry, 13, LMS Lect. Note Ser. 264, CUP, Cambridge, 1999 [arXiv:alg-geom/9502003].

\bibitem{HjPark}
H.~J.~Park,
``Finding the Mirror of the Beauville Manifold'',
arXiv:hep-th/0312056.

\bibitem{Dixon:1985jw}
L.~J.~Dixon, J.~A.~Harvey, C.~Vafa and E.~Witten,
``Strings On Orbifolds,''
Nucl.\ Phys.\  B {\bf 261} (1985) 678.

\bibitem{BrownHiggins}
R.~Brown and P.~J.~Higgins, 
``The fundamental groupoid of the quotient of a Hausdorff space by a discontinuous action of a discrete group 
is the orbit groupoid of the induced action,'' 
math.AT/0212271. 

\end{thebibliography}
\end{document}